\documentclass[ 
	fontsize=12pt,          
	numbers=noenddot,    	
    parskip=half,        	
    listof=totoc,        	
    bibliography=totoc,  	
	headsepline=true,       
	footsepline=false, 		
    DIV=12,                	
    toc=sectionentrywithdots
]{scrartcl}
\usepackage{amsthm}
\usepackage{amsmath, amsfonts, amssymb, bbm, bm}

\usepackage[utf8]{inputenc}
\usepackage[T1]{fontenc}
\usepackage[round]{natbib}
\usepackage{graphicx}
\usepackage{lmodern}
\usepackage{enumerate}
\usepackage{setspace}
\usepackage{xcolor}
\usepackage{soul}
\usepackage{booktabs}
\usepackage{url}
\usepackage{appendix}
\usepackage{lscape}

\makeatletter
\def\namedlabel#1#2{\begingroup
    #2%
    \def\@currentlabel{#2}%
    \phantomsection\label{#1}\endgroup
}
\makeatother

\usepackage[english]{babel}

\usepackage[a4paper, left=1.5cm, right=1.5cm, top=2.5cm, bottom=2.5cm]{geometry}

\theoremstyle{plain}
\newtheorem{defi}{Definition}[section]
\newtheorem{theorem}[defi]{Theorem}
\newtheorem{example}[defi]{Example}

\newtheorem{ass}[defi]{Assumption}
\newtheorem{remark}[defi]{Remark}

\newtheorem{proposition}[defi]{Proposition}

\newcommand{\R}{\mathbb{R}}
\newcommand{\T}{\mathcal{T}}
\newcommand{\E}{\mathbb{E}}

\title{ }
\author{ }
\renewcommand{\title}[1]{ \noindent{\centering \Large \textbf{ #1 } \\} }
\newcommand{\inst}[1]{\textsuperscript{#1}}
\newcommand{\institute}[1]{{\centering \footnotesize{#1}} \vspace{2ex}}

\newcommand{\keywords}{\\[0.2cm] \textbf{Keywords:} }

\date{ }

\begin{document}

\title{ Multiple Comparison Procedures for Simultaneous Inference in Functional MANOVA }
\begin{center}
		Merle Munko\inst{1,*}\let\thefootnote\relax\footnote{*Corresponding author. Email address: \url{merle.munko@ovgu.de}},
		Marc Ditzhaus\inst{1},
  Markus Pauly\inst{2,3} and
  {\L}ukasz Smaga\inst{4}
\end{center}

	\institute{
		\inst{1} Otto-von-Guericke University Magdeburg; Magdeburg (Germany)\newline
        \inst{2} TU Dortmund University; Dortmund (Germany)
\newline
        \inst{3} Research Center Trustworthy Data Science and Security, UA Ruhr; Dortmund (Germany)
        \newline
        \inst{4} Adam Mickiewicz University; Poznań (Poland)
	}

\hrule

\begin{sloppypar}

\abstract{
Functional data analysis is becoming increasingly popular to study data from real-valued random functions. Nevertheless, there is  a lack of multiple testing procedures for such data. These are particularly important in factorial designs to compare different groups or to infer factor effects. We propose a new class of testing procedures for arbitrary linear hypotheses in general factorial designs with functional data. Our methods allow global as well as multiple inference of both, univariate and multivariate mean functions without assuming particular error distributions nor homoscedasticity. That is, we allow for different structures of the covariance functions between groups. To this end, we use point-wise quadratic-form-type test functions that take potential heteroscedasticity into account. Taking the supremum over each test function, we define a class of local test statistics. We analyse their (joint) asymptotic behaviour and propose a resampling approach to approximate the limit distributions. 
The resulting global and multiple testing procedures are asymptotic valid under weak conditions and applicable in general functional MANOVA settings. We evaluate their small-sample performance in extensive simulations and finally illustrate their applicability by analysing a multivariate functional air pollution data set. 
}

\maketitle

\section{Introduction}\label{sec:intro}
Since the appearance of the notion "functional data" in the 1982 article by \cite{ramsay1982data}
and the subsequent seminal 1991 paper 
by \cite{ramsay1991some}, functional data analysis (FDA) has gained in popularity. 
This was already stated in the 2013 review article \cite{ullah2013applications} on applications of FDA with functional ANOVA and MANOVA among the most common inference applications (last two columns of Table 1 therein). A quick Scopus search for 
"functional data" highlights, that its popularity increased even more since then: counting 5,945 hits between 2014 and March 2024 compared to 4,259 hits between 1992 and 2013 as well as 318 until 1991. 

In some of these papers, not only one univariate functional endpoint is of interest but a multi-dimensional vector of functions is observed for each experimental unit or individual. This is, e.g., the case with 
different functional endpoints or with 
longitudinally recorded functional data (special case of functional Repeated Measures). Examples of such data can be found in various disciplines covering longitudinal behavioural monitoring in ecology \citep{chiou2014linear}, angular rotation curves of hip and knee through the gait cycle of children \citep{multiEX}, $\beta$-trace measurements of different gender under longitudinal interventions \citep{sattler2018inference} in medicine, functional EEG and SPECT recordings over time in different disease and control groups in neuroscience \citep{bathke2018testing} or in environmental research. An example of the latter are given by the 
Canadian weather data set available in the R package \texttt{fda} (\citealp{fda}), where daily temperature and precipitation is collected or the data set on air pollution in the U.S. analyzed in 
\cite{ZHU2022105095}, in which the trajectories of three pollutants were recorded over a year. In all these examples, a longitudinal or multivariate functional design is present and the comparison of different groups or experimental conditions over time is of interest. Thus, we need multivariate analysis of variance (MANOVA) methods that can test global (e.g., about group, factor, time or interaction effects) as well as multiple linear hypotheses of mean functions (e.g. all-pairs, many-to-one or factorial change points) in arbitrary factorial designs covering crossed as well as functional Repeated Measures ANOVA (i.e. longitudinal).

There exist various global testing procedures for univariate functional data, see, e.g. \cite{cuevasEtal2004,gorecki2015comparison,Simu,pomann2016two,liebl2023fast} or the monograph by \cite{Zhang2013} on functional ANOVA. In comparison, functional MANOVA seems to be less researched in the literature until recently.
\cite{cao2018simultaneous,smaga2020note,wang2020simultaneous} proposed inference procedures in case of functional Repeated Measures ANOVA with one or two groups. More general, \citeauthor{Gorecki} (\citeyear{Gorecki}) studied permutation tests based on a basis function representation and on random projections for the multivariate analysis of variance (MANOVA) problem for homoscedastic functional data, i.e., functional data with the same covariance structure between the groups. 
\citeauthor{Multiv} (\citeyear{Multiv}) proposed two global tests for multivariate homoscedastic functional data in the two-sample case by integrating the pointwise Hotelling`s $T^2$-test statistic and taking its supremum, respectively.
Moreover, \citeauthor{ZHU2022105095} (\citeyear{ZHU2022105095}) developed another global test for the MANOVA problem for potentially more than two samples by using the functional Lawley–Hotelling trace test statistic still assuming homoscedasticity. 
Homoscedasticity is already a very restrictive assumption in classical ANOVA \citep[e.g.][]{paulyETAL2015} and is rarely fulfilled in real applications with multivariate functional data. Some solutions that allow for heteroscedasticity 
have been published recently 
for the one-way functional MANOVA setting by \cite{ZHU2022105095,ZHUetAl2023}. Moreover, extensions to 
general linear hypotheses can be found in the preprint \cite{zhu2023}. 
The lack of approaches for multiple testing of mean functions is even more serious. There only exist a recent preprint for the univariate case by the authors \citep{FANOVA}. However, there exist no multiple testing procedures for general functional MANOVA (covering functional Repeated Measure ANOVA as special case) to the best of our knowledge. 

Thus, it is the aim of the present paper to develop a new class of global and multiple testing procedures for general linear hypotheses in the functional MANOVA context. The procedures do not postulate heteroscedasticity nor a specific distribution and can be flexibly applied to all kind of multivariate factorial designs covering one-way, crossed two-way or general longitudinal and split-plot settings with functional data. To this end, we consider  Wald-type test statistics for global hypotheses which take the heteroscedastic nature of the functional data into account. Since their limit distributions turn out to be rather complicated, we adapt the parametric bootstrap approach of \cite{cuevasEtal2004} to the present setting to approximate the limit distribution and prove its asymptotic correctness. To also allow for general multiple comparisons with family-wise error rate control, the asymptotic exact dependence structure of the test statistics is incorporated to determine suitable critical values that lead to valid test decisions. 

The remainder of this paper is organised as follows. Section~\ref{sec:methods} is divided into four subsections.
The general setup of the linear hypothesis testing problem for multivariate heteroscedastic functional data is presented in Section~\ref{ssec:setup}. A suitable test statistic for this testing problem is proposed and studied in Section~\ref{ssec:MtestingProc}. In Section~\ref{ssec:MBootstrap}, the parametric bootstrap procedure is extended to multivariate functional data. Furthermore, we construct multiple tests in Section~\ref{ssec:Multiple} for testing multiple hypotheses. In addition, the finite sample performance of the proposed tests is analysed in intensive simulation studies in terms of size control and power in Section~\ref{sec:Simu}. In Section~\ref{sec:Data}, our proposed methodologies are applied to a data example. 
Finally, the results of this paper are discussed in Section~\ref{sec:Discussion}. 
The proofs of all theorems and auxiliary technical results are given in the supplement.

\section{Statistical Methodology}\label{sec:methods}
In this section, we introduce the general functional set-up, a suitable test statistic for the global testing problem and the parametric bootstrap approach for constructing global testing procedures for general linear combinations of multivariate mean functions. Moreover, we develop multiple testing procedures and state the asymptotic validity of all procedures.

\subsection{
General Functional Set-Up}
\label{ssec:setup}
Throughout this paper, let $(\T,\rho)$ denote a compact pseudometric space, e.g., $\T = [0,1]$ with the Euclidean metric. We write $\mathbf{x} \sim \text{GP}_p(\boldsymbol{\eta},\mathit{\Gamma})$ and 
$\mathbf{x} \sim \text{SP}_p(\boldsymbol{\eta},\mathit{\Gamma})$ for a $p$-dimensional Gaussian or general stochastic process $\mathbf{x}$, respectively, over $\T$ with mean function $\boldsymbol{\eta}: \T \to \R^p$, covariance function $\mathit{\Gamma}: \T^2 \to \R^{p \times p}$ and $p\in\mathbb{N}$.  
Suppose we have $k\in\mathbb{N}$ independent multivariate functional samples given by $p$-dimensional stochastic processes
\begin{align}
\label{eq:Msamples}
    \mathbf{x}_{i1},...,\mathbf{x}_{in_i} \sim \text{SP}_p(\boldsymbol{\eta}_i,\mathit{\Gamma}_i)\: \text{ i.i.d. \quad for all } i\in\{1,...,k\}
\end{align} with sample sizes $n_i \in\mathbb{N}, n_i\geq 2, i\in\{1,...,k\},$ and dimension  $p\in\mathbb N$.
 Here, $\boldsymbol{\eta}_i: \T \to \R^p$ denotes the multivariate mean function and \mbox{$\mathit{\Gamma}_i : \T^2 \to \R^{p\times p}$} denotes the covariance function of the $i$th sample for all $i \in \{1,...,k\}$, which are both unknown. 
The total sample size is denoted by $n := \sum_{i=1}^k n_i$. 
In contrast to the multivariate set-ups considered in \citeauthor{Gorecki} (\citeyear{Gorecki}, p.~2173), \citeauthor{Multiv} (\citeyear{Multiv}) or \citeauthor{ZHU2022105095} (\citeyear{ZHU2022105095}), the covariance functions may differ from each other, i.e. we are not assuming homoscedasticity.\\

To infer general linear hypothesis about the mean functions, let $\mathit{H} \in \R^{r \times pk}$ denote a known matrix with $\textrm{rank}(\mathit{H}) \geq 1$.
Furthermore, let $\boldsymbol{\eta} := (\boldsymbol{\eta}_1^{\top},...,\boldsymbol{\eta}_k^{\top})^{\top}: \T \to \R^{pk}$ be the vector of all mean functions and \mbox{$\mathbf{c}: \T\to\R^r$} be a fixed 
function. In many settings, we use the zero function $\mathbf{c}(t) = \mathbf{0}_{r}$ for all $t\in\T$. In the following, we consider the null and alternative hypothesis 
\begin{align}
    \label{eq:global}
    \mathcal H_0: \mathit{H}\boldsymbol{\eta}(t) = \mathbf{c}(t) \text{ for all } t\in \T \quad \text{vs.} \quad \mathcal H_1: \mathit{H}\boldsymbol{\eta}(t) \neq \mathbf{c}(t) \text{ for some } t\in \T.
\end{align}

We note, that this generalizes the setting of \cite{zhu2023} who considered hypotheses of the form $\mathit{G}[\boldsymbol{\eta}_1,...,\boldsymbol{\eta}_k]^{\top} = \mathbf{0}_q$ for a matrix $\mathit{G}\in\mathbb R^{q\times k}$ which can be rewritten in our formulation with $\mathit{H} := \mathit{G} \otimes \mathit{I}_p$  and $\mathbf{c}(t) = \mathbf{0}_{qp}$ for all $t\in\T$. 
We additionally stress, that this set-up is very flexible. In particular, it 
allows for a factorial structure within the pooled vector of mean functions $\boldsymbol{\eta}$ by splitting up indices as in classical ANOVA or MANOVA (\citealp{KONIETSCHKE2015, paulyETAL2015}) 
and thus covers all kind of hypotheses in general factorial designs with functional data. Specific examples are given by 
one- or crossed two-way univariate and multivariate functional analysis of variance problems as well as longitudinal settings as outlined below.

\begin{example}\label{ex2}
\mbox{ }
\begin{enumerate}
    \item[(a)] {\textbf{ One-Way fMANOVA:}} For the one-way functional MANOVA, we may choose $\mathbf{c}(t)=\mathbf{0}_{pk}$ for all $t\in\T$ and $\mathit{H} = \mathit{P}_k\otimes \mathit{I}_p:= (\mathit{I}_k - \mathit{J}_k/k)\otimes \mathit{I}_p$, 
where here and throughout $\mathit{I}_p \in\R^{p\times p}$ denotes the unit matrix and \mbox{$\mathit{J}_k := \mathbf{1}_k\mathbf{1}_k^{\top} \in\R^{k\times k}$} is the matrix of ones. This results in the null hypothesis 
$\mathcal H_0: \boldsymbol\eta_1(t)=\dots=\boldsymbol\eta_k(t)$ for all $t\in \mathcal{T}$.
\item[(b)] {\textbf{ Crossed Two-Way fMANOVA:}} In a functional two-way design with factors A with $a$~levels and B with $b$ levels, we set $k := a b$ and split up the group index $i$ in two subindices \mbox{$(i_1, i_2)\in\{1,\dots,a\}\times\{1,\dots,b\}$.} Thus, the following null hypotheses are covered by our formulation of the testing problem:
\begin{itemize}
    \item No main effect of factor A: $\mathcal{H}_0^A: (\mathit{P}_a \otimes (\mathbf{1}_b^{\top}/b)\otimes \mathit{I}_p) \boldsymbol\eta(t) = \mathbf{0}_{ap} $ for all $t\in\T$,
    \item No main effect of factor B: $\mathcal{H}_0^B: ((\mathbf{1}_a^{\top}/a)  \otimes \mathit{P}_b\otimes \mathit{I}_p) \boldsymbol\eta(t) = \mathbf{0}_{bp} $ for all $t\in\T$,
    \item No interaction effect between factors A and B: $\mathcal{H}_0^{AB}: (\mathit{P}_a  \otimes \mathit{P}_b\otimes \mathit{I}_p) \boldsymbol\eta(t) = \mathbf{0}_{abp} $ for all $t\in\T$.
\end{itemize}
\item[(c)] {\textbf{ Longitudinal fANOVA:}} For the one-way Repeated Measures fANOVA, we set $p := ab$, where $a$ denotes the number of repeated measures and $b$ denotes the dimension. 
   Hence, $\boldsymbol{\eta}_{ij} = ({\eta}_{ij1},...,{\eta}_{ijb})^{\top}$ corresponds to the mean function of an individual in group $i\in\{1,...,k\}$ measured at $j\in\{1,...,a\}$ and $\boldsymbol{\eta} = (\boldsymbol{\eta}_{11},...,\boldsymbol{\eta}_{1a},...,\boldsymbol{\eta}_{ka})^{\top}$.
   Then, the following hypotheses can be formulated in our testing framework:
   \begin{itemize}
           \item No group effect: $\mathcal{H}_0^k: (\mathit{P}_k\otimes (\mathbf{1}_a^{\top}/a)\otimes \mathit{I}_b) \boldsymbol\eta(t) = \mathbf{0}_{kb} $ for all $t\in\T$,
    \item No time effect: $\mathcal{H}_0^a: ((\mathbf{1}_k^{\top}/k) \otimes \mathit{P}_a\otimes \mathit{I}_b) \boldsymbol\eta(t) = \mathbf{0}_{ab} $ for all $t\in\T$,
    \item No interaction effect between time and group: $\mathcal{H}_0^{ka}: (\mathit{P}_k\otimes \mathit{P}_a  \otimes \mathit{I}_b) \boldsymbol\eta(t) = \mathbf{0}_{kab} $ for all $t\in\T$.
   \end{itemize}
   \item[(d)] {\textbf{ Testing for specific functional pattern:}} If it is of interest whether the vector of mean functions $\boldsymbol{\eta}$ has a specific functional pattern $\mathbf{c}:\T\to\R^{pk}$, we can choose $\mathit H = \mathit I_{pk}$ as unit matrix and, thus, receive the null hypothesis  $\mathcal{H}_0: \boldsymbol{\eta}(t) = \mathbf{c}(t)$ for all $t\in\T$.
\end{enumerate}
\end{example}
\subsection{General Assumptions}
In the following, we consider the space of $p$-dimensional continuous functions $C^p(\T)$.
For constructing suitable tests for the general testing problem (\ref{eq:global}) and analyzing their asymptotic properties, the following assumptions are required. 

\begin{ass}
\label{Massumptions}
Consider the $k$ independent samples (\ref{eq:Msamples}). Let $\mathbf{v}_{ij} := \mathbf{x}_{ij} - \boldsymbol{\eta}_i,$ \mbox{$i\in\{1,...,k\},$} $j\in\{1,...,n_i\},$ denote the \emph{subject-effect functions}. Additionally, assume that the following conditions \ref{itm:iid}--\ref{itm:Lipschitz} are fulfilled for all $i\in\{1,...,k\}.$
\begin{enumerate}
    \item[\namedlabel{itm:iid}{(A1)}] The subject-effect functions $\mathbf{v}_{i1},...,\mathbf{v}_{in_i} \sim \text{SP}_p(\mathbf{0}_p,\mathit{\Gamma}_i)$ are i.i.d.~and take values in $C^p(\T)$.
    \item[\namedlabel{itm:J}{(A2)}] There exists a $t^*\in\T, J \geq 2$ such that $\E[||\mathbf{v}_{i1}(t^*)||_2^J] < \infty$, where here and throughout $||.||_2$ denotes the 2-norm of a vector.
    \item[\namedlabel{itm:pd}{(A3)}] 
    The covariance matrix
    $\mathit{\Gamma}_i(t,t)$ is positive definite for all $t\in\T$.  
    \item[\namedlabel{itm:tau}{(A4)}] There exists a  $\tau_i > 0$ such that $\dfrac{n_i}{n} \to \tau_i$ as $n \to \infty$.
    \item[\namedlabel{itm:Lipschitz}{(A5)}] There exists a $C \geq 0$ such that $$ \E\left[ || \mathbf{v}_{i1}(t) - \mathbf{v}_{i1}(s) ||_2^J \right] \leq C (\rho(t,s))^J $$
    holds for all $t,s\in\T$ and
    \begin{align}
        \label{eq:intAss}
        \int_0^{\nu} D(\varepsilon, \rho)^{2/J} \;\mathrm{d}\varepsilon < \infty
    \end{align}
    for some $\nu > 0$. Here and throughout, $J$ is as in \ref{itm:J} and $D(\varepsilon, \rho)$ denotes the packing number, see Definition~2.2.3 in \cite{vaartWellner1996} for details.
\end{enumerate}
\end{ass}

Assumptions \ref{itm:iid}--\ref{itm:tau} are standard assumptions in the functional data setup in the space of continuous functions, see e.g.~\cite{Dette}. Assumptions \ref{itm:iid}, \ref{itm:tau} and \ref{itm:Lipschitz} will guarantee the asymptotic Gaussianity of the mean function estimators and the parametric bootstrap counterparts. Assumptions \ref{itm:iid}, \ref{itm:J} and \ref{itm:Lipschitz} will yield the uniform consistency of the covariance function estimators and the parametric bootstrap counterparts. From this, the convergence of our proposed test statistic will follow by using \ref{itm:pd}.
\\
Assumption \ref{itm:Lipschitz} is motivated by (A3) in \cite{Dette} but less restrictive since we replaced it by a moment condition. 
Moreover, \cite{Dette} focused on $\T = [0,1]$ with the metric $\rho:\T^2 \to [0,\infty), \rho(t,s) := |t-s|^{\theta}$ and $\theta \in (1/J,1]$ while we allow general pseudometric spaces $(\T,\rho)$.
However, for $\T = [0,1]$ and $\rho(t,s) = |t-s|^{\theta}$, (\ref{eq:intAss}) is fulfilled whenever $\theta \in (2/J,1]$ and, thus, the restrictions on $\theta$ are slightly stronger than in their paper. 
Condition (\ref{eq:intAss}) results from empirical process theory (\citealp[Theorem~2.2.4]{vaartWellner1996}) and is required with a higher order than in \cite{Dette} since the parametric bootstrap estimators of the covariance functions are incorporated in our test statistic (\ref{eq:MparaGPH}), see the proof of Lemma~S8 in the supplement for details.

\subsection{Globalising Pointwise Hotelling's $T^2$-Test Statistic}\label{ssec:MtestingProc}
In this section, we present a suitable test statistic for the testing problem (\ref{eq:global}). Therefore, we use a similar test function as in \cite{FANOVA}. However, we globalise this test function by taking the supremum instead of integrating over it since the supremum seemed to outperform the integral in our simulation studies. 
For constructing the test function, we firstly define an unbiased estimator for the mean function of the $i$th group at $t$ by
$$ \widehat{\boldsymbol\eta}_i(t) := \frac{1}{n_i}\sum_{j=1}^{n_i} \mathbf{x}_{ij}(t)$$
and for the covariance function of the $i$th group at $(t,s)$ by
\begin{align}
\label{eq:MhatGamma}
    \widehat{\mathit{\Gamma}}_i(t,s) := \frac{1}{n_i-1}\sum_{j=1}^{n_i} \left(\mathbf{x}_{ij}(t) - \widehat{\boldsymbol\eta}_i(t)\right)\left(\mathbf{x}_{ij}(s) - \widehat{\boldsymbol\eta}_i(s)\right)^{\top}
\end{align}
for all $t,s\in\T, i\in\{1,...,k\}$. 
By the weak law of large numbers, the mean function estimators and covariance function estimators are pointwisely consistent under~\ref{itm:iid}.
Furthermore, we show that the covariance function estimators are uniformly consistent over
$\{(t,t)\in\T^2 \mid t\in\T\}$
under \ref{itm:iid}, \ref{itm:J} and \ref{itm:Lipschitz} in Lemma~S6 in Section~S1 in the supplement.\\
Let
\begin{align*}
	\widehat{\mathit{\Lambda}}(t,s) &:= \bigoplus_{i=1}^k  \left(\frac{n}{n_i}\widehat{\mathit{\Gamma}}_i(t,s)\right)
\end{align*}
be an estimator for
\begin{align*}
	\mathit{\Lambda}(t,s) &:= \bigoplus_{i=1}^k  \left(\frac{1}{\tau_i}{\mathit{\Gamma}}_i(t,s)\right) \in \R^{pk\times pk}
\end{align*}
for all $t,s \in \T$, where $\bigoplus_{i=1}^k  \mathit{A}_i$ denotes the direct sum of the matrices $\mathit{A}_1,...,\mathit{A}_k$. It should be noted that $\mathit{\Lambda}(t,t)$ is invertible under \ref{itm:pd} and \ref{itm:tau} for all $t \in \T$. The pointwise consistency of the covariance function estimators 
provides immediately the pointwise consistency of $\widehat{\mathit{\Lambda}}$ under \ref{itm:iid} and \ref{itm:tau}.
Moreover, the uniform consistency of $\widehat{\mathit{\Lambda}}$ over $\{(t,t)\in\T^2 \mid t\in\T\}$ is proven in Lemma~S6 under the assumptions \ref{itm:iid}, \ref{itm:J}, \ref{itm:tau} and \ref{itm:Lipschitz}.\\
{Then, a reasonable point-wise test statistic } for the multivariate functional data setup is defined by
\begin{align}
\label{eq:Mtestfunc}
	\mathrm{PH}_{n,\mathit{H},\mathbf{c}}(t) &:= n( \mathit{H} \boldsymbol{\widehat\eta}(t) - \mathbf{c}(t) )^{\top} (\mathit{H}\widehat{\mathit{\Lambda}}(t,t) \mathit{H}^{\top})^+ ( \mathit{H} \boldsymbol{\widehat\eta}(t) - \mathbf{c}(t) )
\end{align}
for all $t\in\T$, where $\boldsymbol{\widehat\eta} := (\boldsymbol{\widehat\eta}_1^{\top},...,\boldsymbol{\widehat\eta}_k^{\top})^{\top}$ denotes the vector of all mean function estimators. 

{Following \cite{Multiv} we call $\mathrm{PH}_{n,\mathit{H},\mathbf{c}}$ the \emph{point-wise Hotelling's $T^2$-test statistic}.}
We expect that the point-wise Hotelling's $T^2$-test statistic is small for all $t \in\T$ under the null hypothesis in (\ref{eq:global}) since $\mathit{H} \boldsymbol{\widehat\eta}(t)$ is an estimator for $\mathit{H} \boldsymbol{\eta}(t) = \mathbf{c}(t)$. Moreover, $\mathit{H}\widehat{\mathit{\Lambda}}(t,t) \mathit{H}^{\top}$ can be viewed as an approximation of the covariance matrix of $\sqrt{n} \left(\mathit{H} \boldsymbol{\widehat\eta}(t) - \mathbf{c}(t) \right)$.

Based on the point-wise Hotelling's $T^2$-test statistic (\ref{eq:Mtestfunc}), we can construct a test statistic by taking the supremum over the point-wise test statistics, that is
\begin{align}\label{eq:GPH}
    T_{n}(\mathit{H}, \mathbf{c}) := \sup_{t\in\T} \{ \mathrm{PH}_{n,\mathit{H},\mathbf{c}}(t) \} .
\end{align}
The null hypothesis in (\ref{eq:global}) shall be rejected for large values of the test statistic.

Similarly to the invariance properties of the GPH statistic for the univariate functional data setup in \cite{FANOVA}, one can show the invariance under orthogonal transformations, i.e., $T_n(\mathit{A}\mathit{H},\mathit{A}\mathbf{c}) = T_n(\mathit{H}, \mathbf{c})$ for any orthogonal $\mathit{A}\in\R^{r\times r}$, and the scale-invariance in the sense of \cite{GuoEtAl2019} for the multivariate case.

For the construction of asymptotic tests based on the proposed test statistic, we have to examine its asymptotic behaviour.
The limiting distribution of the test statistic under the null hypothesis is given by the following theorem.

\begin{theorem}
\label{Masy_GPH}
Let $\mathbf{y}
\sim GP_r(\mathbf{0}_r,\mathit{H}\mathit{\Lambda}\mathit{H}^{\top})$.
Under \ref{itm:iid}--\ref{itm:Lipschitz} and the null hypothesis in (\ref{eq:global}), we have
\begin{align}
\label{eq:Masy_GPH_Theo1}
    &T_{n}(\mathit{H}, \mathbf{c}) \xrightarrow{d} \sup_{t\in\T} \{ (\mathbf{y}(t))^{\top}(\mathit{H}\mathit{\Lambda}(t,t)\mathit{H}^{\top})^+\mathbf{y}(t) \}
\end{align}
as $ n \to\infty$,
where here and throughout $\xrightarrow{d}$ denotes convergence in distribution in the sense of \cite{vaartWellner1996}.
\end{theorem}

{The limiting distribution 
depends on the unknown covariance functions $\mathit{\Gamma}_1,\dots,\mathit{\Gamma}_k$ and is thus non-pivotal. To overcome this issue, we below propose a resampling approach to compute adequate critical values.}

\subsection{Parametric Bootstrap}
\label{ssec:MBootstrap}
In this section, we adopt a parametric bootstrap approach as proposed in \cite{FANOVA} to the present set-up. 
Note that a similar bootstrap approach has been proposed earlier by \cite{Dette} in the case of univariate functional time series. 
In this method we substitute $\mathit{\Gamma}_1,...,\mathit{\Gamma}_k$  by their estimators. Then, the parametric bootstrap samples are given by $\mathbf{x}_{i1}^{\mathcal{P}}, ..., \mathbf{x}_{in_i}^{\mathcal{P}} \sim GP_p(\mathbf{0}_p,\widehat{\mathit{\Gamma}}_i), i\in\{1,...,k\},$ conditionally on the functional data $(\mathbf{x}_{i1},...,\mathbf{x}_{in_i})_{i\in\{1,...,k\}}$. 
Note, that we still postulate our general model, i.e., we do not assume a parametric model for our functional data $(\mathbf{x}_{i1},\dots,\mathbf{x}_{in_i})_{i\in\{1,\dots,k\}}$. {The idea, however, is that the bootstrap samples reflect the dependency structure of the true processes; where the distribution is motivated by the functional CLT.}
\\
In the following, we denote the parametric bootstrap counterparts of the statistics defined in Section~\ref{ssec:MtestingProc} with a superscript $\mathcal{P}.$
Thus, we define the \emph{parametric bootstrap pointwise Hotelling's $T^2$-test statistic} at $t$ by
\begin{align*}
    \mathrm{PH}^{\mathcal{P}}_{n,\mathit{H}}(t) := n\left(\mathit{H}\widehat{\boldsymbol\eta}^{\mathcal{P}}(t)\right)^{\top}\left(\mathit{H}\widehat{\mathit{\Lambda}}^{\mathcal{P}}(t,t)\mathit{H}^{\top}\right)^+\mathit{H}\widehat{\boldsymbol\eta}^{\mathcal{P}}(t)
\end{align*} for all $t\in\T.$ 
Taking the supremum 
yields the \emph{parametric bootstrap globalising pointwise Hotelling`s $T^2$-test statistic} for the multivariate functional data setup, that is
\begin{align}\label{eq:MparaGPH}
    T_{n}^{\mathcal{P}}(\mathit{H}) := \sup_{t\in\T} \left\{ \mathrm{PH}^{\mathcal{P}}_{n,\mathit{H}}(t) \right\}.
\end{align}

The consistency of the proposed parametric bootstrap 
is stated in the following theorem.

\begin{theorem}
\label{MParaBS}
Let $\mathbf{y}
\sim GP_r(\mathbf{0}_r,\mathit{H}\mathit{\Lambda}\mathit{H}^{\top})$.
Under \ref{itm:iid}--\ref{itm:Lipschitz}, we have for $ n \to\infty$.
\begin{align}
    &T_{n}^{\mathcal{P}}(\mathit{H}) \xrightarrow{d} \sup_{t\in\T} \{ (\mathbf{y}(t))^{\top}(\mathit{H}\mathit{\Lambda}(t,t)\mathit{H}^{\top})^+\mathbf{y}(t) \}.
\end{align}

\end{theorem}
{
Thus, the parametric bootstrap test statistic $T_{n}^{\mathcal{P}}(\mathit{H})$ always mimics the distributional limit of $T_{n}(\mathit{H},\mathbf{c})$ given in \eqref{eq:Masy_GPH_Theo1}. The distribution of $T_{n}^{\mathcal{P}}(\mathit{H})$} is usually approximated by Monte Carlo with $B\in\mathbb N$ iterations. 
Thus, we obtain the parametric bootstrap test
\begin{align*}
        \varphi_{n,B}^{\mathcal{P}} := \mathbbm{1}\left\{ T_n(\mathit{H}, \mathbf{c}) > Q^{\mathcal{P}}_{n,B}(1-\alpha) \right\},
    \end{align*}
where here and throughout $\alpha\in(0,1)$ denotes the level of significance. Moreover, $Q^{\mathcal{P}}_{n,B}(1-\alpha)$ denotes the empirical $(1-\alpha)$-quantile of $T_{n,1}^{\mathcal{P}}(\mathit{H}),...,T_{n,B}^{\mathcal{P}}(\mathit{H})$, where  $T_{n,1}^{\mathcal{P}}(\mathit{H}),...,T_{n,B}^{\mathcal{P}}(\mathit{H})$ are $B$ independent copies of $T_{n}^{\mathcal{P}}(\mathit{H})$ conditionally on the data $(\mathbf{x}_{i1},...,\mathbf{x}_{in_i})_{i\in\{1,...,k\}}$. It follows that $\varphi_{n,B}^{\mathcal{P}}$ is an asymptotic level-$\alpha$ test {for large $B$ and large $n$.}
    
\begin{proposition}\label{Proposition}
    Let $(B_n)_{n\in\mathbb N}\subset \mathbb N$ be a sequence with $B_n \to \infty$ as $n\to\infty$. Under \ref{itm:iid}--\ref{itm:Lipschitz} and the null hypothesis in (\ref{eq:global}), we have 
    \begin{align}\label{eq:newToShow}
        \lim\limits_{n\to\infty} \E\left[ \varphi_{n,B_n}^{\mathcal{P}} \right] = \alpha.
    \end{align}
\end{proposition}

\begin{remark}\label{RemarkProp}
We note, that the statement (\ref{eq:newToShow}) is generally stronger than the asymptotic result
  \begin{align}\label{eq:consistency:dette}
        \lim\limits_{B\to\infty} \limsup\limits_{n\to\infty} \left|\E\left[ \varphi_{n,B}^{\mathcal{P}} \right]-\alpha\right| = 0
    \end{align}
which is often used to prove consistency, see e.g. (3.23) in \cite{Dette}. In fact, 
if (\ref{eq:newToShow}) holds for all sequences $(B_n)_{n\in\mathbb N}\subset \mathbb N$ with $B_n \to \infty$ as $n\to\infty$, \eqref{eq:consistency:dette} follows.
\end{remark}
{
Our proposed test $\varphi_{n,B}^{\mathcal{P}}$ is a very flexible  tool for inferring different global linear hypotheses of the mean functions of different groups in a general heteroscedastic multivariate functional setting. In the next subsection, we propose extensions for multiple testing that particularly allow for post-hoc testing.}

\subsection{Multiple Tests}\label{ssec:Multiple}
In many applications, it is not only of interest whether there exists a (global) significance but also which specific linear combinations cause the significance.
For example, hypotheses about pairwise comparisons often contain the main research questions in one-way layouts. 
Hence, general multiple tests for heteroscedastic functional data are needed and constructed in this section. To this end, we use the same strategy as in the univariate case (\citealp{FANOVA}) for constructing multiple tests with consonant and coherent test decisions as defined in \cite{gabriel_1969}.
Therefore, the exact asymptotic dependence structure of the local test statistics is taken into account.
In \cite{FANOVA}, simulations have already shown that this procedure yields to a high power and quite accurate family-wise error control for univariate functional data.

In the following, we interpret $\mathit{H}$ as a block matrix, that is $\mathit{H}=\left[\mathit{H}_1^{\top},\dots,\mathit{H}_R^{\top}\right]^{\top}$ for matrices  $\mathit{H}_{\ell} \in \R^{ r_{\ell} \times k }$ with $\mathrm{rank}(\mathit{H}_{\ell}) \geq 1$, and \mbox{$\mathbf{c} = (\mathbf{c}_{\ell})_{\ell\in\{1,\dots,R\}} : \T\to\R^r$} as function with component functions $\mathbf{c}_{\ell}:\T \to \R^{r_{\ell}}$ for all $\ell\in\{1,\dots,R\}$, where $R, r_{\ell}\in\{1,\dots,r\}$ such that $\sum_{\ell=1}^R r_{\ell} = r$. 
 Note that we do not restrict to but also cover the case that $\mathit{H}$ is partitioned in $r$ row vectors $\mathit{H}_1, \dots, \mathit{H}_r \in\R^{1\times k}$.
The main idea of multiple tests is to split up the global null hypothesis in (\ref{eq:global}) with hypothesis matrix
$\mathit{H}= [\mathit{H}_1^{\top}, \dots, \mathit{H}_R^{\top}]^{\top}$ and function $\mathbf{c} = (\mathbf{c}_{\ell})_{\ell\in\{1,\dots,R\}} $ into $R$ single 
tests with
hypothesis matrices $\mathit{H}_1, \dots, \mathit{H}_R$ and functions $\mathbf{c}_1,\dots,\mathbf{c}_R$, respectively.
This leads to the multiple testing problem 
\begin{align}
\label{eq:multiple}
	\mathcal H_{0,{\ell}} : \; \mathit{H}_{\ell} \boldsymbol{\eta}(t) = \mathbf{c}_{\ell}(t) \;\text{ for all }t\in\T, \quad 
	\quad 
	\text{for }\ell\in \{1,\ldots, R\}.
\end{align}

Again, this formulation covers many special cases as will be pointed out in the following example.
\begin{example}
In this example, we focus on all-pairs comparisons and many-to-one comparisons of the mean functions. Further suitable hypothesis matrices, e.g. for trend, can be found in \cite{bretz2001}. 
\begin{enumerate}
    \item[(a)] {\textbf{ All-Pairs Comparisons:}} Pairwise comparisons of the mean functions can be considered by choosing $\mathit{H}_{1},\dots,\mathit{H}_{{k(k-1)}/{2}}$ as the Kronecker products of the rows of the Tukey-type contrast matrix (\citealp{tukey}) and the unit matrix~$\mathit{I}_p$ and $\mathbf{c}(t) = \mathbf{0}_{{k(k-1)}/{2}}$ for all $t\in\T$. This yields the multiple hypotheses $\mathcal H_{0,{\ell_1\ell_2}} : \: \boldsymbol\eta_{\ell_1}(t) = \boldsymbol\eta_{\ell_2}(t)$ for all $t\in\T$ with $\ell_1 < \ell_2$.
    \item[(b)] {\textbf{ Many-To-One Comparisons:}} When there is a reference group, it makes sense to compare the mean functions of all other groups to this reference group.
Here, we assume that the first group is the reference group. Thus, we are interested in the hypotheses
    $\mathcal H_{0,{\ell}} : \: \boldsymbol\eta_{1}(t) = \boldsymbol\eta_{\ell +1}(t)$ for all $t\in\T$ with $\ell\in \{1,\ldots, k-1\}.$ We obtain these hypotheses by choosing $\mathit{H}_{\ell} = \left(-1,\mathbf{e}_{\ell}\right) \otimes \mathit{I}_p$ for all $\ell\in\{1,\dots,k-1\}$ and $\mathbf{c}(t) = \mathbf{0}_{k-1}$ for all $t\in\T$, where $\mathbf{e}_{\ell}\in\R^{k-1}$ denotes the transposed $\ell$th unit vector.
    Here, the row vectors $\left(-1,\mathbf{e}_{\ell}\right), \ell\in\{1,...,k-1\},$ are the rows of the Dunnett-type contrast matrix (\citealp{dunnett_1955}). 
\end{enumerate}
\end{example}
{
To derive a valid multiple testing procedure
, we first investigate the joint distribution of the local test statistics in more detail.} 

\begin{theorem}\label{Multiasy_GPH}
    Let $\mathbf{y} = (\mathbf{y}_{\ell})_{\ell\in\{1,...,R\}} 
\sim GP_r(\mathbf{0}_r,\mathit{H}\mathit{\Lambda}\mathit{H}^{\top})$.
Under \ref{itm:iid}--\ref{itm:Lipschitz} and {the intersection of} the null hypotheses in (\ref{eq:multiple}), we have
\begin{align*}
    \left(T_{n}(\mathit{H}_{\ell}, \mathbf{c}_{\ell}) \right)_{\ell\in\{1,...,R\}} \xrightarrow{d} \left( \sup_{t\in\T} \{ \mathbf{y}^{\top}_{\ell}(t) \left(\mathit{H}_{\ell}  \mathit{\Lambda}(t,t)\mathit{H}_{\ell}^{\top} \right)^+ \mathbf{y}_{\ell}(t) \}  \right)_{\ell\in\{1,...,R\}}
\end{align*} 
as $ n \to\infty$.
\end{theorem}
{
As in the global testing case, the limit distribution is non-pivotal. But even in this case, the vector of the parametric bootstrap counterparts mimic the limit as shown below.
}

\begin{theorem}\label{MultiParaBS}
        Let $\mathbf{y} = (\mathbf{y}_{\ell})_{\ell\in\{1,...,R\}} 
\sim GP_r(\mathbf{0}_r,\mathit{H}\mathit{\Lambda}\mathit{H}^{\top})$.
Under \ref{itm:iid}--\ref{itm:Lipschitz}, we have
\begin{align*}
    \left(T_{n}^{\mathcal{P}}(\mathit{H}_{\ell}) \right)_{\ell\in\{1,...,R\}} \xrightarrow{d} \left( \sup_{t\in\T} \{ \mathbf{y}^{\top}_{\ell}(t) \left(\mathit{H}_{\ell}  \mathit{\Lambda}(t,t)\mathit{H}_{\ell}^{\top} \right)^+ \mathbf{y}_{\ell}(t) \}  \right)_{\ell\in\{1,...,R\}}
\end{align*} 
as $ n \to\infty$.
\end{theorem}
{The theorem can be used to derive a consistent parametric bootstrap procedure for the multiple testing problem. To this end, draw }$B$ independent copies of $(T_{n}^{\mathcal{P}}(\mathit{H}_{\ell}))_{\ell\in\{1,...,R\}}$ conditionally on the data $(\mathbf{x}_{i1},...,\mathbf{x}_{in_i})_{i\in\{1,...,k\}}$ given by 
$$\left(T_{n,1}^{\mathcal{P}}(\mathit{H}_{\ell})\right)_{\ell\in\{1,...,R\}},...,\left(T_{n,B}^{\mathcal{P}}(\mathit{H}_{\ell})\right)_{\ell\in\{1,...,R\}},$$ for some sufficiently large $B\in\mathbb N$.
{Denote the empirical $(1-\beta_{n,B}(\alpha))$-quantile of $T_{n,1}^{\mathcal{P}}(\mathit{H}_{\ell}),...,T_{n,B}^{\mathcal{P}}(\mathit{H}_{\ell})$ for all $\ell\in\{1,...,R\}$
as $Q^{\mathcal{P}}_{n,B,\ell}(1-\alpha)$. Then we define a new class of \emph{multiple parametric bootstrap tests} by}
\begin{align*}
        \varphi_{n,B,\ell}^{\mathcal{P}} := \mathbbm{1}\left\{ T_n(\mathit{H}_{\ell}, \mathbf{c}_{\ell}) > Q^{\mathcal{P}}_{n,B,\ell}(1-\alpha) \right\}, \quad \ell\in\{1,...,R\}.
\end{align*}
 Here,  $\beta_{n,B}(\alpha)$ is chosen such that the approximated family-wise error rate is bounded by the level of significance $\alpha$, see Section~3 in \cite{myRMST} for details. 
Each test statistic $T_{n}(\mathit{H}_{\ell}, \mathbf{c}_{\ell}), \ell\in\{1,\dots,R\},$ is treated in the same way and has the same impact since we use the same local level $\beta_{n,B}(\alpha)$ for each linear combination. 
The following proposition ensures that the proposed multiple testing procedure asymptotically controls the family-wise error rate.

\begin{proposition}\label{Proposition_Multiple}
Let $\mathcal{R} \subset \{1,...,R\}$ denote the subset of indices of true hypotheses, i.e., let $\mathcal{H}_{0,\ell}$ with $\ell\in\mathcal{R}$ from (\ref{eq:multiple}) be true, and let $(B_n)_{n\in\mathbb N} \subset \mathbb N$ be a sequence with $B_n\to\infty$ as $n\to\infty$.
    Under \ref{itm:iid}--\ref{itm:Lipschitz}, 
    we have
    \begin{align}
         \label{eq:consistency MCTP}\lim\limits_{n\to\infty} \E\left[ \max\limits_{\ell\in\mathcal{R}} \varphi_{n,B_n,\ell}^{\mathcal{P}} \right] \leq \alpha.
    \end{align} 
   The inequality becomes an equality if $\mathcal{R} = \{1,...,R\}$.
\end{proposition}

As in Remark~\ref{RemarkProp}, it also holds that
\begin{align*}
    \lim\limits_{B\to\infty} \limsup\limits_{n\to\infty} \left| \E\left[ \max\limits_{\ell\in\mathcal{R}} \varphi_{n,B,\ell}^{\mathcal{P}} \right] - \alpha  \right| = 0.
\end{align*}

\section{Simulations}\label{sec:Simu}
In this section, we investigate the finite sample properties of the new tests and their competitors. We consider the control of the type I error level, power and scale-invariance of the tests.

The new global parametric bootstrap test of Section~\ref{ssec:MBootstrap} will be denoted by SPH (supremum of the pointwise Hotelling's $T^2$-test statistic). On the other hand, the mSPH test is the new multiple parametric bootstrap test (see Section~\ref{ssec:Multiple}). We consider the following competitors for our new tests (see also Section~\ref{sec:intro}):
\begin{itemize}
\item[(a)] W, LH, P, R (\citeauthor{Gorecki}, \citeyear{Gorecki}) - the permutation tests based on a basis function representation of functional data and the Wilks', Lowley-Hotelling's, Pillay's, and Roy's test statistics, respectively,
\item[(b)] Wg, LHg, Pg, Rg (\citeauthor{Gorecki}, \citeyear{Gorecki}) - the tests based on random projections of functional data using gaussian white noise and the Wilks', Lowley-Hotelling's, Pillay's, and Roy's test statistics, respectively,
\item[(c)] Wb, LHb, Pb, Rb (\citeauthor{Gorecki}, \citeyear{Gorecki}) - the tests based on random projections of functional data using Brownian motion and the Wilks', Lowley-Hotelling's, Pillay's, and Roy's test statistics, respectively,
\item[(d)] ZN, ZB \citep{ZHUetAl2023} - the tests for MANOVA problem for heteroscedastic functional data based on naive and bias-reduced estimators, respectively,
\item[(e)] Z \citep{zhu2023} - the test for general linear hypotheses testing problem for heteroscedastic functional data,
\item[(f)] GPH, mGPH \citep{FANOVA} - the global (Section~\ref{ssec:MBootstrap}) and multiple (Section~\ref{ssec:Multiple}) parametric bootstrap globalising pointwise Hotelling's $T^2$-test based on the following test statistic:  $$I_n(\mathit{H},\mathbf{c}):=\int_{\mathcal{T}}\mathrm{PH}_{n,\mathit{H},\mathbf{c}}(t)dt.$$
\item[{(g)}] QI, QS (\citeauthor{Multiv}, \citeyear{Multiv}) - the two-sample tests based on integrating and taking supremum of the pointwise Hotelling's $T^2$-test statistic, respectively. They are used just in the scenario of two-samples for which they are designed. 
\end{itemize}
When we use the global tests (i.e., the test other than mGPH and mSPH) for the multiple testing problem, the Bonferroni correction is used. Note that the tests QI, QS, W, LH, P, R, Wg, LHg, Pg, Rg, Wb, LHb, Pb, and Rb are constructed for homoscedastic functional data.

\subsection{Simulation setup}\label{sec:sim_setup}
{
\cite{Multiv} already conducted a simulation study for the case with two independent groups. For a fair comparability with existing methods, we inspire our models from their study. For ease of presentation, we thereby focus on one-way designs.

\emph{Model 1.}
We first consider a one-way layout with $k=4$ groups and $p=6$ functional variables. Following \cite[Simulation~2]{Multiv}, we consider  the following linear functional model:} $$\mathbf{x}_{ij}(t)=\boldsymbol{\eta}_i(t)+\mathbf{e}_{ij}(t),\ i=1,\dots,4,j=1,\dots,n_i,t\in[0,1].$$ 
{ Below we discuss the different choices for this set-up. 

\emph{Mean Functions.}} The vectors $\boldsymbol{\eta}_i=(\eta_{i1},\dots,\eta_{i6})^{\top}$, $i=1,2,3$, of mean functions in the first three samples are constructed as follows: 
\begin{flalign*}
\eta_{i1}(t)&=(\sin(2\pi t^2))^5,\ \eta_{i2}(t)=(\cos(2\pi t^2))^5,\ \eta_{i3}(t)=t^{1/3}(1-t)-5,\\
\eta_{i4}(t)&=\sqrt{5}t^{2/3}\exp(-7t),\ \eta_{i5}(t)=\sqrt{13t}\exp(-13t/2),\ \eta_{i6}(t)=1+2.3t+3.4t^2+1.5t^3.
\end{flalign*}
In the last sample, we take $\eta_{4m}(t)=\eta_{1m}(t)$ for $m=1,\dots,5$, and 
\begin{equation}
\label{eta46}
\eta_{46}(t)=\left(1+\frac{\delta}{\sqrt{30}}\right)+\left(2.3+\frac{2\delta}{\sqrt{30}}\right)t+\left(3.4+\frac{3\delta}{\sqrt{30}}\right)t^2+\left(1.5+\frac{4\delta}{\sqrt{30}}\right)t^3.
\end{equation}
When the hyperparameter $\delta=0$, the null hypothesis is true, while if $\delta>0$, the alternative holds and $\delta$ controls the difference between the mean functions. We specify its values below. 

{ \emph{Covariance Functions.}
We} set $\mathbf{e}_{ij}(t)=\mathit{A}(z_{ij1}(t),\dots,z_{ij6}(t))^{\top}$, where 
$$\mathit{A}=\left[\begin{array}{rrrrrr}
1&0&-1&-1&0&0\\
0&1&0&0&-1&-1\\
0&0&1&-1&0&-1\\
1&1&0&1&0&-1\\
0&0&1&0&1&-1\\
0&0&0&0&0&1\\
\end{array}
\right]$$
and $z_{ijm}(t)=\mathbf{b}_{ijm}^{\top}\Psi(t)$, $\mathbf{b}_{ijm}=(b_{ijm1},\dots,b_{ijm7})^{\top}$, $b_{ijmr}=\sqrt{\lambda_{ir}}v_{ijmr}$, and $\Psi(t)=(\psi_1(t),\dots,\psi_7(t))^{\top}$, $\psi_1(t)=1$, $\psi_{2s}(t)=\sqrt{2}\sin(2\pi st)$, $\psi_{2s+1}(t)=\sqrt{2}\cos(2\pi st)$ for $s=1,2,3$. We let $\lambda_{ir}=h_i\rho^r$ with the following three scenarios of $\mathit{H}=(h_1,\dots,h_4)^{\top}$: $\mathit{H}_1=(1.5,1.5,1.5,1.5)^{\top},\mathit{H}_2=(1.5, 2, 2.5, 3)^{\top},\mathit{H}_3=(3, 2.5, 2, 1.5)^{\top}$. The homoscedastic case is represented by $\mathit{H}_1$, while $\mathit{H}_2$ and $\mathit{H}_3$ correspond to two heteroscedastic scenarios. Set sample sizes to $n_1=20,n_2=30,n_3=40$, and $n_4=50$. Then, for the heteroscedastic cases, we have the positive and negative pairings scenarios for $\mathit{H}_2$ and $\mathit{H}_3$, respectively, i.e., the variability increases and decreases, respectively, while sample sizes increase (see, for example, \citeauthor{paulyETAL2015}, \citeyear{paulyETAL2015}). As we will see in simulation results, it has an impact on the properties of many of tests considered. The parameter $\rho$ corresponds to the correlation in functional data, namely, the smaller $\rho$ the larger correlation. We set $\rho=0.1,0.3,0.5,0.7,0.9$. Since the power of the tests depends on the amount of correlation in the data, we consider $\delta=0.1,0.2,0.3,0.4,0.5$ for $\rho=0.1,0.3,0.5,0.7,0.9$, respectively, to usually obtain a non-trivial power of the tests. 

{ \emph{Error Distributions.}} The random variables $v_{ijmr}$ are generated independently from three standard distributions: $N(0,1)$, $t_4/\sqrt{2}$, and $(\chi_4^2-4)/\sqrt{8}$. They represent symmetric and skewed distributions as well as the light-tailed and heavy-tailed ones.

{ \emph{Scaling.}} We would also check the impact of scaling functional data on the properties of the tests (see Section~\ref{ssec:MtestingProc}). Thus, in addition to the above scenario without scaling function, we also consider the case of scaling generated functional observations by the function $h(t)=1/(t + 1/50)$ for $t\in[0,1]$. Note that the tests QI, QS, ZN, ZB, Z, GPH, mGPH, SPH, and mSPH are scale-invariant, while the tests by \cite{Gorecki} are not.

\emph{Studied Hypotheses.} For global and multiple testing problem, we consider the Tukey-type contrast \citep{tukey}. Thus, we compare each pair of different groups (each pair just one time), i.e., for $k=4$, we have six contrasts 1-2, 1-3, 1-4, 2-3, 2-4, and 3-4. Then, the hypothesis matrix for testing the global hypothesis 
$\mathcal{H}_0:\boldsymbol\eta_1(t)=\dots=\boldsymbol\eta_4(t)$ for all $t\in\mathcal{T}$ 
in~\eqref{eq:global} is given by
$$\mathit{H}=\mathit{T}_6\otimes I_6=\left[\begin{array}{rrrr}
-1&1&0&0\\
-1&0&1&0\\
-1&0&0&1\\
0&-1&1&0\\
0&-1&0&1\\
0&0&-1&1\\
\end{array}\right]\otimes I_6.$$ For the multiple testing problem, we split this matrix into a block matrix $\mathit{H}=\left[\mathit{H}_1^{\top},\dots,\mathit{H}_6^{\top}\right]^{\top}$, where $\mathit{H}_m=\mathbf{t}_{6m}\otimes I_6$ and $\mathbf{t}_{6m}$ is the $m$-th row of $\mathit{T}_6$, $m=1,\dots,6$.  This corresponds to the individual hypotheses $\mathcal{H}_{0,ij}:\boldsymbol\eta_i(t)=\boldsymbol\eta_j(t)$ for all $t\in\mathcal{T}$ and $1\leq i<j\leq 4$.

\emph{Model 2.} To compare the new test with the two-sample tests by \cite{Multiv}, we also consider the two-sample problem with $k=2$ and $p=2$. The simulation setup is the same as in Simulation~1 in \cite{Multiv}, and it is very similar to that presented above. Let us just point out the differences. The mean functions are $\eta_{11}(t)=\eta_{21}(t)=(\sin(2\pi t^2))^5$, $\eta_{21}(t)=\eta_{16}(t)=1+2.3t+3.4t^2+1.5t^3$, and $\eta_{22}$ is the same as $\eta_{46}$ given in~\eqref{eta46}. The variability is controlled by the $\mathit{H}=(h_1,h_2)^{\top}$ vectors: $\mathit{H}_1=(1.5,1.5)^{\top}$, $\mathit{H}_2=(1.5, 3.5)^{\top}$, and $\mathit{H}_3=(3.5, 1.5)^{\top}$. The sample sizes are $(n_1,n_2)=(20,30),(60,90)$, and $$\mathit{A}=\left[\begin{array}{rr}
1&-1\\
0&1\\
\end{array}
\right].$$

The functional observations are discreetly observed in so-called design time points. Here, their values are calculated in 50 equally spaced points in the interval $[0,1]$. The examples of trajectories of generated functional data in simulation study are presented in Figures~S1-S3 in the supplement.

\emph{Evaluation Criteria and Simulation Runs.} We evaluated the global hypothesis tests by means of their size (type I error rate) and power. To obtain reliable estimates, the empirical size and power of each test is calculated as the proportion of rejections in 1000 repetitions of simulation under the null (size) and alternative hypothesis (power), respectively. 

For the multiple testing procedures, we use the family-wise error rate (FWER) and the local power for evaluation. The empirical FWER is computed as the  proportion of rejecting at least one true null hypothesis. Moreover, the empirical local power is calculated in the same way as for the global hypothesis, but separately for each non true hypothesis.

For the resampling tests, the number of bootstrap or permutation samples equals 1000. The significance level is set to $\alpha=5\%$. The simulation experiments were conducted in the R programming language \citep{Rcore}. The R code for all numerical experiments is given in the supplement.

\subsection{Simulation results}\label{sec:sim_results}
\begin{figure}[t]
\centering
\includegraphics[width=0.99\textwidth,
height=0.4\textheight]{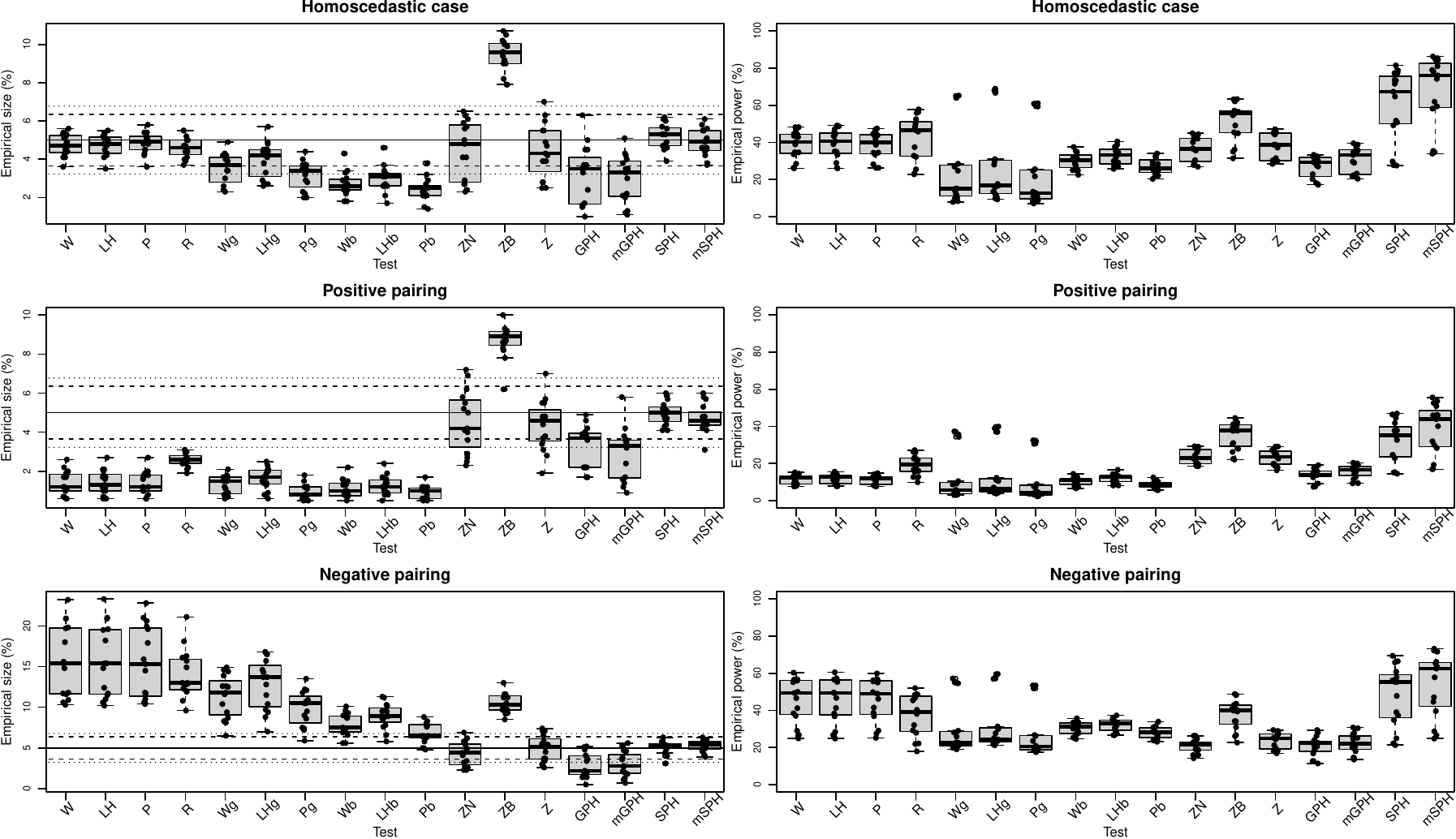}
\caption{Box-and-whisker plots for the empirical sizes and power (as percentages) of all tests (without Rg and Rb tests, since they are always too liberal) obtained in Model~1 under global hypothesis for $k=4$ groups and $p=6$ functional variables and for homoscedastic case, positive pairing, and negative pairing.}
\label{fig_1}
\end{figure}

{In this section we 
discuss our findings in detail by means of the four summarizing Figures~\ref{fig_1}-\ref{fig_3}. Additional details on all simulation results are presented in separate tables and figures in the supplement }(Tables~S1-S10, Figures~S4-S13). 
Figures~\ref{fig_1},~\ref{fig_4} and~\ref{fig_3} show the properties of the tests for homoscedastic and both heteroscedastic cases under different scenarios. On the other hand, Figure~\ref{fig_2} presents the effect of different amounts of correlation in the functional data. 
{We note that we throughout excluded the Rg and Rb tests for ease of presentation as they were extremely liberal in all cases (confer Tables~S1 and~S5 in the supplement).}

\begin{figure}
\centering
\includegraphics[width=0.99\textwidth,
height=0.7\textheight]{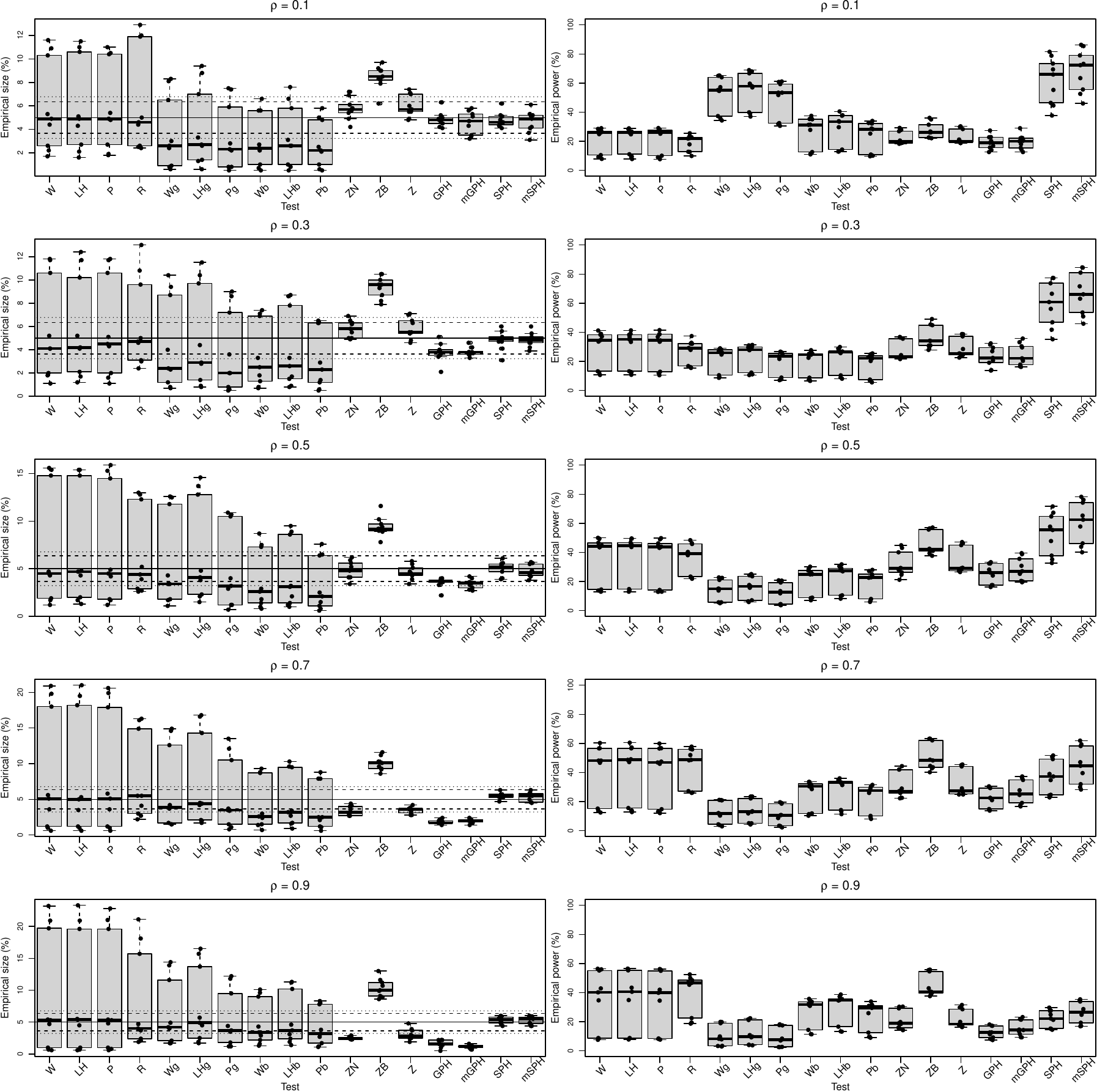}
\caption{Box-and-whisker plots for the empirical size and power (as percentages) of all tests (without Rg and Rb tests, since they are always too liberal) obtained in Model~1 under global hypotheses for $k=4$ groups and $p=6$ functional variables and for different correlation.}
\label{fig_2}
\end{figure}
{
{\textit{ Global Hypothesis Testing.}}
The results for testing the global hypothesis $\mathcal{H}_0:\boldsymbol\eta_1(t)=\dots=\boldsymbol\eta_4(t)$ for all $t\in\mathcal{T}$ 
are given in Figures~\ref{fig_1}-\ref{fig_2}. The mGPH and GPH tests (third and fourth from last boxplot in each plot) by \cite{FANOVA} worked well for one-dimensional functional data. However, for our multivariate settings, they exhibit a rather conservative character (left plots, median empirical sizes always below 5\%), especially for larger negative pairing (last row of Figure~\ref{fig_1}) or larger correlations (last rows of Figure~\ref{fig_2}). This resulted in smaller power (right plots).} 
The ZB test (six last boxplot) is always too liberal. The two other tests of this type, ZN and Z (fifth and seventh last boxplots), control the type I error level much better, but are sometimes a bit too conservative or too liberal (whiskers and parts of boxes  outside the binomial interval). On the other hand, both mSPH and SPH tests (last and second last boxplot in each plot) control the type I error level very well, being neither conservative nor liberal. {
In fact, these are the only two methods that control type I error level  independently of variance pairing or correlations. For describing the results of the remaining ten procedures, we therefore distinguish the settings. 
Under homoscedasticity (first row of Figure~\ref{fig_1}),} the permutation W, LH, P, and R tests control the type I error level well. The other projection tests, i.e. the Wg, LHg, Pg, Wb, LHb, and Pb tests, exhibit a conservative character which is common for this type of test, see, e.g. \citeauthor{Gorecki}, \citeyear{Gorecki}. Under positive pairing, all of these ten tests are conservative, while for negative pairing, they are too liberal. Regarding power, the most accurate procedures, SPH and mSPH, are usually also the most powerful, except for the cases with larger correlations (Figure~\ref{fig_2} with 
$\rho=0.9$ and to some extend also $\rho=0.7$). However, in this case, the new tests are comparable with other tests and control the type I error level well in contrast to most of the competitors. Moreover, the mSPH test is usually better than the SPH test, which shows the improvement of using the multiple testing procedure proposed in Section~\ref{ssec:Multiple} compared to naive Bonferroni. The GPH and mGPH tests are usually much less powerful. In comparison with the other competitors, they are sometimes slightly better, comparable, or worse. However, the GPH and mGPH tests are never too liberal, which is not true for the competitors.

\begin{figure}[t]
\centering
\includegraphics[width=0.99\textwidth,
height=0.4\textheight]{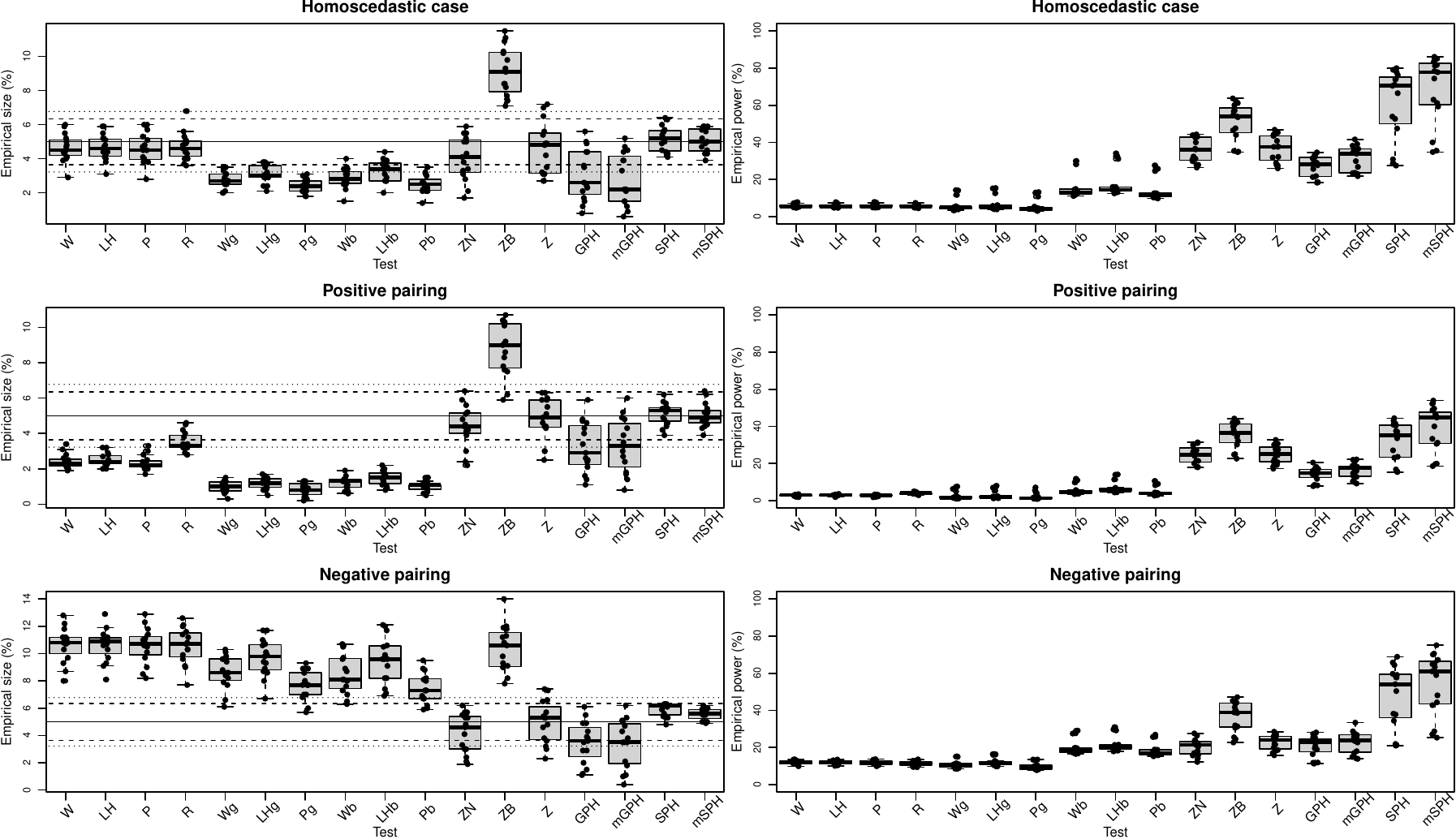}
\caption{Box-and-whisker plots for the empirical sizes and power (as percentages) of all tests (without Rg and Rb tests, since they are always too liberal) obtained in Model~1 under global hypothesis for $k=4$ groups and $p=6$ functional variables with scaling function and for homoscedastic case, positive pairing, and negative pairing.}
\label{fig_4}
\end{figure}

\begin{figure}[t]
\centering
\includegraphics[width=0.99\textwidth,
height=0.4\textheight]{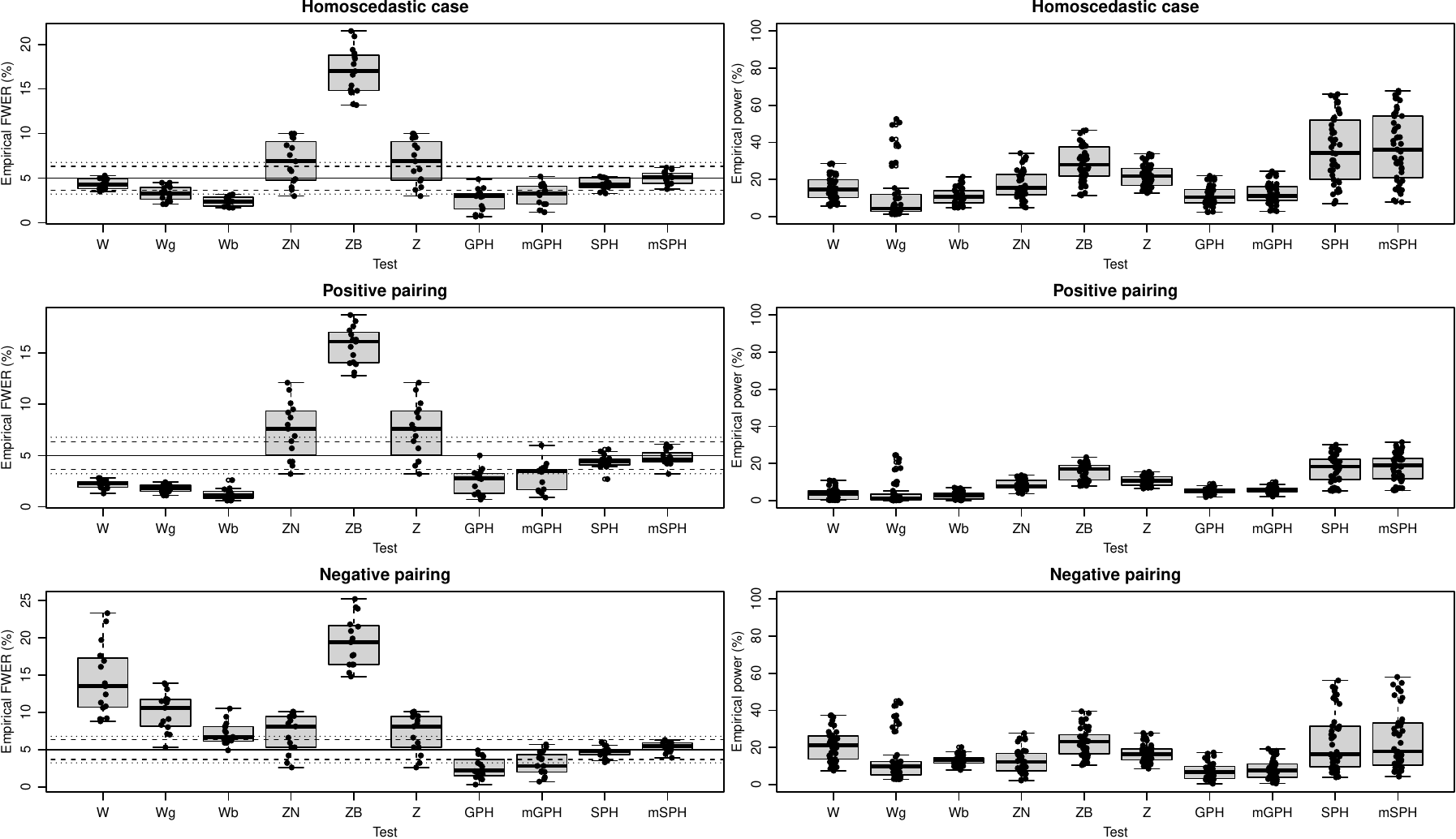}
\caption{Box-and-whisker plots for the empirical FWER and power (as percentages) of all tests obtained in Model~1 under local hypotheses for $k=4$ groups and $p=6$ functional variables and for homoscedastic case, positive pairing, and negative pairing.}
\label{fig_3}
\end{figure}

{
{\textit{ Impact of Scaling.} }}
Let us now discuss the impact of scaling functional data on the properties of the global testing procedures. The results are given in Figure~\ref{fig_4}. The new tests and the GPH, mGPH, ZN, ZB, and Z tests are scale-invariant. Thus, the simulation results are similar to the cases without scaling. On the other hand, {the tests by \cite{Gorecki} (first ten boxplots in all plots) are not scale-invariant, and their empirical power for the present case with scaling is much smaller than for the without scaling case. In fact, they often do not exhibit any power at all for the scaled alternatives. }

{
{\textit{ Multiple Testing.}} 
Turning to the multiple testing problem, Figure~\ref{fig_3} shows the empirical family-wise error rate (FWER) and power (as percentages) of the studied procedures.} {For ease of presentation, we only present the results for the W, Wg, and Wb tests by \cite{Gorecki} using the Wilks test statistic, since these tests are the best representation of that kind of testing procedure.} In general, the conclusions are similar to the above ones for testing the global hypothesis. {In fact, the new mSPH and SPH tests (last and second to last boxplots) exhibit the best results in terms of FWER control and power. }
However, note that the global SPH test with Bonferroni correction is at least slightly more conservative than the mSPH test. Moreover, the ZN, ZB, and Z tests do not control the FWER correctly in many cases, i.e., they are mainly too liberal. 

{
{\textit{ The Classy Two-Sample Problem.}} 
Two-sample comparisons are among the most relevant. That is why we finally study the suitability of the new method for the two-sample testing problem. The detailed results are given in Section~S1.4 in the supplement and we only summarize our findings here. 
Again, the conclusions for all previously considered tests are similar. However, we additionally compare with the recently proposed two sample methods QI and QS by \cite{Multiv} and thus elaborate on them a bit. In many cases, the QI test is too liberal, especially for negative pairing but also for the homoscedastic case. Nevertheless, it is less powerful than the QS test in almost all settings. The QS test is too liberal in negative pairing but otherwise controls type I error level quite well. In contrast, the new SPH test shows the best type I error level control. Moreover, the QS and SPH tests are usually the most powerful, but QS is at least slightly less powerful than SPH.}  

\begin{figure}[t]
\includegraphics[width=0.99\textwidth,
height=0.35\textheight]{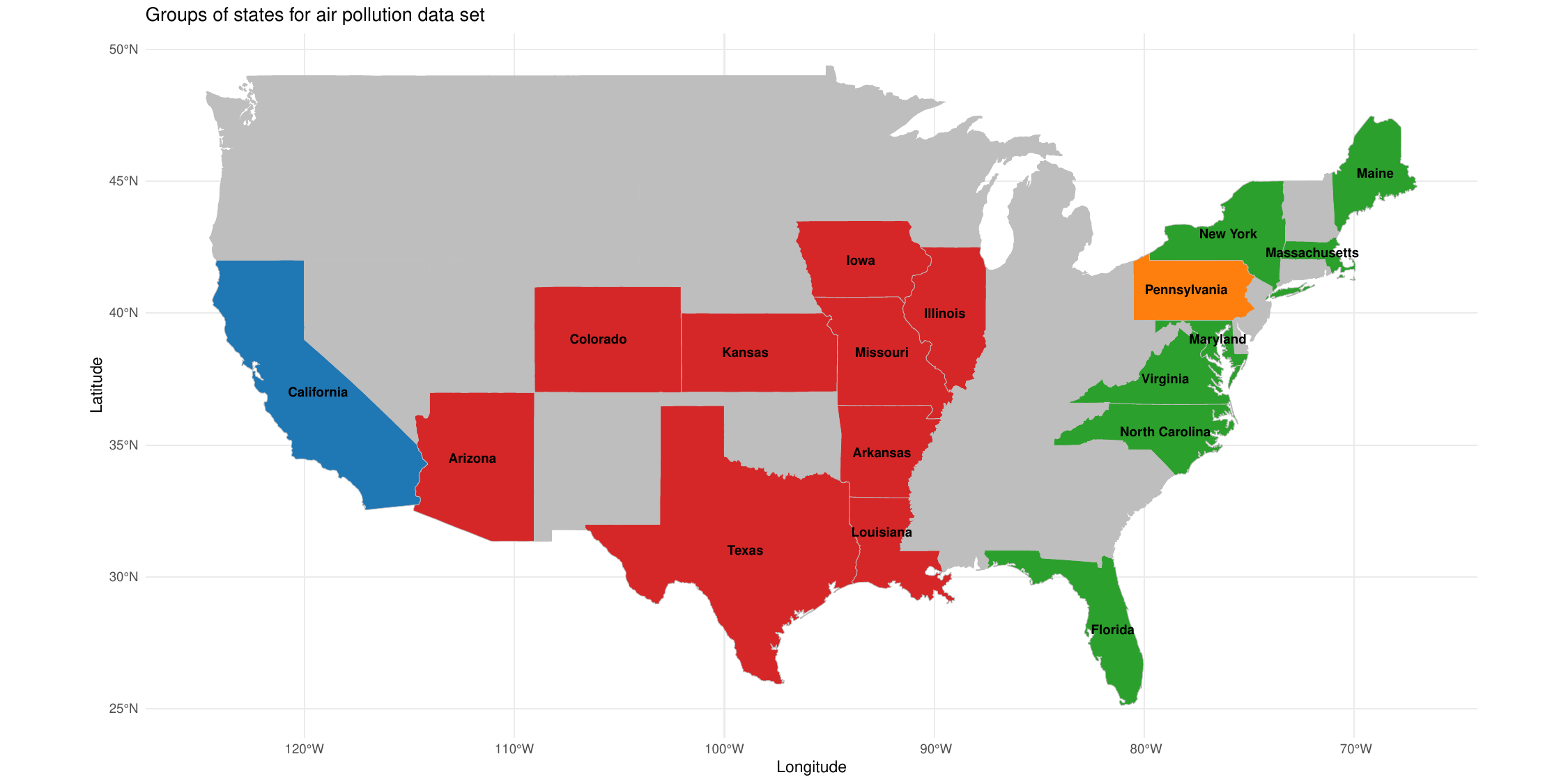}
\caption{Groups of US states considered in our illustrative example: California (blue), Pennsylvania (orange), East states (green), and Central states (red).}
\label{fig_map}
\end{figure}

To sum up, in contrast to the competitors, the new SPH and mSPH tests control the type I error level in all scenarios and usually have the best power. 
From these two tests, the mSPH test using the multiple testing procedure from Section~\ref{ssec:Multiple} is at least slightly better than the global SPH test with Bonferroni correction.
{
\section{Illustrative Data Analysis}\label{sec:Data}
In this section, we illustrate the application on a real data example. It is based on the U.S. air pollution data which is} available from \url{https://www.kaggle.com/datasets/sogun3/uspollution} and was also studied in \cite{ZHU2022105095}. It contains daily concentration measurements of four major pollutants, namely, Nitrogen Dioxide (NO2), Ozone (O3), Sulfur Dioxide (SO2), and Carbon Monoxide (CO) in U.S. cities for many years. For illustrative purposes, we only consider the period from 31 March to 31 October 2008. We used the data for the 2008 year since there are quite a lot of observations for cities in California and Pennsylvania. Moreover, we did not take a whole year, since at its beginning and end, there were a lot of missing values, especially for cities in Pennsylvania. After removing missing values and outliers (detected by visual inspection, see the R code in the supplement for details), we consider the following four samples:
\begin{enumerate}
\item 13 functional observations for cities in California,
\item 15 functional observations for cities in Pennsylvania,
\item 9 functional observations for cities in the East states in the USA,
\item 15 functional observations for cities in the Central states in the USA.
\end{enumerate}
The groups of states are presented in Figure~\ref{fig_map}. Pennsylvania is also located in the eastern part of the USA, but we excluded it from the East states category. The reason for this is that we have data for many cities in Pennsylvania, while for other states, there are data for only one or two cities. As in \cite{ZHU2022105095}, each daily concentration measurement raw curve was first smoothed using a B-spline basis of order 4 and with 20 basis functions. Then, the resulting smoothed concentration measure curve was evaluated over a common grid of time points $t\in\{1,\dots,215\}$ (215 is the number of days between 31 March to 31 October). 

\begin{figure}[t]
\includegraphics[width=0.99\textwidth,
height=0.45\textheight]{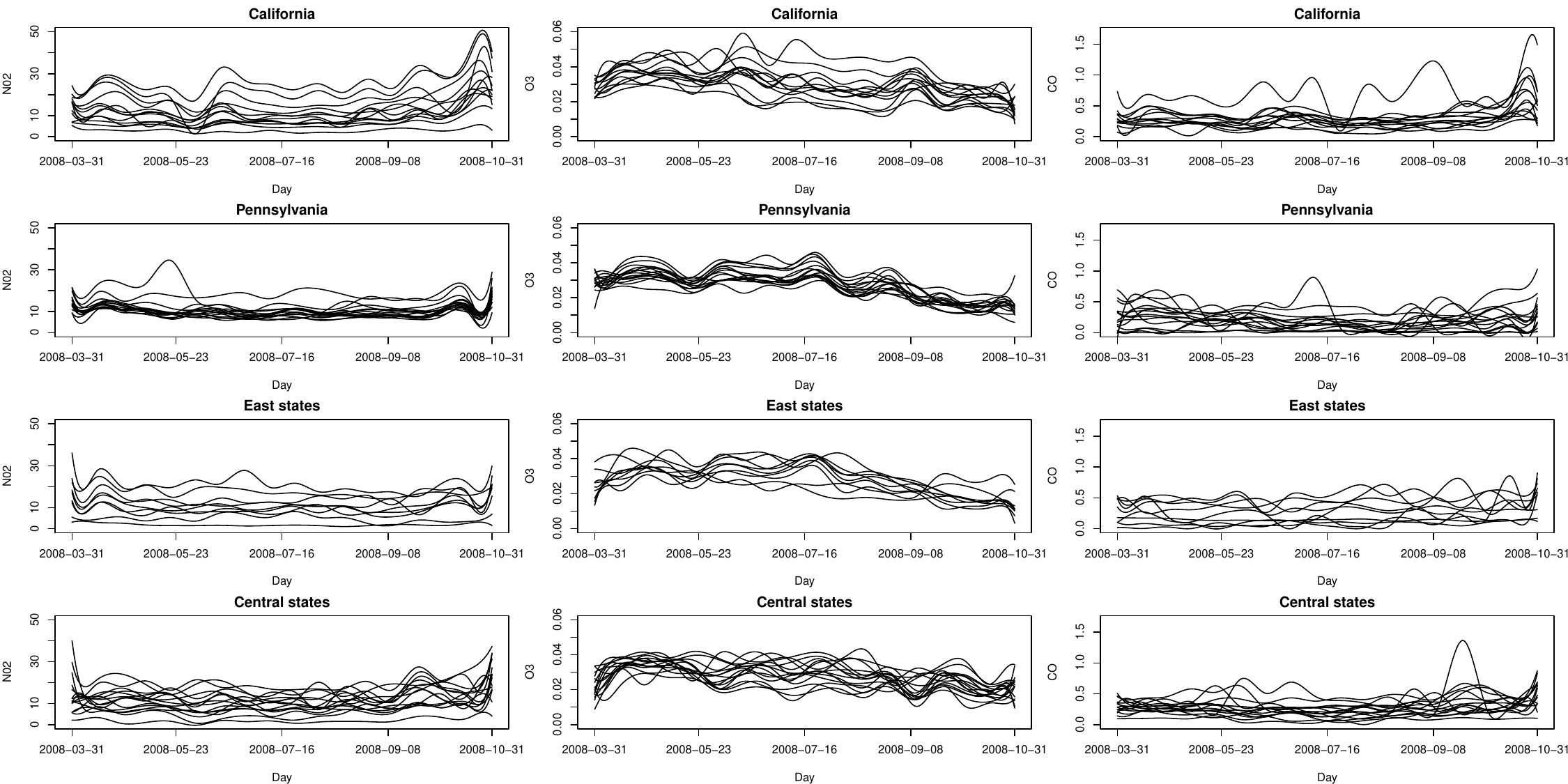}
\caption{Trajectories for the air pollution data set for the three pollutants Nitrogen Dioxide (NO2), Ozone (O3), and Carbon Monoxide (CO) in California, Pennsylvania, East states, and Central states.}
\label{fig_rde_1}
\end{figure}

\begin{figure}[t]
\includegraphics[width=0.99\textwidth,
height=0.35\textheight]{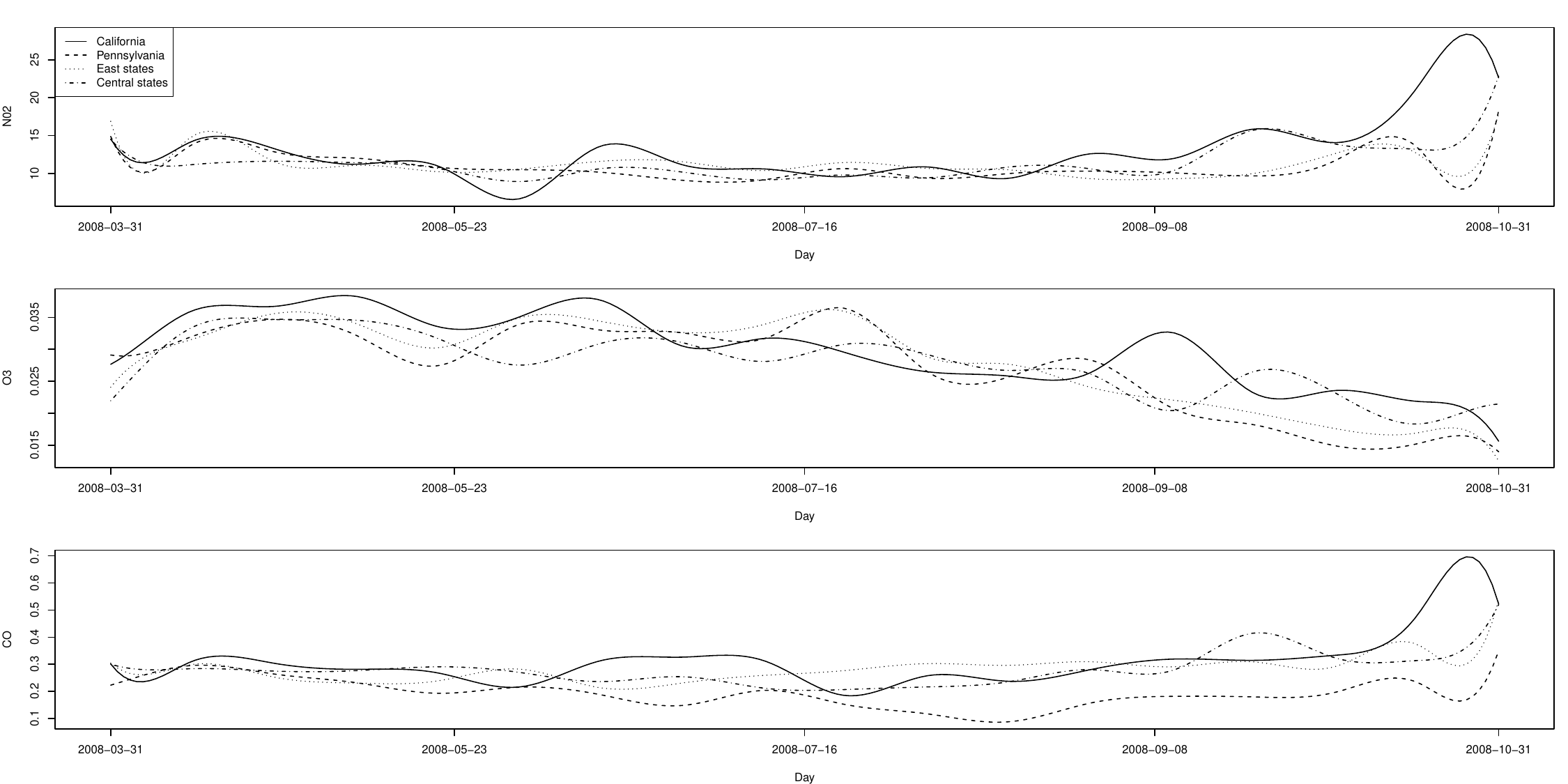}
\caption{Sample mean functions for the air pollution data set for the three pollutants Nitrogen Dioxide (NO2), Ozone (O3) and Carbon Monoxide (CO) in California (solid line), Pennsylvania (dashed), East states (dotted), and Central states (dash-dotted).}
\label{fig_rde_3}
\end{figure}

From the four major pollutants, 
the SO2 variable is very different in the four groups. As this would result in rejecting the null hypothesis by all testing procedures, we omit for our illustrative analysis. Thus, each functional observation 
in this example is a three-dimensional vector of functions corresponding to three pollutants NO2, O3, and CO. The trajectories of our multivariate functional data are presented in Figure~\ref{fig_rde_1}. 
We are interested in comparing the 'location' of the four groups. In particular, we want to test the equality of vectors of group-wise mean functions. The sample mean functions are presented in Figure~\ref{fig_rde_3} (see also Figure~S14 in the supplement). They suggest that there are some differences between the four regional groups. To infer this, we applied all tests considered in Section~\ref{sec:Simu}. The results for the global hypotheses are presented in Table~S12 in the supplement while Table~\ref{tab_rde_2} shows the results for the six local hypotheses (the $P$ columns in both tables). 

In addition, to this illustrative analysis we also conducted a Plasmode-type study, i.e. a simulation study that mimics the present set-up. This allows us to better assess the trustworthiness of the tests' results. To this end, we simulated four groups with the same sample sizes $n_1=13$, $n_2=15$, $n_3=9$, and $n_4=15$ from Gaussian processes with
group-wise covariance functions equal to the sample covariance functions estimated from the data. Moreover, as group-wise mean functions we used (a) the vector of sample mean functions to asses power ('Power' column in the tables) or (b) the pooled sample mean function to infer type-I-error control ('FWER' in Table~\ref{tab_rde_2} and 'Size' in Table~S12). All tests reject the global null hypothesis and their empirical power is close to $100\%$ (Table~S12 in the supplement). However, most tests are too liberal with empirical sizes up to $43\%$ (e.g. for Rg) instead of the desired $5\%$. In fact, only the Pb test and the four PH-tests (GPH, mGPH, SPH and mSPH) exhibit type I error levels smaller than $6\%$  (Table~S12 in the supplement).
The bad properties of some of the tests can be explained by the significant differences in covariance functions in these four groups, which is suggested by the tests for equality of covariance functions for one-dimensional functional data applied to each functional variable separately (see Section~S2.1 in the supplement for more details). Nevertheless, it seems that the vectors of mean functions of the three pollutants are significantly different in the four groups.

As we rejected this global hypothesis, we conducted a post-hoc analysis (Table~\ref{tab_rde_2}) to find out which groups are significantly different and which are not. We can observe that all tests reject the equality for cities in California and Pennsylvania as well as in Pennsylvania and the Central states. On the other hand, in the comparison 'Pennsylvania vs. East states', no test rejects the null hypothesis, which is perhaps caused by the geographical similarity. Finally, we comment the remaining pairs, which are the most interesting cases. Here, we can observe large differences between the methods. Most tests do not reject the null hypotheses which also matches their lower power in the mimicking simulation study. On the other hand, the new SPH (except for the case 'California vs. East states') and mSPH tests as well as the Wg test reject the local null hypotheses and also have larger power in the corresponding simulation study. However, note that the Wg test is a bit too liberal (estimated FWER of 7.3\%) in the mimicking simulation study, perhaps due to the heteroscedastic character of the data. Thus, the most trustworthy results are obtained from the SPH and mSPH tests. Moreover, similar to Section~\ref{sec:Simu}, we also observe a better performance of the mGPH and mSPH tests than the GPH and SPH tests in this simulation study. 

\section{Discussion}\label{sec:Discussion}
{We have proposed a novel class of inference methods for testing (i) global and (ii) local hypotheses about general linear combinations of multivariate mean functions in a functional MANOVA setting. While some methods for (i) were already known (though most under quite restrictive assumptions), we are the first to establish multiple testing procedures with family-wise error rate (FWER) control for (ii) (except for simple Bonferroni adjustments). 
We have proven the asymptotic validity of the novel procedures under weak assumptions, neither requiring Gaussianity nor homoscedasticity, i.e. we do not need to assume equal covariance functions across groups. Moreover, the proposed methods are even scale invariant. To facilitate their application, all new statistical tests are implemented in the R package gmtFD, which will appear on CRAN soon. 

In a comparative simulation study it turned out that both, the new global and multiple testing procedures exhibit accurate type I error rate and FWER control, respectively. In particular, while most existing approaches were either too conservative or too liberal depending on whether positive or negative pairing was present, the new approaches turned out to be robust with respect to changes of covariance functions, dependence structure or sample sizes. In addition, the new methods were also shown to be preferable in terms of power in most studied settings. A reason for this may lie in the form of our test statistics which are based upon globalising pointwise Hotelling-$T^2$-type statistics via the supremum. We note that} it would also be possible to globalise by integrating over it. 
{In fact, our derived theory would in principle also work for this case. 
However, initial simulation results with regard to the procedures'} power were not as convincing as for the supremum-statistic.

The proposed methods are valid for general factorial designs {covering crossed or non-crossed one-, two- and higher-way designs as well as nested layouts. Moreover, they are applicable
with multivariate functional data, longitudinal functional data and even longitudinal multivariate functional data. Hence, many special cases, like the functional MANOVA or functional Repetated Measures ANOVA problem are covered. 
As the current paper focused on the thorough methodological development of the procedures, we could only exemplify the practicability for some special designs. A}  detailed study of the methods' applicability and numerical performance for other important designs, such as one and two group Repeated Measures or complex Split Plot designs, will be done in future research.

\begin{landscape}
\begin{table}[!h]
\caption{$P$-values, empirical FWER, and power (as percentages) of the tests for local hypotheses in the air pollution data set. Too liberal results are presented in bold, i.e., we define the test as too liberal, when its empirical FWER is greater than $6.4\%$, where $6.4\%$ is the value of the upper limit of $95\%$ binomial confidence interval for empirical FWER for 1000 simulation runs \citep{DuchesneFrancq2015}.)}
\centering
\begin{tabular}{l r rr r rr r rr r rr r rr r rr}
\toprule
&&\multicolumn{2}{c}{California vs.}&&\multicolumn{2}{c}{California vs.}&&\multicolumn{2}{c}{California vs.}&&\multicolumn{2}{c}{Pennsylvania vs.}&&\multicolumn{2}{c}{Pennsylvania vs.}&&\multicolumn{2}{c}{East states vs.}\\
&&\multicolumn{2}{c}{Pennsylvania}&&\multicolumn{2}{c}{East states}&&\multicolumn{2}{c}{Central states}&&\multicolumn{2}{c}{East states}&&\multicolumn{2}{c}{Central states}&&\multicolumn{2}{c}{Central states}\\\cmidrule{3-4}\cmidrule{6-7}\cmidrule{9-10}\cmidrule{12-13}\cmidrule{15-16}\cmidrule{18-19}
Test&\multicolumn{1}{c}{FWER}&$P$&\multicolumn{1}{r}{Power}&&$P$&\multicolumn{1}{r}{Power}&&$P$&\multicolumn{1}{r}{Power}&&$P$&\multicolumn{1}{r}{Power}&&$P$&\multicolumn{1}{r}{Power}&&$P$&\multicolumn{1}{r}{Power}\\\midrule
W& \textbf{8.3}& \textbf{0.6}&\textbf{98.4} && \textbf{89.4}&\textbf{19.2} && \textbf{45.6}&\textbf{39.2} && \textbf{100.0}&\textbf{19.8} && \textbf{0.6}&\textbf{97.4} && \textbf{100.0}&\textbf{15.3} \\
Wg& \textbf{7.3}& \textbf{0.0}&\textbf{100.0} && \textbf{0.6}&\textbf{99.5} && \textbf{0.2}&\textbf{98.6} && \textbf{100.0}&\textbf{44.3} && \textbf{0.0}&\textbf{100.0} && \textbf{2.2}&\textbf{71.0} \\
Wb& 5.0& 0.0&99.1 && 19.6&75.3 && 75.7&31.2 && 100.0&19.6 && 0.0&98.8 && 87.8&27.3 \\\addlinespace
ZN& \textbf{12.4}& \textbf{0.0}&\textbf{98.9} && \textbf{55.2}&\textbf{50.6} && \textbf{52.4}&\textbf{50.8} && \textbf{100.0}&\textbf{22.8} && \textbf{0.0}&\textbf{100.0} && \textbf{100.0}&\textbf{36.5} \\
ZB& \textbf{19.2}& \textbf{0.0}&\textbf{99.8} && \textbf{35.6}&\textbf{67.1} && \textbf{34.8}&\textbf{64.0} && \textbf{100.0}&\textbf{37.7} && \textbf{0.0}&\textbf{100.0} && \textbf{75.1}&\textbf{60.8} \\
Z& \textbf{9.5}& \textbf{0.1}&\textbf{98.0} && \textbf{27.5}&\textbf{48.8} && \textbf{44.4}&\textbf{43.2} && \textbf{100.0}&\textbf{19.5} && \textbf{0.0}&\textbf{100.0} && \textbf{63.1}&\textbf{36.1} \\\addlinespace
GPH& 4.7& 1.2&90.8 && 57.6&26.9 && 75.0&25.4 && 100.0&8.0 && 0.0&99.6 && 100.0&14.9 \\
mGPH& 5.8& 1.1&93.1 && 33.0&32.4 && 40.9&30.8 && 91.9&8.9 && 0.0&99.8 && 61.5&18.7 \\\addlinespace
SPH& 5.0& 0.0&99.6 && 5.4&77.6 && 4.8&87.5 && 100.0&19.1 && 0.0&100.0 && 0.0&99.8 \\
mSPH& 5.7& 0.0&99.8 && 4.5&81.2 && 4.1&89.9 && 71.6&22.2 && 0.0&100.0 && 0.0&99.8 \\
\bottomrule
\end{tabular}
\label{tab_rde_2}
\end{table}
\end{landscape}

\newpage
\begin{appendices}

 \section{Remarks on Assumption~\ref{Massumptions}}
In this section, we state remarks about Assumption~\ref{Massumptions}. Here, we see that various other properties follow from these assumptions.

\begin{enumerate}
\item[\namedlabel{itm:224}{(R1)}] By Theorem~2.2.4 in \cite{vaartWellner1996} and Assumption~\ref{itm:Lipschitz}, it follows
\begin{align}
    \label{224}
    \E\left[ \sup\limits_{\rho(t,s) \leq \delta } || \mathbf{v}_{i1}(t) - \mathbf{v}_{i1}(s) ||_2^J \right] \leq K \left\{  \int_0^\nu (D(\varepsilon,\rho))^{1/J} \;\mathrm{d}\varepsilon + \delta (D(\nu,\rho))^{2/J} \right\}^J
\end{align}
 for all $\nu,\delta > 0, i\in\{1,...,k\}$ and some constant $K \geq 0$ depending on $J$ and $C$ only. Since $\nu$ can be chosen arbitrarily small, we have
$$ \E\left[ \sup\limits_{\rho(t,s) \leq \delta } || \mathbf{v}_{i1}(t) - \mathbf{v}_{i1}(s) ||_2^J \right] \to 0 \quad\text{as }\delta\to 0$$ by (\ref{eq:intAss}) for all $i\in\{1,...,k\}.$
\item[\namedlabel{itm:moment}{(R2)}] Assumptions~\ref{itm:J}, \ref{itm:Lipschitz} and (\ref{224}) imply 
\begin{align*}
    \E\left[ \sup\limits_{t\in\T } || \mathbf{v}_{i1}(t) ||_2^J \right] &\leq 2^J\left( \E\left[ \sup\limits_{t\in\T } || \mathbf{v}_{i1}(t) - \mathbf{v}_{i1}(t^*) ||_2^J \right] + \E\left[ || \mathbf{v}_{i1}(t^*) ||_2^J \right]\right)
   \\ &\leq 2^J\left[ K \left\{  \int_0^\nu (D(\varepsilon,\rho))^{1/J} \;\mathrm{d}\varepsilon + \sup\limits_{t\in\T}\rho(t,t^*) (D(\nu,\rho))^{2/J} \right\}^J + \E\left[ || \mathbf{v}_{i1}(t^*) ||_2^J \right]\right] \\& < \infty
\end{align*} for all $i\in\{1,...,k\}$ since $\T$ is compact.
\item[\namedlabel{itm:A7}{(R3)}] Combining~\ref{itm:224} and \ref{itm:moment} yields
\begin{align*}
    &\E\left[ \sup\limits_{\rho(t,s) \leq \delta } ||| \mathbf{v}_{i1}(t)\mathbf{v}^{\top}_{i1}(t) - \mathbf{v}_{i1}(s)\mathbf{v}^{\top}_{i1}(s) ||| \right] 
    \\&\leq \E\left[ \sup\limits_{\rho(t,s) \leq \delta } ||| \mathbf{v}_{i1}(t)\mathbf{v}^{\top}_{i1}(t) -  \mathbf{v}_{i1}(t)\mathbf{v}^{\top}_{i1}(s) |||\right] +  \E\left[\sup\limits_{\rho(t,s) \leq \delta } ||| \mathbf{v}_{i1}(t)\mathbf{v}^{\top}_{i1}(s) - \mathbf{v}_{i1}(s)\mathbf{v}^{\top}_{i1}(s) ||| \right] 
    \\&\leq 2\E\left[ \sup\limits_{t\in\T } || \mathbf{v}_{i1}(t) ||_2 \sup\limits_{\rho(t,s) \leq \delta }||\mathbf{v}_{i1}(t) -  \mathbf{v}_{i1}(s) ||_2\right] 
    \\&\leq 2\sqrt{\E\left[ \sup\limits_{t\in\T } || \mathbf{v}_{i1}(t) ||_2^2\right] \E\left[ \sup\limits_{\rho(t,s) \leq \delta }||\mathbf{v}_{i1}(t) -  \mathbf{v}_{i1}(s) ||_2^2\right] }
    \to 0 
\end{align*} as $\delta\to 0$,
where here and throughout $|||.|||$ denotes the spectral norm.
Thus, it directly follows that the map $$\T \ni t \mapsto \E\left[\mathbf{v}_{i1}(t)\mathbf{v}^{\top}_{i1}(t)\right] = \mathit{\Gamma}_i(t,t) \in \R^{p\times p}$$ is uniformly continuous with respect to $\rho$. 
\\
Analogously, one can show
\begin{align*}
    &\E\left[ \sup\limits_{\rho(t,s) \leq \delta } ||| \mathbf{v}_{i1}(t)\mathbf{v}^{\top}_{i2}(t) - \mathbf{v}_{i1}(s)\mathbf{v}^{\top}_{i2}(s) ||| \right] 
    \to 0 
\end{align*} as $\delta\to 0$.
\end{enumerate}

\end{appendices}

\section*{Competing interests}
No competing interest is declared.

\section*{Data Availability Statement}
The data underlying this article are available in the article and in its online supplementary material, there is the R code for preparing the data set.

\section*{Acknowledgments}
The authors are grateful to Prof. Tianming Zhu for sharing us the code for \cite{zhu2023} and \cite{ZHUetAl2023}. Merle Munko and Marc Ditzhaus gratefully acknowledge funding by the Deutsche Forschungsgemeinschaft (DFG, German Research Foundation) - 314838170, GRK 2297 MathCoRe. A part of calculations for the simulation study and real data example was made at the Pozna\'n Supercomputing and Networking Center (grants 382 and pl0253-02).

\section*{Supplementary materials}
The supplement contains all proofs, all results of simulation studies of Section~3, more details about the analysis of real data examples of Section~4 of the main paper, and the R code for all numerical experiments.


\bibliographystyle{abbrvnat}

\end{sloppypar}
\end{document}


\captionsetup{width=\textwidth}

\author[1]{Merle Munko}
\author[2]{Marc Ditzhaus}
\author[3]{Markus Pauly}
\author[4]{\L ukasz Smaga}

\address[1]{Faculty of Mathematics, Otto-von-Guericke University Magdeburg, Germany}
\address[2]{Faculty of Mathematics, Otto-von-Guericke University Magdeburg, Germany}
\address[3]{TU Dortmund University, Dortmund, Germany; Research Center Trustworthy Data Science and Security, UA Ruhr, Germany}
\address[4]{Faculty of Mathematics and Computer Science, Adam Mickiewicz University, Poland}

\date{Journal: ???}

\title{ Supplementary materials to\\
Multiple Comparison Procedures for Simultaneous Inference \\ in Functional MANOVA }
\begin{center}
		Merle Munko\inst{1},
		Marc Ditzhaus\inst{1},
  Markus Pauly\inst{2,3} and
  {\L}ukasz Smaga\inst{4}
\end{center}

	\institute{
		\inst{1} Otto-von-Guericke University Magdeburg; Magdeburg (Germany)\newline
        \inst{2} TU Dortmund University; Dortmund (Germany)
\newline
        \inst{3} Research Center Trustworthy Data Science and Security, UA Ruhr; Dortmund (Germany)
        \newline
        \inst{4} Adam Mickiewicz University; Poznań (Poland)
	}

\hrule
\vspace{1cm}


This supplement contains all proofs, all results of simulation studies of Section~3 and more details about the analysis of real data examples of Section~4 of the main paper.

\tableofcontents

\listoftables
\listoffigures

\newpage

\section{Proofs}
\label{App3}
In this section, the technical proofs of all stated theorems are presented.

\subsection{Auxiliary Lemmas}
Here, useful auxiliary lemmas are presented.
First of all, we state a lemma which will be applied for proving the weak convergence of our processes.

\begin{lem}
\label{Conv}
Let $\mathbf{X}_1,\mathbf{X}_2,...$ be a sequence of $C^p(\T)$-valued stochastic processes. 
Moreover, assume that there is a $p$-dimensional stochastic process $\mathbf{X}$ such that $$(\mathbf{X}_n(t_1),...,\mathbf{X}_n(t_m)) \xrightarrow{d} (\mathbf{X}(t_1),...,\mathbf{X}(t_m))  \quad\text{as $n\to\infty$}$$ for all finite subsets $\{t_1,...,t_m\}$ of $\T$. 
If there exists a $J \geq 1$ and a constant $C_J\geq 0$ depending on $J$ only such that
\begin{align}
    \label{eq:pequicont}
    \E\left[ ||\mathbf{X}_n(t) - \mathbf{X}_n(s)||_2^J  \right] \leq C_J (\rho(t,s))^J
\end{align}
for all $t,s\in\T, n\in\mathbb N$ and     \begin{align}
        \label{eq:intAss2}
        \int_0^{\nu} D(\varepsilon, \rho)^{1/J} \;\mathrm{d}\varepsilon < \infty
    \end{align} for some $\nu > 0,$ where $D(\varepsilon, \rho)$ denotes the packing number,
then there is a version of $\mathbf{X}$ with continuous sample paths and we have $\mathbf{X}_n \xrightarrow{d} \mathbf{X}$ as $n\to\infty$ in $C^p(\T)$.
\end{lem}
\begin{proof}[{Proof of Lemma~\ref{Conv}}]
Similarly as noted in the supplement of \cite{Dette}, the results of Section 1.5 in \cite{vaartWellner1996} also hold if the space $\ell^{\infty}(T)$ is replaced by $C^p(T)$. We aim to apply Theorem~1.5.4 in \cite{vaartWellner1996}. The convergence of the marginals is given through the assumptions. Thus, the asymptotic tightness remains to show. Therefore, we aim to apply Theorem~1.5.7 in \cite{vaartWellner1996}.\\
For all $t\in\T$, it holds $\mathbf{X}_n(t) \xrightarrow{d} \mathbf{X}(t)$ and, thus, $(\mathbf{X}_n(t))_{n\in\mathbb N}$ is asymptotically tight by Lemma~1.3.8 in \cite{vaartWellner1996}. 
Since $(\T,\rho)$ is compact, it is in particular totally bounded.\\
Now, we apply Theorem~2.2.4 in \cite{vaartWellner1996} with $\psi(x) = x^J$ for all $x\geq 0$, $C = C_J^{1/J}$ and $d = \rho$ to show that $(\mathbf{X}_n)_{n\in\mathbb N}$ is $\rho$-equicontinuous in probability. Note that $\mathbf{X}_n$ is separable for all $n\in\mathbb N$ due to the compactness of $\T$. Then, (\ref{eq:pequicont}) implies
\begin{align*}
    \E\left[ \sup\limits_{\rho(t,s) \leq \delta} ||\mathbf{X}_n(t) - \mathbf{X}_n(s)||_2^J  \right]^{1/J} \leq K_J \left\{ \int_0^\nu (D(\varepsilon,\rho))^{1/J} \;\mathrm{d}\varepsilon + \delta (D(\nu,\rho))^{2/J} \right\}
\end{align*}
 for any $\nu,\delta > 0, n\in\mathbb N$, where $K_J \geq 0$ is a constant depending on $J$ only.
 Thus, it follows
\begin{align*}
   \lim\limits_{\delta \searrow 0}\limsup\limits_{n\to\infty} \E\left[ \sup\limits_{\rho(t,s)<\delta} ||\mathbf{X}_n(t) - \mathbf{X}_n(s)||_2^J  \right] 
   \leq K_J^J \left\{ \int_0^\nu (D(\varepsilon,\rho))^{1/J} \;\mathrm{d}\varepsilon \right\}^J
\end{align*} for any $\nu > 0$. Since $\nu$ can be chosen arbitrarily small, we get
\begin{align*}
   \lim\limits_{\delta \searrow 0}\limsup\limits_{n\to\infty} \E\left[ \sup\limits_{\rho(t,s)<\delta} ||\mathbf{X}_n(t) - \mathbf{X}_n(s)||_2^J  \right] 
   = 0
\end{align*} by (\ref{eq:intAss2}).
By applying Markov's inequality, we see that $(\mathbf{X}_n)_{n\in\mathbb N}$ is $\rho$-equicontinuous in probability.\\
 Hence, Theorem~1.5.7 in \cite{vaartWellner1996} yields that $(\mathbf{X}_n)_{n\in\mathbb N}$ is asymptotically tight. Then, Theorem~1.5.4 in \cite{vaartWellner1996} completes the proof.
\end{proof}

The next lemma follows easily from the Marcinkiewicz-Zygmund inequality (\citealp{Marcinkiewicz1937}) and will help us in the following to show equation (\ref{eq:pequicont}).

\begin{lem}
    \label{MZinequ}
    Let $\mathbf{X}_1,...,\mathbf{X}_n$ denote $n$ i.i.d.~$p$-dimensional random vectors and $J\geq 2$.
    Then, we have
    \begin{align}
        \label{eq:Jensen}
        {{n}^{-J/2}} \E\left[ \left|\left|\left|\sum_{i=1}^n  \mathbf{X}_i\mathbf{X}_i^{\top} \right|\right|\right|^{J/2} \right] \leq \E\left[ \left|\left| \mathbf{X}_1 \right|\right|_2^J \right].
    \end{align}
   If $\E\left[\mathbf{X}_1\right] = \mathbf{0}_p$ and $\E\left[ \left|\left| \mathbf{X}_1 \right|\right|_2^J \right] < \infty$ holds, we have
   \begin{align}
       \label{eq:MZ}
    \E\left[ \left|\left| \frac{1}{\sqrt{n}} \sum_{i=1}^n \mathbf{X}_i \right|\right|_2^J \right] \leq B_J  {{n}^{-J/2}} \E\left[ \left|\left|\left|\sum_{i=1}^n  \mathbf{X}_i\mathbf{X}_i^{\top} \right|\right|\right|^{J/2} \right] \leq B_J \E\left[ \left|\left| \mathbf{X}_1 \right|\right|_2^J \right]   
   \end{align}
     for some constant $B_J > 0$ depending on $J$ and $p$ only.
\end{lem}
\begin{proof}[{Proof of Lemma~\ref{MZinequ}}]
    We start with the proof of (\ref{eq:Jensen}). Note that (\ref{eq:Jensen}) is trivial for $\E\left[ \left|\left| \mathbf{X}_1 \right|\right|_2^J \right] = \infty$ and, thus, we assume $\E\left[ \left|\left| \mathbf{X}_1 \right|\right|_2^J \right] < \infty$. By the triangular inequality and Jensen's inequality, we get
    \begin{align*}
        n^{-J/2}  \E\left[  \left|\left|\left|\sum_{i=1}^n  \mathbf{X}_i\mathbf{X}_i^{\top} \right|\right|\right|^{J/2} \right] &\leq n^{-J/2}  \E\left[  \left(\sum_{i=1}^n  \left|\left|\mathbf{X}_i \right|\right|_2^2\right)^{J/2} \right]
        \\&\leq n^{-J/2} n^{J/2 - 1} \E\left[  \sum_{i=1}^n  \left|\left|\mathbf{X}_i \right|\right|_2^J \right]
        = \E\left[   \left|\left|\mathbf{X}_1 \right|\right|_2^J \right].
    \end{align*}
    For (\ref{eq:MZ}), we denote $\mathbf{X}_i = (X_{i1},...,X_{ip})^{\top}$ for all $i\in\{1,...,n\}.$
    Then, it holds
    \begin{align*}
        \E\left[ \left|\left| \frac{1}{\sqrt{n}} \sum_{i=1}^n \mathbf{X}_i \right|\right|_2^J \right] \leq p^{J/2 - 1}\E\left[  \left|\left| \frac{1}{\sqrt{n}} \sum_{i=1}^n \mathbf{X}_i \right|\right|_J^J \right]
         = p^{J/2 - 1}n^{-J/2}  \E\left[ \sum_{j=1}^p \left|  \sum_{i=1}^n {X}_{ij} \right|^J \right].
    \end{align*}
    Applying the Marcinkiewicz–Zygmund inequality (\citealp{Marcinkiewicz1937}) yields
     \begin{align*}
        \E\left[ \left|\left| \frac{1}{\sqrt{n}} \sum_{i=1}^n \mathbf{X}_i \right|\right|_2^J \right] \leq p^{J/2 - 1} \widetilde{B}_J n^{-J/2}  \E\left[ \sum_{j=1}^p \left(  \sum_{i=1}^n {X}_{ij}^2 \right)^{J/2} \right].
    \end{align*} for some constant $\widetilde{B}_J > 0$ only depending on $J$.
    Since for any symmetric matrix $\mathit{A} = (a_{ij})_{i,j\in\{1,...,p\}}\in\mathbb R^{p\times p},$ we have that the spectral norm is the largest absolute eigenvalue of $\mathit{A}$, it follows easily that $|||\mathit{A}||| \geq |a_{jj}|$ for all $j\in\{1,...,p\}$. Hence, it holds
    $p|||\mathit{A}|||^{J/2} \geq \sum_{j=1}^p |a_{jj}|^{J/2}.$ Thus, we receive
     \begin{align*}
        \E\left[ \left|\left| \frac{1}{\sqrt{n}} \sum_{i=1}^n \mathbf{X}_i \right|\right|_2^J \right] \leq  p^{J/2} \widetilde{B}_J n^{-J/2}  \E\left[  \left|\left|\left|\sum_{i=1}^n  \mathbf{X}_i\mathbf{X}_i^{\top} \right|\right|\right|^{J/2} \right].
    \end{align*} Setting $B_J =  p^{J/2} \widetilde{B}_J$ and applying (\ref{eq:Jensen}) yields (\ref{eq:MZ}).
\end{proof}

The following lemma can be applied to show the stochastic convergence of a bootstrap quantile.

\begin{lem}\label{BSQuant}
    Let $\mathbf{X}_n^{(j)} = ({X}_{n1}^{(j)}, ..., {X}_{nR}^{(j)})^{\top}, j\in\mathbb N,$ be $R$-dimensional random vectors, $B:\mathbb N \to\mathbb N$ with $B(n)\to\infty$ as $n\to\infty$ and $\alpha\in (0,1)$. Furthermore, let $F_{n,\ell}$ denote the empirical distribution function of  ${X}_{n\ell}^{(1)}, ..., {X}_{n\ell}^{(B(n))}$ for all $\ell\in\{1,...,R\}$ and $F_{n}$ denote the empirical distribution function of  $\mathbf{X}_{n}^{(1)}, ..., \mathbf{X}_{n}^{(B(n))}$ for all $n\in\mathbb N$. Additionally, let $F:\R^R \to [0,1]$ denote a distribution function with continuous marginal distribution functions $F_1,...,F_R:\R \to [0,1]$, where $F_{\ell}$ is strictly increasing on $\left[F_{\ell}^{-1}(a), F_{\ell}^{-1}(b)\right]$ for all $\ell\in\{1,...,R\}$  with $a < 1-\alpha, b > 1-\alpha/R$. If 
    \begin{align}
        \label{eq:condition}
        F_n(t_1,...,t_R) \xrightarrow{P} F(t_1,...,t_R) \text{ as } n\to\infty \text{ for all } t_1,...,t_R\in\R,
    \end{align}
     we have
    \begin{align*}
        F_{n,\ell}^{-1}\left( 1-\beta_n(\alpha) \right) \xrightarrow{P} F_{\ell}^{-1}\left( 1- \mathrm{FWER}^{-1}(\alpha)\right)
    \end{align*}
    as $n\to\infty$ for all $\ell\in\{1,...,R\}$,
    where
    \begin{align*}
        &\beta_n(\alpha) := \max\left\{ \beta\in\left\{0,\frac{1}{B(n)},...,\frac{B(n)-1}{B(n)}\right\} \mid \mathrm{FWER}_n(\beta)\leq \alpha\right \},\\ &\mathrm{FWER}_n(\beta) :=  1 - F_n\left( \left(F_{n,\ell}^{-1}(1-\beta)\right)_{\ell\in\{1,...,R\}}  \right)  
    \quad\text{ for all $\beta\in [0,1)$}
    \\\text{ and } \quad&\mathrm{FWER}:\R\to[0,1], \quad \mathrm{FWER}(\beta) := 1- F\left( (F_{\ell}^{-1}(1-\beta))_{\ell\in\{1,...,R\}} \right).
    \end{align*}
\end{lem}\begin{proof}[{Proof of Lemma~\ref{BSQuant}}]
    We proceed similarly as in the proof of Theorem~6 in the supplement of \cite{myRMST}. 
    First of all, we show that $ \beta_n(\alpha) \xrightarrow{P} \mathrm{FWER}^{-1}(\alpha)$  as $n\to\infty$.  It holds
    $$ \mathrm{FWER}_n^{-1}\left(\alpha + \frac{1}{B(n)}\right) \geq  \beta_n(\alpha) \geq \mathrm{FWER}_n^{-1}(\alpha) - \frac{1}{B(n)} .$$ Since $B(n)\to\infty$ as $n\to\infty$, it remains to show 
    \begin{align}\label{eq:FWERQuantil}
        \sup\limits_{r\in [\alpha, c]} \left|\mathrm{FWER}_n^{-1}(r) - \mathrm{FWER}^{-1}(r)\right| \xrightarrow{P} 0\quad \text{ as } n\to\infty
    \end{align} for some $c > \alpha$. Therefore, we aim to apply Lemma~S3 in the supplement of \cite{myRMST}. Note that $\mathrm{FWER}_n$ and $\mathrm{FWER}$ can be seen as distribution functions and $\mathrm{FWER}$ is continuous and strictly increasing on $[1-b,1-a]$. By Lemma~S4 in the supplement of \cite{myRMST}, we have 
    $$\sup\limits_{t_1,...,t_R\in\R} |F_n(t_1,...,t_R) - F(t_1,...,t_R)| \xrightarrow{P} 0 \quad \text{ as } n\to\infty, $$
    which can be shown by applying the subsequence criterion. Moreover, this implies the uniform convergence of the marginal distribution functions, that is
    \begin{align*}
        \sup\limits_{t\in\R} |F_{n,\ell}(t) - F_{\ell}(t)|
        &\leq \sup\limits_{t_1,...,t_R\in\R} |F_n(t_1,...,t_R) - F(t_1,...,t_R)| \xrightarrow{P} 0
    \end{align*} as $n\to\infty$ for all $\ell\in\{1,...,R\}$. Then, Lemma~S3 in the supplement of \cite{myRMST} and the subsequence criterion provide
    \begin{align}\label{eq:MarginalConv}
        \sup\limits_{r\in [p,q]} |F_{n,\ell}^{-1}(r) - F_{\ell}^{-1}(r)|
         \xrightarrow{P} 0
    \end{align} as $n\to\infty$ for all $\ell\in\{1,...,R\}$ and some $p \in (a,1-\alpha), q\in (1-\alpha/R,b)$. 
    Hence, we have
    \begin{align*}
       \mathrm{FWER}_n(\beta) \xrightarrow{P} \mathrm{FWER}(\beta) \quad \text{as $n\to\infty$ for all } \beta\in[1-q,1-p].  
    \end{align*} Applying Lemma~S3 in the supplement of \cite{myRMST} again and the subsequence criterion yield (\ref{eq:FWERQuantil}).
    \\
    Note that $\P\left(1-\beta_n(\alpha) \in [p,q]\right) \to 1$ as $n\to\infty$ holds by (\ref{eq:FWERQuantil}).
    Thus, (\ref{eq:MarginalConv}) completes the proof.
\end{proof}

\begin{remark}\label{RemarkLemma}
    If $\left(\mathbf{X}_n^{(b_1)},\mathbf{X}_n^{(b_2)}\right), b_1, b_2\in\mathbb N$ with $b_1\neq b_2$ are identically distributed for each $n\in\mathbb N$, respectively, and $\left(\mathbf{X}_n^{(1)},\mathbf{X}_n^{(2)}\right) \xrightarrow{d} \left(\mathbf{X}^{(1)},\mathbf{X}^{(2)}\right)$ as $n\to\infty$ for independent random variables $\mathbf{X}^{(1)},\mathbf{X}^{(2)}$ with continuous distribution function $F$,
    (\ref{eq:condition}) in Lemma~\ref{BSQuant} follows. This can be shown by applying Markov's inequality since 
    \begin{align*}
      &\E\left[ \left| F_n(t_1,...,t_R) - F(t_1,...,t_R) \right|^2 \right]
        \\& =  \left(  \E\left[ F_n^2(t_1,...,t_R)\right] -2 \E\left[F_n(t_1,...,t_R) \right]F(t_1,...,t_R) + F^2(t_1,...,t_R) \right)
        \\& =  \frac{1}{B(n)^2}\sum_{b_1,b_2=1}^{B(n)} \P\left( {X}_{n1}^{(b_1)}\leq t_1, ..., {X}_{nR}^{(b_1)} \leq t_R, {X}_{n1}^{(b_2)}\leq t_1, ..., {X}_{nR}^{(b_2)} \leq t_R\right) \\&\quad -2 \P\left({X}_{n1}^{(1)}\leq t_1, ..., {X}_{nR}^{(1)} \leq t_R\right)F(t_1,...,t_R) + F^2(t_1,...,t_R) 
        \\& \to  F^2(t_1,...,t_R) -2 F(t_1,...,t_R)F(t_1,...,t_R) + F^2(t_1,...,t_R) = 0
    \end{align*} as $n\to\infty$ holds for all $t_1,...,t_R\in\R$.
\end{remark}

\begin{lem}\label{ContandMonoton}
    Let $\mathbf{X} \sim GP_r(\mathbf{0}_r, \boldsymbol{\Sigma})$ denote a Gaussian process in $C^r(T)$ with covariance function $\boldsymbol{\Sigma}:\T \to \R^{r\times r}$, where $\boldsymbol{\Sigma}(t_0,t_0)$ is not the zero matrix for some $t_0\in\T$. Furthermore, let $F$ denote the distribution function of $\sup_{t\in\T} || \mathbf{X}(t) ||_2$. Then, $F$ is continuous on $\R$ and strictly increasing on $[0,\infty)$.
\end{lem}
\begin{proof}[{Proof of Lemma~\ref{ContandMonoton}}]
    We proceed similarly as in \cite{ContandMonoton}. By Lemma~1.1 in \cite{ContandMonoton}, it follows that $\P^{\mathbf{X}}$ is log-concave. Then, we define the set
    $$ A_u := \{ \mathbf{x} \in C^r(T) \mid \sup_{t\in\T} || \mathbf{x}(t) ||_2 \leq u\} \quad \text{for all $u > 0$}. $$ It holds $\lambda A_u + (1-\lambda) A_s \subset A_{\lambda u + (1-\lambda)s}$ for all $u,s > 0, \lambda  \in [0,1]$ since for arbitrary $\mathbf{y} \in A_u, \mathbf{z} \in A_s$, we have
    \begin{align*}
        \sup_{t\in\T} || \lambda \mathbf{y}(t) + (1-\lambda) \mathbf{z}(t) ||_2 &\leq  \lambda \sup_{t\in\T}  || \mathbf{y}(t)  ||_2+ (1-\lambda) \sup_{t\in\T} ||\mathbf{z}(t) ||_2 \\&\leq \lambda  u+ (1-\lambda)  s.
    \end{align*}
    By \cite{ledoux1991}, we have $F(u)>0$ for all $u>0$. Hence, it follows that $F:(0,\infty)\to(-\infty,0]$ is log-concave as in the proof of Lemma~1.2 in \cite{ContandMonoton}. Thus, as in Corollary~1.3 in \cite{ContandMonoton}, we get that $F$ is continuous on $\R$ and strictly increasing on $[0,\infty)$.
\end{proof}

\begin{remark}\label{RemContandMonoton}
    Under the notation and conditions as in Lemma~\ref{ContandMonoton}, we can also deduce that the distribution function of $\left(\sup_{t\in\T} || \mathbf{X}(t) ||_2\right)^2 = \sup_{t\in\T} \{ \mathbf{X}^{\top}(t) \mathbf{X}(t) \}$ is continuous on $\R$ and strictly increasing on $[0,\infty)$.
\end{remark}


\subsection{Proof of Theorem~1}

The following lemma ensures that $\mathit{H}\hat{\boldsymbol\eta}$ is asymptotically Gaussian.
\begin{lem}
\label{asy_CLT}
Under (A1), (A4), (A5) and the null hypothesis in (2), we have 
\begin{align*}
    \sqrt{n} (\mathit{H}\hat{\boldsymbol\eta} - \mathbf{c})  \xrightarrow{d}
    \mathbf{y} \sim GP_{r}\left(\mathbf{0}_{r}, \mathit{H}\mathit{\Lambda}\mathit{H}^{\top}\right) \quad\text{as $n\to\infty$}
\end{align*}
 in $C^r(\T)$.
\end{lem}
\begin{proof}[{Proof of Lemma~\ref{asy_CLT}}]
Let $i\in\{1,...,k\}$ be arbitrary but fixed.
Firstly, we apply Lemma~\ref{Conv} to show that $\frac{1}{\sqrt{n_i}} \sum_{j=1}^{n_i} \mathbf{v}_{ij}$ converges weakly to a centered Gaussian process with covariance function $\mathit{\Gamma}_i$ as $n_i\to\infty$  in $C^p(\T)$.
\\
By applying the central limit theorem, we get the convergence of the marginals
$$ \left(\frac{1}{\sqrt{n_i}} \sum_{j=1}^{n_i} \mathbf{v}_{ij}(t_1),..., \frac{1}{\sqrt{n_i}} \sum_{j=1}^{n_i} \mathbf{v}_{ij}(t_m)\right) \xrightarrow{d} \mathcal{N}_{mp}\left(\mathbf{0}_{mp},  (\mathit{\Gamma}_i(t_{j_1},t_{j_2}))_{j_1,j_2\in\{1,...,m\}}  \right)$$
as $n_i \to\infty$ for all $t_1,...,t_m\in\T, m\in\mathbb N$. Hence, it remains to show (\ref{eq:pequicont}) for  $\frac{1}{\sqrt{n_i}} \sum_{j=1}^{n_i} \mathbf{v}_{ij}$.
For $J \geq 2$ as in Assumption~1, it holds 
\begin{align*}
    \E\left[ \left|\left|\frac{1}{\sqrt{n_i}} \sum_{j=1}^{n_i} (\mathbf{v}_{ij}(t) - \mathbf{v}_{ij}(s))\right|\right|_2^J \right] \leq B_J \E\left[ \left|\left|\mathbf{v}_{i1}(t) - \mathbf{v}_{i1}(s)\right|\right|_2^J  \right] \leq B_J C (\rho(t,s))^J
\end{align*} for all $t,s\in\T$ by Lemma~\ref{MZinequ} and (A5).
Hence, Lemma~\ref{Conv} implies 
\begin{align*}
    \frac{1}{\sqrt{n_i}} \sum_{j=1}^{n_i} \mathbf{v}_{ij} \xrightarrow{d} GP_p(\mathbf{0}_p, \mathit{\Gamma}_i)
\end{align*} as $n\to\infty$ in $C^p(\T)$.
\\
Due to the independence of the samples, Slutsky's lemma and (A4), it follows
$$ \sqrt{n}(\widehat{\boldsymbol{\eta}}  - \boldsymbol{\eta}) = \left( \frac{\sqrt{n}}{\sqrt{n_i}}  \frac{1}{\sqrt{n_i}}\sum_{j=1}^{n_i} \mathbf{v}_{ij} \right)_{i\in\{1,...,k\}} \xrightarrow{d} GP_{pk} (\mathbf{0}_{pk}, \mathit{\Lambda}) $$ as $n\to\infty$ in $C^{pk}(\T)$.
Applying the continuous mapping theorem yields the statement of Lemma~\ref{asy_CLT}.
\end{proof}

Now, we state a lemma, which gives us the convergence of the covariance function estimators uniformly over $\{(t,t)\in\T^2 \mid t\in\T\}$ in probability.

\begin{lem}
\label{MUniformConvLemma}
Under (A1), (A2) and (A5), we have
$$  \sup_{t\in\T} \left|\left|\left|\widehat{\mathit{\Gamma}}_i(t,t) - \mathit{\Gamma}_i(t,t)\right|\right|\right| \xrightarrow{P} 0 \quad\text{as $n_i\to\infty$} $$
for all $i\in\{1,...,k\}.$ 
\end{lem}
\begin{proof}[{Proof of Lemma~\ref{MUniformConvLemma}}]
We proceed similarly to the proof of Lemma~S2 in the supplement of \cite{FANOVA} and aim to apply Theorem~2.1 in \cite{newey1991}. 
Let $i\in\{1,...,k\}$ be fixed but arbitrary.
The pointwise convergence
\begin{align}\label{eq:pointw}
    \hat{\mathit{\Gamma}}_i(t,t) \xrightarrow{P} {\mathit{\Gamma}}_i(t,t) \quad\text{as $n\to\infty$}
\end{align}
is well known for all $t\in\T$ under (A1), which can be proved by applying the weak law of large numbers.
Since $\T$ is compact and $\T\ni t\mapsto \mathit{\Gamma}_i(t,t)$ is continuous, see (R3) for details, it is also uniformly continuous and, thus, equicontinuous.
Hence, the stochastic equicontinuity of 
\mbox{$\T\ni t\mapsto \widehat{\mathit{\Gamma}}_i(t,t)$} remains to show. For that purpose, let $\varepsilon, \eta > 0$ and $\delta > 0$ such that 
\begin{align}
\label{eq:Mschranke}
    \E\left[\sup\limits_{\substack{\rho(t,s)<\delta}} |||\mathbf{v}_{i1}(t)\mathbf{v}^{\top}_{ij}(t)-\mathbf{v}_{i1}(s)\mathbf{v}^{\top}_{ij}(s)|||\right] \leq \frac{\eta\varepsilon}{6}
\end{align}
    holds for $j\in\{1,2\}$. By (R3), such a $\delta>0$ exists.
    Inserting the definition of the covariance function estimator (4) and applying the triangular inequality yields
    \begin{align*}
        \sup\limits_{\rho(t,s)<\delta} \left|\left|\left| \hat{\mathit{\Gamma}}_i(t,t) - \hat{\mathit{\Gamma}}_i(s,s) \right|\right|\right| 
        &\leq    \frac{1}{n_i-1}\sum_{j=1}^{n_i} \sup\limits_{\rho(t,s)<\delta} \left| \left| \left|\mathbf{v}_{ij}(t)\mathbf{v}^{\top}_{ij}(t)- \mathbf{v}_{ij}(s)\mathbf{v}^{\top}_{ij}(s)\right|\right|\right|
        \\&\quad + \frac{1}{(n_i-1)n_i}  \sum_{j_1,j_2=1}^{n_i}\sup\limits_{\rho(t,s)<\delta} \left| \left| \left|\mathbf{v}_{ij_1}(t)\mathbf{v}^{\top}_{ij_2}(t)  - \mathbf{v}_{ij_1}(s)\mathbf{v}^{\top}_{ij_2}(s)\right|\right|\right|.
    \end{align*}
By Markov's inequality, (A1) and (\ref{eq:Mschranke}), it follows
\begin{align*}
    \mathbb{P}\left( \sup\limits_{\rho(t,s)<\delta} \left|\left|\left| \hat{\mathit{\Gamma}}_i(t,t) - \hat{\mathit{\Gamma}}_i(s,s) \right|\right|\right|  > \varepsilon \right) &\leq \E\left[  \sup\limits_{\rho(t,s)<\delta} \left|\left|\left| \hat{\mathit{\Gamma}}_i(t,t) - \hat{\mathit{\Gamma}}_i(s,s) \right|\right|\right|   \right]/\varepsilon
   \\& \leq  \frac{n_i}{n_i-1}\frac{\eta}{6} +  \frac{n_i^2}{(n_i-1)n_i}\frac{\eta}{6} \to \frac{\eta}{3} < \eta
\end{align*}
 as $n \to \infty$.
The lemma is proved by applying Theorem~2.1 in \cite{newey1991}.
\end{proof}

By combining the two previous lemmas, we can now prove Theorem~1.

\begin{proof}[{Proof of Theorem~1}]
We proceed similarly as in the proof of Theorem~1 in the supplement of \cite{FANOVA}.
Lemma~\ref{MUniformConvLemma} provides that $\widehat{\mathit{\Lambda}}$ converges uniformly over \mbox{$\{(t,t)\in\T^2 \mid t\in\T\}$} in probability to $\mathit{\Lambda}$.
In the following, we also write $\widehat{\mathit{\Lambda}}(t)$ and ${\mathit{\Lambda}}(t)$ instead of $\widehat{\mathit{\Lambda}}(t,t)$ and ${\mathit{\Lambda}}(t,t)$, respectively, for all $t\in\T$.
Furthermore, Lemma~\ref{asy_CLT} ensures the convergence in distribution of $\sqrt{n} (\mathit{H}\widehat{\boldsymbol\eta} - \mathbf{c})$ to $\mathbf{y}$.\\
Let $\mathbb{D}$ denote the set of all continuous functions $ \widetilde{\mathit{\Lambda}}:\T\to\R^{kp\times kp}$ with 
$\widetilde{\mathit{\Lambda}}(t)$ symmetric and positive semidefinite for all $t\in\T$
and be equipped with the sup-norm.
Then, ${\mathit{\Lambda}}\in\mathbb{D}$ holds due to (R3). Moreover, by (A1),
$\widehat{\mathit{\Lambda}}$ takes values in $\mathbb{D}$.
\\
Furthermore, the map
\begin{align*}
    h: C^r(\T) \times \mathbb{D}
    \ni  (\widetilde{\mathbf{x}}, \widetilde{\mathit{\Lambda}}) \mapsto \sup_{t\in\T} \left\{ \widetilde{\mathbf{x}}^{\top}(t) \left(\mathit{H} \widetilde{\mathit{\Lambda}}(t) \mathit{H}^{\top} \right)^+ \widetilde{\mathbf{x}}(t) \right\} \in \R_0^+
\end{align*}
is continuous on \mbox{$C^r(\T) \times \{\mathit{\Lambda}\}$} under (A3).
This can be shown as in the proof of Theorem~1 in the supplement of \cite{FANOVA}.
Applying the continuous mapping theorem completes the proof.
\end{proof}

\subsection{Proofs of Section~2.3}


The following lemma gives an analogue of Lemma~\ref{asy_CLT} for the parametric bootstrap counterpart of the sample mean function.
\begin{lem}
\label{Para_CLT}
Under (A1), (A4) and (A5), we have 
\begin{align*}
    \sqrt{n} \mathit{H}\hat{\boldsymbol\eta}^{\mathcal{P}} \xrightarrow{d}
    \mathbf{y} \sim GP_{r}\left(\mathbf{0}_{r}, \mathit{H}\mathit{\Lambda}\mathit{H}^{\top}\right) \quad\text{as $n\to\infty$}
\end{align*} 
 in $C^{r}(\T)$.
\end{lem}
\begin{proof}[{Proof of Lemma~\ref{Para_CLT}}]
Let $i\in\{1,...,k\}$ be arbitrary but fixed. We aim to apply Lemma~\ref{Conv} to prove that $\frac{1}{\sqrt{n_i}} \sum_{j=1}^{n_i} \mathbf{x}_{ij}^{\mathcal{P}}$ converges weakly to a centered Gaussian process $\mathbf{z}$ with covariance function $\mathit{\Gamma}_i$ in $C^p(\T)$.\\
In order to see that $\frac{1}{\sqrt{n_i}} \sum_{j=1}^{n_i} \mathbf{x}_{ij}^{\mathcal{P}}$ takes values in $C^p(\T)$, note that \begin{align}\label{eq:representation}
        \mathbf{x}_{ij}^{\mathcal{P}} \overset{d}{=} \frac{1}{\sqrt{n_i-1}} \sum_{k=1}^{n_i} Y_{ijk} (\mathbf{v}_{ik}-\overline{\mathbf{v}}_i)
    \end{align} holds for i.i.d.~standard normal distributed variables $Y_{ijk}, i\in\{1,...,k\}, j\in\{1,...,n_i\}, k\in\{1,...,n_i\},$ with $\overline{\mathbf{v}}_i:= \frac{1}{n_i} \sum_{j=1}^{n_i} \mathbf{v}_{ij}.$
    For proving the convergence of the marginals, let $t_1,...,t_m\in\T, m\in\mathbb N$ be arbitrary.
    Note that $\left(\widehat{\mathit{\Gamma}}_i(t_{j_1},t_{j_2})\right)_{j_1,j_2\in\{1,...,m\}}$ is converging in probability to $\left({\mathit{\Gamma}}_i(t_{j_1},t_{j_2})\right)_{j_1,j_2\in\{1,...,m\}}$ as $n\to\infty$, see (\ref{eq:pointw}). Hence, 
    \begin{align*}
        \left( \frac{1}{\sqrt{n_i}} \sum_{j=1}^{n_i} (\mathbf{x}_{ij}^{\mathcal{P}}(t_1))^{\top},...,\frac{1}{\sqrt{n_i}} \sum_{j=1}^{n_i} (\mathbf{x}_{ij}^{\mathcal{P}}(t_m))^{\top} \right)^{\top} &\overset{d}{=} \left(\widehat{\mathit{\Gamma}}_i(t_{j_1},t_{j_2})\right)_{j_1,j_2\in\{1,...,m\}}^{1/2} \widetilde{\mathbf{Z}} \\&\xrightarrow{d} \left( \mathbf{z}^{\top}(t_1),...,\mathbf{z}^{\top}(t_m) \right)^{\top}
    \end{align*}
    as $n\to\infty$ for $\widetilde{\mathbf{Z}}\sim\mathcal{N}_{pm}(\mathbf{0}_{pm},\mathit{I}_{pm})$ independent of the data \begin{align}\label{eq:funcData}
    (\mathbf{x}_{i1},\mathbf{x}_{i2},...)_{i\in\{1,...,k\}}
\end{align}  and $\mathbf{z}\sim GP_p(\mathbf{0}_p,\mathit{\Gamma}_i).$ Thus,
    the convergence of the marginals follows.\\
    It remains to show (\ref{eq:pequicont}) for $\frac{1}{\sqrt{n_i}} \sum_{j=1}^{n_i} \mathbf{x}_{ij}^{\mathcal{P}}$. Therefore, let $J\geq 2$ be as in Assumption~1, $\mathbf{Z}\sim\mathcal{N}_p(\mathbf{0}_p,\mathit{I}_p)$ independent of the data (\ref{eq:funcData}) and $A_{J} := \E\left[ ||\mathbf{Z}||_2^J  \right] < \infty$ a constant depending on $J$ and $p$ only. 
    Note that $$\frac{1}{\sqrt{n_i}} \sum_{j=1}^{n_i} \mathbf{x}_{ij}^{\mathcal{P}}(t) - \frac{1}{\sqrt{n_i}} \sum_{j=1}^{n_i}\mathbf{x}_{ij}^{\mathcal{P}}(s) \overset{d}{=} \left( \widehat{\mathit{\Gamma}}_i(t,t) + \widehat{\mathit{\Gamma}}_i(s,s) -\widehat{\mathit{\Gamma}}_i(t,s)-\widehat{\mathit{\Gamma}}_i(s,t) \right)^{1/2}\mathbf{Z} $$ holds for all $t,s\in\T$.
    Thus, we have
    \begin{align*}
        &\E\left[ \left|\left| \frac{1}{\sqrt{n_i}} \sum_{j=1}^{n_i} \mathbf{x}_{ij}^{\mathcal{P}}(t) - \frac{1}{\sqrt{n_i}} \sum_{j=1}^{n_i}\mathbf{x}_{ij}^{\mathcal{P}}(s) \right|\right|_2^J \right] 
        \\&=\E\left[\left|\left| \left( \widehat{\mathit{\Gamma}}_i(t,t) + \widehat{\mathit{\Gamma}}_i(s,s) -\widehat{\mathit{\Gamma}}_i(t,s)-\widehat{\mathit{\Gamma}}_i(s,t) \right)^{1/2}\mathbf{Z} \right|\right|_2^J \right] 
        \\&\leq 
        A_{J} \E\left[\left|\left|\left| \widehat{\mathit{\Gamma}}_i(t,t) + \widehat{\mathit{\Gamma}}_i(s,s) -\widehat{\mathit{\Gamma}}_i(t,s)-\widehat{\mathit{\Gamma}}_i(s,t) \right|\right| \right|^{J/2} \right]
    \end{align*}
    for all $t,s\in\T$.
    Let us write $\widehat{\mathit{\Gamma}}_i(t,t) + \widehat{\mathit{\Gamma}}_i(s,s) -\widehat{\mathit{\Gamma}}_i(t,s)-\widehat{\mathit{\Gamma}}_i(s,t)$ as
    \begin{align*}
         \frac{1}{n_i-1} \sum_{j=1}^{n_i} (\mathbf{v}_{ij}(t) - \mathbf{v}_{ij}(s))(\mathbf{v}_{ij}(t) - \mathbf{v}_{ij}(s))^{\top} - \frac{n_i}{n_i-1} (\overline{\mathbf{v}}_i(t) - \overline{\mathbf{v}}_i(s))(\overline{\mathbf{v}}_i(t) - \overline{\mathbf{v}}_i(s))^{\top}
    \end{align*} with $\overline{\mathbf{v}}_i := \frac{1}{n_i} \sum_{j=1}^{n_i} \mathbf{v}_{ij}$.
    For the first part, Lemma~\ref{MZinequ} and (A5) provide
    \begin{align*}
        \E\left[\left|\left|\left| \frac{1}{n_i-1} \sum_{j=1}^{n_i} (\mathbf{v}_{ij}(t) - \mathbf{v}_{ij}(s))(\mathbf{v}_{ij}(t) - \mathbf{v}_{ij}(s))^{\top} \right|\right| \right|^{J/2} \right] 
        &\leq 
        2 \E\left[\left|\left| \mathbf{v}_{i1}(t) - \mathbf{v}_{i1}(s) \right|\right|_2^{J} \right]
        \\&\leq 2 C (\rho(t,s))^J
    \end{align*} for all $t,s\in\T$.
    For the second part, Lemma~\ref{MZinequ} and (A5) yields
    \begin{align*}
        \E\left[\left|\left|\left| \frac{n_i}{n_i-1} (\overline{\mathbf{v}}_i(t) - \overline{\mathbf{v}}_i(s))(\overline{\mathbf{v}}_i(t) - \overline{\mathbf{v}}_i(s))^{\top} \right|\right| \right|^{J/2} \right] 
        &= \left(\frac{n_i}{n_i-1}\right)^{J/2} \E\left[||\overline{\mathbf{v}}_i(t) - \overline{\mathbf{v}}_i(s)||_2^{J} \right] 
       \\&\leq 
        \left({n_i-1}\right)^{-J/2} B_J \E\left[\left|\left| \mathbf{v}_{i1}(t) - \mathbf{v}_{i1}(s) \right|\right|_2^{J} \right]
        \\&\leq B_J C (\rho(t,s))^J
    \end{align*} for all $t,s\in\T$.
    Thus, it follows
    \begin{align}\label{eq:absJmoment}
        \E\left[ \left|\left| \frac{1}{\sqrt{n_i}} \sum_{j=1}^{n_i} \mathbf{x}_{ij}^{\mathcal{P}}(t) - \frac{1}{\sqrt{n_i}} \sum_{j=1}^{n_i}\mathbf{x}_{ij}^{\mathcal{P}}(s) \right|\right|_2^J \right] 
        \leq A_{J} 2^{J/2} (2 + B_J) C (\rho(t,s))^J
    \end{align} for all $t,s\in\T$
    by the triangular inequality. Hence, (\ref{eq:pequicont}) is fulfilled and Lemma~\ref{Conv} completes the proof of
    \begin{align}\label{eq:onlyEtai}
        \sqrt{n_i}\widehat{\boldsymbol{\eta}}_i^{\mathcal{P}} = \frac{1}{\sqrt{n_i}} \sum_{j=1}^{n_i} \mathbf{x}_{ij}^{\mathcal{P}} \xrightarrow{d} GP_p(\mathbf{0}_p,\mathit{\Gamma}_i)
    \end{align}
    as $n\to\infty$ in $C^p(\T)$ for all $i\in\{1,...,k\}$.
    \\
Due to the independence of the samples, Slutsky's lemma and (A4), it follows
$$ \sqrt{n}\widehat{\boldsymbol{\eta}}^{\mathcal{P}} = \left( \frac{\sqrt{n}}{\sqrt{n_i}}  \frac{1}{\sqrt{n_i}}\sum_{j=1}^{n_i} \mathbf{x}_{ij}^{\mathcal{P}} \right)_{i\in\{1,...,k\}} \xrightarrow{d} GP_{pk} (\mathbf{0}_{pk}, \mathit{\Lambda}) $$ as $n\to\infty$ in $C^{pk}(\T)$.
Applying the continuous mapping theorem yields the statement of the lemma.
\end{proof}

The next lemma gives us the uniform consistency of the parametric bootstrap covariance function estimator.

\begin{lem}\label{MUnifCovPara}
    Under (A1), (A2) and (A5), we have
    $$ \sup_{t\in\T} \left|\left|\left|\widehat{\mathit{\Gamma}}_i^{\mathcal{P}}(t,t) - \mathit{\Gamma}_i(t,t)\right|\right|\right| \xrightarrow{P} 0 \quad\text{as $n_i\to\infty$} $$
    for all $i\in\{1,...,k\}$.
\end{lem}
\begin{proof}[{Proof of Lemma~\ref{MUnifCovPara}}]
    We proceed similarly as in Lemma~\ref{MUniformConvLemma}. Let $i\in\{1,...,k\}$ be arbitrary but fixed. We write
    \begin{align*}
        \widehat{\mathit{\Gamma}}_i^{\mathcal{P}}(t,t) = \frac{1}{n_i-1} \sum_{j=1}^{n_i} \mathbf{x}_{ij}^{\mathcal{P}}(t)(\mathbf{x}_{ij}^{\mathcal{P}}(t))^{\top} - \frac{n_i}{n_i - 1} \widehat{\boldsymbol{\eta}}_i^{\mathcal{P}}(t) (\widehat{\boldsymbol{\eta}}_i^{\mathcal{P}}(t))^{\top}
    \end{align*} for all $t\in\T.$ Due to (\ref{eq:onlyEtai}), $\widehat{\boldsymbol{\eta}}_i^{\mathcal{P}}$ vanishes uniformly in probability as $n_i\to\infty$. Hence, it follows
    \begin{align*}
        \sup_{t\in\T} \left|\left|\left| \frac{n_i}{n_i - 1} \widehat{\boldsymbol{\eta}}_i^{\mathcal{P}}(t) (\widehat{\boldsymbol{\eta}}_i^{\mathcal{P}}(t))^{\top} \right|\right|\right| = \frac{n_i}{n_i - 1} \sup_{t\in\T} \left|\left|  \widehat{\boldsymbol{\eta}}_i^{\mathcal{P}}(t) \right|\right|_2^2 \xrightarrow{P} 0 \quad\text{as $n_i\to\infty$}.
    \end{align*}
    Thus, the first term remains to be investigated. We aim to apply Theorem~2.1 in \cite{newey1991}. 
    For showing the pointwise convergence, let $q_1,q_2\in\{1,...,p\}$ be arbitrary. 
    Then, we consider the conditional expectation and variance of $\frac{1}{n_i-1} \sum_{j=1}^{n_i} \mathbf{x}_{ij,q_1}^{\mathcal{P}}(t)\mathbf{x}_{ij,q_2}^{\mathcal{P}}(t)$ for all $t\in\T$.
    In the following, we denote the conditional expectation given the data (\ref{eq:funcData}) by $\E^*$ and the conditional variance given the data (\ref{eq:funcData}) by $\mathbb Var^*$.
    We have that 
    \begin{align*}
        \E^*\left[ \frac{1}{n_i-1} \sum_{j=1}^{n_i} \mathbf{x}_{ij,q_1}^{\mathcal{P}}(t)\mathbf{x}_{ij,q_2}^{\mathcal{P}}(t) \right] = \frac{n_i}{n_i -1 } \widehat{\mathit{\Gamma}}_{i,q_1q_2}(t,t) \xrightarrow{P} {\mathit{\Gamma}}_{i,q_1q_2}(t,t)
    \end{align*} as $n_i \to\infty$ for all $t\in\T$.
    Additionally, it holds
    \begin{align*}
        &\mathbb Var^*\left(  \frac{1}{n_i-1} \sum_{j=1}^{n_i} \mathbf{x}_{ij,q_1}^{\mathcal{P}}(t)\mathbf{x}_{ij,q_2}^{\mathcal{P}}(t) \right)
        \\&=  \frac{n_i}{(n_i-1)^2} \mathbb Var^*\left(    \mathbf{x}_{i1,q_1}^{\mathcal{P}}(t)\mathbf{x}_{i1,q_2}^{\mathcal{P}}(t) \right)
        \\&=\frac{n_i}{(n_i-1)^2} \left\{\E^*\left[\left(   \mathbf{x}_{i1,q_1}^{\mathcal{P}}(t)\mathbf{x}_{i1,q_2}^{\mathcal{P}}(t) \right)^2\right] -  ( \widehat{\mathit{\Gamma}}_{i,q_1q_2}(t,t) )^2\right\}
        \\&\leq\frac{n_i}{(n_i-1)^2} \left\{\sqrt{\E^*\left[(   \mathbf{x}_{i1,q_1}^{\mathcal{P}}(t))^4\right]\E^*\left[(\mathbf{x}_{i1,q_2}^{\mathcal{P}}(t))^4\right]} -  ( \widehat{\mathit{\Gamma}}_{i,q_1q_2}(t,t) )^2\right\}
        \\&=\frac{n_i}{(n_i-1)^2} \left\{3\widehat{\mathit{\Gamma}}_{i,q_1q_1}(t,t)\widehat{\mathit{\Gamma}}_{i,q_2q_2}(t,t) -  ( \widehat{\mathit{\Gamma}}_{i,q_1q_2}(t,t) )^2\right\}
        \\&\xrightarrow{P} 0
    \end{align*} as $n_i \to\infty$ for all $t\in\T$ by the Cauchy-Schwarz inequality. Thus, we get for all $\varepsilon > 0$
    \begin{align*}
        \P\left( \left| \frac{1}{n_i-1} \sum_{j=1}^{n_i} \mathbf{x}_{ij,q_1}^{\mathcal{P}}(t)\mathbf{x}_{ij,q_2}^{\mathcal{P}}(t) - {\mathit{\Gamma}}_{i,q_1q_2}(t,t) \right| > \varepsilon  \:\bigg \vert\: (\mathbf{x}_{i1},\mathbf{x}_{i2},...)_{i\in\{1,...,k\}} \right) \xrightarrow{P} 0
    \end{align*} as $n_i\to\infty$ for all $t\in\T$.
    By the dominated convergence theorem, it follows
    \begin{align*}
        &\P\left( \left| \frac{1}{n_i-1} \sum_{j=1}^{n_i} \mathbf{x}_{ij,q_1}^{\mathcal{P}}(t)\mathbf{x}_{ij,q_2}^{\mathcal{P}}(t) - {\mathit{\Gamma}}_{i,q_1q_2}(t,t) \right| > \varepsilon  \right)
        \\& =
        \E\left[ \P\left( \left| \frac{1}{n_i-1} \sum_{j=1}^{n_i} \mathbf{x}_{ij,q_1}^{\mathcal{P}}(t)\mathbf{x}_{ij,q_2}^{\mathcal{P}}(t) - {\mathit{\Gamma}}_{i,q_1q_2}(t,t) \right| > \varepsilon  \:\bigg \vert\: (\mathbf{x}_{i1},\mathbf{x}_{i2},...)_{i\in\{1,...,k\}} \right) \right]
        \to 0
    \end{align*} as $n_i\to\infty$ for all $\varepsilon > 0, t\in\T$
    and, hence, the pointwise convergence
    \begin{align*}
        \frac{1}{n_i-1} \sum_{j=1}^{n_i} \mathbf{x}_{ij,q_1}^{\mathcal{P}}(t)\mathbf{x}_{ij,q_2}^{\mathcal{P}}(t) \xrightarrow{P} {\mathit{\Gamma}}_{i,q_1,q_2}(t,t)
    \end{align*} as $n_i\to\infty$ for all $t\in\T$.
    Since $q_1,q_2\in\{1,...,p\}$ were arbitrary, we also have
    the pointwise convergence
    \begin{align*}
        \frac{1}{n_i-1} \sum_{j=1}^{n_i} \mathbf{x}_{ij}^{\mathcal{P}}(t)(\mathbf{x}_{ij}^{\mathcal{P}}(t))^{\top} \xrightarrow{P} {\mathit{\Gamma}}_{i}(t,t)
    \end{align*} as $n_i\to\infty$ for all $t\in\T$.\\
    It remains to show the stochastic equicontinuity, i.e., that 
    \begin{align*}
       \lim\limits_{\delta \searrow 0} \limsup_{n_i\to\infty} \P\left( \sup\limits_{\rho(t,s)<\delta}\left|\left|\left|  \frac{1}{n_i-1} \sum_{j=1}^{n_i} \left\{\mathbf{x}_{ij}^{\mathcal{P}}(t)(\mathbf{x}_{ij}^{\mathcal{P}}(t))^{\top} -  \mathbf{x}_{ij}^{\mathcal{P}}(s)(\mathbf{x}_{ij}^{\mathcal{P}}(s))^{\top}\right\}  \right|\right|\right| > \varepsilon \right) = 0
    \end{align*} holds for all $\varepsilon > 0$.
    By applying Markov's inequality, we get the upper bound
    \begin{align*}
       \varepsilon^{-J/2} \lim\limits_{\delta \searrow 0} \limsup_{n_i\to\infty} \E\left[ \sup\limits_{\rho(t,s)<\delta}\left|\left|\left|  \frac{1}{n_i-1} \sum_{j=1}^{n_i} \left\{\mathbf{x}_{ij}^{\mathcal{P}}(t)(\mathbf{x}_{ij}^{\mathcal{P}}(t))^{\top} -  \mathbf{x}_{ij}^{\mathcal{P}}(s)(\mathbf{x}_{ij}^{\mathcal{P}}(s))^{\top}\right\}  \right|\right|\right|^{J/2} \right]
    \end{align*} for all $\varepsilon > 0$.
    We aim to bound this term by applying Theorem~2.2.4 in \cite{vaartWellner1996} with $\psi(x)=x^{J/2}$ for all $x\geq 0$ and $d=\rho$. Note that $\left(\frac{1}{n_i-1} \sum_{j=1}^{n_i} \mathbf{x}_{ij}^{\mathcal{P}}(t)(\mathbf{x}_{ij}^{\mathcal{P}}(t))^{\top}\right)_{t\in\T}$ is separable since it takes values in the continuous functions  under (A1). In order to see this, consider the representation in (\ref{eq:representation}). \\
    Moreover, one can show
    \begin{align}\label{eq:together}\begin{split}
        &\E\left[ \left|\left|\left|  \frac{1}{n_i-1} \sum_{j=1}^{n_i} \left\{\mathbf{x}_{ij}^{\mathcal{P}}(t)(\mathbf{x}_{ij}^{\mathcal{P}}(t))^{\top} -  \mathbf{x}_{ij}^{\mathcal{P}}(s)(\mathbf{x}_{ij}^{\mathcal{P}}(s))^{\top}\right\}  \right|\right|\right|^{J/2} \right] 
        \\&\leq (n_i-1)^{-J/2} n_i^{J/2 - 1} \E\left[    \sum_{j=1}^{n_i} \left|\left|\left|  \mathbf{x}_{ij}^{\mathcal{P}}(t)(\mathbf{x}_{ij}^{\mathcal{P}}(t))^{\top} -  \mathbf{x}_{ij}^{\mathcal{P}}(s)(\mathbf{x}_{ij}^{\mathcal{P}}(s))^{\top} \right|\right|\right|^{J/2} \right]
        \\&\leq 2^{J/2}\E\left[     \left|\left|\left|  \mathbf{x}_{i1}^{\mathcal{P}}(t)(\mathbf{x}_{i1}^{\mathcal{P}}(t))^{\top} -  \mathbf{x}_{i1}^{\mathcal{P}}(s)(\mathbf{x}_{i1}^{\mathcal{P}}(s))^{\top} \right|\right|\right|^{J/2} \right]
        \\&\leq 
        2^{J+1}\sqrt{ \E\left[  \left|\left|  \mathbf{x}_{i1}^{\mathcal{P}}(t) \right|\right|_2^{J} + \left|\left|  \mathbf{x}_{i1}^{\mathcal{P}}(s) \right|\right|_2^{J} \right] \E\left[\left|\left| \mathbf{x}_{i1}^{\mathcal{P}}(t) -  \mathbf{x}_{i1}^{\mathcal{P}}(s) \right|\right|_2^{J} \right] }
    \end{split}\end{align} for all $t,s\in\T$ by applying the triangular inequality and norm equivalence for p-norms 
    in the first step, $n_i\geq 2$ and (A1) in the second step and the triangular inequality and the Cauchy-Schwarz inequality in the last step.
    Since $$\mathbf{x}_{i1}^{\mathcal{P}}(t) -  \mathbf{x}_{i1}^{\mathcal{P}}(s) \overset{d}{=} \frac{1}{\sqrt{n_i}} \sum_{j=1}^{n_i} \mathbf{x}_{ij}^{\mathcal{P}}(t) - \frac{1}{\sqrt{n_i}} \sum_{j=1}^{n_i}\mathbf{x}_{ij}^{\mathcal{P}}(s)$$ holds, we already showed
    \begin{align}\label{eq:ExtxsJ}
        \E\left[ \left|\left| \mathbf{x}_{i1}^{\mathcal{P}}(t) -  \mathbf{x}_{i1}^{\mathcal{P}}(s) \right|\right|_2^J \right] 
        \leq A_J 2^{J/2} (2 + B_J) C (\rho(t,s))^J
    \end{align} for all $t,s\in\T$ in (\ref{eq:absJmoment}). Hence, the first factor remains to consider.
    Therefore, we write $\mathbf{x}_{i1}^{\mathcal{P}}(t) \overset{d}{=} (\widehat{\mathit{\Gamma}}_i(t,t))^{1/2} \mathbf{Z}$ for $t\in\T$, where $\mathbf{Z}\sim\mathcal{N}_p(\mathbf{0}_p,\mathit{I}_p)$ independent of the data (\ref{eq:funcData}) and $A_J := \E\left[ ||\mathbf{Z}||_2^J  \right] < \infty$. Then, it holds
    \begin{align*}
        \E\left[  \left|\left|  \mathbf{x}_{i1}^{\mathcal{P}}(t) \right|\right|_2^{J}\right] &=  \E\left[  \left|\left|  (\widehat{\mathit{\Gamma}}_i(t,t))^{1/2} \mathbf{Z} \right|\right|_2^{J}\right]
        \leq \E\left[  \left|\left|\left| (\widehat{\mathit{\Gamma}}_i(t,t))^{1/2} \right|\right|\right|^J || \mathbf{Z} ||_2^{J}\right]
        = A_J \E\left[  \left|\left|\left| \widehat{\mathit{\Gamma}}_i(t,t) \right|\right|\right|^{J/2} \right]
    \end{align*} for all $t\in\T$. Inserting
    \begin{align*}
        \widehat{\mathit{\Gamma}}_i(t,t) = \frac{1}{n_i-1} \sum_{j=1}^{n_i} \mathbf{v}_{ij}(t)(\mathbf{v}_{ij}(t))^{\top}  - \frac{n_i}{n_i-1} \overline{\mathbf{v}}_i(t)( \overline{\mathbf{v}}_i(t))^{\top}
    \end{align*} results in 
    \begin{align*}
        &\E\left[  \left|\left|  \mathbf{x}_{i1}^{\mathcal{P}}(t) \right|\right|_2^{J}\right] \\&\leq A_J 2^{J/2} \left(\E\left[  \left|\left|\left| \frac{1}{n_i-1} \sum_{j=1}^{n_i} \mathbf{v}_{ij}(t)(\mathbf{v}_{ij}(t))^{\top}  \right|\right|\right|^{J/2} \right] + \E\left[  \left|\left|\left|\frac{n_i}{n_i-1} \overline{\mathbf{v}}_i(t)( \overline{\mathbf{v}}_i(t))^{\top} \right|\right|\right|^{J/2} \right]\right)
        \\&\leq A_J 2^{J/2} \left\{(n_i-1)^{-J/2}\E\left[\left( \sum_{j=1}^{n_i} \left|\left|\left|  \mathbf{v}_{ij}(t)(\mathbf{v}_{ij}(t))^{\top}  \right|\right|\right|\right)^{J/2} \right] + \left(\frac{n_i}{n_i-1}\right)^{J/2}\E\left[  \left|\left| \overline{\mathbf{v}}_i(t)\right|\right|_2^{J} \right]\right\}
    \end{align*} for all $t\in\T$. Thus, Lemma~\ref{MZinequ} implies
    \begin{align}\label{eq:ExtJ}\begin{split}
        \E\left[  \left|\left|  \mathbf{x}_{i1}^{\mathcal{P}}(t) \right|\right|_2^{J}\right] &\leq  A_J 2^{J/2} \left\{\left(\frac{n_i}{n_i-1}\right)^{J/2}\E\left[\left|\left|  \mathbf{v}_{i1}(t)\right|\right|_2^{J} \right] + \left(\frac{n_i}{n_i-1}\right)^{J/2}n_i^{-J/2}B_J \E\left[  \left|\left| {\mathbf{v}}_{i1}(t)\right|\right|_2^{J} \right]\right\}
        \\&\leq A_J2^{J/2}(2^{J/2}+B_J) \E\left[ \sup\limits_{s\in\T} \left|\left| {\mathbf{v}}_{i1}(s)\right|\right|_2^{J} \right]
    \end{split}\end{align} for all $t\in\T$. By (R2), we have $D_J:=\E\left[ \sup\limits_{s\in\T} \left|\left| {\mathbf{v}}_{i1}(s)\right|\right|_2^{J} \right] <\infty$.
    Inserting (\ref{eq:ExtxsJ}) and (\ref{eq:ExtJ}) in (\ref{eq:together}) 
    yields
    \begin{align*}
        &\E\left[ \left|\left|\left|  \frac{1}{n_i-1} \sum_{j=1}^{n_i} \left\{\mathbf{x}_{ij}^{\mathcal{P}}(t)(\mathbf{x}_{ij}^{\mathcal{P}}(t))^{\top} -  \mathbf{x}_{ij}^{\mathcal{P}}(s)(\mathbf{x}_{ij}^{\mathcal{P}}(s))^{\top}\right\}  \right|\right|\right|^{J/2} \right] 
        \leq \widetilde{C}_J (\rho(t,s))^{J/2} 
    \end{align*} for all $t,s\in\T$ with $$\widetilde{C}_J = 2^{3(J+1)/2} A_J\sqrt{(2^{J/2}+B_J)   (2 + B_J)  C D_J } \geq 0$$ depending on $J, C$ and $D_J$ only.
    Thus, Theorem~2.2.4 in \cite{vaartWellner1996} provides that there exists a constant $K$ depending on $J, C$ and $D_J$ only such that
    \begin{align*}
        &\E\left[ \sup\limits_{\rho(t,s) \leq \delta } \left|\left|\left|  \frac{1}{n_i-1} \sum_{j=1}^{n_i} \left\{\mathbf{x}_{ij}^{\mathcal{P}}(t)(\mathbf{x}_{ij}^{\mathcal{P}}(t))^{\top} -  \mathbf{x}_{ij}^{\mathcal{P}}(s)(\mathbf{x}_{ij}^{\mathcal{P}}(s))^{\top}\right\}  \right|\right|\right|^{J/2} \right] \\& \leq K \left\{  \int_0^\nu (D(\varepsilon,\rho))^{2/J} \;\mathrm{d}\varepsilon + \delta (D(\nu,\rho))^{4/J} \right\}^{J/2}
    \end{align*} holds  for all $\nu,\delta > 0$.
    Since $\nu$ can be chosen arbitrarily small, we have
$$\lim\limits_{\delta\searrow 0}\limsup\limits_{n_i\to\infty} \E\left[ \sup\limits_{\rho(t,s) \leq \delta } \left|\left|\left|  \frac{1}{n_i-1} \sum_{j=1}^{n_i} \left\{\mathbf{x}_{ij}^{\mathcal{P}}(t)(\mathbf{x}_{ij}^{\mathcal{P}}(t))^{\top} -  \mathbf{x}_{ij}^{\mathcal{P}}(s)(\mathbf{x}_{ij}^{\mathcal{P}}(s))^{\top}\right\}  \right|\right|\right|^{J/2} \right] = 0 $$ due to (3). Hence, the stochastic equicontinuity follows and Theorem~2.1 in \cite{newey1991} completes the proof.
\end{proof}

Theorem~2 can now be proved by using Lemma~\ref{Para_CLT} and \ref{MUnifCovPara}.

\begin{proof}[{Proof of Theorem~2}]
Here, we proceed similarly as in the proof of Theorem~1.
Lemma~\ref{MUnifCovPara} provides that $\widehat{\mathit{\Lambda}}^{\mathcal{P}}$ converges uniformly over \mbox{$\{(t,t)\in\T^2 \mid t\in\T\}$} in probability to $\mathit{\Lambda}$.
Furthermore, Lemma~\ref{Para_CLT} ensures the convergence in distribution of $\sqrt{n} \mathit{H}\widehat{\boldsymbol\eta}^{\mathcal{P}}$ to $\mathbf{y}$ in $C^r(\T)$.\\
Let $\mathbb{D}$ denote the set of all continuous functions $ \widetilde{\mathit{\Lambda}}:\T\to\R^{kp\times kp}$ with 
$\widetilde{\mathit{\Lambda}}(t)$ symmetric and positive semidefinite for all $t\in\T$
and be equipped with the sup-norm.
Then, ${\mathit{\Lambda}}\in\mathbb{D}$ holds due to (R3). Moreover, by (A1),
$\widehat{\mathit{\Lambda}}^{\mathcal{P}}$ takes values in $\mathbb{D}$, which can be seen through the representation in (\ref{eq:representation}).
\\
Furthermore, the map
$h$ in the proof of Theorem~1
is continuous on \mbox{$C^r(\T) \times \{\mathit{\Lambda}\}$} under (A3).
Thus, applying the continuous mapping theorem completes the proof.
 \end{proof}

\begin{proof}[{Proof of Proposition~1}]
We aim to show the assumtions of Lemma~\ref{BSQuant} with $\mathbf{X}_n^{(b)} := T_{n,b}^{\mathcal{P}}(\mathit{H})$ for $b\in\mathbb N$, $R=1$ and $F$ the distribution function of $T:= \sup_{t\in\T} \{ (\mathbf{y}(t))^{\top}(\mathit{H}\mathit{\Lambda}(t,t)\mathit{H}^{\top})^+\mathbf{y}(t) \}$, where $\mathbf{y}\sim GP_r(\mathbf{0}_r,\mathit{H}\mathit{\Lambda}\mathit{H}^{\top})$. By setting $\mathbf{X}(t) := \left((\mathit{H}\mathit{\Lambda}(t,t)\mathit{H}^{\top})^+\right)^{1/2}\mathbf{y}(t)$ for all $t\in\T$,  $F$ is continuous and strictly increasing on $[0,\infty)$ by Remark~\ref{RemContandMonoton}.
Now, we verify the conditions in Remark~\ref{RemarkLemma}. The identical distribution holds due to the definition of the random variables. 
In order to prove the convergence in distribution, let
    $$ \left(\mathbf{x}_{i1}^{\mathcal{P},(b)},...,\mathbf{x}_{in_i}^{\mathcal{P},(b)}\right)_{i\in\{1,...,k\}}, \quad b\in\{1,2\}, $$
    be independent copies of  $(\mathbf{x}_{i1}^{\mathcal{P}},...,\mathbf{x}_{in_i}^{\mathcal{P}})_{i\in\{1,...,k\}}$ conditionally on the data (\ref{eq:funcData}).
    Moreover, let
    $\widehat{\boldsymbol{\eta}}^{\mathcal{P},(b)}$ and $\widehat{\mathit{\Lambda}}^{\mathcal{P},(b)}$ denote the corresponding parametric bootstrap counterparts of $\widehat{\boldsymbol{\eta}}$ and $\widehat{\mathit{\Lambda}}$ based on the data $ \left(\mathbf{x}_{i1}^{\mathcal{P},(b)},...,\mathbf{x}_{in_i}^{\mathcal{P},(b)}\right)_{i\in\{1,...,k\}}$ for all $b\in\{1,2\}$.
    Firstly, we aim to show that 
    \begin{align}\label{eq:jointConv}
        \left(\sqrt{n}\mathit{H}\widehat{\boldsymbol{\eta}}^{\mathcal{P},(1)}, \sqrt{n}\mathit{H}\widehat{\boldsymbol{\eta}}^{\mathcal{P},(2)}\right) \xrightarrow{d} (\mathbf{y}^{(1)},\mathbf{y}^{(2)}) 
    \end{align} as $n\to\infty$ in $C^{2r}(\T)$, where $\mathbf{y}^{(1)}, \mathbf{y}^{(2)}$ are independent copies of $\mathbf{y}$. The asymptotic tightness of $\sqrt{n}\mathit{H}\widehat{\boldsymbol{\eta}}^{\mathcal{P},(1)}$, $ \sqrt{n}\mathit{H}\widehat{\boldsymbol{\eta}}^{\mathcal{P},(2)}$ follows easily by Theorem~1.5.7 in \cite{vaartWellner1996} due to the uniform $\rho$-equicontinuity of the processes as shown in the proofs of Lemma~\ref{Conv} and~\ref{Para_CLT}. Thus, the vector
    $$ \left(\sqrt{n}\mathit{H}\widehat{\boldsymbol{\eta}}^{\mathcal{P},(1)},  \sqrt{n}\mathit{H}\widehat{\boldsymbol{\eta}}^{\mathcal{P},(2)}\right) $$ is also asymptotically tight by Lemma~1.4.3 in \cite{vaartWellner1996}. In order to show the convergence of the marginals, 
    let $t_1,...,t_m\in\T, m\in\mathbb N$ be arbitrary and $\mathbf{Z}^{(1)}, \mathbf{Z}^{(2)} \sim \mathcal{N}_r(\mathbf{0}_r, \mathit{I}_r )$ i.i.d.~and independent of the data (\ref{eq:funcData}). Then,
    it holds
    \begin{align*}
        &\left((\sqrt{n}\mathit{H}\widehat{\boldsymbol{\eta}}^{\mathcal{P},(1)}(t_i))_{i\in\{1,...,m\}},  (\sqrt{n}\mathit{H}\widehat{\boldsymbol{\eta}}^{\mathcal{P},(2)}(t_i))_{i\in\{1,...,m\}}\right)^{\top} \\&\overset{d}{=} 
        \left(\mathit{I}_{2} \otimes  (\mathit{H}\widehat{\mathit{\Lambda}}(t_i,t_j)\mathit{H}^{\top})_{i,j\in\{1,...,m\}}^{1/2}  \right)
\begin{pmatrix}
 \mathbf{Z}^{(1)}  \\ \mathbf{Z}^{(2)}
\end{pmatrix}.
    \end{align*}
    By Lemma~\ref{MUniformConvLemma} and (A4), we have that
    \begin{align*}
        \mathit{I}_{2} \otimes  (\mathit{H}\widehat{\mathit{\Lambda}}(t_i,t_j)\mathit{H}^{\top})_{i,j\in\{1,...,m\}}^{1/2}  
\xrightarrow{P}  \mathit{I}_{2} \otimes  (\mathit{H}{\mathit{\Lambda}}(t_i,t_j)\mathit{H}^{\top})_{i,j\in\{1,...,m\}}^{1/2}  
    \end{align*} as $n\to\infty$. Hence, the convergence of the marginals, that is
    \begin{align*}
        &\left( (\sqrt{n}\mathit{H}\widehat{\boldsymbol{\eta}}^{\mathcal{P},(1)}(t_i))_{i\in\{1,...,m\}},  (\sqrt{n}\mathit{H}\widehat{\boldsymbol{\eta}}^{\mathcal{P},(2)}(t_i))_{i\in\{1,...,m\}}\right) \xrightarrow{d} \left( (\mathbf{y}^{(1)}(t_i))_{i\in\{1,...,m\}}, (\mathbf{y}^{(2)}(t_i))_{i\in\{1,...,m\}}\right)
    \end{align*} as $n\to\infty$,
    follows by Slutsky's lemma. By Theorem~1.5.4 in \cite{vaartWellner1996}, (\ref{eq:jointConv}) follows. \\
    Furthermore, it holds 
    \begin{align*}
      \sup\limits_{t\in\T} ||| \widehat{\mathit{\Lambda}}^{\mathcal{P},(b)}(t,t) - {\mathit{\Lambda}}(t,t) ||| \xrightarrow{P} 0
    \end{align*} as $n\to\infty$ for $b\in\{1,2\}$ due to Lemma~
    \ref{MUnifCovPara}.
    \\
    Let $\mathbb{D}$ denote the set of all continuous functions $ \widetilde{\mathit{\Lambda}}:\T\to\R^{kp\times kp}$ with 
$\widetilde{\mathit{\Lambda}}(t)$ symmetric and positive semidefinite for all $t\in\T$
and be equipped with the sup-norm.
Then, ${\mathit{\Lambda}}\in\mathbb{D}$ holds due to (R3). Moreover, by (\ref{eq:representation}),
$\widehat{\mathit{\Lambda}}^{\mathcal{P},(1)},\widehat{\mathit{\Lambda}}^{\mathcal{P},(2)}$ take values in $\mathbb{D}$.
Furthermore, the map \begin{align*}
    h:\: & C^{2r}(\T) \times \mathbb{D}^{2}
    \to (\R_0^+)^{2},\\
      &(\widetilde{\mathbf{x}}_1,\widetilde{\mathbf{x}}_2, \widetilde{\mathit{\Lambda}}_1, \widetilde{\mathit{\Lambda}}_2) \mapsto \left(\sup_{t\in\T} \left\{\widetilde{\mathbf{x}}_b^{\top}(t) \left(\mathit{H} \widetilde{\mathit{\Lambda}}_b(t) \mathit{H}^{\top} \right)^+ \widetilde{\mathbf{x}}_b(t) \right\}\right)_{b\in\{1,2\}} 
\end{align*}
is continuous on \mbox{$C^{2r}(\T) \times \{\mathit{\Lambda}\}^{2}$} under (A3) similarly as in the proof of Theorem~1.
Applying the continuous mapping theorem yields
\begin{align*}
    \left(  T_{n,1}^{\mathcal{P}}(\mathit{H}),T_{n,2}^{\mathcal{P}}(\mathit{H}) \right)  \xrightarrow{d} \left( T^{(1)},  T^{(2)} \right) 
\end{align*} as $n\to\infty$, where $T^{(1)},T^{(2)}$ are independent copies of $T.$
Thus, all assumptions of Lemma~\ref{BSQuant} are satisfied. 
Let $(B_n)_{n\in\mathbb N}$ be a sequence with $B_n \to\infty$ as $n\to\infty$. Then, (\ref{eq:MarginalConv}) provides
\begin{align*}
    Q_{n,B_n}^{\mathcal{P}}(1-\alpha) \xrightarrow{P} F^{-1}(1-\alpha) \quad\text{as $n\to\infty$}
\end{align*} and, thus,
\begin{align*}
   \left(T_n(\mathit{H}, \mathbf{c}), Q_{n,B_n}^{\mathcal{P}}(1-\alpha)\right) \xrightarrow{P} \left(T, F^{-1}(1-\alpha) \right)\quad\text{as $n\to\infty$}.
\end{align*}
Together with Theorem~1, it follows
\begin{align*}
    \E\left[ \varphi_{n,B_n}^{\mathcal{P}} \right] &= 1 - \P\left( T_n(\mathit{H}, \mathbf{c}) \leq Q^{\mathcal{P}}_{n,B}(1-\alpha) \right)
    \\&\to 1 - \P\left( T \leq F^{-1}(1-\alpha) \right) = \alpha
\end{align*} as $n\to\infty$.
\end{proof}

\begin{proof}[Proof of Remark~1]
    Firstly, we prove that (10) implies that for all $\varepsilon > 0$ there exists an $M\in\mathbb N$ such that 
    \begin{align}\label{eq:Implication}
        \left|\E\left[ \varphi_{n,B}^{\mathcal{P}} \right] - \alpha\right| < \varepsilon \quad\text{for all $n,B \geq M$} 
    \end{align}
     by contradiction. Therefore fix $\varepsilon > 0$ and assume that for all $M\in\mathbb N$ there exist $n_M,B_M\geq M$ such that $\left|\E\left[ \varphi_{n_M,B_M}^{\mathcal{P}} \right] - \alpha\right| \geq \varepsilon$. Then, it holds $n_M \to\infty, B_M\to\infty$ as $M\to\infty$. Hence, (10) implies the contradiction $$\varepsilon \leq \lim\limits_{M\to\infty} \left|\E\left[ \varphi_{n_M,B_M}^{\mathcal{P}} \right] - \alpha\right| =  0 .$$
    Thus, (\ref{eq:Implication}) follows. Now, fix again $\varepsilon>0$ and choose $M\in\mathbb N$ such that (\ref{eq:Implication}) holds. Then, we have 
    \begin{align*}
        \limsup\limits_{n\to\infty} \left| \E\left[ \varphi_{n,B}^{\mathcal{P}} \right] - \alpha \right| \leq \varepsilon
    \end{align*} for all $B\geq M$ and, thus, the statement follows.
\end{proof}

\subsection{Proofs of Section~2.4}

With minor adjustments, the proofs of Theorem~3 and 4 can be obtained similarly to the proofs of Theorem~3 and 4 in the supplement of \cite{FANOVA}, respectively.

\begin{proof}[{Proof of Theorem~3}]
Lemma~\ref{MUniformConvLemma} provides that $\widehat{\mathit{\Lambda}}$ converges uniformly over \mbox{$\{(t,t)\in\T^2 \mid t\in\T\}$} in probability to $\mathit{\Lambda}$.
Furthermore, Lemma~\ref{asy_CLT} ensures the convergence in distribution of $\sqrt{n} (\mathit{H}\hat{\boldsymbol\eta} - \mathbf{c})$ to $\mathbf{z}$.
    By the continuous mapping theorem, it suffices to show that the map 
    $$ \mathbf{g}: \mathcal{L}_2^r(\T) \times \mathbb{D} \to (\R_0^+ \cup \{\infty\})^R, \quad \mathbf{g}(\widetilde{\mathbf{x}}, \widetilde{\mathit{\Lambda}}) := \left(\sup\limits_{t\in\T} \left\{\widetilde{\mathbf{x}}_{\ell}^{\top}(t) \left(\mathit{H}_{\ell} \widetilde{\mathit{\Lambda}}(t) \mathit{H}_{\ell}^{\top} \right)^+ \widetilde{\mathbf{x}}_{\ell}(t) \right\}\right)_{\ell\in\{1,...,R\}} $$ is continuous on $ \mathcal{L}_2^r(\T) \times \{\mathit{\Lambda} \}$ with $\mathbb{D}$ as in the proof of Theorem~1.
    This holds whenever its component functions are continuous. The continuity of the component functions follows analogously as in the proof of Theorem~1.
\end{proof}

\begin{proof}[{Proof of Theorem~4}]
By Lemma~\ref{Para_CLT}, 
$ \sqrt{n}\mathit{H}\widehat{\boldsymbol\eta}^{\mathcal{P}}$ converges in distribution to $\mathbf{z}$ in $C^r(\T)$. Applying the continuous mapping theorem as in the proof of Theorem~3 
yields 
that 
\begin{align*}
    &\left(\sup\limits_{t\in\T} \left\{ n\left(\mathit{H}_{\ell}\widehat{\boldsymbol\eta}^{\mathcal{P}}(t)\right)^{\top}\left(\mathit{H}_{\ell}\widehat{\mathit{\Lambda}}^{\mathcal{P}}(t,t)\mathit{H}_{\ell}^{\top}\right)^+\mathit{H}_{\ell}\widehat{\boldsymbol\eta}^{\mathcal{P}}(t)\right\} \right)_{\ell\in\{1,...,R\}} \\&\xrightarrow{d} \left(\sup\limits_{t\in\T} \{ \mathbf{z}_{\ell}^{\top}(t)(\mathit{H}_{\ell}\mathit{\Lambda}(t,t)\mathit{H}_{\ell}^{\top})^+\mathbf{z}_{\ell}(t) \}\right)_{\ell\in\{1,...,R\}}
\end{align*} as $n\to\infty$.
\end{proof}

\begin{proof}[{Proof of Proposition~2}]
Let $(B_n)_{n\in\mathbb N}$ be a sequence with $B_n\to\infty$ as $n\to\infty$.
We aim to apply Lemma~\ref{BSQuant} with $X_{n\ell}^{(b)} := T_{n,b}^{\mathcal{P}}(\mathit{H}_{\ell})$ for all $b\in\mathbb N, \ell\in\{1,...,R\}$. Let ${F}_\ell$ denote the distribution function of $$T_\ell := \sup_{t\in\T} \{ (\mathbf{y}_\ell(t))^{\top}(\mathit{H}_\ell\mathit{\Lambda}(t,t)\mathit{H}_{\ell}^{\top})^+\mathbf{y}_\ell(t) \}$$ for all $\ell\in\{1,...,R\}$ and ${F}$ denote the joint distribution function of $(T_\ell)_{\ell\in\{1,...,R\}}$.
The condition of identical distributions in Remark~\ref{RemarkLemma} follows due to the definition of $T_{n,b}^{\mathcal{P}}(\mathit{H}_{\ell}), b\in\mathbb N, \ell\in\{1,...,R\}.$
Analogously as in the proof of Proposition~1, one can show that 
\begin{align*}
    \left( \left(T_{n,1}^{\mathcal{P}}(\mathit{H}_{\ell})\right)_{\ell\in\{1,...,R\}} ,  \left(T_{n,2}^{\mathcal{P}}(\mathit{H}_{\ell})\right)_{\ell\in\{1,...,R\}}\right) \xrightarrow{d} \left( \left(T_{\ell}^{(1)}\right)_{\ell\in\{1,...,R\}} ,  \left(T_{\ell}^{(2)}\right)_{\ell\in\{1,...,R\}}\right)
\end{align*}
as $n\to\infty$ holds, where $\left(T_{\ell}^{(1)}\right)_{\ell\in\{1,...,R\}}, \left(T_{\ell}^{(2)}\right)_{\ell\in\{1,...,R\}}$ are independent copies of $(T_{\ell})_{\ell\in\{1,...,R\}}$. Thus, Remark~\ref{RemarkLemma} implies (\ref{eq:condition}).
Additionally, for each $\ell\in\{1,...,R\}$, ${F}_\ell$ is continuous and strictly increasing on $[0,\infty)$ due to $\mathrm{rank}(\mathit{H}_\ell) \geq 1$, which can be shown by Remark~\ref{RemContandMonoton} with $\mathbf{X}(t) := \left((\mathit{H}_{\ell}\mathit{\Lambda}(t,t)\mathit{H}_{\ell}^{\top})^+\right)^{1/2}\mathbf{y}(t)$ for all $t\in\T$.
Then, Lemma~\ref{BSQuant} implies that 
\begin{align*}
    Q_{n,B_n,\ell}^{\mathcal{P}}(1-\alpha) \xrightarrow{P} F_{\ell}^{-1} \left( 1 - \mathrm{FWER}^{-1}(\alpha) \right)
\end{align*} as $n\to\infty$ for all $\ell\in\{1,...,R\}$, where
$$ \mathrm{FWER}(\beta) := 1 - {F}\left( \left( {F}_\ell^{-1} (1-\beta) \right)_{\ell\in\{1,...,R\}} \right)  $$  for $\beta\in(0,1).$
Hence, it follows
\begin{align*}
   \left(\left(  T_{n}(\mathit{H}_{\ell}, \mathbf{c}_{\ell})\right)_{\ell\in\mathcal{R}}, \left( Q_{n,B_n,\ell}^{\mathcal{P}}(1-\alpha)\right)_{\ell\in\mathcal{R}} \right) \xrightarrow{d} \left(\left(  T_{\ell}\right)_{\ell\in\mathcal{R}}, \left( F_{\ell}^{-1} \left( 1 - \mathrm{FWER}^{-1}(\alpha) \right)\right)_{\ell\in\mathcal{R}} \right)
\end{align*} as $n\to\infty$ under $\mathcal{H}_{0,\ell}, \ell\in\mathcal{R},$ in (12) by Theorem~3.
Thus, we have
\begin{align*}
   \E\left[ \max\limits_{\ell\in\mathcal{R}} \varphi_{n,B_n,\ell}^{\mathcal{P}} \right] &= 1- \P\left( \forall \ell\in\mathcal{R}:\:  T_{n}(\mathit{H}_{\ell}, \mathbf{c}_{\ell}) \leq Q_{n,B_n,\ell}^{\mathcal{P}}(1-\alpha)\right)
   \\&\to 1- \P\left( \forall \ell\in\mathcal{R}:\:  T_{\ell} \leq F_{\ell}^{-1} \left( 1 - \mathrm{FWER}^{-1}(\alpha) \right)\right)
   \\& \leq 1- \P\left( \forall \ell\in\{1,...,R\}:\:  T_{\ell} \leq F_{\ell}^{-1} \left( 1 - \mathrm{FWER}^{-1}(\alpha) \right)\right)
   \\& = \alpha
\end{align*}
as $n\to\infty$ under $\mathcal{H}_{0,\ell}, \ell\in\mathcal{R}.$ Note that the inequality is an equality if $\mathcal{R} = \{1,...,R\}.$
\end{proof}

\section{Simulation Results}\label{simulation-results}
In this section, we present all simulation results in tables and summarize them in box-and-whisker plots. We also present the example of trajectories of generated functional data.

\subsection{Examples of generated functional data}
\begin{figure}[h]
\centering
\includegraphics[width=0.99\textwidth,
height=0.6\textheight]{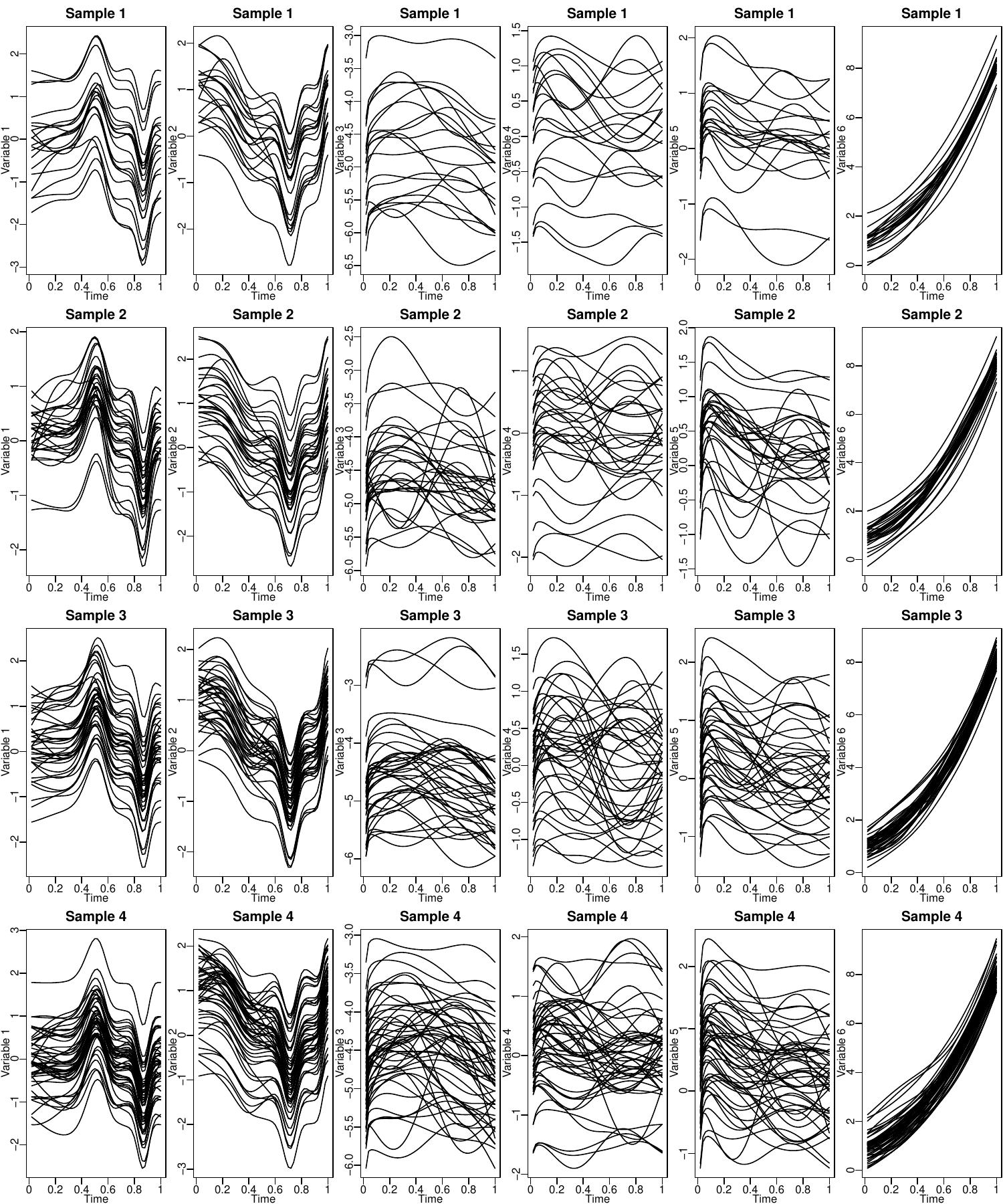}
\caption[Trajectories of functional data generated in simulation studies with $\rho=0.1$.]{Trajectories of functional data generated in simulation studies in Model~1 under the null hypothesis for $k=4$ groups and $p=6$ functional variables with $\rho=0.1$.}
\label{fig_s_01}
\end{figure}

\begin{figure}[h]
\centering
\includegraphics[width=0.99\textwidth,
height=0.6\textheight]{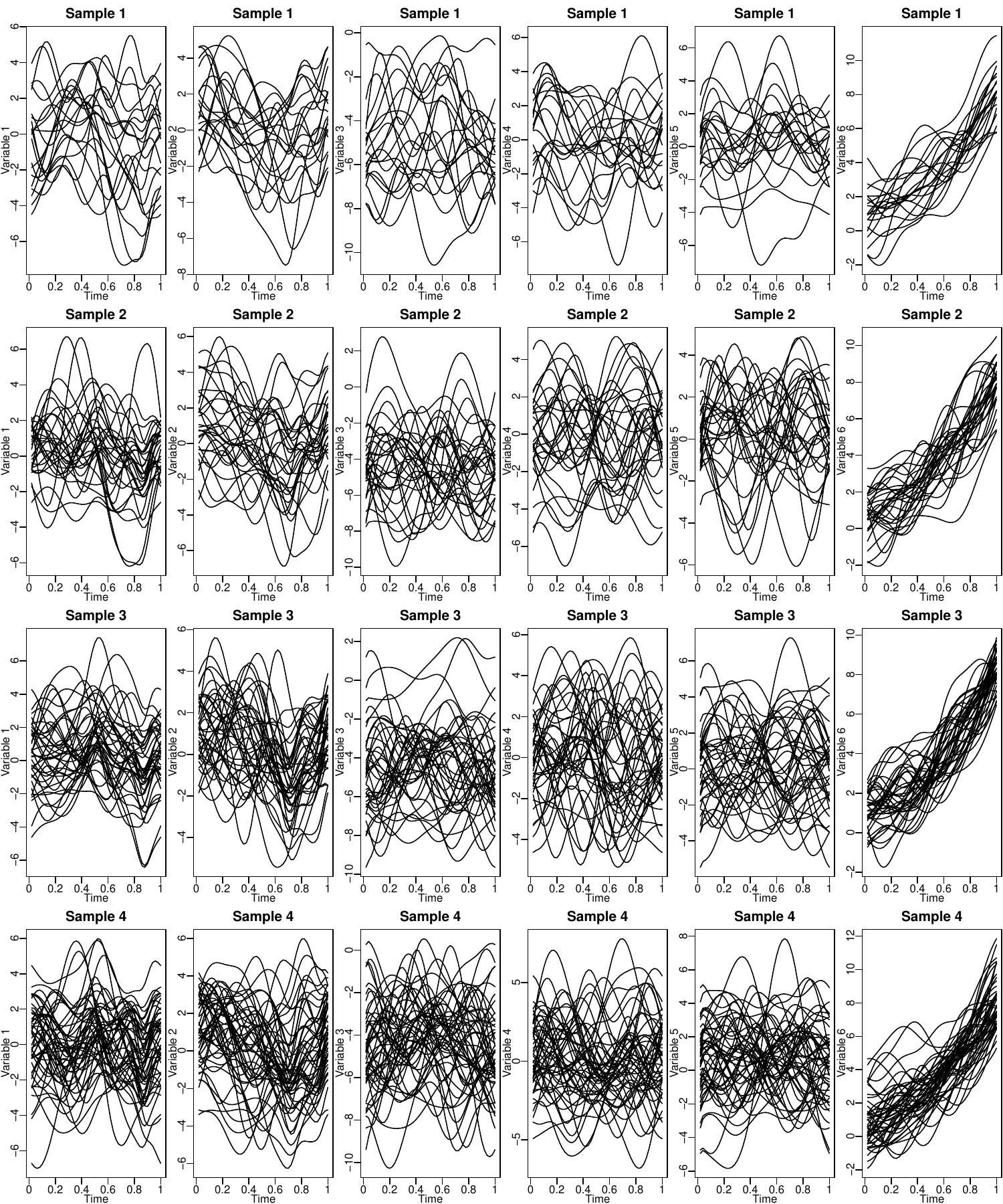}
\caption[Trajectories of functional data generated in simulation studies with $\rho=0.5$.]{Trajectories of functional data generated in simulation studies in Model~1 under the null hypothesis for $k=4$ groups and $p=6$ functional variables with $\rho=0.5$.}
\label{fig_s_02}
\end{figure}

\begin{figure}[h]
\centering
\includegraphics[width=0.99\textwidth,
height=0.6\textheight]{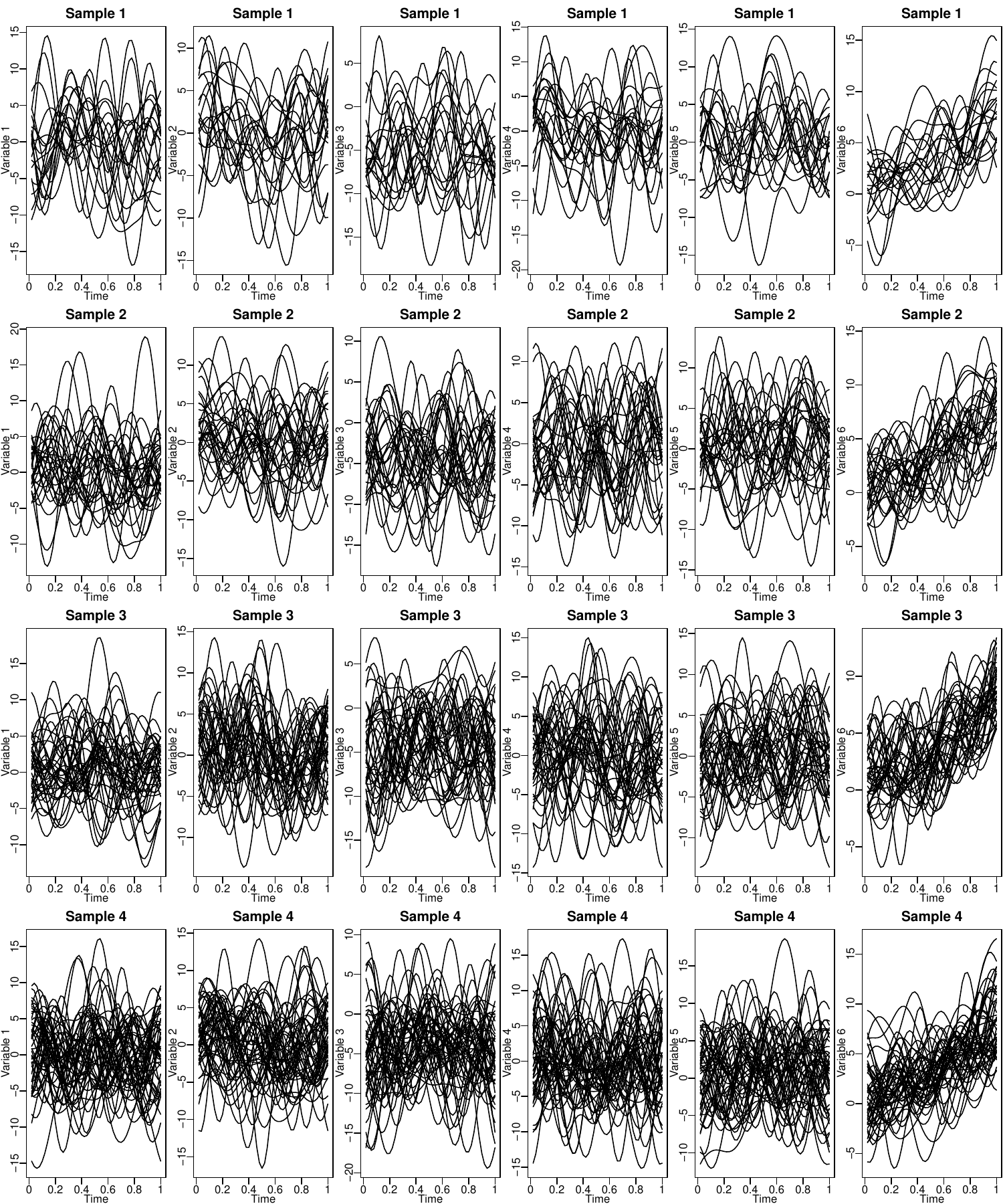}
\caption[Trajectories of functional data generated in simulation studies with $\rho=0.9$.]{Trajectories of functional data generated in simulation studies in Model~1 under the null hypothesis for $k=4$ groups and $p=6$ functional variables with $\rho=0.9$.}
\label{fig_s_03}
\end{figure}

Figures~\ref{fig_s_01}-\ref{fig_s_03} present the trajectories of functional data generated in simulation studies. We can observe that the data are more noisely for less correlation (greater $\rho$).

\clearpage

\subsection{Results without Scaling Function}\label{results-without-scaling-function}
\begin{figure}
\centering
\includegraphics[width=0.99\textwidth,
height=0.4\textheight]{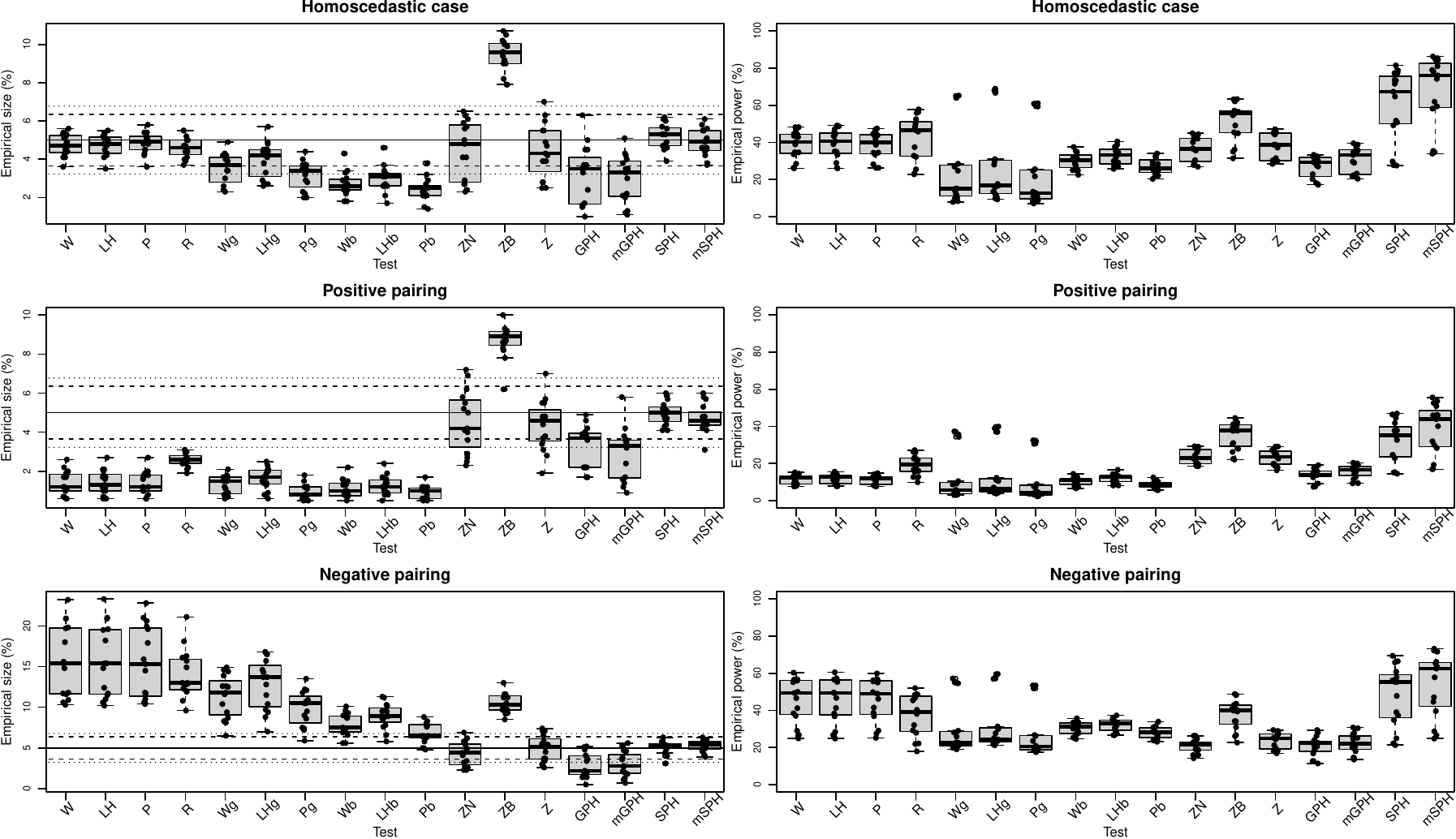}
\caption[Box-and-whisker plots for the empirical sizes and power in Model~1 under global hypothesis for $k=4$ groups and $p=6$ functional variables and for homoscedastic case, positive pairing, and negative pairing]{Box-and-whisker plots for the empirical sizes and power (as percentages) of all tests (without projection tests based on Roy test, i.e., Rg and Rb, since they are always too liberal) obtained in Model~1 under global hypothesis for $k=4$ groups and $p=6$ functional variables and for homoscedastic case, positive pairing, and negative pairing.}
\label{fig_s_1}
\end{figure}

\begin{figure}
\centering
\includegraphics[width=0.99\textwidth,
height=0.8\textheight]{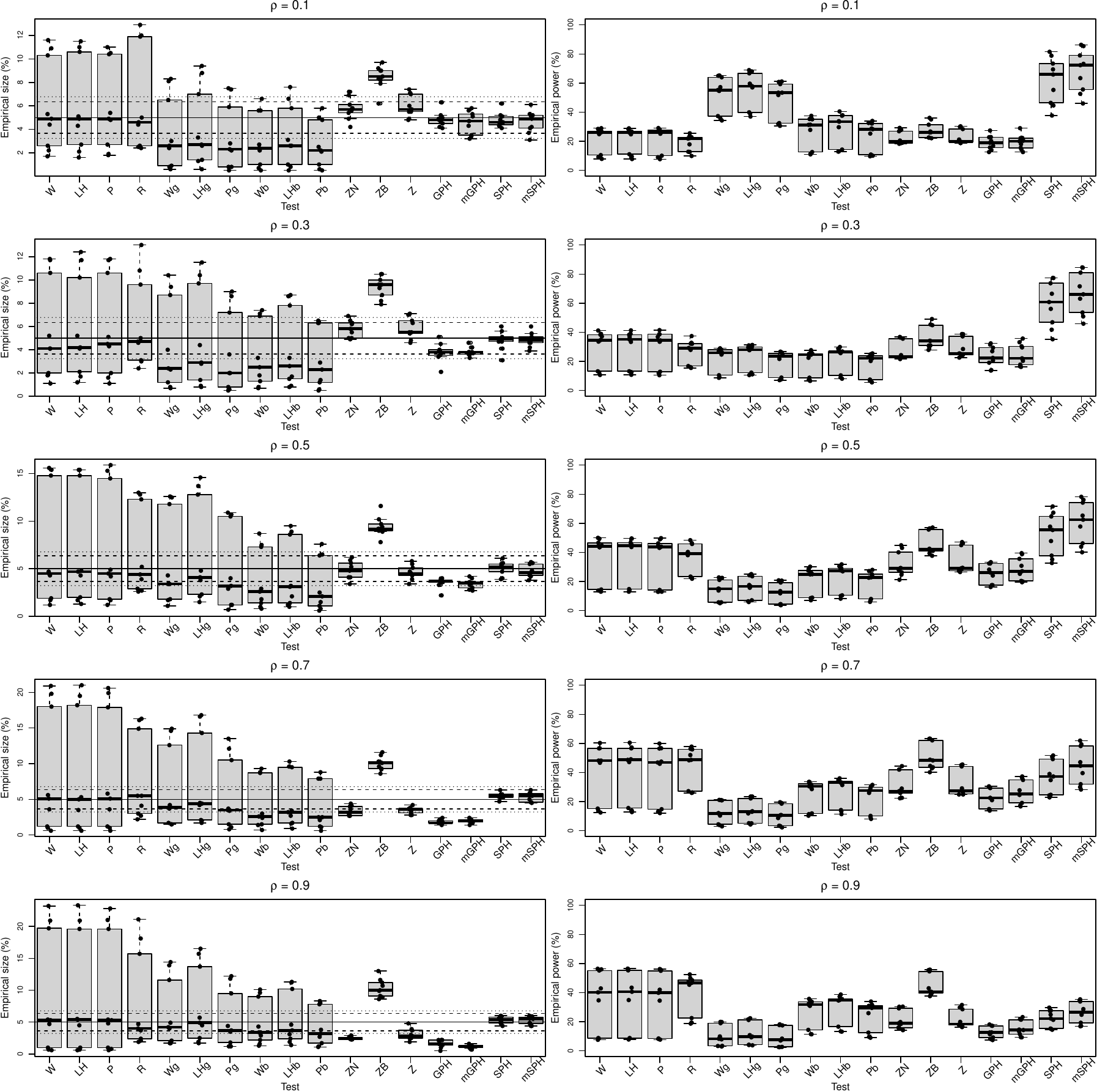}
\caption[Box-and-whisker plots for the empirical size and power in Model~1 under global hypothesis for $k=4$ groups and $p=6$ functional variables and for different correlation]{Box-and-whisker plots for the empirical size and power (as percentages) of all tests (without projection tests based on Roy test, i.e., Rg and Rb, since they are always too liberal) obtained in Model~1 under global hypotheses for $k=4$ groups and $p=6$ functional variables and for different correlation.}
\label{fig_s_2}
\end{figure}

\begin{figure}
\centering
\includegraphics[width=0.99\textwidth,
height=0.4\textheight]{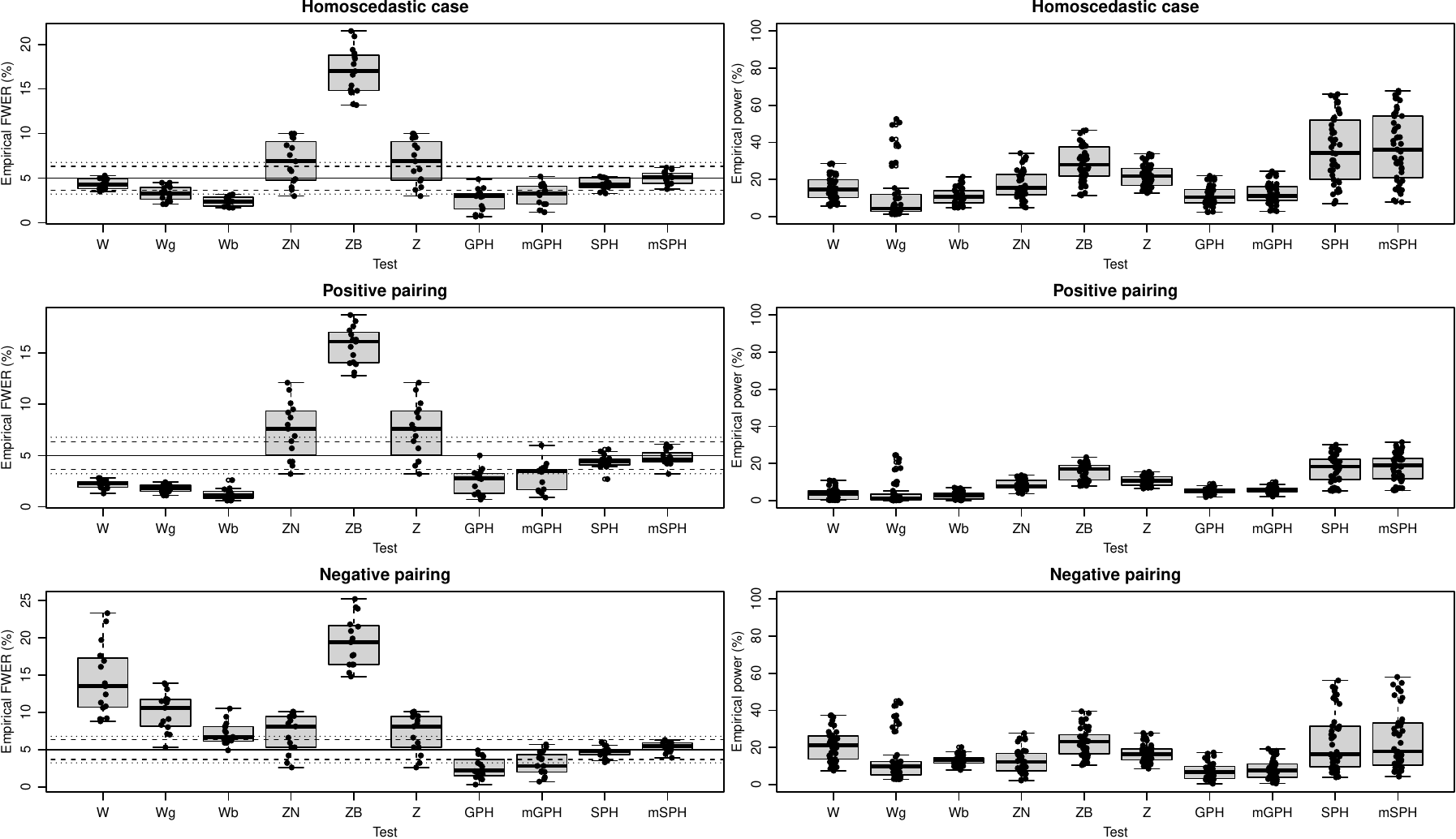}
\caption[Box-and-whisker plots for the empirical FWER and power in Model~1 under local hypotheses for $k=4$ groups and $p=6$ functional variables and for homoscedastic case, positive pairing, and negative pairing.]{Box-and-whisker plots for the empirical FWER and power (as percentages) of all tests obtained in Model~1 under local hypotheses for $k=4$ groups and $p=6$ functional variables and for homoscedastic case, positive pairing, and negative
pairing.}
\label{fig_s_3}
\end{figure}

\begin{figure}
\centering
\includegraphics[width=0.99\textwidth,
height=0.8\textheight]{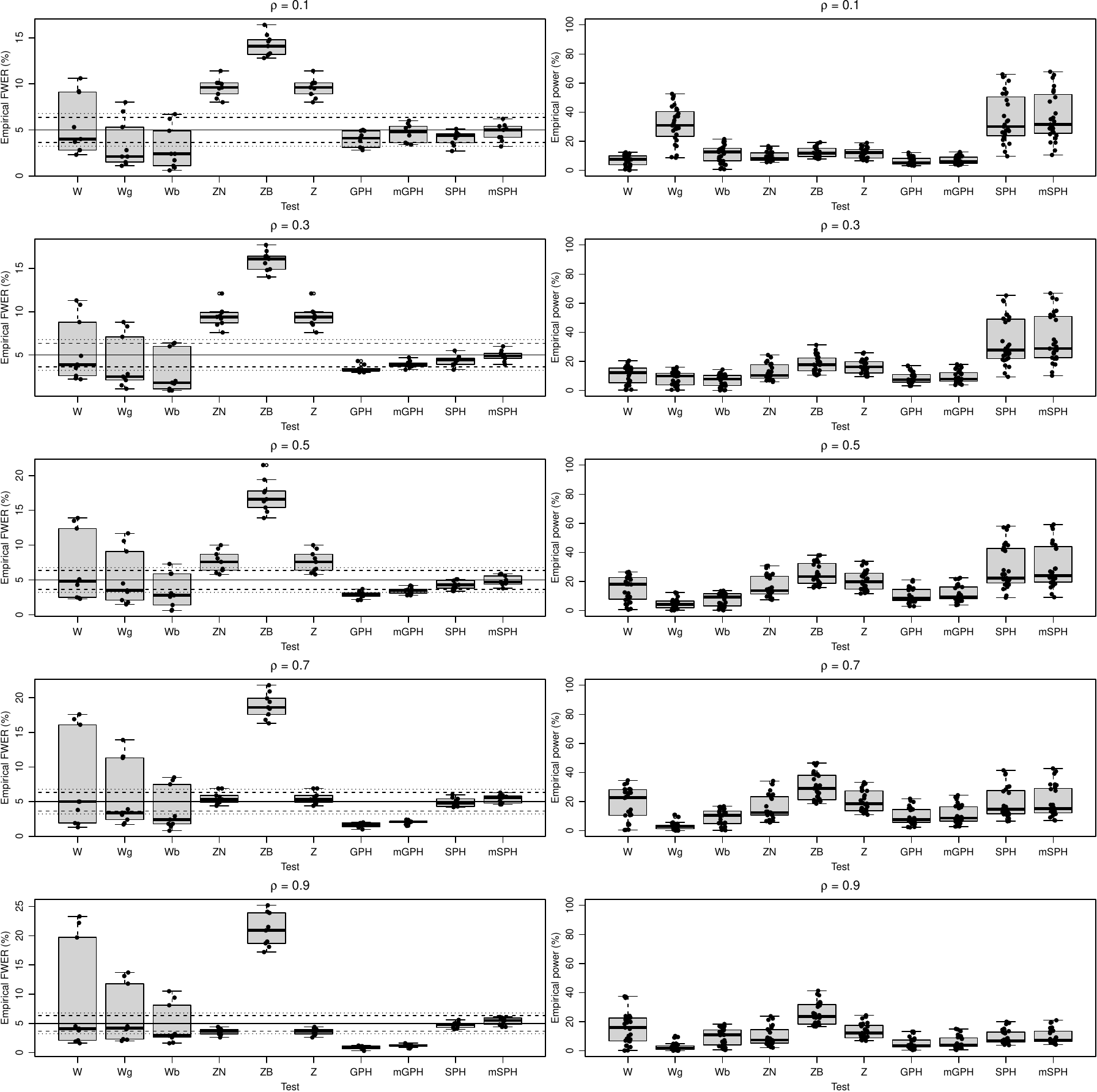}
\caption[Box-and-whisker plots for the empirical FWER and power in Model~1 under local hypotheses for $k=4$ groups and $p=6$ functional variables and for different correlation]{Box-and-whisker plots for the empirical FWER and power (as percentages) of all tests obtained in Model~1 under local hypotheses for $k=4$ groups and $p=6$ functional variables and for different correlation.}
\label{fig_s_4}
\end{figure}

Tables~\ref{tab:unnamed-chunk-3}-\ref{tab:unnamed-chunk-8} and Figures~\ref{fig_s_1}-\ref{fig_s_4} contain and summarize the results of simulation studies for the global and local hypotheses for $k=4$ samples and $p=6$ functional variables.

\begin{landscape}\begingroup\fontsize{10}{12}\selectfont
\begin{longtable}[t]{lllrrrrrrrrrrrrrrrrrrr}
\caption[Empirical sizes in Model~1 under global hypothesis for $k=4$ groups and $p=6$ functional variables]{\label{tab:unnamed-chunk-3}Empirical sizes (as percentages) in Model~1 under global hypothesis for $k=4$ groups and $p=6$ functional variables. Columns: D - distribution ($N$ - normal, $t_4$ - t-Student, $\chi_4^2$ - $\chi_4^2$-squared), HH - homoscedasticity or heteroscedasticity (hom - homoscedasticity, pos - positive pairing, neg - negative pairing), $\mathbf{n}=(20,30,40,50)$ - vector of sample sizes.}\\
\hline
D & $\rho$ & HH & W & LH & P & R & Wg & LHg & Pg & Rg & Wb & LHb & Pb & Rb & ZN & ZB & Z & GPH & mGPH & SPH & mSPH\\
\hline
\endfirsthead
\caption[]{Empirical sizes (as percentages) in Model~1 under global hypothesis for $k=4$ groups and $p=6$ functional variables. Columns: D - distribution ($N$ - normal, $t_4$ - t-Student, $\chi_4^2$ - $\chi_4^2$-squared), HH - homoscedasticity or heteroscedasticity (hom - homoscedasticity, pos - positive pairing, neg - negative pairing), $\mathbf{n}=(20,30,40,50)$ - vector of sample sizes. \textit{(continued)}}\\
\hline
D & $\rho$ & HH & W & LH & P & R & Wg & LHg & Pg & Rg & Wb & LHb & Pb & Rb & ZN & ZB & Z & GPH & mGPH & SPH & mSPH\\
\hline
\endhead
$N$ & 0.1 & hom & 5.3 & 5.1 & 5.4 & 4.4 & 2.4 & 2.6 & 2.2 & 26.6 & 2.4 & 2.6 & 2.2 & 24.0 & 5.7 & 7.9 & 7.0 & 4.5 & 4.3 & 4.6 & 4.2\\
&& pos & 2.2 & 2.1 & 1.9 & 2.6 & 0.6 & 0.6 & 0.5 & 19.5 & 0.7 & 0.8 & 0.6 & 18.8 & 7.2 & 9.0 & 5.7 & 4.9 & 5.8 & 5.2 & 5.3\\
&& neg & 11.6 & 11.5 & 11.0 & 12.9 & 8.3 & 9.4 & 7.4 & 45.6 & 5.6 & 6.6 & 5.0 & 42.1 & 6.9 & 9.7 & 7.4 & 4.8 & 5.6 & 5.1 & 5.2\\
& 0.3 & hom & 4.1 & 4.1 & 4.5 & 4.6 & 2.4 & 2.8 & 2.0 & 30.8 & 2.5 & 2.6 & 2.3 & 21.8 & 6.5 & 9.6 & 5.5 & 3.5 & 3.3 & 4.7 & 4.6\\
&& pos & 2.0 & 2.1 & 2.0 & 2.4 & 1.2 & 1.4 & 0.8 & 21.9 & 1.3 & 1.5 & 1.2 & 17.1 & 6.9 & 10.0 & 7.0 & 4.0 & 3.6 & 4.9 & 4.6\\
&& neg & 11.8 & 12.4 & 11.8 & 13.0 & 8.7 & 9.7 & 7.2 & 54.9 & 6.9 & 7.8 & 6.3 & 45.4 & 6.2 & 9.6 & 7.1 & 5.1 & 4.6 & 5.6 & 6.0\\
& 0.5 & hom & 4.3 & 4.3 & 4.2 & 4.4 & 3.4 & 4.1 & 3.2 & 38.1 & 1.8 & 2.1 & 1.5 & 28.2 & 4.8 & 9.2 & 4.3 & 3.7 & 4.0 & 3.9 & 4.5\\
&& pos & 1.2 & 1.3 & 1.2 & 2.6 & 1.1 & 1.5 & 0.7 & 29.3 & 1.0 & 1.2 & 0.8 & 21.7 & 5.2 & 8.9 & 5.5 & 3.6 & 4.2 & 5.1 & 5.7\\
&& neg & 14.8 & 14.8 & 14.5 & 12.3 & 12.6 & 13.7 & 10.9 & 60.4 & 7.5 & 8.9 & 6.4 & 45.4 & 5.6 & 11.6 & 4.5 & 2.2 & 2.8 & 4.0 & 4.2\\
& 0.7 & hom & 5.6 & 5.3 & 5.8 & 5.5 & 4.2 & 4.5 & 3.7 & 41.8 & 3.0 & 3.4 & 2.5 & 29.3 & 4.1 & 10.1 & 4.2 & 2.4 & 2.1 & 5.7 & 5.0\\
&& pos & 0.6 & 0.6 & 0.6 & 2.2 & 1.5 & 2.0 & 0.8 & 31.0 & 0.7 & 0.9 & 0.6 & 23.9 & 4.0 & 9.3 & 3.8 & 1.7 & 2.4 & 5.7 & 4.6\\
&& neg & 19.8 & 19.5 & 19.9 & 14.9 & 14.9 & 16.6 & 13.5 & 66.6 & 9.3 & 10.3 & 8.8 & 49.2 & 4.4 & 11.2 & 3.4 & 2.0 & 2.1 & 5.4 & 5.9\\
& 0.9 & hom & 5.3 & 5.5 & 5.3 & 4.7 & 4.2 & 4.9 & 3.7 & 46.0 & 4.3 & 4.6 & 3.8 & 31.8 & 2.7 & 10.0 & 2.5 & 1.5 & 1.3 & 5.6 & 5.5\\
&& pos & 0.7 & 0.7 & 0.8 & 2.4 & 1.7 & 1.7 & 1.2 & 31.7 & 1.3 & 1.4 & 1.1 & 21.1 & 2.9 & 8.6 & 4.8 & 2.2 & 1.6 & 5.9 & 6.0\\
&& neg & 20.9 & 20.9 & 21.0 & 18.1 & 13.9 & 15.7 & 11.8 & 70.5 & 10.1 & 11.2 & 8.3 & 56.0 & 2.6 & 11.2 & 2.6 & 1.4 & 1.1 & 5.8 & 5.7\\\addlinespace
$t_4$ & 0.1 & hom & 4.4 & 4.3 & 4.8 & 4.6 & 3.0 & 3.3 & 2.8 & 25.5 & 2.7 & 3.1 & 2.5 & 23.5 & 6.1 & 8.2 & 4.9 & 5.0 & 3.5 & 4.5 & 3.7\\
&& pos & 2.6 & 2.7 & 2.7 & 2.5 & 0.8 & 1.4 & 0.8 & 18.1 & 1.0 & 1.0 & 1.0 & 16.6 & 4.2 & 6.2 & 5.5 & 4.2 & 3.4 & 4.1 & 3.1\\
&& neg & 10.9 & 11.0 & 10.4 & 12.0 & 8.1 & 8.8 & 7.5 & 44.6 & 6.6 & 7.6 & 5.8 & 39.2 & 5.4 & 9.2 & 7.1 & 4.1 & 4.7 & 4.4 & 4.9\\
& 0.3 & hom & 5.2 & 5.2 & 5.1 & 5.0 & 4.0 & 4.4 & 3.6 & 31.4 & 3.3 & 3.3 & 2.9 & 25.6 & 6.3 & 10.5 & 6.3 & 3.6 & 3.8 & 4.7 & 5.1\\
&& pos & 1.8 & 1.7 & 1.6 & 3.1 & 0.8 & 0.8 & 0.6 & 24.0 & 0.8 & 0.9 & 0.6 & 19.0 & 5.0 & 8.7 & 4.8 & 3.9 & 3.6 & 5.0 & 4.8\\
&& neg & 10.6 & 10.2 & 10.6 & 10.8 & 10.4 & 11.5 & 9.0 & 53.4 & 7.4 & 8.6 & 6.5 & 42.6 & 4.9 & 9.3 & 5.4 & 2.1 & 3.7 & 3.1 & 3.9\\
& 0.5 & hom & 4.5 & 4.7 & 4.5 & 3.9 & 3.4 & 3.8 & 2.9 & 38.6 & 2.6 & 3.1 & 2.1 & 25.3 & 4.1 & 9.0 & 3.9 & 3.3 & 3.0 & 6.1 & 5.5\\
&& pos & 1.7 & 1.7 & 1.7 & 2.7 & 1.7 & 2.1 & 1.2 & 25.9 & 0.8 & 1.0 & 0.6 & 19.8 & 4.1 & 7.8 & 4.4 & 3.9 & 3.5 & 4.7 & 4.3\\
&& neg & 15.6 & 15.4 & 15.9 & 12.8 & 11.8 & 12.8 & 10.5 & 60.0 & 7.3 & 8.6 & 6.5 & 46.4 & 4.1 & 10.2 & 5.1 & 3.6 & 2.7 & 5.6 & 4.6\\
& 0.7 & hom & 5.1 & 5.0 & 5.1 & 5.5 & 3.7 & 4.4 & 3.4 & 41.1 & 2.6 & 3.2 & 2.5 & 28.8 & 2.7 & 10.2 & 2.8 & 1.7 & 2.2 & 5.3 & 5.6\\
&& pos & 1.0 & 1.0 & 1.0 & 3.0 & 1.7 & 2.1 & 1.1 & 30.3 & 1.5 & 1.7 & 1.2 & 21.3 & 2.7 & 8.6 & 3.1 & 1.7 & 1.7 & 4.7 & 4.5\\
&& neg & 20.9 & 21.0 & 20.6 & 16.3 & 14.6 & 16.8 & 12.1 & 65.0 & 8.8 & 9.5 & 7.9 & 50.4 & 2.7 & 10.3 & 3.6 & 1.4 & 2.0 & 5.3 & 5.5\\
& 0.9 & hom & 5.4 & 5.4 & 5.4 & 3.7 & 4.0 & 4.5 & 3.5 & 42.8 & 2.9 & 3.1 & 2.7 & 30.8 & 2.3 & 10.7 & 2.5 & 1.0 & 1.1 & 4.8 & 4.4\\
&& pos & 0.6 & 0.6 & 0.6 & 1.9 & 1.7 & 1.9 & 1.2 & 31.1 & 2.2 & 2.4 & 1.7 & 22.9 & 2.3 & 8.9 & 1.9 & 2.2 & 1.2 & 4.4 & 4.8\\
&& neg & 19.7 & 19.6 & 19.6 & 15.7 & 11.6 & 13.7 & 9.5 & 66.5 & 9.0 & 10.2 & 8.0 & 51.9 & 2.3 & 13.0 & 3.7 & 0.5 & 1.3 & 5.3 & 4.9\\\addlinespace
$\chi_4^2$ & 0.1 & hom & 4.9 & 4.9 & 4.9 & 5.0 & 2.6 & 2.7 & 2.3 & 25.9 & 2.2 & 2.6 & 2.1 & 24.4 & 5.6 & 9.0 & 6.0 & 6.3 & 5.1 & 6.2 & 6.1\\
&& pos & 1.7 & 1.6 & 1.8 & 2.4 & 0.9 & 1.3 & 0.8 & 19.1 & 0.5 & 0.5 & 0.5 & 16.9 & 5.8 & 8.3 & 4.8 & 4.6 & 3.2 & 4.3 & 4.1\\
&& neg & 10.3 & 10.6 & 10.5 & 11.9 & 6.5 & 7.0 & 5.9 & 43.0 & 5.6 & 5.8 & 4.8 & 39.5 & 4.9 & 8.5 & 5.5 & 5.2 & 5.3 & 4.8 & 5.1\\
& 0.3 & hom & 4.1 & 4.2 & 4.3 & 4.7 & 2.3 & 2.9 & 2.0 & 32.0 & 1.8 & 1.7 & 1.4 & 24.4 & 5.0 & 7.9 & 5.5 & 4.5 & 4.2 & 6.0 & 4.9\\
&& pos & 1.1 & 1.2 & 1.1 & 3.0 & 0.7 & 0.9 & 0.5 & 23.8 & 0.7 & 0.8 & 0.5 & 17.8 & 5.5 & 8.2 & 4.6 & 3.8 & 3.8 & 4.1 & 4.2\\
&& neg & 11.7 & 11.7 & 11.7 & 9.6 & 9.4 & 10.4 & 8.6 & 52.3 & 7.1 & 8.7 & 6.3 & 42.3 & 5.8 & 10.5 & 6.5 & 3.4 & 3.8 & 5.1 & 5.4\\
& 0.5 & hom & 4.7 & 4.8 & 4.9 & 5.2 & 4.3 & 4.8 & 4.0 & 38.7 & 2.9 & 3.2 & 2.5 & 27.5 & 5.9 & 9.4 & 4.7 & 3.7 & 3.5 & 5.3 & 3.8\\
&& pos & 1.9 & 2.0 & 1.8 & 2.9 & 1.8 & 2.3 & 1.2 & 29.2 & 1.4 & 1.4 & 1.1 & 19.6 & 6.2 & 9.2 & 3.4 & 3.7 & 3.3 & 5.2 & 4.8\\
&& neg & 15.4 & 15.4 & 15.3 & 13.0 & 12.5 & 14.6 & 10.7 & 60.9 & 8.7 & 9.5 & 7.6 & 48.3 & 3.4 & 9.7 & 5.8 & 4.0 & 3.6 & 5.5 & 5.6\\
& 0.7 & hom & 3.6 & 3.5 & 3.6 & 4.1 & 3.9 & 4.2 & 3.5 & 40.1 & 2.4 & 2.7 & 2.1 & 26.8 & 2.9 & 9.6 & 2.8 & 1.6 & 2.0 & 5.3 & 4.6\\
&& pos & 1.2 & 1.2 & 1.2 & 2.7 & 1.5 & 1.7 & 1.5 & 28.9 & 1.4 & 1.6 & 1.1 & 20.8 & 3.6 & 9.2 & 3.7 & 2.2 & 1.4 & 5.4 & 5.8\\
&& neg & 18.0 & 18.2 & 17.9 & 16.1 & 12.6 & 14.3 & 10.5 & 65.8 & 8.7 & 9.6 & 7.9 & 48.6 & 3.2 & 11.6 & 3.7 & 1.9 & 1.8 & 6.3 & 6.3\\
& 0.9 & hom & 4.7 & 4.5 & 4.8 & 4.0 & 4.9 & 5.7 & 4.4 & 43.9 & 3.4 & 3.7 & 3.2 & 31.2 & 2.4 & 9.9 & 3.9 & 1.6 & 1.2 & 4.9 & 5.8\\
&& pos & 1.0 & 1.0 & 1.0 & 2.1 & 2.1 & 2.5 & 1.8 & 33.5 & 1.6 & 1.9 & 1.2 & 24.0 & 2.5 & 9.1 & 2.8 & 2.2 & 0.9 & 6.0 & 4.4\\
&& neg & 23.2 & 23.3 & 22.8 & 21.1 & 14.4 & 16.5 & 12.2 & 69.0 & 9.6 & 11.3 & 7.8 & 54.1 & 2.3 & 11.6 & 3.0 & 1.6 & 0.7 & 5.4 & 5.9\\
\hline
\end{longtable}
\endgroup{}
\end{landscape}

\newpage

\begin{landscape}\begingroup\fontsize{10}{12}\selectfont

\begin{longtable}[t]{llllrrrrrrrrrrrrrrrrr}
\caption[Empirical power in Model~1 under global hypothesis for $k=4$ groups and $p=6$ functional variables]{\label{tab:unnamed-chunk-4}Empirical power (as percentages) in Model~1 under global hypothesis for $k=4$ groups and $p=6$ functional variables (without projection tests based on Roy test, i.e., Rg and Rb, since they are always too liberal). Columns: D - distribution ($N$ - normal, $t_4$ - t-Student, $\chi_4^2$ - $\chi_4^2$-squared), HH - homoscedasticity or heteroscedasticity (hom - homoscedasticity, pos - positive pairing, neg - negative pairing), $\mathbf{n}=(20,30,40,50)$ - vector of sample sizes, $d$ - hyperparameter in scenario for alternative hypothesis.}\\
\hline
D & $\rho$ & $d$ & HH & W & LH & P & R & Wg & LHg & Pg & Wb & LHb & Pb & ZN & ZB & Z & GPH & mGPH & SPH & mSPH\\
\hline
\endfirsthead
\caption[]{Empirical power (as percentages) in Model~1 under global hypothesis for $k=4$ groups and $p=6$ functional variables (without projection tests based on Roy test, i.e., Rg and Rb, since they are always too liberal). Columns: D - distribution ($N$ - normal, $t_4$ - t-Student, $\chi_4^2$ - $\chi_4^2$-squared), HH - homoscedasticity or heteroscedasticity (hom - homoscedasticity, pos - positive pairing, neg - negative pairing), $\mathbf{n}=(20,30,40,50)$ - vector of sample sizes, $d$ - hyperparameter in scenario for alternative hypothesis. \textit{(continued)}}\\
\hline
D & $\rho$ & $d$ & HH & W & LH & P & R & Wg & LHg & Pg & Wb & LHb & Pb & ZN & ZB & Z & GPH & mGPH & SPH & mSPH\\
\hline
\endhead
$N$ & 0.1 & 0.1 & hom & 26.0 & 26.0 & 26.2 & 23.3 & 65.0 & 68.9 & 61.1 & 35.3 & 37.9 & 32.6 & 26.9 & 31.4 & 30.3 & 20.2 & 20.4 & 73.4 & 79.0\\
&&& pos & 7.9 & 7.8 & 7.7 & 9.9 & 37.3 & 39.9 & 32.5 & 11.6 & 13.0 & 10.1 & 19.8 & 22.5 & 18.8 & 12.6 & 12.6 & 37.7 & 46.0\\
&&& neg & 24.9 & 24.8 & 25.2 & 17.9 & 57.2 & 59.6 & 53.4 & 27.8 & 29.7 & 25.3 & 18.3 & 22.6 & 19.4 & 16.8 & 18.0 & 55.3 & 63.4\\
& 0.3 & 0.2 & hom & 33.6 & 33.4 & 33.5 & 32.2 & 24.7 & 28.0 & 21.7 & 22.5 & 25.7 & 20.3 & 36.8 & 45.4 & 37.2 & 27.0 & 29.6 & 73.7 & 81.0\\
&&& pos & 10.7 & 10.8 & 10.5 & 16.0 & 9.2 & 11.1 & 7.7 & 6.6 & 8.1 & 5.6 & 24.3 & 30.5 & 25.5 & 13.7 & 16.2 & 35.2 & 45.9\\
&&& neg & 37.1 & 36.8 & 36.9 & 27.9 & 25.9 & 27.7 & 23.5 & 24.6 & 26.6 & 23.1 & 22.7 & 30.9 & 24.4 & 20.6 & 20.4 & 56.1 & 63.1\\
& 0.5 & 0.3 & hom & 44.2 & 44.7 & 43.9 & 45.8 & 15.0 & 16.7 & 13.4 & 24.9 & 27.7 & 22.9 & 44.9 & 57.1 & 47.1 & 32.4 & 38.8 & 64.7 & 74.2\\
&&& pos & 14.5 & 14.8 & 14.2 & 22.6 & 5.1 & 5.9 & 3.7 & 8.6 & 10.1 & 7.3 & 29.0 & 37.7 & 29.1 & 17.5 & 20.0 & 32.7 & 40.1\\
&&& neg & 46.6 & 46.7 & 46.2 & 38.1 & 21.1 & 23.6 & 19.3 & 25.9 & 27.5 & 24.0 & 24.7 & 39.4 & 27.3 & 24.1 & 25.3 & 47.8 & 57.8\\
& 0.7 & 0.4 & hom & 46.7 & 47.2 & 46.0 & 56.3 & 10.4 & 11.9 & 8.7 & 28.6 & 31.4 & 26.0 & 44.4 & 62.1 & 44.2 & 30.0 & 36.5 & 49.2 & 58.2\\
&&& pos & 15.2 & 15.6 & 14.8 & 27.2 & 4.4 & 5.2 & 3.4 & 11.9 & 14.2 & 10.2 & 29.0 & 42.9 & 29.1 & 14.9 & 19.1 & 24.3 & 30.0\\
&&& neg & 56.7 & 57.2 & 56.5 & 48.7 & 20.9 & 22.2 & 19.6 & 32.2 & 33.1 & 29.9 & 22.3 & 43.6 & 24.9 & 21.0 & 25.2 & 34.8 & 39.6\\
& 0.9 & 0.5 & hom & 40.2 & 40.6 & 39.9 & 50.0 & 7.9 & 9.8 & 7.1 & 31.8 & 35.1 & 29.5 & 30.3 & 55.8 & 31.5 & 17.2 & 23.2 & 27.7 & 33.9\\
&&& pos & 7.6 & 7.8 & 7.7 & 22.5 & 3.4 & 4.2 & 2.6 & 14.4 & 16.6 & 12.5 & 19.8 & 39.0 & 19.9 & 8.2 & 11.5 & 15.3 & 19.3\\
&&& neg & 55.2 & 55.3 & 55.0 & 45.4 & 19.8 & 22.4 & 18.2 & 33.7 & 35.6 & 30.9 & 15.9 & 39.9 & 18.0 & 12.6 & 13.5 & 21.3 & 24.8\\\addlinespace
$t_4$ & 0.1 & 0.1 & hom & 28.5 & 28.9 & 28.2 & 25.5 & 65.2 & 68.0 & 60.8 & 37.5 & 40.5 & 34.0 & 27.8 & 36.1 & 28.4 & 22.9 & 22.3 & 78.8 & 85.4\\
&&& pos & 10.5 & 11.1 & 10.1 & 12.7 & 35.7 & 39.0 & 31.9 & 12.6 & 14.3 & 10.9 & 20.5 & 26.2 & 19.7 & 14.0 & 14.4 & 44.6 & 52.6\\
&&& neg & 26.8 & 26.7 & 27.5 & 22.2 & 54.6 & 57.0 & 51.8 & 31.4 & 33.5 & 28.3 & 19.0 & 26.7 & 18.8 & 19.0 & 20.0 & 65.9 & 72.3\\
& 0.3 & 0.2 & hom & 38.9 & 39.0 & 38.7 & 37.4 & 28.5 & 31.0 & 25.6 & 27.7 & 29.9 & 24.9 & 35.6 & 49.1 & 38.8 & 31.5 & 33.2 & 77.3 & 84.6\\
&&& pos & 12.5 & 12.6 & 12.0 & 16.8 & 10.5 & 12.3 & 8.9 & 8.8 & 10.4 & 7.5 & 22.8 & 31.0 & 23.6 & 14.0 & 16.6 & 41.8 & 50.9\\
&&& neg & 38.4 & 38.3 & 38.7 & 29.2 & 28.1 & 30.0 & 26.2 & 25.3 & 26.7 & 23.8 & 21.6 & 34.2 & 25.2 & 22.4 & 22.1 & 60.8 & 66.1\\
& 0.5 & 0.3 & hom & 43.5 & 43.7 & 43.1 & 46.4 & 15.1 & 17.6 & 12.7 & 27.5 & 29.1 & 25.0 & 40.1 & 55.7 & 45.0 & 32.4 & 35.6 & 67.3 & 76.1\\
&&& pos & 13.8 & 14.1 & 13.0 & 23.4 & 5.7 & 7.1 & 4.2 & 8.9 & 10.5 & 8.0 & 28.1 & 42.1 & 27.0 & 16.2 & 19.5 & 35.2 & 43.9\\
&&& neg & 49.4 & 49.4 & 49.0 & 40.3 & 22.8 & 25.1 & 20.8 & 30.2 & 31.8 & 28.0 & 21.2 & 42.5 & 29.4 & 26.2 & 27.1 & 55.4 & 62.5\\
& 0.7 & 0.4 & hom & 48.2 & 48.7 & 46.9 & 57.8 & 11.9 & 13.4 & 10.7 & 30.5 & 33.2 & 27.6 & 42.0 & 63.3 & 45.1 & 29.3 & 34.9 & 51.7 & 61.9\\
&&& pos & 13.6 & 13.9 & 13.2 & 26.3 & 3.0 & 4.3 & 2.3 & 10.3 & 11.4 & 8.1 & 27.2 & 44.5 & 24.9 & 13.8 & 16.7 & 22.9 & 28.3\\
&&& neg & 60.3 & 60.5 & 59.9 & 52.0 & 21.0 & 23.5 & 18.6 & 33.6 & 36.0 & 30.6 & 22.6 & 48.3 & 25.3 & 22.5 & 24.6 & 37.3 & 44.6\\
& 0.9 & 0.5 & hom & 42.9 & 43.4 & 42.0 & 52.5 & 9.9 & 10.8 & 8.4 & 34.8 & 38.7 & 31.6 & 29.2 & 55.5 & 28.6 & 17.3 & 21.4 & 27.5 & 34.7\\
&&& pos & 8.5 & 8.6 & 8.2 & 18.7 & 3.3 & 4.0 & 2.7 & 11.4 & 13.4 & 9.6 & 18.9 & 40.6 & 16.3 & 7.4 & 9.4 & 14.5 & 16.9\\
&&& neg & 56.0 & 56.1 & 56.2 & 48.5 & 19.9 & 21.7 & 17.7 & 35.7 & 37.3 & 33.9 & 15.9 & 43.0 & 16.9 & 11.4 & 14.3 & 22.2 & 26.6\\\addlinespace
$\chi_4^2$ & 0.1 & 0.1 & hom & 26.3 & 26.3 & 26.3 & 22.8 & 64.0 & 66.7 & 59.3 & 35.1 & 37.6 & 32.2 & 29.2 & 35.0 & 29.9 & 27.4 & 29.0 & 81.5 & 86.3\\
&&& pos & 9.2 & 9.6 & 9.1 & 12.8 & 34.3 & 37.0 & 30.6 & 11.0 & 12.9 & 9.7 & 18.6 & 22.1 & 19.4 & 15.7 & 15.5 & 46.4 & 55.6\\
&&& neg & 29.0 & 28.4 & 29.1 & 21.9 & 55.2 & 57.9 & 53.4 & 31.3 & 33.8 & 28.8 & 18.7 & 26.1 & 21.2 & 22.8 & 22.1 & 69.4 & 73.2\\
& 0.3 & 0.2 & hom & 34.6 & 35.2 & 34.4 & 32.8 & 26.9 & 29.8 & 24.7 & 25.2 & 26.8 & 22.1 & 36.4 & 44.9 & 38.7 & 32.4 & 35.7 & 77.4 & 84.4\\
&&& pos & 13.2 & 13.2 & 12.9 & 15.5 & 8.7 & 10.6 & 7.1 & 7.3 & 9.0 & 6.7 & 22.5 & 27.9 & 22.6 & 19.1 & 17.8 & 46.9 & 53.6\\
&&& neg & 41.2 & 41.3 & 41.5 & 32.1 & 29.0 & 31.3 & 26.7 & 27.1 & 28.8 & 25.5 & 23.1 & 34.1 & 28.5 & 29.4 & 30.4 & 66.5 & 71.6\\
& 0.5 & 0.3 & hom & 44.7 & 45.1 & 44.5 & 48.4 & 13.8 & 15.7 & 11.9 & 25.0 & 27.0 & 22.6 & 43.1 & 56.5 & 46.3 & 33.3 & 39.5 & 71.6 & 78.2\\
&&& pos & 13.0 & 13.0 & 12.9 & 21.9 & 5.7 & 6.4 & 4.5 & 7.1 & 8.2 & 6.0 & 29.3 & 41.1 & 26.5 & 17.5 & 20.4 & 37.8 & 46.2\\
&&& neg & 50.0 & 49.6 & 49.8 & 39.2 & 22.5 & 24.3 & 20.6 & 28.8 & 30.6 & 27.2 & 26.3 & 41.3 & 28.9 & 28.2 & 30.8 & 57.7 & 65.3\\
& 0.7 & 0.4 & hom & 48.2 & 49.1 & 47.5 & 55.9 & 12.1 & 13.0 & 10.8 & 31.7 & 33.9 & 28.4 & 42.2 & 63.2 & 45.5 & 30.5 & 37.3 & 50.8 & 59.4\\
&&& pos & 12.5 & 12.8 & 12.1 & 26.0 & 3.9 & 5.1 & 3.2 & 11.4 & 13.2 & 9.4 & 26.0 & 40.2 & 27.3 & 14.7 & 17.4 & 24.7 & 32.0\\
&&& neg & 56.6 & 56.6 & 56.7 & 48.4 & 21.2 & 23.5 & 19.0 & 33.5 & 34.9 & 31.5 & 26.3 & 48.8 & 27.2 & 22.6 & 27.8 & 39.0 & 47.1\\
& 0.9 & 0.5 & hom & 34.7 & 34.9 & 34.5 & 47.1 & 8.3 & 9.3 & 7.7 & 30.9 & 34.6 & 25.9 & 30.4 & 54.6 & 29.8 & 18.1 & 21.8 & 29.7 & 35.4\\
&&& pos & 8.8 & 8.8 & 8.5 & 19.4 & 3.0 & 3.8 & 2.5 & 11.5 & 13.2 & 9.0 & 18.5 & 37.6 & 17.8 & 9.2 & 9.3 & 15.1 & 17.1\\
&&& neg & 56.4 & 56.5 & 55.6 & 46.8 & 19.0 & 21.2 & 17.5 & 32.7 & 34.7 & 30.6 & 14.2 & 40.5 & 18.4 & 12.7 & 14.6 & 24.9 & 28.7\\
\hline
\end{longtable}
\endgroup{}
\end{landscape}

\begin{longtable}[t]{lllrrrrrrrrrr}
\caption[Empirical FWER in Model~1 for $k=4$ groups and $p=6$ functional variables]{\label{tab:unnamed-chunk-7}Empirical FWER (as percentages) in Model~1 for $k=4$ groups and $p=6$ functional variables. Columns: D - distribution ($N$ - normal, $t_4$ - t-Student, $\chi_4^2$ - chi-squared), HH - homoscedasticity or heteroscedasticity (hom - homoscedasticity, pos - positive pairing, neg - negative pairing), $\mathbf{n}=(20,30,40,50)$ - vector of sample sizes.}\\
\hline
D & $\rho$ & HH & W & Wg & Wb & ZN & ZB & Z & GPH & mGPH & SPH & mSPH\\
\hline
\endfirsthead
\caption[]{Empirical FWER (as percentages) in Model~1 for $k=4$ groups and $p=6$ functional variables. Columns: D - distribution ($N$ - normal, $t_4$ - t-Student, $\chi_4^2$ - chi-squared), HH - homoscedasticity or heteroscedasticity (hom - homoscedasticity, pos - positive pairing, neg - negative pairing), $\mathbf{n}=(20,30,40,50)$ - vector of sample sizes. \textit{(continued)}}\\
\hline
D & $\rho$ & HH & W & Wg & Wb & ZN & ZB & Z & GPH & mGPH & SPH & mSPH\\
\hline
\endhead
$N$ & 0.1 & hom & 4.0 & 2.1 & 2.4 & 9.6 & 13.3 & 9.6 & 3.8 & 4.4 & 3.6 & 4.2\\
&& pos & 2.8 & 1.5 & 0.9 & 11.4 & 14.1 & 11.4 & 5.0 & 6.0 & 4.7 & 5.5\\
&& neg & 10.6 & 8.0 & 6.7 & 10.1 & 14.8 & 10.1 & 4.9 & 5.7 & 4.6 & 5.4\\
& 0.3 & hom & 3.5 & 2.4 & 2.0 & 9.5 & 14.9 & 9.5 & 3.0 & 3.3 & 3.9 & 4.8\\
&& pos & 2.2 & 1.5 & 0.9 & 12.1 & 16.1 & 12.1 & 3.2 & 3.7 & 4.4 & 4.6\\
&& neg & 10.8 & 8.8 & 6.3 & 9.4 & 16.4 & 9.4 & 4.3 & 4.7 & 5.5 & 6.0\\
& 0.5 & hom & 4.3 & 4.5 & 2.8 & 6.0 & 15.4 & 6.0 & 3.4 & 4.1 & 4.4 & 4.5\\
&& pos & 2.5 & 1.8 & 1.4 & 7.6 & 14.8 & 7.6 & 3.7 & 4.2 & 5.1 & 5.8\\
&& neg & 12.4 & 10.6 & 7.3 & 10.0 & 21.5 & 10.0 & 2.1 & 2.8 & 3.5 & 4.4\\
& 0.7 & hom & 5.0 & 3.9 & 2.9 & 6.9 & 18.6 & 6.9 & 1.6 & 2.1 & 4.3 & 5.1\\
&& pos & 1.3 & 2.4 & 1.8 & 6.9 & 17.6 & 6.9 & 2.0 & 2.4 & 4.3 & 4.6\\
&& neg & 16.9 & 13.9 & 8.5 & 5.9 & 19.9 & 5.9 & 1.9 & 2.1 & 5.6 & 5.9\\
& 0.9 & hom & 4.1 & 4.5 & 2.9 & 3.7 & 19.0 & 3.7 & 0.8 & 1.4 & 5.0 & 5.5\\
&& pos & 2.1 & 2.0 & 1.6 & 4.4 & 17.2 & 4.4 & 1.2 & 1.6 & 5.6 & 6.1\\
&& neg & 19.7 & 13.1 & 10.5 & 4.2 & 24.1 & 4.2 & 1.1 & 1.1 & 5.0 & 5.9\\\addlinespace
$t_4$ & 0.1 & hom & 3.7 & 2.1 & 1.7 & 8.4 & 13.2 & 8.4 & 3.1 & 3.6 & 3.3 & 3.8\\
&& pos & 2.3 & 1.1 & 1.1 & 8.0 & 12.8 & 8.0 & 3.0 & 3.4 & 2.7 & 3.2\\
&& neg & 9.1 & 7.0 & 6.2 & 8.9 & 15.3 & 8.9 & 4.1 & 4.8 & 4.4 & 5.0\\
& 0.3 & hom & 4.9 & 2.5 & 1.8 & 10.0 & 17.0 & 10.0 & 3.2 & 4.1 & 4.5 & 5.2\\
&& pos & 2.3 & 1.1 & 0.9 & 9.2 & 15.6 & 9.2 & 3.4 & 3.6 & 4.6 & 4.9\\
&& neg & 11.3 & 7.1 & 6.4 & 8.5 & 17.7 & 8.5 & 3.0 & 3.8 & 3.3 & 3.9\\
& 0.5 & hom & 4.8 & 3.3 & 3.1 & 5.8 & 16.6 & 5.8 & 2.9 & 3.1 & 5.1 & 5.6\\
&& pos & 2.3 & 1.5 & 0.6 & 6.4 & 13.9 & 6.4 & 3.1 & 3.5 & 3.9 & 4.4\\
&& neg & 13.9 & 9.1 & 5.9 & 6.6 & 17.6 & 6.6 & 2.2 & 2.8 & 3.8 & 4.7\\
& 0.7 & hom & 5.0 & 3.1 & 2.4 & 4.7 & 18.4 & 4.7 & 1.9 & 2.3 & 5.2 & 5.7\\
&& pos & 1.8 & 1.9 & 0.8 & 4.4 & 16.3 & 4.4 & 1.4 & 1.7 & 4.4 & 4.6\\
&& neg & 17.6 & 11.3 & 7.5 & 5.3 & 21.8 & 5.3 & 1.3 & 2.0 & 4.8 & 5.6\\
& 0.9 & hom & 4.5 & 4.0 & 2.9 & 3.0 & 21.5 & 3.0 & 0.7 & 1.2 & 4.0 & 4.4\\
&& pos & 1.9 & 2.2 & 2.6 & 3.2 & 18.1 & 3.2 & 1.1 & 1.3 & 4.6 & 4.8\\
&& neg & 22.2 & 11.8 & 8.1 & 2.6 & 23.9 & 2.6 & 1.0 & 1.3 & 4.3 & 5.0\\\addlinespace
$\chi_4^2$ & 0.1 & hom & 5.3 & 2.8 & 2.4 & 10.0 & 14.6 & 10.0 & 4.9 & 5.2 & 5.1 & 6.2\\
&& pos & 2.8 & 1.3 & 0.6 & 10.1 & 13.1 & 10.1 & 2.8 & 3.5 & 4.1 & 4.2\\
&& neg & 9.2 & 5.3 & 4.9 & 9.5 & 16.4 & 9.5 & 4.4 & 5.4 & 4.5 & 5.1\\
& 0.3 & hom & 3.9 & 2.8 & 1.7 & 7.6 & 14.8 & 7.6 & 3.9 & 4.3 & 4.1 & 5.1\\
&& pos & 2.6 & 2.1 & 1.1 & 8.7 & 14.0 & 8.9 & 3.3 & 3.9 & 3.9 & 4.2\\
&& neg & 8.8 & 8.3 & 6.0 & 9.9 & 16.4 & 9.9 & 3.2 & 3.9 & 4.8 & 5.5\\
& 0.5 & hom & 5.1 & 3.5 & 2.6 & 8.7 & 17.8 & 8.7 & 3.0 & 3.6 & 3.4 & 3.8\\
&& pos & 2.4 & 2.1 & 0.7 & 9.5 & 16.3 & 9.5 & 2.7 & 3.5 & 4.3 & 5.0\\
&& neg & 13.5 & 11.7 & 5.9 & 8.1 & 19.4 & 8.1 & 2.7 & 3.7 & 4.9 & 5.9\\
& 0.7 & hom & 3.8 & 3.4 & 1.8 & 4.9 & 19.4 & 4.9 & 1.5 & 2.1 & 4.2 & 4.8\\
&& pos & 1.9 & 1.7 & 1.4 & 5.7 & 16.8 & 5.7 & 1.0 & 1.5 & 5.4 & 5.8\\
&& neg & 16.1 & 11.5 & 8.1 & 5.3 & 20.9 & 5.3 & 1.7 & 2.0 & 6.0 & 6.3\\
& 0.9 & hom & 3.8 & 4.2 & 3.2 & 4.0 & 20.9 & 4.0 & 0.9 & 1.2 & 5.1 & 6.0\\
&& pos & 1.6 & 2.3 & 1.7 & 4.0 & 18.7 & 4.0 & 0.7 & 0.9 & 4.1 & 4.4\\
&& neg & 23.3 & 13.7 & 9.4 & 3.2 & 25.2 & 3.2 & 0.3 & 0.7 & 4.8 & 5.9\\
\hline
\end{longtable}

\newpage

\begin{longtable}[t]{lllllrrrrrrrrrr}
\caption[Empirical power in Model~1 under local hypotheses for $k=4$ groups and $p=6$ functional variables]{\label{tab:unnamed-chunk-8}Empirical power (as percentages) in Model~1 under local hypotheses for $k=4$ groups and $p=6$ functional variables. Columns: D - distribution ($N$ - normal, $t_4$ - t-Student, $\chi_4^2$ - chi-squared), HH - homoscedasticity or heteroscedasticity (hom - homoscedasticity, pos - positive pairing, neg - negative pairing), $\mathbf{n}=(20,30,40,50)$ - vector of sample sizes, $d$ - hyperparameter in scenario for alternative hypothesis, Con - contrast.}\\
\hline
D & $\rho$ & $d$ & HH & Con & W & Wg & Wb & ZN & ZB & Z & GPH & mGPH & SPH & mSPH\\
\hline
\endfirsthead
\caption[]{Empirical power (as percentages) in Model~1 under local hypotheses for $k=4$ groups and $p=6$ functional variables. Columns: D - distribution ($N$ - normal, $t_4$ - t-Student, $\chi_4^2$ - chi-squared), HH - homoscedasticity or heteroscedasticity (hom - homoscedasticity, pos - positive pairing, neg - negative pairing), $\mathbf{n}=(20,30,40,50)$ - vector of sample sizes, $d$ - hyperparameter in scenario for alternative hypothesis, Con - contrast. \textit{(continued)}}\\
\hline
D & $\rho$ & $\delta$ & HH & Con & W & Wg & Wb & ZN & ZB & Z & GPH & mGPH & SPH & mSPH\\
\hline
\endhead
$N$ & 0.1 & 0.1 & hom & 1-4 & 6.3 & 27.3 & 9.7 & 8.1 & 11.4 & 13.9 & 5.2 & 6.0 & 27.1 & 28.8\\
&&&& 2-4 & 7.7 & 39.1 & 15.1 & 9.6 & 11.8 & 13.3 & 6.7 & 7.8 & 45.3 & 47.3\\
&&&& 3-4 & 10.9 & 52.6 & 21.5 & 15.1 & 16.3 & 16.5 & 9.7 & 10.2 & 61.6 & 63.7\\
&&& pos & 1-4 & 0.7 & 8.7 & 1.0 & 6.6 & 8.7 & 8.4 & 3.4 & 4.0 & 18.3 & 19.1\\
&&&& 2-4 & 1.8 & 16.7 & 3.8 & 7.1 & 8.1 & 7.9 & 3.9 & 4.3 & 21.0 & 22.1\\
&&&& 3-4 & 3.6 & 24.5 & 6.4 & 7.0 & 7.9 & 6.5 & 4.4 & 4.6 & 24.8 & 26.0\\
&&& neg & 1-4 & 10.0 & 28.6 & 13.2 & 6.7 & 10.5 & 11.8 & 3.2 & 3.5 & 9.7 & 10.6\\
&&&& 2-4 & 7.5 & 34.4 & 11.5 & 8.2 & 10.9 & 10.3 & 4.1 & 5.0 & 25.5 & 27.9\\
&&&& 3-4 & 8.5 & 43.9 & 15.9 & 12.6 & 14.7 & 12.4 & 8.0 & 9.0 & 48.3 & 50.3\\
& 0.3 & 0.2 & hom & 1-4 & 8.0 & 6.4 & 5.0 & 10.4 & 17.3 & 19.8 & 7.4 & 7.9 & 25.1 & 25.7\\
&&&& 2-4 & 11.4 & 10.2 & 8.7 & 15.6 & 22.0 & 18.6 & 9.7 & 11.1 & 47.0 & 48.5\\
&&&& 3-4 & 14.7 & 12.1 & 10.9 & 20.4 & 24.5 & 21.3 & 14.4 & 15.7 & 61.3 & 62.7\\
&&& pos & 1-4 & 0.4 & 0.5 & 0.3 & 6.8 & 10.5 & 10.6 & 4.1 & 4.7 & 17.2 & 18.5\\
&&&& 2-4 & 3.5 & 2.2 & 1.6 & 8.8 & 11.2 & 11.7 & 5.3 & 6.4 & 20.8 & 21.9\\
&&&& 3-4 & 5.3 & 3.6 & 3.2 & 9.9 & 13.2 & 9.6 & 6.1 & 6.5 & 23.9 & 25.4\\
&&& neg & 1-4 & 15.6 & 14.9 & 12.4 & 7.4 & 13.7 & 14.1 & 3.2 & 3.8 & 9.4 & 10.2\\
&&&& 2-4 & 12.8 & 11.4 & 7.9 & 9.5 & 14.2 & 15.4 & 6.6 & 7.5 & 25.6 & 26.5\\
&&&& 3-4 & 12.2 & 11.5 & 9.0 & 17.4 & 20.3 & 15.8 & 10.9 & 11.6 & 48.5 & 51.4\\
& 0.5 & 0.3 & hom & 1-4 & 12.5 & 2.4 & 4.9 & 14.1 & 25.6 & 23.3 & 9.3 & 10.5 & 22.3 & 23.9\\
&&&& 2-4 & 18.7 & 3.6 & 11.2 & 24.1 & 33.4 & 25.5 & 15.3 & 16.8 & 41.9 & 42.8\\
&&&& 3-4 & 22.3 & 5.7 & 11.6 & 30.6 & 38.1 & 33.8 & 20.0 & 21.8 & 55.6 & 56.2\\
&&& pos & 1-4 & 1.2 & 0.4 & 0.8 & 9.6 & 16.2 & 14.8 & 6.2 & 6.6 & 17.8 & 18.3\\
&&&& 2-4 & 4.5 & 1.1 & 2.9 & 13.7 & 19.5 & 12.1 & 7.5 & 8.4 & 17.9 & 18.5\\
&&&& 3-4 & 7.3 & 1.9 & 3.6 & 10.8 & 15.8 & 14.4 & 7.0 & 8.1 & 19.2 & 20.0\\
&&& neg & 1-4 & 24.8 & 9.7 & 12.0 & 8.3 & 16.7 & 14.8 & 3.0 & 3.8 & 8.9 & 9.2\\
&&&& 2-4 & 18.1 & 6.0 & 10.4 & 15.9 & 23.4 & 17.0 & 8.4 & 9.1 & 21.5 & 22.5\\
&&&& 3-4 & 19.9 & 4.3 & 10.5 & 23.2 & 30.3 & 23.8 & 14.4 & 16.2 & 43.5 & 45.0\\
& 0.7 & 0.4 & hom & 1-4 & 13.1 & 1.7 & 5.1 & 10.7 & 29.1 & 22.2 & 6.0 & 7.4 & 14.2 & 14.7\\
&&&& 2-4 & 22.8 & 3.2 & 10.5 & 22.6 & 37.7 & 27.4 & 14.2 & 15.5 & 27.1 & 28.6\\
&&&& 3-4 & 28.7 & 3.1 & 13.9 & 34.2 & 46.3 & 32.4 & 22.0 & 24.4 & 39.2 & 40.8\\
&&& pos & 1-4 & 0.7 & 0.3 & 0.6 & 8.9 & 20.8 & 12.9 & 5.1 & 5.6 & 11.3 & 11.7\\
&&&& 2-4 & 4.2 & 0.7 & 2.2 & 11.8 & 19.7 & 13.5 & 5.8 & 6.9 & 11.3 & 11.9\\
&&&& 3-4 & 10.7 & 1.4 & 5.4 & 12.6 & 20.2 & 13.5 & 9.1 & 10.0 & 14.7 & 15.3\\
&&& neg & 1-4 & 31.9 & 10.9 & 15.5 & 6.6 & 23.8 & 15.6 & 2.3 & 3.0 & 6.5 & 7.3\\
&&&& 2-4 & 25.0 & 5.7 & 12.2 & 12.5 & 26.9 & 18.2 & 6.6 & 7.3 & 12.2 & 13.3\\
&&&& 3-4 & 26.1 & 2.9 & 11.6 & 24.4 & 35.4 & 24.4 & 15.5 & 17.6 & 26.0 & 27.5\\
& 0.9 & 0.5 & hom & 1-4 & 11.2 & 1.3 & 7.3 & 5.2 & 22.7 & 14.7 & 2.5 & 3.5 & 7.7 & 8.1\\
&&&& 2-4 & 16.0 & 1.6 & 10.8 & 14.8 & 32.9 & 16.9 & 7.3 & 8.6 & 12.0 & 12.2\\
&&&& 3-4 & 23.9 & 3.7 & 16.4 & 21.3 & 38.2 & 23.4 & 13.2 & 15.1 & 19.2 & 19.6\\
&&& pos & 1-4 & 0.3 & 0.1 & 1.3 & 5.6 & 17.7 & 8.3 & 3.4 & 3.7 & 5.4 & 5.7\\
&&&& 2-4 & 3.5 & 0.7 & 2.6 & 7.5 & 16.7 & 9.7 & 3.4 & 4.2 & 6.5 & 6.8\\
&&&& 3-4 & 7.0 & 0.6 & 5.9 & 9.1 & 17.4 & 10.0 & 4.9 & 5.1 & 6.6 & 6.8\\
&&& neg & 1-4 & 36.4 & 9.5 & 17.1 & 2.2 & 19.7 & 9.9 & 0.5 & 0.6 & 4.2 & 4.8\\
&&&& 2-4 & 21.4 & 4.2 & 11.0 & 8.0 & 24.5 & 12.2 & 3.0 & 3.3 & 7.0 & 7.3\\
&&&& 3-4 & 18.5 & 3.0 & 14.4 & 16.9 & 30.3 & 16.8 & 8.5 & 10.1 & 12.6 & 13.3\\\addlinespace
$t_4$& 0.1 & 0.1 & hom & 1-4 & 7.2 & 29.1 & 9.2 & 8.8 & 12.8 & 15.0 & 5.1 & 5.7 & 34.3 & 36.2\\
&&&& 2-4 & 9.5 & 41.7 & 15.3 & 11.8 & 16.9 & 13.7 & 8.3 & 9.3 & 53.4 & 55.2\\
&&&& 3-4 & 11.9 & 50.7 & 19.3 & 15.4 & 18.5 & 15.8 & 10.7 & 11.8 & 66.0 & 67.8\\
&&& pos & 1-4 & 0.3 & 10.2 & 0.8 & 6.9 & 10.9 & 9.8 & 4.3 & 4.7 & 22.7 & 24.0\\
&&&& 2-4 & 3.2 & 17.5 & 4.6 & 6.5 & 8.8 & 8.3 & 4.6 & 5.0 & 28.3 & 29.8\\
&&&& 3-4 & 4.0 & 24.5 & 7.0 & 7.7 & 9.6 & 7.5 & 4.8 & 5.2 & 30.1 & 31.5\\
&&& neg & 1-4 & 12.4 & 32.7 & 13.8 & 7.0 & 11.7 & 11.4 & 3.8 & 4.2 & 12.8 & 13.7\\
&&&& 2-4 & 11.0 & 37.2 & 13.5 & 10.3 & 14.6 & 11.5 & 5.6 & 6.9 & 31.4 & 33.6\\
&&&& 3-4 & 10.4 & 45.1 & 20.2 & 12.7 & 15.9 & 13.3 & 8.6 & 9.8 & 56.1 & 58.0\\
& 0.3 & 0.2 & hom & 1-4 & 10.2 & 6.8 & 7.3 & 11.5 & 21.9 & 18.9 & 7.7 & 8.8 & 30.3 & 31.7\\
&&&& 2-4 & 15.2 & 9.9 & 10.2 & 19.4 & 25.7 & 20.0 & 13.6 & 14.9 & 52.1 & 54.1\\
&&&& 3-4 & 19.2 & 13.3 & 11.9 & 24.4 & 31.3 & 25.5 & 17.1 & 18.0 & 65.3 & 66.8\\
&&& pos & 1-4 & 0.6 & 0.7 & 0.2 & 6.9 & 13.3 & 12.6 & 5.0 & 5.5 & 21.2 & 21.9\\
&&&& 2-4 & 3.7 & 2.8 & 2.9 & 10.4 & 15.0 & 10.4 & 6.0 & 6.8 & 24.6 & 26.0\\
&&&& 3-4 & 5.4 & 5.3 & 3.5 & 9.6 & 13.8 & 11.4 & 5.6 & 6.2 & 26.6 & 27.7\\
&&& neg & 1-4 & 17.3 & 12.4 & 10.7 & 7.0 & 17.6 & 16.2 & 3.2 & 4.0 & 11.8 & 12.5\\
&&&& 2-4 & 14.4 & 10.1 & 9.0 & 12.9 & 19.1 & 14.4 & 7.5 & 8.4 & 28.8 & 30.1\\
&&&& 3-4 & 15.7 & 9.9 & 9.8 & 18.0 & 24.1 & 19.8 & 11.2 & 12.7 & 50.4 & 52.0\\
& 0.5 & 0.3 & hom & 1-4 & 11.9 & 3.4 & 5.6 & 13.0 & 26.5 & 22.0 & 8.2 & 9.4 & 24.9 & 26.3\\
&&&& 2-4 & 18.2 & 5.2 & 9.4 & 22.1 & 31.6 & 27.8 & 14.1 & 15.5 & 41.0 & 42.5\\
&&&& 3-4 & 23.9 & 6.3 & 12.5 & 29.3 & 38.0 & 31.1 & 20.1 & 21.7 & 58.0 & 59.2\\
&&& pos & 1-4 & 1.9 & 0.2 & 0.7 & 9.8 & 18.4 & 15.0 & 6.2 & 6.7 & 16.3 & 17.1\\
&&&& 2-4 & 5.6 & 0.9 & 2.2 & 12.9 & 20.3 & 14.8 & 7.6 & 8.6 & 20.9 & 22.0\\
&&&& 3-4 & 9.4 & 2.1 & 3.0 & 13.1 & 19.0 & 12.9 & 7.3 & 8.2 & 21.5 & 22.5\\
&&& neg & 1-4 & 26.2 & 12.5 & 13.4 & 7.4 & 21.9 & 17.1 & 3.1 & 4.1 & 10.3 & 11.3\\
&&&& 2-4 & 22.6 & 7.3 & 11.6 & 14.5 & 25.8 & 19.8 & 9.2 & 10.5 & 24.9 & 26.9\\
&&&& 3-4 & 21.3 & 6.8 & 11.6 & 25.2 & 35.1 & 24.9 & 14.8 & 16.5 & 46.6 & 48.2\\
& 0.7 & 0.4 & hom & 1-4 & 14.5 & 1.9 & 7.7 & 11.0 & 30.4 & 20.8 & 6.2 & 7.2 & 14.9 & 15.9\\
&&&& 2-4 & 23.4 & 3.2 & 11.1 & 25.0 & 40.8 & 27.7 & 16.0 & 17.8 & 30.4 & 31.3\\
&&&& 3-4 & 28.3 & 3.4 & 16.8 & 31.9 & 46.5 & 31.6 & 19.8 & 21.7 & 41.5 & 42.8\\
&&& pos & 1-4 & 0.5 & 0.1 & 0.8 & 6.9 & 18.6 & 11.9 & 4.6 & 5.0 & 10.0 & 10.3\\
&&&& 2-4 & 5.0 & 0.5 & 1.7 & 11.5 & 21.7 & 13.8 & 5.7 & 6.2 & 11.9 & 12.7\\
&&&& 3-4 & 10.9 & 1.7 & 4.7 & 13.1 & 23.4 & 11.0 & 7.7 & 8.5 & 13.2 & 13.5\\
&&& neg & 1-4 & 34.6 & 10.9 & 16.0 & 5.8 & 25.9 & 14.7 & 2.6 & 3.1 & 6.7 & 7.0\\
&&&& 2-4 & 26.5 & 4.5 & 12.2 & 12.2 & 29.3 & 16.5 & 6.8 & 7.8 & 14.1 & 15.2\\
&&&& 3-4 & 28.4 & 3.0 & 13.3 & 24.9 & 38.5 & 27.8 & 14.7 & 16.8 & 30.0 & 31.8\\
& 0.9 & 0.5 & hom & 1-4 & 7.9 & 1.2 & 6.3 & 6.4 & 25.3 & 12.7 & 2.4 & 3.1 & 7.0 & 7.8\\
&&&& 2-4 & 19.5 & 2.9 & 13.0 & 12.6 & 33.4 & 18.7 & 7.5 & 9.3 & 13.5 & 14.3\\
&&&& 3-4 & 23.4 & 1.8 & 18.4 & 23.8 & 41.3 & 24.4 & 13.2 & 14.7 & 18.4 & 19.3\\
&&& pos & 1-4 & 0.2 & 0.1 & 0.7 & 4.2 & 17.7 & 6.9 & 1.9 & 2.2 & 5.6 & 5.8\\
&&&& 2-4 & 3.6 & 0.6 & 3.2 & 7.4 & 19.4 & 7.2 & 2.9 & 3.4 & 5.7 & 5.9\\
&&&& 3-4 & 6.5 & 1.2 & 3.8 & 6.9 & 16.8 & 7.7 & 3.5 & 4.0 & 6.4 & 6.7\\
&&& neg & 1-4 & 37.4 & 9.1 & 17.0 & 2.6 & 23.7 & 8.6 & 0.9 & 1.0 & 3.9 & 4.3\\
&&&& 2-4 & 24.0 & 5.0 & 14.5 & 7.4 & 25.9 & 12.2 & 2.8 & 3.2 & 6.4 & 7.4\\
&&&& 3-4 & 19.9 & 3.3 & 14.4 & 17.7 & 33.4 & 18.0 & 9.8 & 11.1 & 15.2 & 16.0\\\addlinespace
$\chi_4^2$ & 0.1 & 0.1 & hom & 1-4 & 5.7 & 28.7 & 8.2 & 8.5 & 12.0 & 18.9 & 8.1 & 9.0 & 37.4 & 39.2\\
&&&& 2-4 & 7.4 & 37.9 & 14.4 & 12.4 & 15.8 & 13.9 & 9.9 & 10.9 & 52.6 & 54.2\\
&&&& 3-4 & 9.8 & 49.4 & 20.0 & 16.7 & 19.3 & 16.6 & 12.2 & 12.6 & 64.2 & 65.6\\
&&& pos & 1-4 & 0.2 & 9.1 & 1.7 & 5.6 & 8.3 & 8.7 & 4.8 & 5.1 & 23.0 & 24.9\\
&&&& 2-4 & 2.6 & 20.7 & 4.1 & 5.9 & 8.0 & 7.0 & 5.1 & 5.9 & 26.1 & 27.3\\
&&&& 3-4 & 4.2 & 22.5 & 6.7 & 7.4 & 9.3 & 7.8 & 4.8 & 5.2 & 27.3 & 28.8\\
&&& neg & 1-4 & 11.7 & 30.9 & 12.6 & 7.0 & 11.7 & 17.5 & 5.9 & 6.7 & 17.6 & 19.2\\
&&&& 2-4 & 9.7 & 36.8 & 14.6 & 9.6 & 12.6 & 13.9 & 6.8 & 7.7 & 33.7 & 35.2\\
&&&& 3-4 & 9.4 & 42.4 & 17.6 & 12.5 & 15.9 & 15.5 & 9.3 & 10.0 & 52.7 & 53.9\\
& 0.3 & 0.2 & hom & 1-4 & 6.6 & 4.4 & 4.8 & 12.3 & 20.0 & 20.8 & 10.5 & 11.2 & 33.9 & 35.0\\
&&&& 2-4 & 12.8 & 9.7 & 7.2 & 19.9 & 25.4 & 21.8 & 14.6 & 15.9 & 49.4 & 50.5\\
&&&& 3-4 & 15.6 & 15.4 & 11.1 & 22.8 & 27.9 & 26.0 & 16.7 & 17.2 & 61.9 & 63.7\\
&&& pos & 1-4 & 0.7 & 0.4 & 0.1 & 5.9 & 10.5 & 12.9 & 5.7 & 6.4 & 22.3 & 22.8\\
&&&& 2-4 & 3.3 & 1.8 & 1.8 & 7.4 & 11.1 & 10.5 & 5.3 & 6.0 & 21.5 & 22.5\\
&&&& 3-4 & 5.3 & 4.2 & 3.5 & 8.2 & 11.8 & 9.6 & 6.2 & 6.5 & 27.9 & 28.8\\
&&& neg & 1-4 & 20.5 & 16.0 & 14.4 & 8.9 & 17.7 & 21.1 & 6.8 & 7.7 & 16.2 & 17.9\\
&&&& 2-4 & 16.8 & 12.3 & 11.0 & 13.2 & 19.4 & 17.3 & 10.3 & 11.9 & 31.5 & 33.3\\
&&&& 3-4 & 13.7 & 11.3 & 10.0 & 18.3 & 22.8 & 19.4 & 12.9 & 14.2 & 52.1 & 54.8\\
& 0.5 & 0.3 & hom & 1-4 & 9.9 & 3.5 & 5.2 & 13.0 & 24.8 & 25.8 & 9.6 & 10.7 & 27.8 & 28.8\\
&&&& 2-4 & 18.0 & 4.5 & 7.7 & 25.5 & 34.4 & 26.7 & 16.2 & 17.6 & 44.9 & 46.0\\
&&&& 3-4 & 23.2 & 5.4 & 11.4 & 30.8 & 37.9 & 32.4 & 21.0 & 22.5 & 57.2 & 58.0\\
&&& pos & 1-4 & 0.6 & 0.2 & 0.2 & 8.5 & 17.1 & 15.4 & 5.8 & 6.6 & 18.8 & 19.3\\
&&&& 2-4 & 4.4 & 1.1 & 1.9 & 11.9 & 17.8 & 11.7 & 7.2 & 8.2 & 22.1 & 22.7\\
&&&& 3-4 & 8.2 & 2.4 & 3.9 & 12.7 & 18.1 & 13.3 & 8.1 & 8.5 & 21.1 & 22.2\\
&&& neg & 1-4 & 26.6 & 12.1 & 13.7 & 7.9 & 21.3 & 21.2 & 4.9 & 5.7 & 14.8 & 16.0\\
&&&& 2-4 & 23.2 & 7.4 & 11.5 & 15.0 & 23.4 & 18.5 & 10.7 & 12.0 & 27.0 & 28.9\\
&&&& 3-4 & 22.8 & 6.7 & 11.3 & 24.6 & 33.8 & 25.9 & 16.7 & 18.2 & 45.5 & 46.8\\
& 0.7 & 0.4 & hom & 1-4 & 12.7 & 1.3 & 5.4 & 12.4 & 31.1 & 23.7 & 8.7 & 9.5 & 17.1 & 17.8\\
&&&& 2-4 & 22.6 & 3.2 & 10.0 & 21.9 & 38.8 & 29.6 & 14.4 & 16.2 & 28.3 & 29.5\\
&&&& 3-4 & 28.5 & 3.1 & 16.3 & 32.1 & 44.9 & 33.3 & 20.6 & 23.1 & 38.7 & 39.2\\
&&& pos & 1-4 & 0.6 & 0.1 & 0.3 & 7.4 & 19.7 & 15.5 & 4.6 & 5.1 & 11.0 & 11.5\\
&&&& 2-4 & 4.0 & 0.9 & 2.6 & 13.1 & 21.1 & 13.8 & 7.9 & 8.7 & 15.2 & 15.7\\
&&&& 3-4 & 10.5 & 1.4 & 5.2 & 11.4 & 20.9 & 12.9 & 7.6 & 8.8 & 12.9 & 13.3\\
&&& neg & 1-4 & 32.6 & 9.5 & 15.1 & 5.7 & 25.9 & 18.6 & 2.5 & 2.9 & 8.4 & 9.1\\
&&&& 2-4 & 25.9 & 4.8 & 12.7 & 16.8 & 31.3 & 19.0 & 7.6 & 9.4 & 16.4 & 17.8\\
&&&& 3-4 & 28.3 & 5.3 & 16.3 & 27.7 & 39.5 & 27.4 & 17.3 & 19.3 & 30.4 & 31.8\\
& 0.9 & 0.5 & hom & 1-4 & 10.3 & 2.3 & 6.0 & 4.9 & 24.8 & 13.8 & 2.6 & 2.9 & 8.5 & 8.8\\
&&&& 2-4 & 15.1 & 1.6 & 11.2 & 14.2 & 32.3 & 18.3 & 7.5 & 8.2 & 13.1 & 13.8\\
&&&& 3-4 & 19.9 & 2.8 & 16.5 & 22.2 & 39.0 & 22.3 & 12.4 & 13.5 & 20.0 & 21.0\\
&&& pos & 1-4 & 0.2 & 0.2 & 0.6 & 3.7 & 19.0 & 9.2 & 2.6 & 3.4 & 5.4 & 5.5\\
&&&& 2-4 & 2.6 & 0.5 & 2.2 & 6.4 & 17.3 & 8.2 & 3.1 & 3.3 & 5.2 & 5.7\\
&&&& 3-4 & 8.4 & 1.8 & 4.5 & 6.9 & 17.6 & 7.7 & 3.8 & 4.0 & 7.0 & 7.3\\
&&& neg & 1-4 & 37.2 & 10.2 & 17.4 & 3.1 & 22.8 & 10.9 & 1.4 & 1.8 & 6.4 & 6.8\\
&&&& 2-4 & 22.9 & 4.9 & 14.0 & 8.5 & 23.1 & 12.1 & 3.8 & 4.2 & 8.7 & 9.2\\
&&&& 3-4 & 22.2 & 2.9 & 13.1 & 16.2 & 31.2 & 19.2 & 7.8 & 9.9 & 14.5 & 15.7\\
\hline
\end{longtable}

\newpage

\clearpage

\subsection{Results with Scaling
Function}\label{results-with-scaling-function}
\begin{figure}
\centering
\includegraphics[width=0.99\textwidth,
height=0.4\textheight]{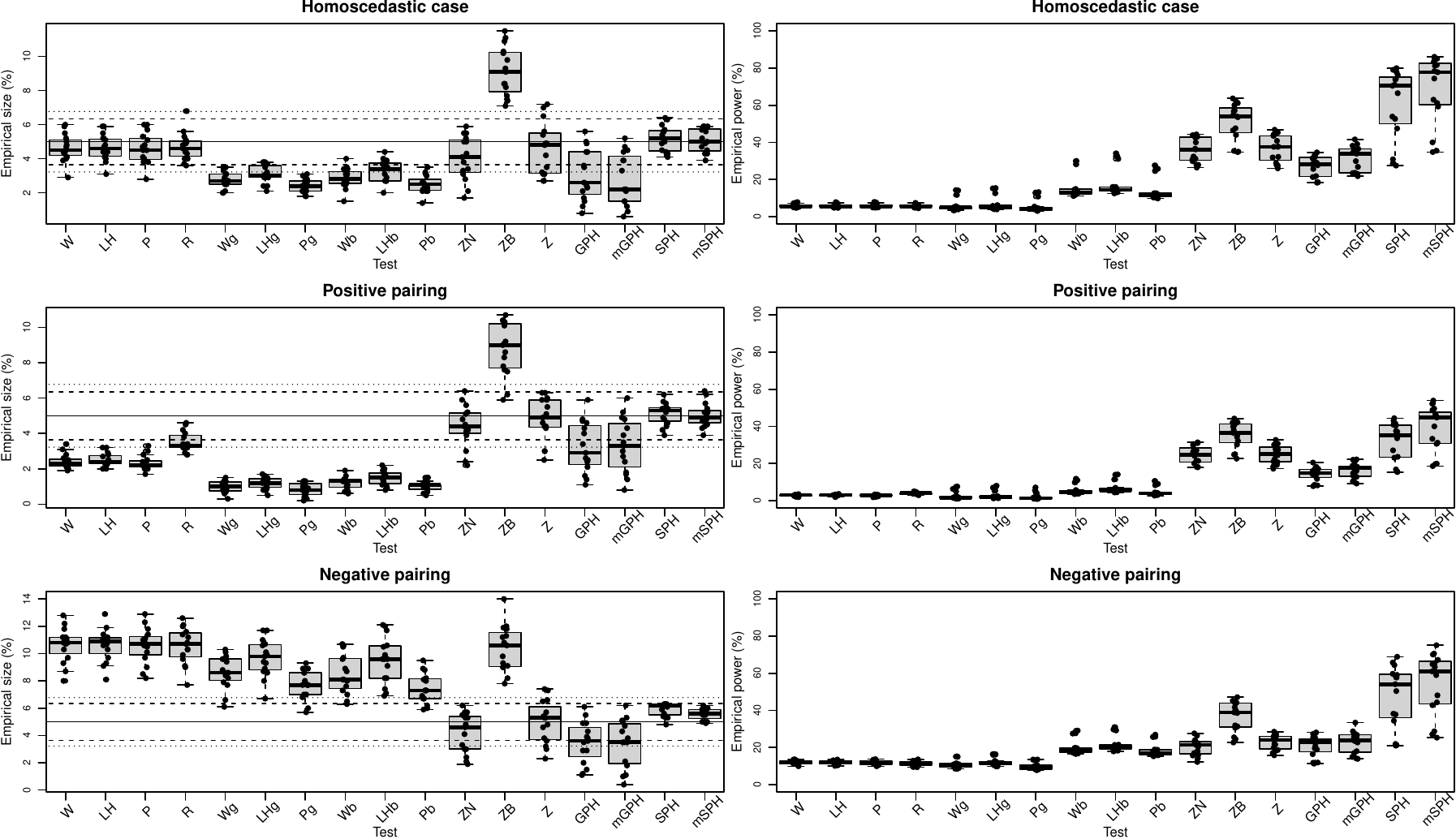}
\caption[Box-and-whisker plots for the empirical sizes and power in Model~1 under global hypothesis for $k=4$ groups and $p=6$ functional variables with scaling function and for homoscedastic case, positive pairing, and negative pairing.]{Box-and-whisker plots for the empirical sizes and power (as percentages) of all tests (without projection tests based on Roy test, i.e., Rg and Rb, since they are always too liberal) obtained in Model~1 under global hypothesis for $k=4$ groups and $p=6$ functional variables with scaling function and for homoscedastic case, positive pairing, and negative pairing.}
\label{fig_s_5}
\end{figure}

\begin{figure}
\centering
\includegraphics[width=0.99\textwidth,
height=0.8\textheight]{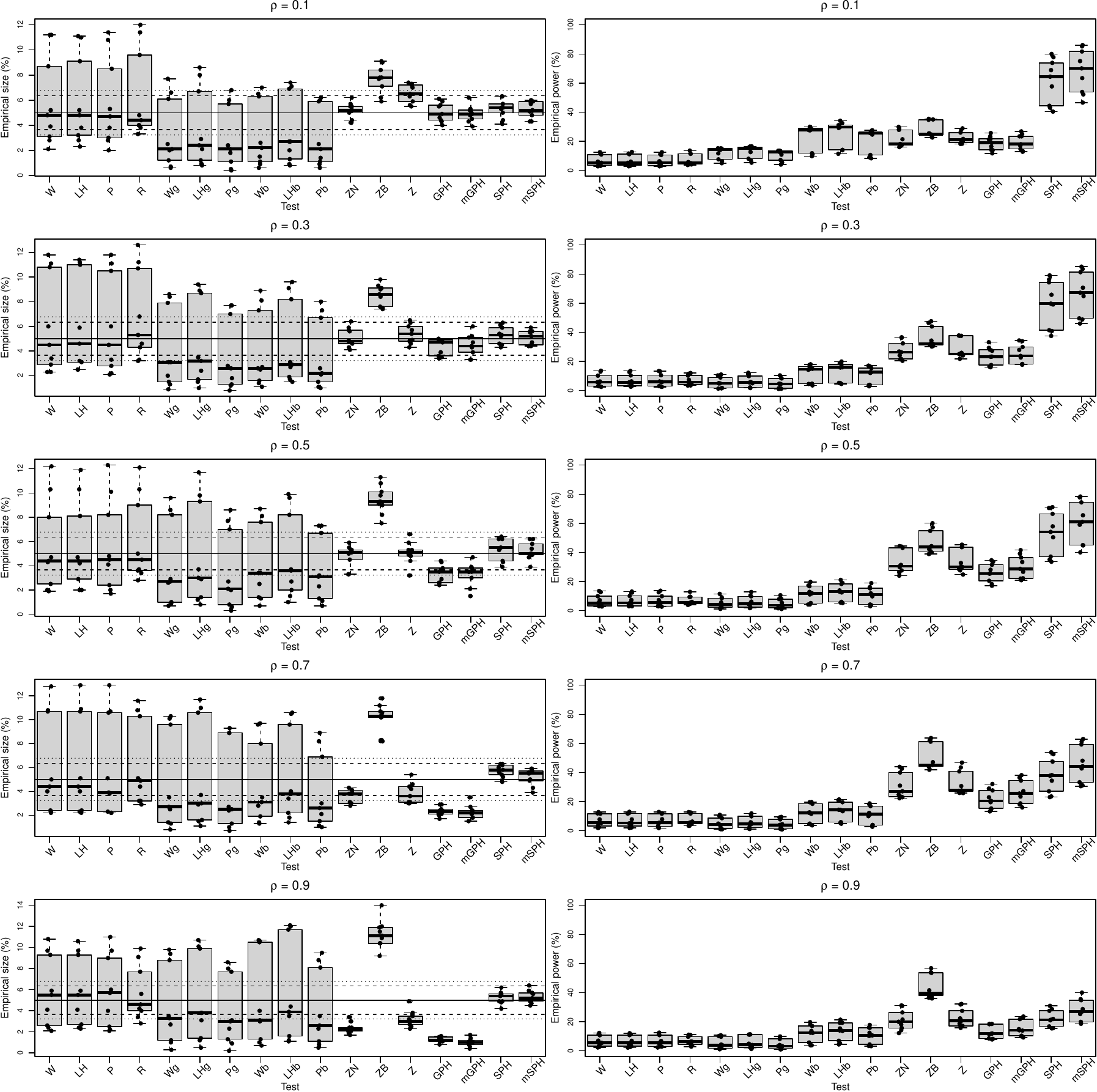}
\caption[Box-and-whisker plots for the empirical size and power in Model~1 under global hypothesis for $k=4$ groups and $p=6$ functional variables with scaling function and for different correlation]{Box-and-whisker plots for the empirical size and power (as percentages) of all tests (without projection tests based on Roy test, i.e., Rg and Rb, since they are always too liberal) obtained in Model~1 under global hypothesis for $k=4$ groups and $p=6$ functional variables with scaling function and for different correlation.}
\label{fig_s_6}
\end{figure}

\begin{figure}
\centering
\includegraphics[width=0.99\textwidth,
height=0.4\textheight]{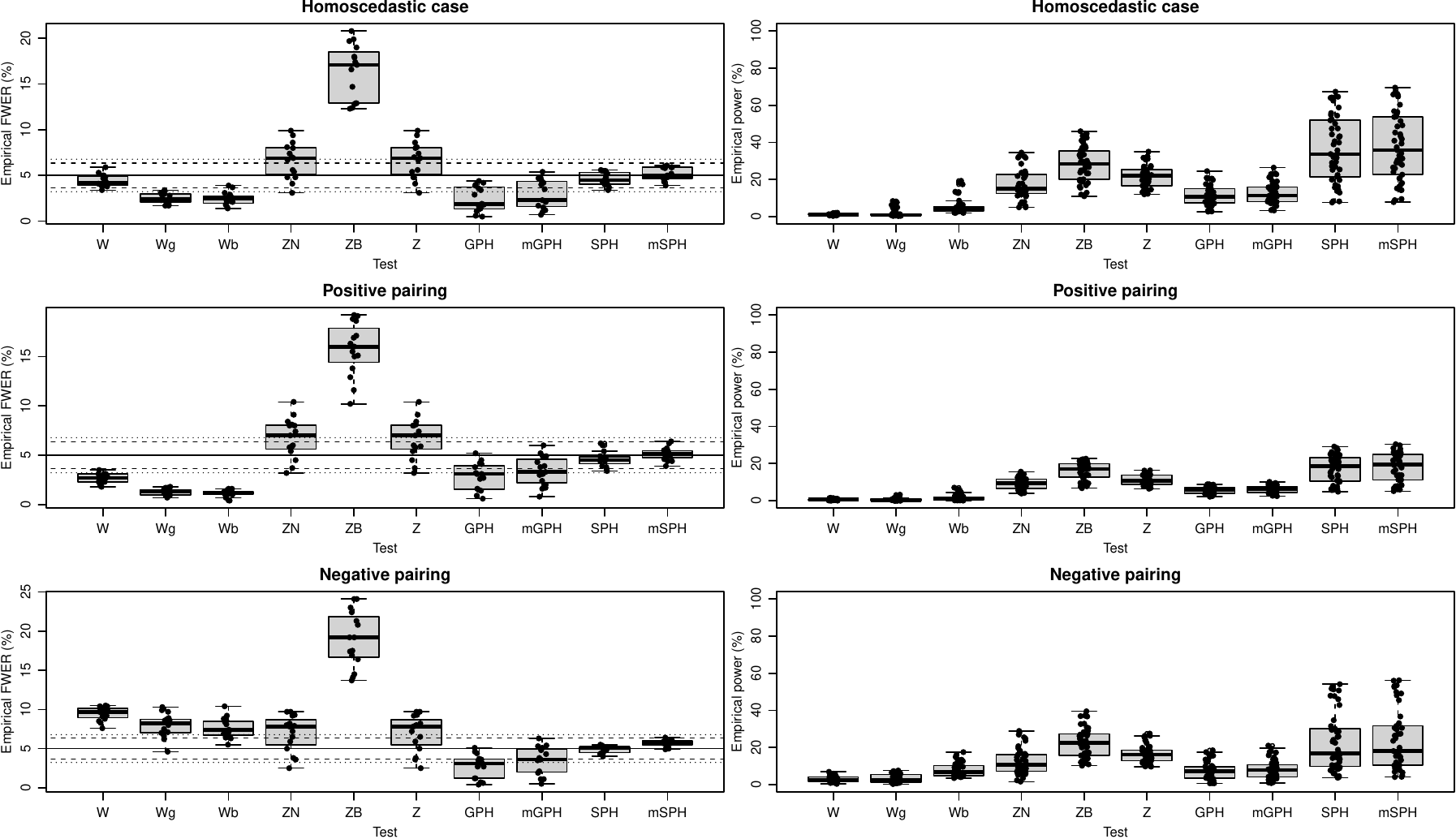}
\caption[Box-and-whisker plots for the empirical FWER and power in Model~1 under local hypotheses for $k=4$ groups and $p=6$ functional variables with scaling function and for homoscedastic case, positive pairing, and negative pairing]{Box-and-whisker plots for the empirical FWER and power (as percentages) of all tests obtained in Model~1 under local hypotheses for $k=4$ groups and $p=6$ functional variables with scaling function and for homoscedastic case, positive pairing, and negative pairing.}
\label{fig_s_7}
\end{figure}

\begin{figure}
\centering
\includegraphics[width=0.99\textwidth,
height=0.8\textheight]{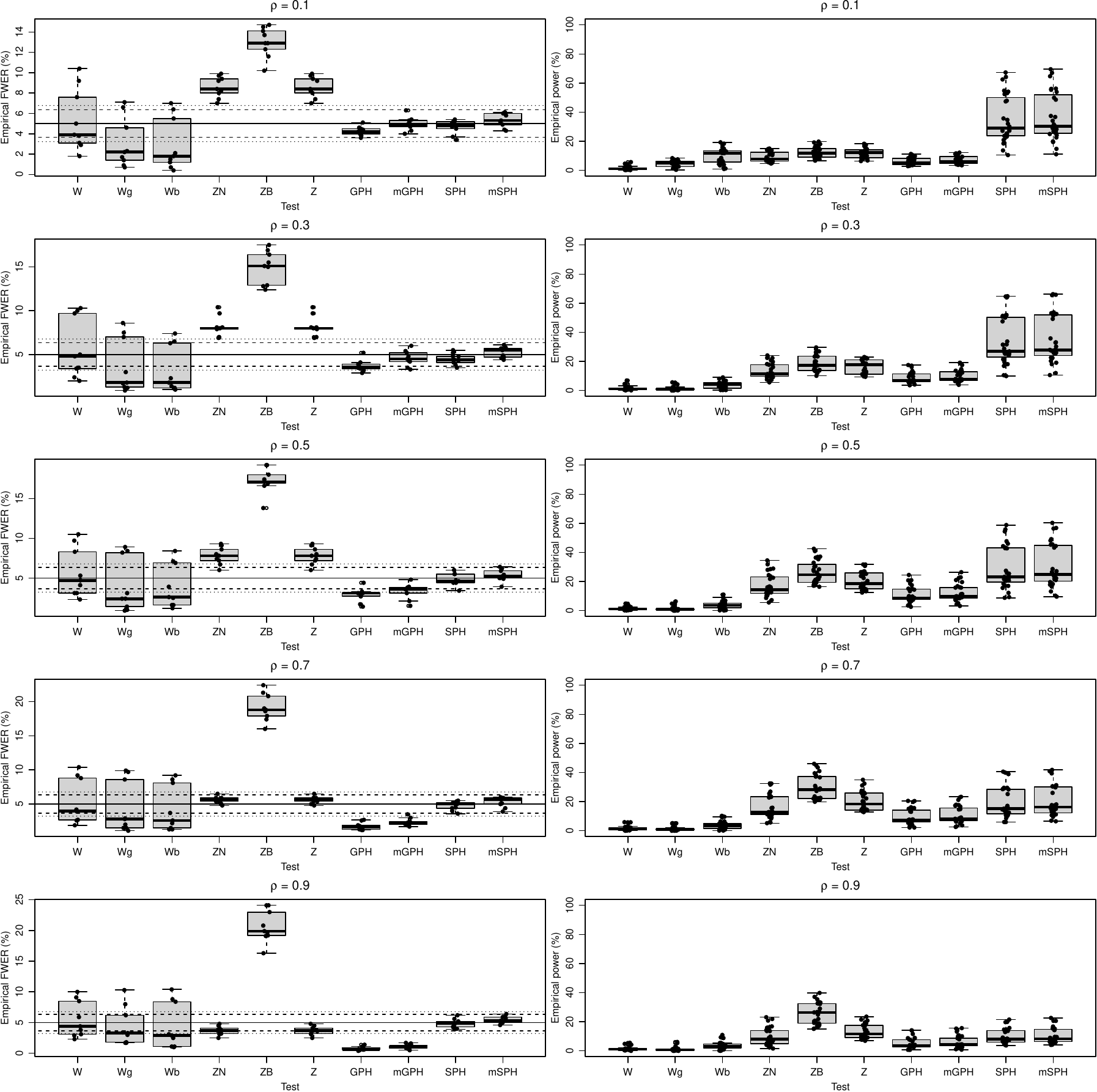}
\caption[Box-and-whisker plots for the empirical FWER and power in Model~1 under local hypotheses for $k=4$ groups and $p=6$ functional variables with scaling function and for different correlation]{Box-and-whisker plots for the empirical FWER and power (as percentages) of all tests obtained in Model~1 under local hypotheses for $k=4$ groups and $p=6$ functional variables with scaling function and for different correlation.}
\label{fig_s_8}
\end{figure}

Tables~\ref{tab:unnamed-chunk-12}-\ref{tab:unnamed-chunk-17} and Figures~\ref{fig_s_5}-\ref{fig_s_8} contain and summarize the results of simulation studies in Model~1 for the global and local hypotheses for $k=4$ samples and $p=6$ functional variables with scaling function.

\begin{landscape}\begingroup\fontsize{10}{12}\selectfont

\begin{longtable}[t]{lllrrrrrrrrrrrrrrrrrrr}
\caption[Empirical sizes in Model~1 under global hypothesis for $k=4$ groups and $p=6$ functional variables with scaling function]{\label{tab:unnamed-chunk-12}Empirical sizes (as percentages) in Model~1 under global hypothesis for $k=4$ groups and $p=6$ functional variables with scaling function. Columns: D - distribution ($N$ - normal, $t_4$ - t-Student, $\chi_4^2$ - chi-squared), HH - homoscedasticity or heteroscedasticity (hom - homoscedasticity, pos - positive pairing, neg - negative pairing), $\mathbf{n}=(20,30,40,50)$ - vector of sample sizes.}\\
\hline
D & $\rho$ & HH & W & LH & P & R & Wg & LHg & Pg & Rg & Wb & LHb & Pb & Rb & ZN & ZB & Z & GPH & mGPH & SPH & mSPH\\
\hline
\endfirsthead
\caption[]{Empirical sizes (as percentages) in Model~1 under global hypothesis for $k=4$ groups and $p=6$ functional variables with scaling function. Columns: D - distribution ($N$ - normal, $t_4$ - t-Student, $\chi_4^2$ - chi-squared), HH - homoscedasticity or heteroscedasticity (hom - homoscedasticity, pos - positive pairing, neg - negative pairing), $\mathbf{n}=(20,30,40,50)$ - vector of sample sizes. \textit{(continued)}}\\
\hline
D & $\rho$ & HH & W & LH & P & R & Wg & LHg & Pg & Rg & Wb & LHb & Pb & Rb & ZN & ZB & Z & GPH & mGPH & SPH & mSPH\\
\hline
\endhead
$N$ & 0.1 & hom & 4.8 & 4.8 & 4.7 & 4.4 & 2.5 & 2.9 & 2.4 & 24.5 & 2.6 & 2.7 & 2.5 & 26.0 & 5.0 & 7.1 & 7.0 & 4.9 & 4.5 & 5.0 & 5.2\\
&& pos & 2.8 & 2.8 & 2.8 & 4.0 & 0.6 & 0.9 & 0.4 & 20.9 & 0.9 & 0.9 & 0.9 & 21.9 & 4.4 & 6.2 & 5.9 & 4.6 & 4.8 & 5.4 & 5.1\\
&& neg & 11.2 & 11.0 & 11.4 & 11.4 & 7.7 & 8.6 & 6.8 & 42.9 & 7.0 & 7.4 & 6.2 & 42.7 & 5.5 & 7.8 & 7.4 & 5.5 & 5.3 & 5.4 & 5.9\\
& 0.3 & hom & 2.9 & 3.1 & 2.8 & 4.3 & 2.0 & 2.4 & 1.8 & 26.1 & 2.5 & 2.7 & 2.1 & 24.7 & 5.9 & 8.4 & 6.5 & 3.6 & 3.9 & 4.5 & 4.5\\
&& pos & 3.4 & 3.2 & 3.3 & 4.6 & 1.3 & 1.5 & 1.2 & 21.7 & 1.6 & 1.9 & 1.5 & 19.4 & 6.4 & 9.1 & 6.0 & 4.8 & 6.0 & 5.2 & 4.7\\
&& neg & 10.8 & 11.0 & 10.5 & 10.7 & 8.6 & 9.4 & 7.7 & 45.9 & 8.1 & 9.1 & 7.3 & 45.9 & 5.7 & 9.8 & 5.7 & 3.9 & 4.3 & 6.1 & 5.6\\
& 0.5 & hom & 4.3 & 4.4 & 4.1 & 5.0 & 2.7 & 3.0 & 2.0 & 30.6 & 3.4 & 3.7 & 3.1 & 30.2 & 5.5 & 9.8 & 4.8 & 4.4 & 4.7 & 6.4 & 5.8\\
&& pos & 2.0 & 2.0 & 2.0 & 2.8 & 1.0 & 1.4 & 0.8 & 20.1 & 1.3 & 1.5 & 1.1 & 20.6 & 5.9 & 10.1 & 5.9 & 4.3 & 3.5 & 6.2 & 6.2\\
&& neg & 8.0 & 8.1 & 8.2 & 9.0 & 8.2 & 9.8 & 7.0 & 47.8 & 7.6 & 8.2 & 6.7 & 48.2 & 5.3 & 10.8 & 6.6 & 3.6 & 3.5 & 5.3 & 6.2\\
& 0.7 & hom & 4.0 & 4.0 & 3.9 & 4.4 & 3.5 & 3.7 & 2.7 & 32.4 & 3.5 & 4.0 & 3.0 & 31.7 & 3.8 & 10.3 & 4.2 & 2.1 & 2.2 & 5.2 & 5.6\\
&& pos & 2.2 & 2.3 & 2.2 & 3.2 & 0.8 & 1.1 & 0.7 & 21.2 & 1.3 & 1.4 & 1.1 & 20.7 & 3.9 & 8.3 & 4.6 & 2.1 & 2.7 & 5.8 & 5.7\\
&& neg & 10.8 & 10.9 & 10.7 & 10.8 & 10.1 & 11.0 & 9.3 & 50.8 & 9.6 & 10.5 & 8.2 & 51.5 & 4.1 & 11.2 & 5.4 & 2.9 & 3.5 & 6.3 & 5.9\\
& 0.9 & hom & 5.5 & 5.5 & 6.0 & 4.0 & 3.3 & 3.8 & 3.0 & 32.8 & 3.0 & 3.5 & 2.3 & 34.3 & 3.4 & 11.1 & 3.5 & 1.5 & 1.2 & 4.9 & 4.7\\
&& pos & 2.4 & 2.4 & 2.1 & 4.2 & 1.0 & 1.2 & 0.9 & 22.8 & 0.7 & 1.2 & 0.5 & 23.0 & 2.2 & 9.2 & 4.9 & 1.6 & 1.4 & 4.9 & 5.2\\
&& neg & 10.8 & 10.6 & 11.0 & 9.9 & 9.8 & 10.7 & 8.6 & 53.5 & 10.6 & 12.0 & 8.1 & 56.4 & 2.1 & 14.0 & 3.8 & 1.2 & 0.4 & 4.8 & 5.7\\\addlinespace
$t_4$& 0.1 & hom & 5.2 & 5.2 & 5.3 & 4.8 & 2.1 & 2.4 & 2.1 & 22.7 & 2.2 & 2.7 & 2.1 & 23.2 & 5.2 & 7.7 & 5.6 & 4.4 & 3.9 & 4.1 & 4.3\\
&& pos & 3.1 & 3.2 & 3.0 & 3.8 & 1.2 & 1.2 & 1.1 & 18.1 & 1.1 & 1.3 & 1.1 & 19.4 & 5.2 & 7.8 & 6.3 & 5.9 & 4.9 & 5.7 & 4.8\\
&& neg & 11.2 & 11.1 & 10.8 & 12.0 & 6.6 & 8.0 & 6.0 & 40.4 & 6.5 & 7.1 & 6.0 & 40.4 & 5.7 & 9.0 & 7.3 & 4.3 & 5.0 & 5.3 & 6.0\\
& 0.3 & hom & 4.5 & 4.6 & 4.5 & 5.3 & 3.1 & 3.5 & 2.8 & 26.0 & 2.7 & 2.9 & 2.6 & 26.2 & 4.1 & 7.5 & 4.9 & 3.5 & 3.3 & 4.3 & 4.4\\
&& pos & 2.3 & 2.5 & 2.1 & 3.2 & 1.5 & 1.7 & 1.3 & 17.7 & 1.1 & 1.5 & 1.0 & 19.4 & 4.8 & 8.6 & 4.3 & 3.4 & 3.3 & 4.6 & 4.4\\
&& neg & 11.1 & 11.2 & 11.1 & 12.6 & 7.9 & 8.9 & 7.0 & 43.7 & 8.9 & 9.6 & 8.0 & 44.0 & 4.6 & 9.3 & 4.8 & 4.9 & 4.7 & 5.9 & 5.6\\
& 0.5 & hom & 4.7 & 4.7 & 4.8 & 3.6 & 2.9 & 3.7 & 2.7 & 29.2 & 3.4 & 3.6 & 3.2 & 27.1 & 3.3 & 9.1 & 3.2 & 2.4 & 2.1 & 4.4 & 5.0\\
&& pos & 2.5 & 2.9 & 2.4 & 4.5 & 0.7 & 0.8 & 0.6 & 21.3 & 1.4 & 2.0 & 1.3 & 20.3 & 5.1 & 9.0 & 4.4 & 2.9 & 3.9 & 3.9 & 4.7\\
&& neg & 10.3 & 10.3 & 10.1 & 10.3 & 8.6 & 9.3 & 7.7 & 49.0 & 8.7 & 9.9 & 7.3 & 46.5 & 3.3 & 8.2 & 5.3 & 3.5 & 3.5 & 6.3 & 4.9\\
& 0.7 & hom & 5.0 & 5.1 & 5.1 & 4.9 & 2.7 & 3.0 & 2.5 & 33.5 & 2.8 & 3.4 & 2.1 & 31.7 & 3.1 & 10.2 & 3.1 & 1.7 & 1.5 & 6.0 & 4.3\\
&& pos & 2.3 & 2.2 & 2.3 & 3.2 & 1.4 & 1.6 & 1.3 & 21.4 & 1.4 & 1.8 & 1.0 & 22.4 & 4.1 & 10.7 & 3.0 & 2.5 & 1.8 & 4.8 & 4.9\\
&& neg & 10.7 & 10.7 & 10.6 & 10.3 & 10.3 & 11.7 & 8.9 & 51.6 & 8.0 & 9.6 & 6.9 & 48.3 & 3.0 & 11.8 & 3.0 & 2.0 & 2.1 & 6.3 & 5.0\\
& 0.9 & hom & 5.9 & 5.9 & 5.7 & 5.6 & 3.5 & 3.8 & 3.1 & 31.1 & 3.1 & 3.9 & 2.6 & 33.9 & 1.7 & 10.9 & 2.7 & 0.8 & 0.6 & 5.7 & 5.9\\
&& pos & 2.1 & 2.3 & 2.2 & 3.4 & 0.3 & 0.5 & 0.2 & 23.2 & 1.3 & 1.6 & 1.1 & 23.7 & 2.4 & 10.4 & 2.5 & 1.1 & 1.7 & 4.2 & 4.5\\
&& neg & 9.3 & 9.3 & 9.0 & 7.7 & 9.4 & 10.1 & 8.0 & 51.1 & 10.5 & 12.1 & 9.5 & 50.7 & 1.9 & 12.0 & 2.3 & 1.1 & 1.1 & 5.6 & 5.0\\\addlinespace
$\chi_4^2$ & 0.1 & hom & 3.9 & 3.8 & 3.8 & 4.0 & 2.0 & 2.1 & 1.8 & 23.5 & 1.5 & 2.0 & 1.4 & 25.1 & 5.5 & 8.4 & 7.2 & 5.6 & 5.2 & 6.3 & 5.7\\
&& pos & 2.1 & 2.3 & 2.0 & 3.3 & 0.7 & 0.8 & 0.5 & 16.6 & 0.6 & 0.8 & 0.6 & 17.5 & 4.2 & 5.9 & 5.5 & 4.0 & 4.3 & 4.4 & 4.3\\
&& neg & 8.7 & 9.1 & 8.5 & 9.6 & 6.1 & 6.7 & 5.7 & 39.5 & 6.3 & 6.9 & 5.9 & 38.5 & 6.2 & 9.1 & 6.5 & 6.1 & 6.2 & 6.3 & 5.9\\
& 0.3 & hom & 6.0 & 5.9 & 6.0 & 6.8 & 3.1 & 3.2 & 2.6 & 28.7 & 2.6 & 3.1 & 2.2 & 26.0 & 4.3 & 7.4 & 5.4 & 5.0 & 4.4 & 5.4 & 5.9\\
&& pos & 2.3 & 2.5 & 2.2 & 3.3 & 0.9 & 1.0 & 0.8 & 18.6 & 1.4 & 1.7 & 1.1 & 18.8 & 5.6 & 7.6 & 6.3 & 4.7 & 5.2 & 5.3 & 5.2\\
&& neg & 11.8 & 11.4 & 11.8 & 11.2 & 8.4 & 8.7 & 7.7 & 46.2 & 7.3 & 8.2 & 6.7 & 45.4 & 4.8 & 9.0 & 4.6 & 4.9 & 5.1 & 6.3 & 5.5\\
& 0.5 & hom & 4.4 & 4.2 & 4.5 & 3.7 & 2.6 & 2.9 & 2.1 & 28.4 & 2.5 & 2.7 & 2.3 & 26.5 & 5.0 & 9.3 & 4.9 & 2.6 & 1.5 & 4.3 & 3.9\\
&& pos & 1.9 & 2.0 & 1.7 & 3.4 & 0.9 & 1.2 & 0.3 & 21.5 & 0.7 & 1.0 & 0.7 & 19.8 & 4.5 & 7.5 & 5.1 & 2.6 & 3.0 & 5.5 & 5.4\\
&& neg & 12.2 & 11.9 & 12.3 & 12.1 & 9.6 & 11.7 & 8.6 & 51.8 & 8.1 & 9.6 & 7.3 & 48.7 & 5.3 & 11.3 & 5.3 & 3.8 & 3.8 & 6.2 & 5.0\\
& 0.7 & hom & 4.4 & 4.4 & 3.8 & 5.1 & 2.5 & 2.9 & 2.4 & 29.5 & 3.1 & 3.8 & 2.6 & 29.5 & 2.8 & 8.2 & 3.1 & 2.3 & 2.2 & 5.6 & 5.9\\
&& pos & 2.4 & 2.4 & 2.2 & 2.9 & 1.3 & 1.5 & 1.0 & 23.4 & 1.9 & 2.2 & 1.5 & 20.3 & 4.3 & 10.3 & 4.4 & 2.4 & 2.4 & 5.4 & 3.9\\
&& neg & 12.8 & 12.9 & 12.9 & 11.6 & 9.6 & 10.6 & 8.9 & 52.2 & 9.7 & 10.6 & 8.9 & 52.5 & 3.0 & 10.6 & 3.6 & 2.9 & 1.8 & 6.2 & 5.5\\
& 0.9 & hom & 4.1 & 4.1 & 4.0 & 4.6 & 2.7 & 3.1 & 2.3 & 32.5 & 4.0 & 4.4 & 3.5 & 33.9 & 2.1 & 11.5 & 2.7 & 1.2 & 0.9 & 5.6 & 4.9\\
&& pos & 2.6 & 2.7 & 2.5 & 2.8 & 1.2 & 1.4 & 1.3 & 22.1 & 1.0 & 1.1 & 0.8 & 26.5 & 3.0 & 10.4 & 3.0 & 1.4 & 0.8 & 5.4 & 6.4\\
&& neg & 9.7 & 9.7 & 9.7 & 9.1 & 8.8 & 9.9 & 7.7 & 51.7 & 10.7 & 11.7 & 8.8 & 53.1 & 2.4 & 11.9 & 3.2 & 1.5 & 1.0 & 6.2 & 5.6\\
\hline
\end{longtable}
\endgroup{}
\end{landscape}

\newpage

\begin{landscape}\begingroup\fontsize{10}{12}\selectfont

\begin{longtable}[t]{llllrrrrrrrrrrrrrrrrr}
\caption[Empirical power in Model~1 under global hypothesis for $k=4$ groups and $p=6$ functional variables with scaling function]{\label{tab:unnamed-chunk-13}Empirical power (as percentages) in Model~1 under global hypothesis for $k=4$ groups and $p=6$ functional variables with scaling function (without projection tests based on Roy test, i.e., Rg and Rb, since they are always too liberal). Columns: D - distribution ($N$ - normal, $t_4$ - t-Student, $\chi_4^2$ - chi-squared), HH - homoscedasticity or heteroscedasticity (hom - homoscedasticity, pos - positive pairing, neg - negative pairing), $\mathbf{n}=(20,30,40,50)$ - vector of sample sizes, $\delta$ - hyperparameter in scenario for alternative hypothesis.}\\
\hline
D & $\rho$ & $\delta$ & HH & W & LH & P & R & Wg & LHg & Pg & Wb & LHb & Pb & ZN & ZB & Z & GPH & mGPH & SPH & mSPH\\
\hline
\endfirsthead
\caption[]{Empirical power (as percentages) in Model~1 under global hypothesis for $k=4$ groups and $p=6$ functional variables with scaling function (without projection tests based on Roy test, i.e., Rg and Rb, since they are always too liberal). Columns: D - distribution ($N$ - normal, $t_4$ - t-Student, $\chi_4^2$ - chi-squared), HH - homoscedasticity or heteroscedasticity (hom - homoscedasticity, pos - positive pairing, neg - negative pairing), $\mathbf{n}=(20,30,40,50)$ - vector of sample sizes, $\delta$ - hyperparameter in scenario for alternative hypothesis. \textit{(continued)}}\\
\hline
D & $\rho$ & $\delta$ & HH & W & LH & P & R & Wg & LHg & Pg & Wb & LHb & Pb & ZN & ZB & Z & GPH & mGPH & SPH & mSPH\\
\hline
\endhead
$N$ & 0.1 & 0.1 & hom & 5.2 & 5.2 & 5.3 & 5.7 & 14.1 & 15.2 & 12.8 & 30.0 & 32.5 & 27.6 & 29.9 & 34.8 & 26.0 & 21.3 & 23.4 & 73.9 & 81.7\\
&&& pos & 3.1 & 3.1 & 3.2 & 3.7 & 6.3 & 7.0 & 5.5 & 11.8 & 14.1 & 10.6 & 21.4 & 24.8 & 19.2 & 14.2 & 14.7 & 40.4 & 46.6\\
&&& neg & 10.9 & 11.1 & 10.6 & 11.6 & 15.0 & 16.0 & 13.5 & 27.5 & 29.2 & 25.7 & 17.0 & 22.7 & 18.2 & 16.4 & 17.0 & 57.7 & 63.4\\
& 0.3 & 0.2 & hom & 6.2 & 6.3 & 6.2 & 7.4 & 5.0 & 5.8 & 4.5 & 14.3 & 15.8 & 12.6 & 36.2 & 46.5 & 37.6 & 28.1 & 29.9 & 74.3 & 81.3\\
&&& pos & 3.0 & 3.2 & 2.8 & 3.5 & 1.5 & 1.7 & 1.2 & 4.5 & 5.1 & 3.9 & 26.8 & 32.2 & 25.9 & 16.0 & 18.0 & 37.5 & 46.0\\
&&& neg & 13.0 & 12.7 & 13.1 & 10.3 & 10.4 & 11.3 & 8.6 & 17.6 & 18.9 & 16.7 & 21.7 & 31.9 & 24.1 & 23.3 & 22.7 & 59.1 & 64.9\\
& 0.5 & 0.3 & hom & 4.7 & 4.7 & 4.9 & 4.8 & 4.3 & 4.8 & 3.5 & 11.1 & 12.5 & 9.8 & 43.1 & 54.8 & 45.3 & 31.5 & 36.5 & 66.5 & 74.4\\
&&& pos & 3.2 & 3.3 & 3.0 & 4.7 & 2.3 & 2.4 & 1.9 & 4.2 & 5.5 & 3.0 & 31.4 & 40.8 & 32.7 & 17.1 & 20.9 & 33.6 & 40.0\\
&&& neg & 13.5 & 13.1 & 13.7 & 12.9 & 11.5 & 12.8 & 10.6 & 19.7 & 21.1 & 18.9 & 24.0 & 38.9 & 28.5 & 25.4 & 26.7 & 50.4 & 59.0\\
& 0.7 & 0.4 & hom & 5.2 & 5.3 & 5.0 & 5.1 & 3.4 & 3.9 & 3.1 & 11.6 & 13.4 & 10.0 & 42.6 & 62.0 & 46.8 & 27.7 & 34.5 & 47.5 & 59.3\\
&&& pos & 2.3 & 2.2 & 2.2 & 3.4 & 1.5 & 2.2 & 1.3 & 4.6 & 5.4 & 3.9 & 31.0 & 44.2 & 27.4 & 15.3 & 18.0 & 23.6 & 31.2\\
&&& neg & 12.8 & 13.0 & 12.9 & 12.5 & 10.9 & 11.9 & 9.5 & 19.6 & 21.0 & 18.8 & 23.0 & 44.2 & 25.9 & 20.5 & 25.6 & 34.4 & 42.8\\
& 0.9 & 0.5 & hom & 5.5 & 5.5 & 5.6 & 6.3 & 3.9 & 4.4 & 3.2 & 12.5 & 14.0 & 10.3 & 31.2 & 53.6 & 32.2 & 18.5 & 23.0 & 27.5 & 35.5\\
&&& pos & 2.8 & 2.7 & 2.9 & 3.7 & 1.8 & 2.3 & 1.8 & 3.8 & 4.4 & 3.4 & 21.5 & 36.0 & 22.3 & 8.5 & 10.0 & 16.8 & 18.6\\
&&& neg & 11.5 & 11.3 & 11.4 & 10.9 & 9.7 & 11.2 & 8.2 & 18.1 & 19.3 & 15.8 & 15.7 & 41.6 & 17.0 & 11.8 & 13.9 & 21.0 & 25.3\\\addlinespace
$t_4$& 0.1 & 0.1 & hom & 5.1 & 4.9 & 5.2 & 5.3 & 11.7 & 12.6 & 10.8 & 29.8 & 34.1 & 25.8 & 27.9 & 35.4 & 28.9 & 21.8 & 23.4 & 77.8 & 85.2\\
&&& pos & 2.8 & 2.8 & 2.8 & 4.2 & 4.9 & 5.3 & 4.0 & 9.8 & 11.4 & 8.3 & 18.1 & 25.0 & 21.1 & 11.7 & 12.8 & 43.1 & 52.2\\
&&& neg & 11.5 & 11.6 & 11.2 & 11.4 & 15.2 & 16.4 & 13.5 & 27.7 & 29.7 & 25.7 & 16.0 & 25.5 & 19.9 & 18.9 & 18.0 & 64.4 & 70.1\\
& 0.3 & 0.2 & hom & 5.7 & 5.6 & 5.9 & 5.8 & 4.9 & 5.0 & 4.7 & 14.6 & 16.0 & 12.1 & 32.4 & 44.0 & 37.6 & 29.3 & 33.9 & 76.3 & 83.5\\
&&& pos & 2.5 & 2.6 & 2.5 & 4.1 & 1.3 & 1.9 & 1.1 & 3.5 & 4.5 & 2.9 & 26.3 & 34.3 & 24.5 & 16.4 & 17.7 & 40.9 & 49.6\\
&&& neg & 10.2 & 10.4 & 10.5 & 11.5 & 8.9 & 9.9 & 8.3 & 16.5 & 17.9 & 15.4 & 20.4 & 31.3 & 21.8 & 23.2 & 23.8 & 59.7 & 67.2\\
& 0.5 & 0.3 & hom & 5.2 & 5.3 & 5.4 & 5.5 & 4.1 & 4.4 & 3.7 & 11.9 & 13.1 & 10.9 & 44.0 & 60.1 & 44.1 & 33.6 & 39.1 & 71.1 & 77.8\\
&&& pos & 3.0 & 3.0 & 3.0 & 4.1 & 1.8 & 2.0 & 1.6 & 4.2 & 5.2 & 3.8 & 29.9 & 42.1 & 30.0 & 18.2 & 22.0 & 35.1 & 44.8\\
&&& neg & 9.9 & 10.1 & 9.9 & 9.5 & 8.5 & 9.8 & 7.9 & 16.8 & 18.3 & 15.3 & 26.4 & 44.3 & 24.8 & 25.5 & 28.7 & 54.1 & 61.0\\
& 0.7 & 0.4 & hom & 5.4 & 5.1 & 5.5 & 5.6 & 4.4 & 4.8 & 4.0 & 12.6 & 14.3 & 12.2 & 40.0 & 61.2 & 43.5 & 29.0 & 36.5 & 53.9 & 61.2\\
&&& pos & 1.9 & 1.9 & 1.8 & 3.4 & 1.0 & 1.2 & 1.0 & 4.9 & 5.9 & 3.9 & 24.8 & 42.2 & 27.7 & 13.3 & 16.0 & 23.1 & 30.6\\
&&& neg & 11.7 & 11.8 & 11.7 & 11.8 & 10.2 & 11.2 & 9.1 & 18.5 & 19.6 & 16.9 & 23.3 & 47.1 & 26.0 & 19.6 & 23.5 & 37.8 & 44.3\\
& 0.9 & 0.5 & hom & 7.0 & 6.9 & 6.7 & 6.3 & 3.4 & 4.0 & 3.2 & 13.7 & 15.4 & 12.4 & 26.4 & 53.9 & 27.5 & 18.3 & 21.8 & 28.6 & 34.8\\
&&& pos & 3.2 & 3.2 & 3.1 & 4.6 & 1.1 & 1.3 & 1.0 & 5.7 & 6.9 & 4.4 & 18.1 & 36.5 & 17.2 & 7.8 & 9.2 & 15.4 & 19.7\\
&&& neg & 10.7 & 10.8 & 10.8 & 9.7 & 10.2 & 11.2 & 9.5 & 19.7 & 21.4 & 17.8 & 12.2 & 39.7 & 15.8 & 11.3 & 14.1 & 21.3 & 27.1\\\addlinespace
$\chi_4^2$ & 0.1 & 0.1 & hom & 6.0 & 6.0 & 6.3 & 4.9 & 14.3 & 15.4 & 13.2 & 28.2 & 31.1 & 24.6 & 28.0 & 35.0 & 27.7 & 25.7 & 26.8 & 80.0 & 86.1\\
&&& pos & 3.5 & 3.6 & 3.3 & 4.1 & 7.6 & 8.0 & 7.1 & 11.0 & 13.9 & 9.0 & 18.0 & 22.6 & 18.3 & 13.9 & 13.3 & 44.4 & 53.9\\
&&& neg & 12.5 & 12.7 & 12.5 & 13.6 & 15.1 & 16.5 & 13.3 & 29.1 & 31.0 & 26.9 & 18.1 & 24.5 & 21.6 & 23.4 & 25.0 & 68.9 & 75.1\\
& 0.3 & 0.2 & hom & 5.7 & 5.4 & 5.8 & 4.5 & 4.9 & 5.5 & 4.1 & 14.9 & 16.0 & 12.6 & 36.6 & 47.6 & 37.8 & 33.2 & 34.4 & 79.1 & 85.1\\
&&& pos & 3.0 & 3.2 & 3.0 & 3.4 & 1.7 & 2.1 & 1.5 & 4.7 & 5.2 & 3.6 & 24.1 & 30.2 & 25.1 & 17.5 & 18.2 & 41.4 & 48.8\\
&&& neg & 13.6 & 13.5 & 13.4 & 12.0 & 11.0 & 12.2 & 10.4 & 18.0 & 19.9 & 17.1 & 21.4 & 30.6 & 25.0 & 27.3 & 27.9 & 65.8 & 70.6\\
& 0.5 & 0.3 & hom & 7.2 & 7.3 & 7.3 & 6.0 & 5.6 & 6.2 & 5.4 & 12.9 & 13.9 & 11.9 & 44.3 & 57.2 & 43.6 & 34.6 & 41.6 & 70.5 & 78.4\\
&&& pos & 2.8 & 3.0 & 2.7 & 4.0 & 1.2 & 1.7 & 1.1 & 5.2 & 6.0 & 4.2 & 30.5 & 39.9 & 30.1 & 20.5 & 22.2 & 37.1 & 45.0\\
&&& neg & 12.7 & 12.4 & 12.6 & 9.4 & 9.2 & 10.4 & 8.1 & 17.3 & 19.0 & 16.1 & 27.6 & 43.8 & 27.7 & 28.1 & 33.4 & 57.9 & 65.5\\
& 0.7 & 0.4 & hom & 7.8 & 7.8 & 7.9 & 6.3 & 4.9 & 5.8 & 4.5 & 12.4 & 14.4 & 11.3 & 43.9 & 63.7 & 40.8 & 32.1 & 38.2 & 52.7 & 63.0\\
&&& pos & 3.2 & 3.3 & 3.1 & 5.0 & 1.1 & 1.5 & 0.9 & 3.7 & 4.8 & 2.7 & 27.1 & 41.9 & 30.7 & 14.9 & 18.8 & 27.1 & 33.3\\
&&& neg & 11.9 & 12.1 & 12.1 & 12.8 & 8.6 & 9.9 & 7.8 & 19.7 & 21.5 & 18.0 & 22.3 & 45.2 & 27.1 & 23.0 & 27.2 & 38.2 & 48.1\\
& 0.9 & 0.5 & hom & 5.0 & 4.8 & 4.8 & 5.5 & 4.3 & 4.9 & 4.0 & 12.3 & 13.7 & 10.6 & 30.9 & 56.8 & 31.9 & 18.5 & 23.6 & 30.9 & 40.0\\
&&& pos & 2.3 & 2.2 & 2.3 & 2.9 & 1.4 & 1.8 & 1.2 & 3.8 & 5.2 & 3.2 & 20.0 & 38.1 & 20.7 & 8.2 & 10.4 & 15.6 & 19.9\\
&&& neg & 12.3 & 12.1 & 12.5 & 11.1 & 10.7 & 11.3 & 9.9 & 17.1 & 19.3 & 16.0 & 14.2 & 38.6 & 18.6 & 11.9 & 14.9 & 21.9 & 28.7\\
\hline
\end{longtable}
\endgroup{}
\end{landscape}

\begin{longtable}[t]{lllrrrrrrrrrr}
\caption[Empirical FWER in Model~1 for $k=4$ groups and $p=6$ functional variables with scaling function]{\label{tab:unnamed-chunk-16}Empirical FWER (as percentages) in Model~1 for $k=4$ groups and $p=6$ functional variables with scaling function. Columns: D - distribution ($N$ - normal, $t_4$ - t-Student, $\chi_4^2$ - chi-squared), HH - homoscedasticity or heteroscedasticity (hom - homoscedasticity, pos - positive pairing, neg - negative pairing), $\mathbf{n}=(20,30,40,50)$ - vector of sample sizes.}\\
\hline
D & $\rho$ & HH & W & Wg & Wb & ZN & ZB & Z & GPH & mGPH & SPH & mSPH\\
\hline
\endfirsthead
\caption[]{Empirical FWER (as percentages) in Model~1 for $k=4$ groups and $p=6$ functional variables with scaling function. Columns: D - distribution ($N$ - normal, $t_4$ - t-Student, $\chi_4^2$ - chi-squared), HH - homoscedasticity or heteroscedasticity (hom - homoscedasticity, pos - positive pairing, neg - negative pairing), $\mathbf{n}=(20,30,40,50)$ - vector of sample sizes. \textit{(continued)}}\\
\hline
D & $\rho$ & HH & W & Wg & Wb & ZN & ZB & Z & GPH & mGPH & SPH & mSPH\\
\hline
\endhead
$N$ & 0.1 & hom & 5.0 & 2.3 & 2.1 & 9.9 & 12.3 & 9.9 & 4.2 & 4.7 & 4.8 & 5.2\\
&& pos & 1.8 & 0.9 & 0.4 & 7.0 & 10.2 & 7.0 & 4.5 & 4.9 & 4.7 & 5.3\\
&& neg & 9.2 & 7.1 & 7.0 & 9.7 & 14.1 & 9.7 & 4.6 & 5.3 & 5.1 & 6.1\\
& 0.3 & hom & 3.4 & 1.8 & 1.4 & 6.9 & 12.9 & 6.9 & 3.4 & 4.2 & 4.1 & 4.6\\
&& pos & 3.5 & 1.3 & 1.2 & 10.4 & 15.5 & 10.4 & 5.2 & 6.0 & 4.4 & 4.9\\
&& neg & 9.7 & 8.6 & 6.5 & 9.7 & 16.9 & 9.7 & 3.5 & 4.3 & 4.3 & 5.7\\
& 0.5 & hom & 5.3 & 3.1 & 3.9 & 8.6 & 17.1 & 8.6 & 4.4 & 4.8 & 5.5 & 5.9\\
&& pos & 2.3 & 1.4 & 1.6 & 9.1 & 16.9 & 9.1 & 3.1 & 3.8 & 6.0 & 6.2\\
&& neg & 8.3 & 8.2 & 6.9 & 9.3 & 19.2 & 9.3 & 3.1 & 3.6 & 5.5 & 6.4\\
& 0.7 & hom & 3.9 & 2.8 & 3.7 & 5.4 & 19.0 & 5.4 & 1.9 & 2.3 & 5.0 & 5.9\\
&& pos & 2.6 & 1.5 & 1.3 & 5.8 & 16.0 & 5.8 & 2.6 & 3.0 & 5.4 & 5.7\\
&& neg & 10.4 & 9.9 & 8.6 & 6.5 & 20.8 & 6.5 & 2.7 & 3.5 & 5.2 & 6.0\\
& 0.9 & hom & 3.8 & 3.0 & 2.5 & 4.8 & 19.7 & 4.8 & 1.1 & 1.2 & 4.4 & 4.8\\
&& pos & 3.1 & 1.7 & 1.0 & 3.7 & 16.3 & 3.7 & 0.9 & 1.6 & 4.9 & 5.3\\
&& neg & 10.0 & 10.3 & 8.8 & 3.8 & 24.1 & 3.8 & 0.4 & 0.5 & 5.2 & 5.9\\\addlinespace
$t_4$& 0.1 & hom & 3.8 & 2.2 & 1.8 & 8.0 & 12.9 & 8.0 & 3.6 & 4.0 & 3.7 & 4.3\\
&& pos & 3.1 & 0.7 & 0.7 & 7.4 & 12.9 & 7.4 & 4.2 & 4.9 & 4.5 & 4.9\\
&& neg & 10.4 & 6.6 & 6.4 & 8.2 & 13.7 & 8.2 & 4.4 & 5.1 & 5.2 & 6.1\\
& 0.3 & hom & 4.8 & 1.7 & 1.8 & 7.0 & 12.4 & 7.0 & 2.9 & 3.5 & 3.5 & 4.7\\
&& pos & 2.4 & 0.9 & 1.0 & 8.1 & 15.1 & 8.1 & 3.2 & 3.3 & 3.8 & 4.4\\
&& neg & 10.3 & 7.5 & 6.3 & 7.9 & 17.5 & 7.9 & 3.6 & 4.8 & 5.2 & 5.8\\
& 0.5 & hom & 4.7 & 2.4 & 2.5 & 6.7 & 18.0 & 6.7 & 1.7 & 2.1 & 4.5 & 5.0\\
&& pos & 3.1 & 0.9 & 1.2 & 8.0 & 17.1 & 8.0 & 3.1 & 3.9 & 4.4 & 5.0\\
&& neg & 10.5 & 8.4 & 8.4 & 7.2 & 17.4 & 7.2 & 2.7 & 3.6 & 4.6 & 4.9\\
& 0.7 & hom & 4.2 & 2.9 & 2.6 & 4.8 & 17.9 & 4.8 & 1.3 & 1.7 & 4.0 & 4.4\\
&& pos & 1.9 & 1.5 & 1.3 & 5.4 & 18.8 & 5.4 & 1.5 & 2.0 & 4.5 & 5.1\\
&& neg & 8.8 & 9.7 & 8.1 & 5.0 & 21.3 & 5.0 & 1.2 & 2.1 & 4.4 & 5.0\\
& 0.9 & hom & 5.9 & 3.4 & 3.1 & 3.1 & 19.9 & 3.1 & 0.5 & 0.7 & 5.6 & 6.0\\
&& pos & 2.3 & 1.7 & 1.1 & 3.2 & 19.1 & 3.2 & 1.4 & 1.7 & 3.9 & 4.6\\
&& neg & 8.5 & 6.2 & 8.4 & 2.5 & 23.0 & 2.5 & 0.7 & 1.1 & 4.0 & 5.3\\\addlinespace
$\chi_4^2$ & 0.1 & hom & 3.9 & 1.7 & 1.5 & 9.4 & 14.7 & 9.4 & 4.0 & 5.4 & 5.2 & 5.8\\
&& pos & 2.9 & 1.4 & 1.2 & 8.4 & 11.6 & 8.4 & 3.8 & 4.3 & 3.4 & 4.4\\
&& neg & 7.6 & 4.6 & 5.5 & 9.2 & 14.5 & 9.2 & 5.1 & 6.3 & 5.4 & 6.0\\
& 0.3 & hom & 5.0 & 3.0 & 2.3 & 8.1 & 12.8 & 8.1 & 3.9 & 4.5 & 5.5 & 6.1\\
&& pos & 2.0 & 1.2 & 1.1 & 8.0 & 15.0 & 8.0 & 4.1 & 5.2 & 4.8 & 5.5\\
&& neg & 10.0 & 7.0 & 7.4 & 8.0 & 16.4 & 8.0 & 3.7 & 5.4 & 4.8 & 5.6\\
& 0.5 & hom & 4.1 & 2.4 & 2.6 & 7.4 & 16.6 & 7.4 & 1.4 & 1.5 & 3.4 & 3.9\\
&& pos & 3.1 & 1.0 & 1.6 & 6.0 & 13.8 & 6.0 & 2.7 & 3.1 & 4.7 & 5.4\\
&& neg & 9.7 & 8.9 & 7.1 & 7.8 & 19.2 & 7.8 & 3.3 & 3.8 & 4.4 & 5.2\\
& 0.7 & hom & 4.0 & 2.0 & 2.2 & 5.6 & 17.4 & 5.6 & 1.8 & 2.2 & 5.5 & 5.9\\
&& pos & 2.7 & 1.1 & 1.5 & 5.9 & 18.6 & 5.9 & 1.6 & 2.4 & 3.6 & 3.9\\
&& neg & 9.2 & 8.6 & 9.2 & 5.9 & 22.4 & 5.9 & 1.2 & 1.9 & 5.0 & 5.7\\
& 0.9 & hom & 4.4 & 3.3 & 2.9 & 4.1 & 20.8 & 4.1 & 0.6 & 1.0 & 4.3 & 5.0\\
&& pos & 2.9 & 1.8 & 1.1 & 4.5 & 19.2 & 4.5 & 0.6 & 0.8 & 6.2 & 6.4\\
&& neg & 9.1 & 8.0 & 10.4 & 3.6 & 24.1 & 3.6 & 0.6 & 1.1 & 5.1 & 5.7\\
\hline
\end{longtable}

\newpage

\begin{longtable}[t]{lllllrrrrrrrrrr}
\caption[Empirical power under local hypotheses in Model~1 for $k=4$ groups and $p=6$ functional variables with scaling function]{\label{tab:unnamed-chunk-17}Empirical power (as percentages) in Model~1 under local hypotheses for $k=4$ groups and $p=6$ functional variables with scaling function. Columns: D - distribution ($N$ - normal, $t_4$ - t-Student, $\chi_4^2$ - chi-squared), HH - homoscedasticity or heteroscedasticity (hom - homoscedasticity, pos - positive pairing, neg - negative pairing), $\mathbf{n}=(20,30,40,50)$ - vector of sample sizes, $\delta$ - hyperparameter in scenario for alternative hypothesis, Con - contrast.}\\
\hline
D & $\rho$ & $\delta$ & HH & Con & W & Wg & Wb & ZN & ZB & Z & GPH & mGPH & SPH & mSPH\\
\hline
\endfirsthead
\caption[]{Empirical power (as percentages) in Model~1 under local hypotheses for $k=4$ groups and $p=6$ functional variables with scaling function. Columns: D - distribution ($N$ - normal, $t_4$ - t-Student, $\chi_4^2$ - chi-squared), HH - homoscedasticity or heteroscedasticity (hom - homoscedasticity, pos - positive pairing, neg - negative pairing), $\mathbf{n}=(20,30,40,50)$ - vector of sample sizes, $\delta$ - hyperparameter in scenario for alternative hypothesis, Con - contrast. \textit{(continued)}}\\
\hline
D & $\rho$ & $\delta$ & HH & Con & W & Wg & Wb & ZN & ZB & Z & GPH & mGPH & SPH & mSPH\\
\hline
\endhead
$N$ & 0.1 & 0.1 & hom & 1-4 & 1.1 & 2.8 & 6.8 & 7.8 & 11.0 & 12.0 & 5.1 & 5.7 & 27.2 & 29.0\\
&&&& 2-4 & 1.2 & 5.0 & 13.4 & 13.8 & 15.5 & 12.3 & 8.5 & 9.3 & 47.4 & 48.6\\
&&&& 3-4 & 1.1 & 7.0 & 17.4 & 14.2 & 16.2 & 12.9 & 9.9 & 10.9 & 62.5 & 64.7\\
&&& pos & 1-4 & 0.2 & 0.9 & 1.0 & 5.5 & 7.8 & 8.7 & 3.9 & 4.3 & 18.5 & 19.5\\
&&&& 2-4 & 0.4 & 2.1 & 3.4 & 6.5 & 7.7 & 7.8 & 4.0 & 4.5 & 23.2 & 24.2\\
&&&& 3-4 & 1.0 & 3.2 & 7.0 & 7.8 & 8.7 & 6.4 & 4.8 & 5.8 & 25.4 & 26.9\\
&&& neg & 1-4 & 4.2 & 5.7 & 12.6 & 6.2 & 10.2 & 12.8 & 2.9 & 3.5 & 10.5 & 11.1\\
&&&& 2-4 & 1.7 & 5.3 & 11.9 & 9.0 & 11.3 & 9.7 & 4.9 & 5.7 & 24.1 & 26.4\\
&&&& 3-4 & 1.5 & 6.5 & 15.9 & 10.5 & 12.7 & 11.4 & 8.4 & 9.5 & 47.9 & 49.9\\
& 0.3 & 0.2 & hom & 1-4 & 1.2 & 1.0 & 2.8 & 12.7 & 18.8 & 18.4 & 7.5 & 8.1 & 26.9 & 27.9\\
&&&& 2-4 & 1.1 & 1.1 & 5.1 & 17.7 & 23.8 & 19.9 & 10.6 & 11.6 & 47.5 & 49.2\\
&&&& 3-4 & 1.1 & 1.3 & 4.6 & 22.4 & 26.3 & 22.0 & 16.2 & 17.5 & 64.0 & 65.5\\
&&& pos & 1-4 & 0.1 & 0.3 & 0.3 & 9.2 & 13.5 & 11.1 & 6.1 & 6.4 & 18.1 & 19.1\\
&&&& 2-4 & 0.6 & 0.3 & 1.3 & 10.2 & 13.7 & 9.7 & 6.0 & 7.1 & 21.2 & 22.4\\
&&&& 3-4 & 0.8 & 0.5 & 1.2 & 10.2 & 12.6 & 11.1 & 5.7 & 6.2 & 24.6 & 25.8\\
&&& neg & 1-4 & 3.7 & 5.0 & 8.1 & 7.3 & 14.1 & 13.8 & 4.1 & 4.6 & 9.9 & 10.6\\
&&&& 2-4 & 2.4 & 1.8 & 5.0 & 10.8 & 14.8 & 13.1 & 6.2 & 7.0 & 25.7 & 27.4\\
&&&& 3-4 & 1.8 & 0.9 & 5.0 & 19.7 & 23.4 & 18.1 & 11.9 & 13.5 & 51.3 & 53.1\\
& 0.5 & 0.3 & hom & 1-4 & 1.2 & 0.9 & 2.0 & 12.5 & 23.3 & 25.3 & 8.6 & 9.5 & 22.4 & 24.0\\
&&&& 2-4 & 0.8 & 0.7 & 3.6 & 21.8 & 30.1 & 24.3 & 15.1 & 16.0 & 42.8 & 43.3\\
&&&& 3-4 & 1.0 & 0.7 & 4.9 & 28.5 & 36.4 & 30.7 & 20.8 & 22.4 & 54.4 & 56.4\\
&&& pos & 1-4 & 0.2 & 0.1 & 0.3 & 10.2 & 17.1 & 16.3 & 6.5 & 7.1 & 15.6 & 16.5\\
&&&& 2-4 & 0.5 & 0.1 & 1.3 & 12.1 & 16.4 & 12.4 & 7.4 & 8.5 & 17.0 & 17.8\\
&&&& 3-4 & 0.5 & 0.8 & 2.0 & 14.3 & 19.4 & 13.5 & 7.6 & 8.7 & 20.5 & 21.0\\
&&& neg & 1-4 & 4.8 & 6.3 & 11.0 & 7.2 & 17.3 & 18.4 & 3.5 & 4.3 & 9.3 & 10.3\\
&&&& 2-4 & 2.5 & 2.3 & 5.7 & 16.6 & 24.7 & 18.4 & 8.5 & 9.6 & 22.5 & 23.6\\
&&&& 3-4 & 1.7 & 1.0 & 3.5 & 23.6 & 30.5 & 23.8 & 13.4 & 15.8 & 42.9 & 45.5\\
& 0.7 & 0.4 & hom & 1-4 & 0.5 & 0.7 & 2.5 & 11.9 & 29.0 & 23.5 & 6.3 & 8.1 & 15.7 & 16.3\\
&&&& 2-4 & 1.2 & 1.0 & 3.3 & 22.8 & 37.2 & 26.2 & 13.2 & 14.8 & 25.2 & 26.0\\
&&&& 3-4 & 1.2 & 0.8 & 5.0 & 32.4 & 43.5 & 35.0 & 19.8 & 21.2 & 38.9 & 39.4\\
&&& pos & 1-4 & 0.1 & 0.0 & 0.4 & 9.4 & 20.5 & 13.8 & 5.4 & 6.2 & 10.1 & 10.8\\
&&&& 2-4 & 0.3 & 0.2 & 1.2 & 11.6 & 21.5 & 14.1 & 6.6 & 7.1 & 12.5 & 13.4\\
&&&& 3-4 & 0.4 & 0.7 & 1.6 & 15.6 & 22.3 & 14.6 & 7.6 & 8.5 & 13.3 & 13.6\\
&&& neg & 1-4 & 5.9 & 5.2 & 10.2 & 5.7 & 23.6 & 13.8 & 2.2 & 3.0 & 6.0 & 6.9\\
&&&& 2-4 & 3.1 & 3.1 & 5.7 & 13.3 & 27.3 & 19.8 & 6.9 & 8.2 & 14.5 & 15.0\\
&&&& 3-4 & 1.7 & 1.0 & 4.5 & 25.5 & 36.7 & 26.3 & 15.1 & 17.6 & 28.4 & 29.8\\
& 0.9 & 0.5 & hom & 1-4 & 0.6 & 0.9 & 2.0 & 5.0 & 24.9 & 15.5 & 2.7 & 3.3 & 7.6 & 7.8\\
&&&& 2-4 & 0.9 & 0.7 & 3.2 & 15.9 & 33.4 & 18.8 & 7.8 & 8.8 & 13.7 & 14.3\\
&&&& 3-4 & 1.3 & 0.8 & 4.5 & 20.8 & 35.4 & 23.5 & 13.9 & 15.3 & 19.9 & 20.6\\
&&& pos & 1-4 & 0.0 & 0.0 & 0.2 & 3.8 & 15.1 & 9.1 & 2.7 & 3.2 & 5.7 & 5.9\\
&&&& 2-4 & 0.5 & 0.3 & 0.9 & 8.0 & 17.1 & 9.1 & 3.6 & 4.3 & 5.8 & 6.1\\
&&&& 3-4 & 1.0 & 0.7 & 2.1 & 8.8 & 17.9 & 9.5 & 3.6 & 4.2 & 7.3 & 7.7\\
&&& neg & 1-4 & 4.8 & 5.4 & 10.9 & 3.4 & 21.9 & 9.8 & 0.7 & 0.8 & 3.7 & 4.1\\
&&&& 2-4 & 1.1 & 1.2 & 4.5 & 9.0 & 26.7 & 11.6 & 3.1 & 3.5 & 7.7 & 7.8\\
&&&& 3-4 & 0.4 & 0.3 & 4.0 & 16.8 & 32.6 & 17.6 & 8.7 & 10.3 & 14.5 & 15.4\\\addlinespace
$t_4$& 0.1 & 0.1 & hom & 1-4 & 1.0 & 4.8 & 8.6 & 7.9 & 12.9 & 13.1 & 6.5 & 7.2 & 32.1 & 34.1\\
&&&& 2-4 & 0.8 & 4.7 & 13.1 & 14.1 & 19.2 & 16.7 & 8.0 & 9.9 & 52.4 & 53.9\\
&&&& 3-4 & 1.3 & 7.7 & 18.4 & 15.0 & 19.8 & 14.6 & 10.2 & 10.7 & 67.3 & 69.5\\
&&& pos & 1-4 & 0.3 & 1.3 & 1.3 & 5.2 & 8.1 & 8.7 & 3.2 & 3.8 & 20.9 & 22.6\\
&&&& 2-4 & 0.4 & 1.5 & 3.9 & 6.2 & 9.5 & 8.7 & 3.7 & 4.2 & 27.3 & 28.3\\
&&&& 3-4 & 0.5 & 2.6 & 5.4 & 7.4 & 9.3 & 7.8 & 4.5 & 5.0 & 29.1 & 30.4\\
&&& neg & 1-4 & 4.6 & 6.1 & 11.9 & 6.3 & 12.0 & 12.9 & 3.6 & 3.9 & 13.7 & 14.8\\
&&&& 2-4 & 2.0 & 5.6 & 12.0 & 8.3 & 10.9 & 9.9 & 5.1 & 5.9 & 30.1 & 31.6\\
&&&& 3-4 & 1.7 & 6.6 & 17.5 & 11.6 & 14.2 & 12.5 & 8.3 & 9.4 & 53.1 & 56.2\\
& 0.3 & 0.2 & hom & 1-4 & 1.1 & 1.2 & 2.9 & 11.8 & 20.1 & 21.2 & 7.0 & 7.9 & 31.7 & 32.9\\
&&&& 2-4 & 1.1 & 1.1 & 4.3 & 16.1 & 23.3 & 21.0 & 11.7 & 13.4 & 50.1 & 51.5\\
&&&& 3-4 & 2.0 & 1.3 & 6.1 & 22.3 & 28.5 & 23.0 & 17.1 & 19.1 & 64.6 & 66.2\\
&&& pos & 1-4 & 0.4 & 0.0 & 0.3 & 7.5 & 12.7 & 11.2 & 5.3 & 5.8 & 19.4 & 20.3\\
&&&& 2-4 & 0.7 & 0.3 & 1.0 & 9.0 & 13.6 & 10.8 & 6.3 & 7.0 & 24.4 & 25.8\\
&&&& 3-4 & 0.8 & 0.1 & 1.9 & 10.2 & 14.1 & 10.3 & 5.7 & 6.4 & 26.2 & 27.9\\
&&& neg & 1-4 & 4.5 & 4.2 & 7.7 & 5.6 & 14.3 & 12.4 & 3.6 & 3.9 & 10.5 & 12.0\\
&&&& 2-4 & 1.8 & 1.4 & 4.6 & 11.6 & 17.2 & 15.2 & 6.5 & 7.7 & 28.8 & 30.2\\
&&&& 3-4 & 0.9 & 1.3 & 6.0 & 20.7 & 26.9 & 18.8 & 13.5 & 14.7 & 51.7 & 53.5\\
& 0.5 & 0.3 & hom & 1-4 & 1.2 & 0.9 & 2.2 & 14.2 & 28.0 & 22.4 & 8.7 & 9.1 & 25.4 & 26.5\\
&&&& 2-4 & 1.1 & 1.2 & 4.3 & 24.4 & 35.3 & 26.4 & 15.0 & 16.8 & 44.3 & 45.3\\
&&&& 3-4 & 1.2 & 0.6 & 4.2 & 31.8 & 40.9 & 31.7 & 20.9 & 22.9 & 58.8 & 60.3\\
&&& pos & 1-4 & 0.3 & 0.0 & 0.2 & 10.4 & 19.4 & 14.7 & 7.1 & 7.6 & 17.3 & 18.3\\
&&&& 2-4 & 0.6 & 0.3 & 0.6 & 13.9 & 21.2 & 13.3 & 8.1 & 8.6 & 20.7 & 21.9\\
&&&& 3-4 & 1.4 & 0.6 & 2.0 & 14.9 & 21.6 & 13.3 & 8.8 & 9.5 & 23.2 & 24.8\\
&&& neg & 1-4 & 4.8 & 4.8 & 8.7 & 5.6 & 20.2 & 16.1 & 2.6 & 3.2 & 8.8 & 9.6\\
&&&& 2-4 & 1.6 & 1.6 & 5.4 & 16.1 & 26.2 & 15.2 & 8.5 & 9.9 & 23.2 & 25.0\\
&&&& 3-4 & 1.0 & 0.8 & 4.7 & 28.0 & 36.9 & 23.8 & 17.5 & 19.6 & 46.0 & 48.1\\
& 0.7 & 0.4 & hom & 1-4 & 1.3 & 0.5 & 2.8 & 10.9 & 28.4 & 22.2 & 6.1 & 7.1 & 16.4 & 17.5\\
&&&& 2-4 & 1.1 & 0.9 & 3.2 & 23.7 & 41.0 & 27.3 & 13.0 & 14.6 & 29.0 & 31.3\\
&&&& 3-4 & 1.8 & 1.1 & 6.4 & 32.5 & 46.0 & 32.4 & 20.5 & 23.3 & 40.6 & 41.8\\
&&& pos & 1-4 & 0.0 & 0.0 & 0.1 & 8.4 & 22.0 & 13.8 & 4.0 & 4.9 & 10.1 & 10.7\\
&&&& 2-4 & 0.3 & 0.1 & 1.0 & 11.3 & 22.0 & 13.1 & 6.1 & 6.9 & 10.8 & 11.3\\
&&&& 3-4 & 0.9 & 0.3 & 1.6 & 12.8 & 22.6 & 14.0 & 6.9 & 7.8 & 15.1 & 15.6\\
&&& neg & 1-4 & 5.4 & 5.0 & 9.6 & 5.2 & 22.3 & 14.9 & 2.2 & 2.6 & 6.0 & 6.5\\
&&&& 2-4 & 2.6 & 2.1 & 4.8 & 12.8 & 28.3 & 16.0 & 7.4 & 7.8 & 14.6 & 16.2\\
&&&& 3-4 & 1.3 & 0.8 & 4.9 & 24.8 & 37.5 & 25.6 & 15.6 & 17.3 & 28.6 & 30.6\\
& 0.9 & 0.5 & hom & 1-4 & 1.1 & 0.7 & 3.0 & 5.0 & 26.4 & 14.0 & 3.3 & 3.9 & 8.3 & 9.5\\
&&&& 2-4 & 1.1 & 0.3 & 3.3 & 14.2 & 32.2 & 16.7 & 7.5 & 8.7 & 12.9 & 14.3\\
&&&& 3-4 & 1.5 & 0.6 & 5.4 & 21.9 & 37.1 & 21.5 & 11.6 & 13.2 & 19.7 & 20.6\\
&&& pos & 1-4 & 0.4 & 0.0 & 0.2 & 4.1 & 15.3 & 7.7 & 2.1 & 2.4 & 6.3 & 6.3\\
&&&& 2-4 & 0.5 & 0.3 & 0.6 & 6.4 & 17.0 & 6.9 & 3.3 & 3.7 & 6.4 & 6.7\\
&&&& 3-4 & 1.5 & 0.8 & 2.6 & 7.9 & 18.8 & 7.8 & 4.1 & 4.5 & 8.0 & 8.3\\
&&& neg & 1-4 & 4.2 & 3.0 & 8.8 & 1.6 & 20.7 & 10.2 & 0.7 & 1.1 & 4.6 & 5.0\\
&&&& 2-4 & 2.1 & 1.2 & 4.7 & 8.7 & 26.5 & 11.1 & 3.6 & 4.2 & 8.3 & 8.7\\
&&&& 3-4 & 1.0 & 0.8 & 4.7 & 16.2 & 32.8 & 17.2 & 8.9 & 10.0 & 14.3 & 15.5\\\addlinespace
$\chi_4^2$ & 0.1 & 0.1 & hom & 1-4 & 0.4 & 2.6 & 7.4 & 6.7 & 11.9 & 15.1 & 7.1 & 7.6 & 35.7 & 37.4\\
&&&& 2-4 & 1.0 & 6.3 & 12.7 & 13.4 & 16.9 & 17.5 & 11.3 & 12.2 & 54.8 & 55.4\\
&&&& 3-4 & 1.8 & 8.4 & 19.3 & 14.8 & 18.1 & 15.1 & 10.8 & 11.3 & 64.2 & 67.0\\
&&& pos & 1-4 & 0.3 & 0.4 & 1.8 & 4.7 & 7.5 & 10.0 & 4.2 & 4.8 & 23.7 & 25.0\\
&&&& 2-4 & 0.4 & 2.8 & 4.3 & 6.6 & 8.2 & 8.4 & 4.8 & 5.3 & 25.0 & 25.9\\
&&&& 3-4 & 0.8 & 2.0 & 6.1 & 5.4 & 6.7 & 6.8 & 3.9 & 4.4 & 28.4 & 29.8\\
&&& neg & 1-4 & 5.7 & 7.6 & 13.6 & 7.0 & 12.2 & 18.3 & 7.1 & 7.7 & 18.8 & 19.9\\
&&&& 2-4 & 2.8 & 6.2 & 13.4 & 10.8 & 14.4 & 14.2 & 7.4 & 7.8 & 34.7 & 36.7\\
&&&& 3-4 & 1.1 & 7.3 & 14.9 & 14.8 & 17.0 & 14.3 & 10.1 & 11.7 & 54.1 & 56.0\\
& 0.3 & 0.2 & hom & 1-4 & 0.8 & 0.8 & 2.5 & 11.3 & 19.7 & 22.0 & 9.6 & 10.6 & 33.7 & 35.8\\
&&&& 2-4 & 0.6 & 0.8 & 4.3 & 17.8 & 24.9 & 22.4 & 11.1 & 12.3 & 51.9 & 53.7\\
&&&& 3-4 & 1.1 & 0.8 & 5.0 & 24.1 & 29.6 & 22.3 & 17.5 & 19.0 & 64.8 & 65.8\\
&&& pos & 1-4 & 0.1 & 0.0 & 0.3 & 6.5 & 10.1 & 11.9 & 6.4 & 6.8 & 21.7 & 22.6\\
&&&& 2-4 & 0.7 & 0.4 & 1.0 & 10.2 & 13.7 & 9.4 & 7.0 & 7.4 & 25.2 & 26.1\\
&&&& 3-4 & 0.9 & 1.0 & 2.6 & 10.2 & 12.0 & 10.6 & 6.2 & 7.0 & 25.0 & 25.4\\
&&& neg & 1-4 & 6.9 & 5.6 & 9.0 & 9.0 & 15.7 & 17.7 & 6.3 & 7.2 & 16.9 & 18.1\\
&&&& 2-4 & 3.2 & 2.5 & 5.4 & 14.6 & 20.0 & 18.6 & 9.5 & 10.7 & 31.5 & 33.1\\
&&&& 3-4 & 1.8 & 1.9 & 6.7 & 18.2 & 23.9 & 22.4 & 12.4 & 13.6 & 50.7 & 52.6\\
& 0.5 & 0.3 & hom & 1-4 & 1.0 & 1.1 & 2.5 & 13.8 & 28.4 & 27.0 & 10.0 & 11.3 & 27.4 & 28.6\\
&&&& 2-4 & 1.1 & 0.4 & 4.3 & 22.6 & 32.9 & 26.9 & 14.4 & 15.8 & 43.3 & 44.4\\
&&&& 3-4 & 0.8 & 1.0 & 5.0 & 34.5 & 42.5 & 31.7 & 24.5 & 26.4 & 55.8 & 56.8\\
&&& pos & 1-4 & 0.1 & 0.0 & 0.1 & 11.5 & 17.8 & 13.9 & 7.5 & 8.1 & 18.5 & 19.6\\
&&&& 2-4 & 0.6 & 0.1 & 0.8 & 11.6 & 18.0 & 16.2 & 8.8 & 10.1 & 21.2 & 21.7\\
&&&& 3-4 & 0.6 & 0.9 & 2.9 & 13.4 & 18.9 & 13.7 & 8.6 & 9.3 & 21.6 & 22.3\\
&&& neg & 1-4 & 3.6 & 4.1 & 9.2 & 8.7 & 22.0 & 18.7 & 4.6 & 5.1 & 12.2 & 13.2\\
&&&& 2-4 & 2.4 & 2.4 & 6.4 & 16.0 & 27.2 & 20.4 & 9.7 & 11.4 & 25.9 & 27.1\\
&&&& 3-4 & 1.8 & 1.2 & 6.9 & 28.9 & 37.1 & 26.3 & 18.5 & 21.0 & 47.7 & 49.1\\
& 0.7 & 0.4 & hom & 1-4 & 1.2 & 0.4 & 2.1 & 11.3 & 29.6 & 20.3 & 6.5 & 7.8 & 17.1 & 18.1\\
&&&& 2-4 & 2.1 & 1.5 & 4.7 & 23.3 & 38.2 & 25.3 & 15.1 & 16.6 & 29.5 & 30.4\\
&&&& 3-4 & 1.1 & 0.8 & 4.7 & 31.7 & 44.7 & 31.1 & 20.4 & 22.8 & 39.9 & 40.9\\
&&& pos & 1-4 & 0.3 & 0.1 & 0.1 & 9.3 & 21.3 & 12.8 & 5.5 & 6.5 & 10.6 & 11.2\\
&&&& 2-4 & 0.6 & 0.5 & 0.8 & 12.5 & 21.4 & 14.0 & 7.1 & 7.9 & 13.7 & 13.8\\
&&&& 3-4 & 1.1 & 0.5 & 1.9 & 12.7 & 19.9 & 14.6 & 7.7 & 8.4 & 13.3 & 13.7\\
&&& neg & 1-4 & 5.8 & 5.1 & 9.2 & 5.9 & 22.5 & 16.5 & 3.2 & 3.7 & 8.6 & 9.6\\
&&&& 2-4 & 2.6 & 2.0 & 6.3 & 15.8 & 30.8 & 18.3 & 7.7 & 9.1 & 15.9 & 17.6\\
&&&& 3-4 & 1.8 & 1.0 & 4.0 & 26.9 & 39.5 & 27.5 & 16.1 & 17.9 & 29.5 & 30.6\\
& 0.9 & 0.5 & hom & 1-4 & 0.9 & 0.8 & 2.9 & 5.7 & 26.7 & 14.2 & 2.8 & 3.5 & 8.2 & 8.9\\
&&&& 2-4 & 1.1 & 0.7 & 3.0 & 13.6 & 33.2 & 19.6 & 6.6 & 7.9 & 14.6 & 15.2\\
&&&& 3-4 & 1.3 & 1.2 & 5.4 & 23.1 & 39.8 & 20.8 & 14.1 & 15.6 & 21.5 & 22.6\\
&&& pos & 1-4 & 0.3 & 0.0 & 0.0 & 4.3 & 16.8 & 8.8 & 2.4 & 2.6 & 4.8 & 5.1\\
&&&& 2-4 & 0.5 & 0.4 & 0.9 & 7.1 & 19.8 & 7.7 & 3.9 & 4.3 & 5.8 & 6.3\\
&&&& 3-4 & 0.9 & 0.5 & 2.2 & 9.8 & 19.2 & 8.4 & 3.9 & 4.3 & 7.5 & 7.9\\
&&& neg & 1-4 & 5.5 & 6.1 & 9.3 & 2.3 & 20.3 & 10.8 & 0.8 & 1.2 & 4.0 & 4.2\\
&&&& 2-4 & 1.7 & 1.4 & 5.3 & 7.7 & 26.4 & 12.6 & 3.6 & 4.3 & 9.1 & 9.6\\
&&&& 3-4 & 1.5 & 0.9 & 4.2 & 13.6 & 29.0 & 17.9 & 8.6 & 10.6 & 15.7 & 16.7\\
\hline
\end{longtable}

\newpage

\clearpage

\subsection{Results for Two-Sample Problem}\label{results-for-the-two-sample-problem}
\begin{figure}
\centering
\includegraphics[width=0.99\textwidth,
height=0.4\textheight]{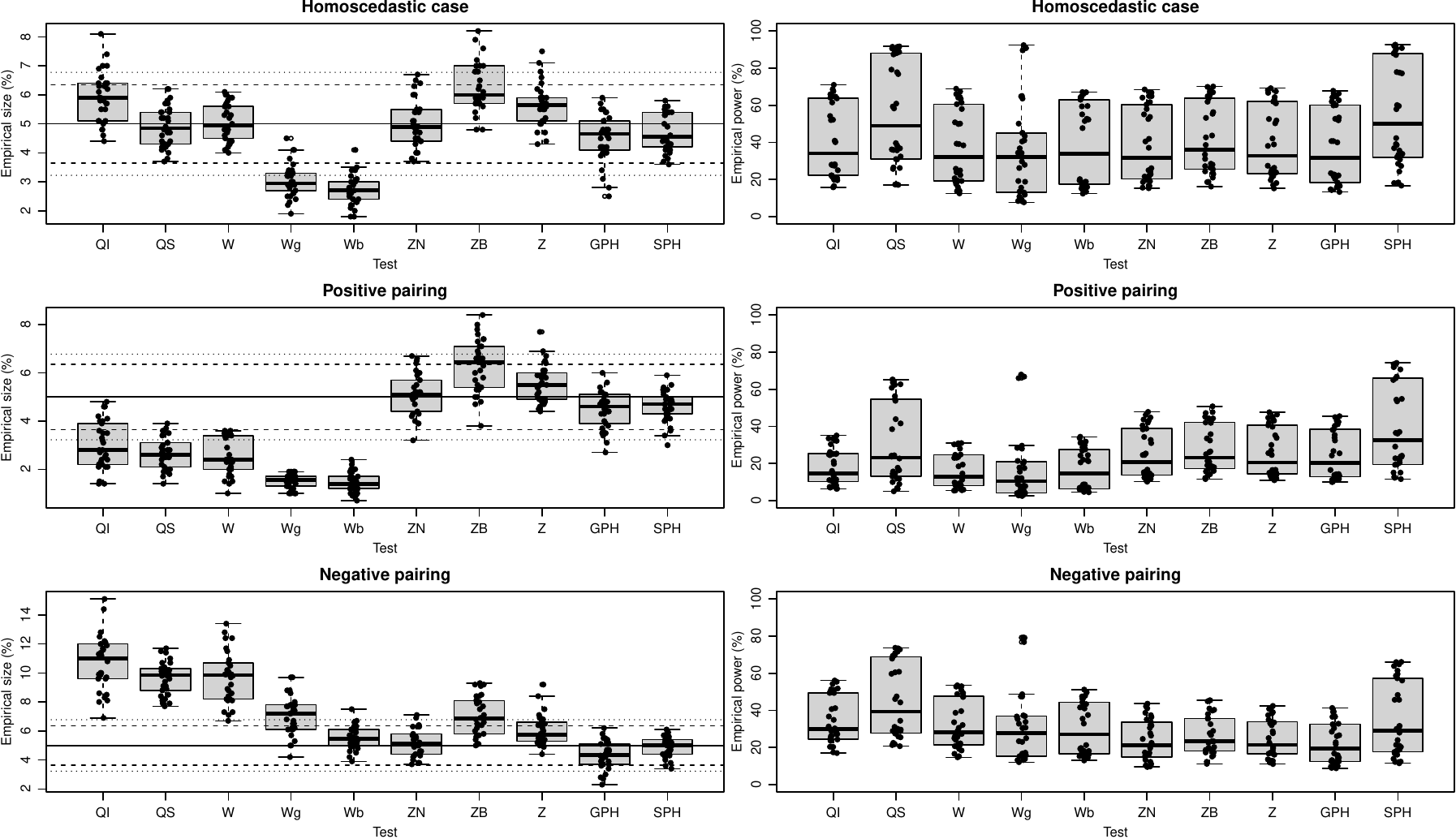}
\caption[Box-and-whisker plots for the empirical sizes and power in Model~2 under global hypothesis for $k=2$ groups and $p=2$ functional variables and for homoscedastic case, positive pairing, and negative pairing]{Box-and-whisker plots for the empirical sizes and power (as percentages) of all tests obtained in Model~2 under global hypothesis for $k=2$ groups and $p=2$ functional variables and for homoscedastic case, positive pairing, and negative pairing.}
\label{fig_s_9}
\end{figure}

\begin{figure}
\centering
\includegraphics[width=0.99\textwidth,
height=0.8\textheight]{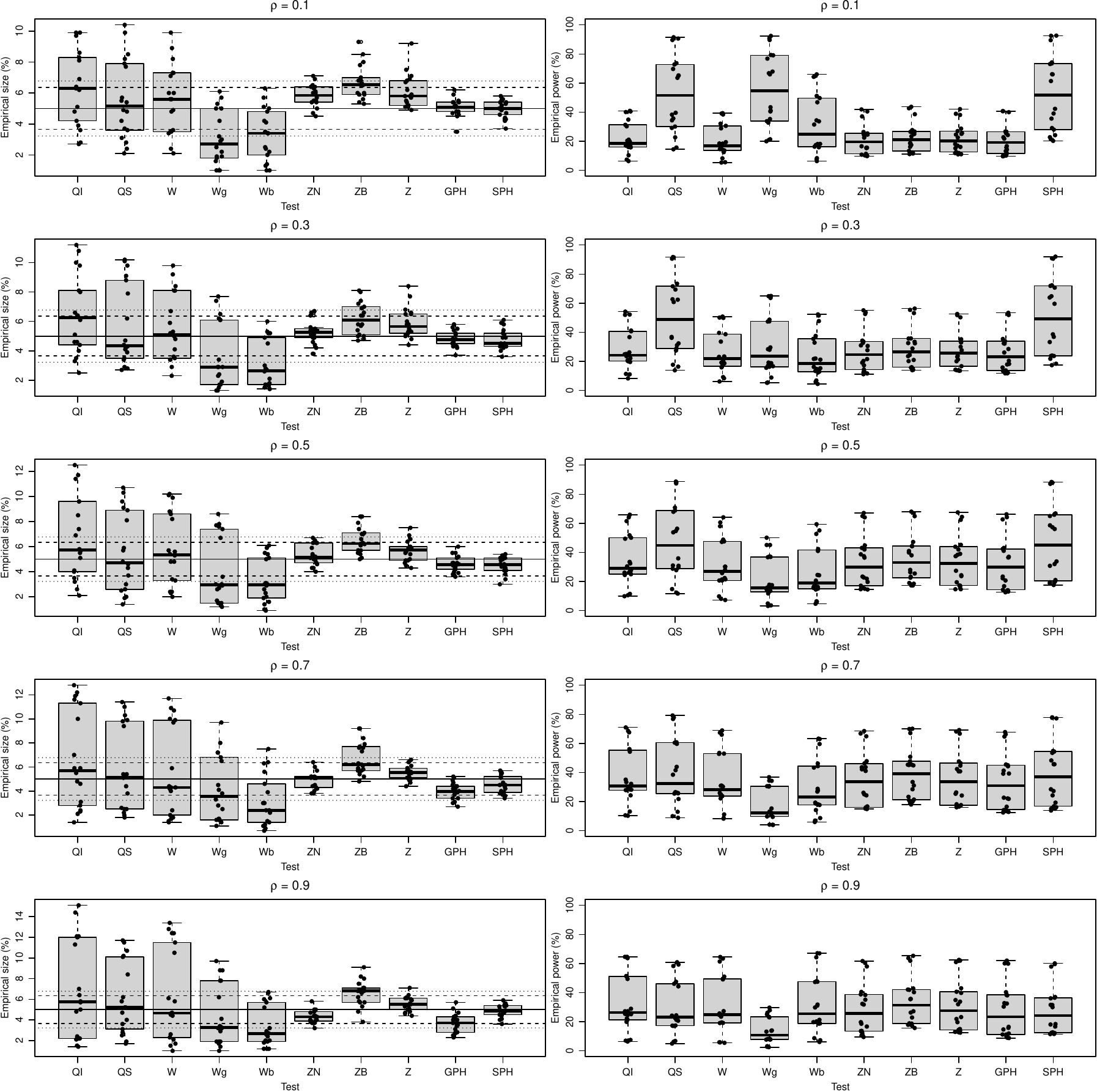}
\caption[Box-and-whisker plots for the empirical size and power in Model~2 under global hypothesis for $k=2$ groups and $p=2$ functional variables and for different correlation]{Box-and-whisker plots for the empirical size and power (as percentages) of all tests obtained in Model~2 under global hypothesis for $k=2$ groups and $p=2$ functional variables and for different correlation.}
\label{fig_s_10}
\end{figure}

Tables~\ref{tab:unnamed-chunk-20}-\ref{tab:unnamed-chunk-21} and Figures~\ref{fig_s_9}-\ref{fig_s_10} contain and summarize the results of simulation studies in Model~2 for the two-sample problem ($k=2$) and $p=2$ functional variables. In this case, the results of tests by \cite{Gorecki} based on different multivariate tests, i.e., W, LH, P, R, Wg, LHg, Pg, Rg, Wb, LHb, Pb, and Rb, are very similar, so we just report the results for the tests based on Wilks' test statistic (i.e., W, Wg, Wb).

\begingroup\fontsize{10}{12}\selectfont

\begin{longtable}[t]{llllrrrrrrrrrr}
\caption[Empirical sizes in Model~2 under global hypothesis for $k=2$ groups and $p=2$ functional variables]{\label{tab:unnamed-chunk-20}Empirical sizes (as percentages) in Model~2 under global hypothesis for $k=2$ groups and $p=2$ functional variables. Columns: D - distribution ($N$ - normal, $t_4$ - t-Student, $\chi_4^2$ - chi-squared), HH - homoscedasticity or heteroscedasticity (hom - homoscedasticity, pos - positive pairing, neg - negative pairing), $\mathbf{n}=(n_1,n_2)$ - vector of sample sizes ($\mathbf{n}_1=(20,30)$; $\mathbf{n}_2=(60,90)$).}\\
\hline
D & $\rho$ & HH & $\mathbf{n}$ & QI & QS & W & Wg & Wb & ZN & ZB & Z & GPH & SPH\\
\hline
\endfirsthead
\caption[]{Empirical sizes (as percentages) in Model~2 under global hypothesis for $k=2$ groups and $p=2$ functional variables. Columns: D - distribution ($N$ - normal, $t_4$ - t-Student, $\chi_4^2$ - chi-squared), HH - homoscedasticity or heteroscedasticity (hom - homoscedasticity, pos - positive pairing, neg - negative pairing), $\mathbf{n}=(n_1,n_2)$ - vector of sample sizes ($\mathbf{n}_1=(20,30)$; $\mathbf{n}_2=(60,90)$). \textit{(continued)}}\\
\hline
D & $\rho$ & HH & $\mathbf{n}$ & QI & QS & W & Wg & Wb & ZN & ZB & Z & GPH & SPH\\
\hline
\endhead
$N$ & 0.1 & hom & $\mathbf{n}_1$ & 8.1 & 4.8 & 6.0 & 3.2 & 4.1 & 6.0 & 6.5 & 5.8 & 5.9 & 5.0\\
&&& $\mathbf{n}_2$ & 5.1 & 4.9 & 4.8 & 3.0 & 3.5 & 5.5 & 5.9 & 5.1 & 4.8 & 4.3\\
&& pos & $\mathbf{n}_1$ & 3.9 & 3.1 & 3.5 & 1.8 & 1.4 & 5.5 & 6.0 & 5.4 & 4.7 & 5.5\\
&&& $\mathbf{n}_2$ & 2.8 & 2.1 & 2.4 & 1.0 & 1.0 & 5.4 & 5.4 & 4.9 & 4.8 & 4.7\\
&& neg & $\mathbf{n}_1$ & 8.3 & 7.9 & 7.3 & 4.2 & 5.0 & 6.3 & 8.0 & 7.5 & 5.1 & 4.6\\
&&& $\mathbf{n}_2$ & 9.7 & 9.9 & 8.9 & 5.7 & 5.7 & 6.4 & 6.5 & 5.8 & 5.4 & 5.0\\
& 0.3 & hom & $\mathbf{n}_1$ & 6.4 & 4.2 & 5.0 & 2.3 & 2.7 & 6.4 & 7.0 & 6.6 & 5.2 & 4.3\\
&&& $\mathbf{n}_2$ & 4.4 & 4.3 & 4.1 & 2.9 & 1.8 & 5.4 & 5.8 & 5.2 & 4.2 & 4.3\\
&& pos & $\mathbf{n}_1$ & 4.0 & 2.9 & 2.9 & 1.3 & 1.7 & 6.7 & 7.4 & 7.7 & 4.7 & 5.2\\
&&& $\mathbf{n}_2$ & 3.5 & 3.5 & 3.4 & 1.9 & 1.7 & 4.9 & 5.0 & 6.0 & 4.8 & 4.9\\
&& neg & $\mathbf{n}_1$ & 10.0 & 9.8 & 8.1 & 6.5 & 4.5 & 4.9 & 6.4 & 6.5 & 4.3 & 5.4\\
&&& $\mathbf{n}_2$ & 8.0 & 8.8 & 6.7 & 6.2 & 5.2 & 4.6 & 5.1 & 6.5 & 3.7 & 4.3\\
& 0.5 & hom & $\mathbf{n}_1$ & 7.4 & 5.9 & 5.7 & 3.6 & 3.0 & 6.3 & 7.0 & 7.5 & 5.5 & 5.3\\
&&& $\mathbf{n}_2$ & 5.5 & 3.7 & 5.4 & 3.0 & 3.1 & 5.4 & 5.7 & 4.3 & 4.5 & 4.1\\
&& pos & $\mathbf{n}_1$ & 3.2 & 3.0 & 2.4 & 1.5 & 1.9 & 6.5 & 7.4 & 6.9 & 4.7 & 4.1\\
&&& $\mathbf{n}_2$ & 2.1 & 1.4 & 2.0 & 1.2 & 1.6 & 5.9 & 6.3 & 5.9 & 3.8 & 4.3\\
&& neg & $\mathbf{n}_1$ & 12.5 & 10.7 & 9.9 & 8.6 & 5.9 & 6.3 & 8.4 & 6.6 & 4.6 & 5.1\\
&&& $\mathbf{n}_2$ & 9.6 & 9.1 & 8.2 & 7.8 & 6.1 & 4.6 & 5.1 & 5.4 & 5.2 & 4.9\\
& 0.7 & hom & $\mathbf{n}_1$ & 5.9 & 5.4 & 4.3 & 4.1 & 3.0 & 6.0 & 7.9 & 6.4 & 4.0 & 5.5\\
&&& $\mathbf{n}_2$ & 5.5 & 5.2 & 4.2 & 3.8 & 2.3 & 5.1 & 5.9 & 4.4 & 4.7 & 5.6\\
&& pos & $\mathbf{n}_1$ & 2.8 & 2.5 & 1.8 & 1.6 & 1.1 & 6.4 & 7.6 & 6.1 & 4.4 & 4.5\\
&&& $\mathbf{n}_2$ & 3.1 & 2.0 & 2.4 & 1.4 & 0.7 & 5.1 & 5.7 & 6.5 & 4.9 & 4.5\\
&& neg & $\mathbf{n}_1$ & 12.8 & 11.4 & 10.0 & 9.7 & 7.5 & 5.7 & 9.2 & 5.1 & 4.2 & 5.7\\
&&& $\mathbf{n}_2$ & 11.9 & 9.8 & 10.9 & 7.2 & 6.4 & 5.1 & 5.6 & 6.6 & 3.9 & 4.2\\
& 0.9 & hom & $\mathbf{n}_1$ & 6.4 & 5.8 & 4.5 & 3.2 & 2.9 & 4.7 & 6.8 & 5.9 & 3.1 & 5.4\\
&&& $\mathbf{n}_2$ & 7.0 & 4.0 & 6.1 & 4.1 & 3.2 & 4.9 & 5.8 & 5.6 & 5.7 & 4.1\\
&& pos & $\mathbf{n}_1$ & 2.2 & 3.1 & 2.1 & 1.9 & 1.9 & 5.0 & 6.9 & 6.4 & 3.7 & 5.3\\
&&& $\mathbf{n}_2$ & 1.5 & 1.7 & 1.7 & 1.7 & 2.0 & 3.9 & 4.8 & 6.1 & 4.4 & 4.7\\
&& neg & $\mathbf{n}_1$ & 15.1 & 10.7 & 12.4 & 8.8 & 6.6 & 4.3 & 7.5 & 6.2 & 2.8 & 5.4\\
&&& $\mathbf{n}_2$ & 12.0 & 8.4 & 11.5 & 7.8 & 6.0 & 5.8 & 7.1 & 7.1 & 3.7 & 3.6\\\addlinespace
$t_4$& 0.1 & hom & $\mathbf{n}_1$ & 6.3 & 5.7 & 6.0 & 2.9 & 3.4 & 4.5 & 5.9 & 6.2 & 5.1 & 5.6\\
&&& $\mathbf{n}_2$ & 6.3 & 5.4 & 5.6 & 2.2 & 3.4 & 5.0 & 5.6 & 5.5 & 5.4 & 5.4\\
&& pos & $\mathbf{n}_1$ & 2.7 & 2.4 & 2.1 & 1.0 & 1.3 & 6.1 & 7.8 & 5.2 & 3.5 & 3.7\\
&&& $\mathbf{n}_2$ & 4.2 & 3.6 & 3.6 & 1.6 & 2.4 & 4.7 & 5.3 & 5.1 & 5.2 & 5.3\\
&& neg & $\mathbf{n}_1$ & 9.9 & 8.5 & 8.2 & 5.3 & 5.1 & 7.1 & 9.3 & 5.2 & 5.2 & 5.8\\
&&& $\mathbf{n}_2$ & 6.9 & 8.2 & 7.3 & 5.0 & 4.8 & 5.4 & 6.0 & 6.9 & 4.5 & 4.4\\
& 0.3 & hom & $\mathbf{n}_1$ & 6.1 & 4.1 & 5.3 & 2.4 & 2.6 & 5.4 & 7.2 & 5.5 & 4.8 & 3.6\\
&&& $\mathbf{n}_2$ & 5.1 & 4.7 & 4.8 & 2.9 & 3.0 & 5.1 & 5.7 & 5.1 & 4.7 & 4.2\\
&& pos & $\mathbf{n}_1$ & 4.6 & 3.9 & 3.6 & 1.7 & 1.6 & 5.1 & 6.6 & 5.8 & 5.6 & 5.9\\
&&& $\mathbf{n}_2$ & 2.5 & 2.8 & 2.3 & 1.3 & 1.4 & 4.2 & 5.0 & 5.8 & 4.3 & 4.3\\
&& neg & $\mathbf{n}_1$ & 10.8 & 10.2 & 9.2 & 7.4 & 5.2 & 5.6 & 8.1 & 5.3 & 5.5 & 5.1\\
&&& $\mathbf{n}_2$ & 8.1 & 9.1 & 8.1 & 6.1 & 4.9 & 5.4 & 6.4 & 4.4 & 4.5 & 4.0\\
& 0.5 & hom & $\mathbf{n}_1$ & 6.9 & 5.7 & 4.8 & 2.9 & 2.3 & 4.9 & 6.2 & 5.9 & 4.7 & 4.6\\
&&& $\mathbf{n}_2$ & 5.7 & 4.6 & 5.6 & 3.0 & 3.0 & 4.7 & 5.7 & 5.0 & 5.1 & 4.4\\
&& pos & $\mathbf{n}_1$ & 3.5 & 2.0 & 2.4 & 1.5 & 0.9 & 5.0 & 7.1 & 4.9 & 3.9 & 3.0\\
&&& $\mathbf{n}_2$ & 2.6 & 1.9 & 2.3 & 1.5 & 1.0 & 4.3 & 5.4 & 5.6 & 4.5 & 4.1\\
&& neg & $\mathbf{n}_1$ & 11.4 & 9.6 & 10.2 & 6.7 & 5.4 & 5.2 & 7.9 & 5.5 & 5.5 & 5.1\\
&&& $\mathbf{n}_2$ & 8.5 & 8.1 & 8.6 & 7.4 & 5.5 & 4.0 & 5.0 & 6.0 & 3.6 & 3.4\\
& 0.7 & hom & $\mathbf{n}_1$ & 7.0 & 4.4 & 5.9 & 2.8 & 2.4 & 4.5 & 7.6 & 5.6 & 4.5 & 3.9\\
&&& $\mathbf{n}_2$ & 4.6 & 3.8 & 4.3 & 3.3 & 2.2 & 3.8 & 4.8 & 5.9 & 3.4 & 3.7\\
&& pos & $\mathbf{n}_1$ & 2.1 & 1.8 & 1.4 & 1.1 & 0.9 & 5.2 & 8.4 & 5.5 & 3.1 & 3.6\\
&&& $\mathbf{n}_2$ & 1.4 & 2.2 & 1.5 & 1.6 & 1.2 & 5.2 & 6.1 & 4.9 & 4.3 & 4.5\\
&& neg & $\mathbf{n}_1$ & 11.3 & 9.9 & 9.9 & 6.8 & 3.9 & 5.1 & 9.2 & 5.1 & 3.0 & 3.8\\
&&& $\mathbf{n}_2$ & 11.6 & 10.3 & 10.7 & 8.0 & 5.6 & 4.2 & 5.4 & 5.0 & 3.4 & 4.4\\
& 0.9 & hom & $\mathbf{n}_1$ & 5.0 & 6.2 & 4.4 & 3.3 & 2.7 & 3.7 & 6.8 & 5.1 & 2.5 & 5.4\\
&&& $\mathbf{n}_2$ & 5.7 & 5.2 & 5.8 & 2.9 & 1.8 & 4.4 & 5.7 & 4.7 & 4.1 & 4.9\\
&& pos & $\mathbf{n}_1$ & 2.2 & 2.8 & 2.3 & 1.4 & 1.2 & 4.3 & 7.1 & 4.6 & 3.5 & 4.8\\
&&& $\mathbf{n}_2$ & 1.4 & 1.9 & 1.5 & 1.0 & 1.2 & 4.0 & 5.3 & 5.0 & 4.0 & 4.0\\
&& neg & $\mathbf{n}_1$ & 11.3 & 10.2 & 10.5 & 6.2 & 5.2 & 3.7 & 9.1 & 5.4 & 2.3 & 4.4\\
&&& $\mathbf{n}_2$ & 12.1 & 10.1 & 12.8 & 7.8 & 5.7 & 4.4 & 6.6 & 5.3 & 4.7 & 4.5\\\addlinespace
$\chi_4^2$ & 0.1 & hom & $\mathbf{n}_1$ & 6.4 & 4.2 & 4.9 & 1.9 & 2.0 & 5.5 & 6.8 & 7.1 & 4.6 & 4.2\\
&&& $\mathbf{n}_2$ & 6.2 & 5.5 & 5.6 & 2.5 & 2.5 & 6.5 & 7.0 & 5.7 & 5.5 & 5.4\\
&& pos & $\mathbf{n}_1$ & 4.8 & 3.7 & 3.5 & 1.9 & 2.2 & 5.7 & 6.6 & 6.7 & 5.1 & 4.9\\
&&& $\mathbf{n}_2$ & 3.6 & 2.8 & 3.4 & 1.0 & 1.0 & 6.0 & 6.6 & 5.9 & 5.4 & 4.9\\
&& neg & $\mathbf{n}_1$ & 8.6 & 7.7 & 7.1 & 5.0 & 4.2 & 6.9 & 8.5 & 9.2 & 5.1 & 5.2\\
&&& $\mathbf{n}_2$ & 9.9 & 10.4 & 9.9 & 6.1 & 6.3 & 6.4 & 6.8 & 6.8 & 6.2 & 5.6\\
& 0.3 & hom & $\mathbf{n}_1$ & 6.4 & 4.4 & 5.2 & 2.9 & 2.1 & 3.8 & 4.8 & 6.8 & 4.8 & 4.4\\
&&& $\mathbf{n}_2$ & 6.6 & 6.2 & 5.9 & 3.4 & 2.8 & 4.9 & 5.4 & 5.0 & 5.4 & 5.8\\
&& pos & $\mathbf{n}_1$ & 4.6 & 3.4 & 3.5 & 1.5 & 1.7 & 6.6 & 8.0 & 5.4 & 4.5 & 4.5\\
&&& $\mathbf{n}_2$ & 3.3 & 2.7 & 3.4 & 1.7 & 1.5 & 4.4 & 4.7 & 4.8 & 5.1 & 4.9\\
&& neg & $\mathbf{n}_1$ & 9.8 & 7.9 & 8.4 & 6.1 & 5.3 & 5.5 & 6.9 & 8.4 & 4.7 & 4.5\\
&&& $\mathbf{n}_2$ & 11.2 & 10.1 & 9.8 & 7.7 & 6.0 & 5.5 & 5.8 & 5.3 & 5.8 & 6.1\\
& 0.5 & hom & $\mathbf{n}_1$ & 5.8 & 4.8 & 5.3 & 2.6 & 2.6 & 4.3 & 6.1 & 5.9 & 4.2 & 3.8\\
&&& $\mathbf{n}_2$ & 5.1 & 4.3 & 4.8 & 2.7 & 2.9 & 6.7 & 7.0 & 4.7 & 4.2 & 4.6\\
&& pos & $\mathbf{n}_1$ & 4.1 & 2.6 & 3.3 & 1.3 & 2.0 & 5.2 & 6.6 & 5.9 & 4.7 & 4.5\\
&&& $\mathbf{n}_2$ & 4.0 & 2.5 & 3.4 & 1.7 & 1.4 & 5.1 & 5.5 & 4.5 & 6.0 & 5.4\\
&& neg & $\mathbf{n}_1$ & 11.7 & 8.9 & 10.1 & 7.7 & 5.1 & 6.3 & 8.4 & 6.4 & 4.4 & 5.0\\
&&& $\mathbf{n}_2$ & 9.6 & 10.3 & 8.8 & 7.6 & 5.0 & 5.0 & 5.7 & 4.9 & 4.0 & 5.2\\
& 0.7 & hom & $\mathbf{n}_1$ & 5.9 & 5.1 & 4.4 & 4.5 & 3.0 & 4.0 & 6.3 & 5.9 & 4.2 & 4.0\\
&&& $\mathbf{n}_2$ & 4.8 & 5.4 & 4.0 & 2.5 & 2.4 & 4.4 & 5.2 & 5.7 & 3.9 & 5.4\\
&& pos & $\mathbf{n}_1$ & 2.3 & 2.6 & 1.7 & 1.6 & 1.4 & 5.2 & 7.3 & 5.5 & 2.7 & 3.4\\
&&& $\mathbf{n}_2$ & 2.4 & 2.5 & 2.0 & 1.7 & 1.5 & 5.1 & 5.8 & 4.7 & 5.2 & 5.1\\
&& neg & $\mathbf{n}_1$ & 12.2 & 11.0 & 11.7 & 7.2 & 6.3 & 4.3 & 7.7 & 5.7 & 3.7 & 5.2\\
&&& $\mathbf{n}_2$ & 10.0 & 9.4 & 9.7 & 6.5 & 4.6 & 3.8 & 5.7 & 5.4 & 3.9 & 4.7\\
& 0.9 & hom & $\mathbf{n}_1$ & 5.8 & 4.7 & 4.6 & 3.2 & 2.4 & 4.8 & 8.2 & 5.5 & 2.8 & 4.5\\
&&& $\mathbf{n}_2$ & 4.8 & 5.2 & 4.7 & 3.4 & 2.6 & 4.0 & 5.7 & 6.1 & 4.0 & 5.3\\
&& pos & $\mathbf{n}_1$ & 2.1 & 3.6 & 1.0 & 1.9 & 2.0 & 3.9 & 6.8 & 4.4 & 3.4 & 5.0\\
&&& $\mathbf{n}_2$ & 2.4 & 2.5 & 2.6 & 1.9 & 1.2 & 3.2 & 3.8 & 4.8 & 5.1 & 4.8\\
&& neg & $\mathbf{n}_1$ & 14.4 & 11.5 & 13.4 & 8.8 & 6.1 & 3.8 & 8.0 & 5.5 & 2.7 & 5.6\\
&&& $\mathbf{n}_2$ & 12.1 & 11.7 & 12.4 & 9.7 & 6.7 & 5.0 & 6.2 & 6.0 & 4.3 & 5.9\\
\hline
\end{longtable}
\endgroup{}

\newpage

\begingroup\fontsize{10}{12}\selectfont

\begin{longtable}[t]{lllllrrrrrrrrrr}
\caption[Empirical power in Model~2 under global hypothesis for $k=2$ groups and $p=2$ functional variables]{\label{tab:unnamed-chunk-21}Empirical power (as percentages) in Model~2 under global hypothesis for $k=2$ groups and $p=2$ functional variables. Columns: D - distribution ($N$ - normal, $t_4$ - t-Student, $\chi_4^2$ - chi-squared), HH - homoscedasticity or heteroscedasticity (hom - homoscedasticity, pos - positive pairing, neg - negative pairing), $\mathbf{n}=(n_1,n_2)$ - vector of sample sizes ($\mathbf{n}_1=(20,30)$; $\mathbf{n}_2=(60,90)$), $\delta$ - hyperparameter in scenario for alternative hypothesis.}\\
\hline
D & $\rho$ & $\delta$ & HH & $\mathbf{n}$ & QI & QS & W & Wg & Wb & ZN & ZB & Z & GPH & SPH\\
\hline
\endfirsthead
\caption[]{Empirical power (as percentages) in Model~2 under global hypothesis for $k=2$ groups and $p=2$ functional variables. Columns: D - distribution ($N$ - normal, $t_4$ - t-Student, $\chi_4^2$ - chi-squared), HH - homoscedasticity or heteroscedasticity (hom - homoscedasticity, pos - positive pairing, neg - negative pairing), $\mathbf{n}=(n_1,n_2)$ - vector of sample sizes ($\mathbf{n}_1=(20,30)$; $\mathbf{n}_2=(60,90)$), $\delta$ - hyperparameter in scenario for alternative hypothesis. \textit{(continued)}}\\
\hline
D & $\rho$ & $\delta$ & HH & $\mathbf{n}$ & QI & QS & W & Wg & Wb & ZN & ZB & Z & GPH & SPH\\
\hline
\endhead
$N$ & 0.1 & 0.1 & hom & $\mathbf{n}_1$ & 15.8 & 35.7 & 12.5 & 40.5 & 17.2 & 15.3 & 16.2 & 15.2 & 14.3 & 35.6\\
&&&& $\mathbf{n}_2$ & 40.1 & 91.5 & 38.4 & 90.8 & 66.0 & 41.9 & 42.7 & 38.9 & 40.7 & 92.5\\
&&& pos & $\mathbf{n}_1$ & 7.4 & 15.7 & 5.6 & 20.0 & 6.4 & 10.8 & 12.2 & 11.0 & 10.2 & 24.2\\
&&&& $\mathbf{n}_2$ & 19.8 & 63.6 & 18.5 & 66.8 & 33.7 & 24.6 & 25.0 & 24.6 & 24.3 & 73.5\\
&&& neg & $\mathbf{n}_1$ & 17.0 & 29.2 & 14.7 & 35.3 & 17.4 & 10.3 & 11.2 & 11.1 & 9.9 & 20.3\\
&&&& $\mathbf{n}_2$ & 30.3 & 69.7 & 28.9 & 76.8 & 46.8 & 22.5 & 23.4 & 25.1 & 21.8 & 61.4\\
& 0.3 & 0.2 & hom & $\mathbf{n}_1$ & 19.6 & 32.4 & 15.9 & 15.6 & 12.4 & 19.1 & 21.0 & 21.3 & 17.9 & 33.3\\
&&&& $\mathbf{n}_2$ & 54.2 & 91.5 & 49.9 & 64.8 & 52.3 & 52.6 & 53.2 & 50.3 & 51.9 & 90.8\\
&&& pos & $\mathbf{n}_1$ & 11.4 & 16.4 & 8.4 & 5.3 & 5.9 & 14.4 & 16.0 & 14.3 & 13.4 & 23.7\\
&&&& $\mathbf{n}_2$ & 26.2 & 60.8 & 23.4 & 28.4 & 22.0 & 32.7 & 33.5 & 30.0 & 32.4 & 70.7\\
&&& neg & $\mathbf{n}_1$ & 20.3 & 27.5 & 16.6 & 16.7 & 13.0 & 13.3 & 15.1 & 16.7 & 11.8 & 17.4\\
&&&& $\mathbf{n}_2$ & 36.7 & 69.4 & 33.6 & 47.5 & 32.9 & 27.7 & 28.9 & 27.7 & 26.1 & 59.6\\
& 0.5 & 0.3 & hom & $\mathbf{n}_1$ & 27.4 & 31.0 & 22.4 & 13.0 & 16.3 & 21.9 & 25.5 & 24.3 & 22.2 & 30.8\\
&&&& $\mathbf{n}_2$ & 61.5 & 87.4 & 58.0 & 45.1 & 52.7 & 64.1 & 64.9 & 62.4 & 62.7 & 87.1\\
&&& pos & $\mathbf{n}_1$ & 10.6 & 11.7 & 8.1 & 3.2 & 6.1 & 16.2 & 18.5 & 14.7 & 14.3 & 19.5\\
&&&& $\mathbf{n}_2$ & 33.5 & 53.8 & 30.2 & 17.5 & 24.5 & 43.1 & 44.3 & 45.2 & 42.4 & 65.0\\
&&& neg & $\mathbf{n}_1$ & 26.1 & 28.8 & 21.5 & 15.9 & 14.5 & 15.0 & 17.3 & 17.3 & 12.6 & 17.6\\
&&&& $\mathbf{n}_2$ & 51.8 & 68.8 & 48.8 & 36.9 & 42.3 & 36.4 & 38.0 & 38.4 & 36.5 & 56.1\\
& 0.7 & 0.4 & hom & $\mathbf{n}_1$ & 27.9 & 25.7 & 23.8 & 9.8 & 17.7 & 24.3 & 28.5 & 24.2 & 21.9 & 24.3\\
&&&& $\mathbf{n}_2$ & 71.0 & 77.8 & 68.9 & 34.3 & 63.0 & 68.5 & 69.9 & 69.2 & 67.8 & 77.9\\
&&& pos & $\mathbf{n}_1$ & 10.3 & 10.0 & 8.3 & 4.1 & 7.0 & 15.3 & 18.0 & 16.6 & 12.9 & 15.1\\
&&&& $\mathbf{n}_2$ & 35.1 & 41.6 & 30.9 & 12.0 & 27.6 & 43.7 & 45.2 & 46.4 & 45.0 & 54.4\\
&&& neg & $\mathbf{n}_1$ & 28.4 & 25.5 & 24.1 & 13.2 & 18.1 & 16.0 & 19.9 & 18.2 & 13.2 & 16.1\\
&&&& $\mathbf{n}_2$ & 54.3 & 59.7 & 53.0 & 30.6 & 43.4 & 43.6 & 45.4 & 40.8 & 41.3 & 45.7\\
& 0.9 & 0.5 & hom & $\mathbf{n}_1$ & 22.2 & 17.3 & 19.2 & 8.4 & 20.2 & 16.7 & 22.9 & 23.2 & 14.7 & 16.7\\
&&&& $\mathbf{n}_2$ & 63.7 & 58.3 & 61.4 & 26.2 & 64.3 & 61.7 & 63.7 & 62.5 & 60.1 & 57.9\\
&&& pos & $\mathbf{n}_1$ & 6.4 & 6.3 & 5.9 & 2.5 & 6.1 & 13.2 & 15.7 & 14.4 & 11.2 & 12.3\\
&&&& $\mathbf{n}_2$ & 25.4 & 24.4 & 24.7 & 7.8 & 31.1 & 40.5 & 42.9 & 40.6 & 40.3 & 36.5\\
&&& neg & $\mathbf{n}_1$ & 27.2 & 21.3 & 25.0 & 14.0 & 19.7 & 12.3 & 18.1 & 13.0 & 10.0 & 12.4\\
&&&& $\mathbf{n}_2$ & 51.1 & 44.3 & 49.4 & 23.0 & 47.5 & 31.8 & 35.5 & 32.0 & 30.4 & 30.0\\\addlinespace
$t_4$& 0.1 & 0.1 & hom & $\mathbf{n}_1$ & 16.1 & 39.6 & 13.7 & 43.4 & 18.1 & 15.6 & 18.6 & 17.0 & 13.3 & 39.1\\
&&&& $\mathbf{n}_2$ & 40.5 & 89.6 & 39.1 & 89.6 & 64.3 & 40.5 & 43.7 & 42.1 & 40.0 & 92.6\\
&&& pos & $\mathbf{n}_1$ & 10.4 & 22.7 & 8.3 & 21.4 & 8.4 & 11.4 & 13.4 & 12.6 & 11.7 & 29.1\\
&&&& $\mathbf{n}_2$ & 21.1 & 65.2 & 19.4 & 67.9 & 34.4 & 24.6 & 26.5 & 27.0 & 26.7 & 74.2\\
&&& neg & $\mathbf{n}_1$ & 17.4 & 30.1 & 15.2 & 32.8 & 16.3 & 9.8 & 12.3 & 11.9 & 9.7 & 21.4\\
&&&& $\mathbf{n}_2$ & 31.3 & 73.7 & 30.6 & 79.3 & 49.7 & 24.6 & 26.7 & 22.6 & 22.8 & 66.1\\
& 0.3 & 0.2 & hom & $\mathbf{n}_1$ & 22.2 & 36.8 & 18.7 & 19.2 & 14.8 & 20.3 & 24.4 & 23.1 & 18.3 & 36.9\\
&&&& $\mathbf{n}_2$ & 52.0 & 91.7 & 50.8 & 63.6 & 51.5 & 55.1 & 56.3 & 51.8 & 52.7 & 92.0\\
&&& pos & $\mathbf{n}_1$ & 11.1 & 17.7 & 9.2 & 8.9 & 6.9 & 11.7 & 14.2 & 14.0 & 13.8 & 23.7\\
&&&& $\mathbf{n}_2$ & 25.1 & 62.4 & 23.1 & 28.0 & 21.7 & 34.3 & 35.8 & 35.2 & 33.9 & 72.3\\
&&& neg & $\mathbf{n}_1$ & 20.4 & 28.9 & 19.2 & 16.3 & 15.3 & 11.2 & 15.7 & 13.6 & 12.2 & 18.4\\
&&&& $\mathbf{n}_2$ & 40.7 & 71.6 & 38.7 & 48.7 & 37.5 & 30.6 & 32.6 & 33.0 & 28.5 & 63.9\\
& 0.5 & 0.3 & hom & $\mathbf{n}_1$ & 27.7 & 35.9 & 25.8 & 13.5 & 17.5 & 23.4 & 28.0 & 23.7 & 23.4 & 33.8\\
&&&& $\mathbf{n}_2$ & 65.9 & 88.7 & 64.1 & 50.1 & 59.4 & 65.0 & 67.0 & 67.5 & 64.2 & 88.4\\
&&& pos & $\mathbf{n}_1$ & 11.5 & 14.8 & 9.8 & 4.9 & 6.4 & 14.7 & 19.0 & 14.7 & 13.9 & 20.4\\
&&&& $\mathbf{n}_2$ & 32.4 & 56.3 & 30.2 & 17.8 & 24.3 & 44.8 & 47.2 & 43.9 & 43.1 & 66.8\\
&&& neg & $\mathbf{n}_1$ & 24.5 & 29.6 & 21.4 & 14.6 & 15.6 & 17.1 & 22.6 & 17.1 & 14.0 & 18.8\\
&&&& $\mathbf{n}_2$ & 49.4 & 70.8 & 47.6 & 37.3 & 41.6 & 37.4 & 39.9 & 37.4 & 36.6 & 58.6\\
& 0.7 & 0.4 & hom & $\mathbf{n}_1$ & 28.0 & 26.2 & 24.4 & 9.5 & 18.7 & 25.0 & 33.5 & 26.3 & 22.2 & 27.3\\
&&&& $\mathbf{n}_2$ & 68.6 & 79.2 & 67.4 & 36.8 & 63.3 & 66.8 & 70.1 & 68.2 & 66.3 & 77.6\\
&&& pos & $\mathbf{n}_1$ & 13.2 & 13.0 & 11.2 & 4.2 & 8.7 & 15.2 & 21.4 & 16.5 & 16.5 & 19.3\\
&&&& $\mathbf{n}_2$ & 33.4 & 43.9 & 31.0 & 12.1 & 28.8 & 47.8 & 50.8 & 47.6 & 45.5 & 54.8\\
&&& neg & $\mathbf{n}_1$ & 24.3 & 21.8 & 22.2 & 12.1 & 14.2 & 15.9 & 23.3 & 16.9 & 12.5 & 13.9\\
&&&& $\mathbf{n}_2$ & 55.4 & 60.5 & 52.8 & 30.4 & 44.4 & 42.2 & 45.4 & 40.8 & 39.2 & 45.7\\
& 0.9 & 0.5 & hom & $\mathbf{n}_1$ & 19.9 & 17.4 & 19.5 & 7.7 & 18.8 & 19.7 & 27.2 & 21.9 & 15.8 & 18.1\\
&&&& $\mathbf{n}_2$ & 64.6 & 59.3 & 64.5 & 29.7 & 67.0 & 60.4 & 65.3 & 62.0 & 62.0 & 59.6\\
&&& pos & $\mathbf{n}_1$ & 7.0 & 6.8 & 5.8 & 3.2 & 8.7 & 13.6 & 18.9 & 15.0 & 11.9 & 12.7\\
&&&& $\mathbf{n}_2$ & 25.2 & 23.7 & 24.8 & 8.1 & 31.8 & 38.8 & 42.1 & 39.8 & 38.5 & 37.0\\
&&& neg & $\mathbf{n}_1$ & 28.9 & 22.5 & 27.3 & 13.2 & 18.7 & 9.5 & 18.7 & 12.5 & 9.0 & 12.2\\
&&&& $\mathbf{n}_2$ & 49.4 & 46.1 & 49.1 & 23.4 & 47.3 & 35.1 & 40.3 & 34.9 & 32.8 & 31.6\\\addlinespace
$\chi_4^2$ & 0.1 & 0.1 & hom & $\mathbf{n}_1$ & 16.5 & 38.3 & 13.9 & 41.1 & 15.9 & 16.6 & 18.7 & 18.1 & 16.7 & 42.3\\
&&&& $\mathbf{n}_2$ & 40.9 & 90.6 & 39.3 & 92.4 & 64.8 & 37.1 & 38.6 & 38.1 & 40.7 & 89.5\\
&&& pos & $\mathbf{n}_1$ & 6.4 & 14.5 & 5.4 & 21.0 & 7.6 & 10.2 & 11.6 & 11.9 & 10.0 & 22.8\\
&&&& $\mathbf{n}_2$ & 16.0 & 65.2 & 14.6 & 66.0 & 31.5 & 25.4 & 26.1 & 25.6 & 25.5 & 73.1\\
&&& neg & $\mathbf{n}_1$ & 20.4 & 34.4 & 18.4 & 33.8 & 18.2 & 15.6 & 17.5 & 16.1 & 15.7 & 28.0\\
&&&& $\mathbf{n}_2$ & 34.9 & 72.8 & 32.1 & 79.1 & 51.2 & 26.4 & 27.7 & 28.7 & 26.6 & 66.0\\
& 0.3 & 0.2 & hom & $\mathbf{n}_1$ & 20.8 & 36.2 & 17.6 & 18.8 & 13.4 & 21.6 & 23.6 & 23.8 & 20.4 & 38.6\\
&&&& $\mathbf{n}_2$ & 52.2 & 90.6 & 50.1 & 65.0 & 47.7 & 53.7 & 55.4 & 52.6 & 53.4 & 90.6\\
&&& pos & $\mathbf{n}_1$ & 8.3 & 13.9 & 6.2 & 5.5 & 4.5 & 12.4 & 14.0 & 13.6 & 11.8 & 21.8\\
&&&& $\mathbf{n}_2$ & 24.9 & 62.7 & 23.0 & 29.6 & 21.2 & 31.0 & 32.4 & 33.8 & 35.4 & 71.9\\
&&& neg & $\mathbf{n}_1$ & 23.6 & 31.1 & 20.7 & 17.2 & 16.2 & 17.3 & 20.2 & 20.1 & 15.8 & 24.0\\
&&&& $\mathbf{n}_2$ & 41.2 & 73.4 & 39.6 & 47.4 & 35.5 & 33.6 & 35.7 & 33.5 & 32.2 & 64.8\\
& 0.5 & 0.3 & hom & $\mathbf{n}_1$ & 25.0 & 30.8 & 20.7 & 11.8 & 15.1 & 22.8 & 26.4 & 27.4 & 20.7 & 32.0\\
&&&& $\mathbf{n}_2$ & 64.1 & 88.1 & 60.6 & 45.1 & 55.1 & 67.1 & 68.0 & 64.4 & 66.2 & 87.9\\
&&& pos & $\mathbf{n}_1$ & 9.9 & 12.4 & 7.3 & 3.6 & 4.7 & 14.5 & 17.3 & 16.4 & 13.1 & 19.4\\
&&&& $\mathbf{n}_2$ & 30.4 & 54.7 & 28.0 & 15.6 & 20.6 & 42.0 & 44.3 & 42.0 & 41.1 & 65.9\\
&&& neg & $\mathbf{n}_1$ & 26.1 & 27.8 & 20.7 & 14.6 & 16.5 & 19.7 & 23.3 & 19.5 & 16.8 & 21.0\\
&&&& $\mathbf{n}_2$ & 50.1 & 67.9 & 47.2 & 36.7 & 41.0 & 38.7 & 41.0 & 39.0 & 37.0 & 57.3\\
& 0.7 & 0.4 & hom & $\mathbf{n}_1$ & 28.2 & 26.5 & 24.9 & 10.8 & 18.1 & 26.1 & 31.1 & 25.4 & 22.7 & 28.5\\
&&&& $\mathbf{n}_2$ & 67.6 & 76.8 & 66.0 & 34.4 & 59.6 & 64.9 & 67.1 & 66.7 & 64.7 & 77.2\\
&&& pos & $\mathbf{n}_1$ & 10.6 & 8.9 & 8.7 & 4.0 & 6.0 & 16.7 & 20.9 & 16.1 & 14.2 & 15.1\\
&&&& $\mathbf{n}_2$ & 32.2 & 38.5 & 30.3 & 12.2 & 26.9 & 46.2 & 47.7 & 44.7 & 44.3 & 53.5\\
&&& neg & $\mathbf{n}_1$ & 29.4 & 25.9 & 25.9 & 15.3 & 19.3 & 14.9 & 19.0 & 17.6 & 14.5 & 16.8\\
&&&& $\mathbf{n}_2$ & 56.1 & 60.8 & 53.5 & 30.7 & 46.2 & 41.1 & 44.9 & 42.4 & 39.3 & 48.3\\
& 0.9 & 0.5 & hom & $\mathbf{n}_1$ & 21.4 & 17.1 & 19.1 & 8.1 & 19.5 & 18.6 & 26.3 & 19.7 & 16.5 & 18.4\\
&&&& $\mathbf{n}_2$ & 64.6 & 60.9 & 63.4 & 27.9 & 67.1 & 58.0 & 61.9 & 61.0 & 59.9 & 60.1\\
&&& pos & $\mathbf{n}_1$ & 7.5 & 5.0 & 5.5 & 2.8 & 7.3 & 13.7 & 17.9 & 13.9 & 11.2 & 11.6\\
&&&& $\mathbf{n}_2$ & 24.5 & 24.2 & 23.5 & 7.8 & 29.8 & 38.4 & 41.9 & 43.0 & 38.1 & 36.1\\
&&& neg & $\mathbf{n}_1$ & 27.2 & 20.7 & 25.7 & 13.8 & 21.3 & 10.5 & 17.4 & 12.8 & 8.8 & 11.5\\
&&&& $\mathbf{n}_2$ & 51.4 & 47.5 & 51.0 & 25.0 & 47.9 & 32.3 & 35.7 & 33.8 & 32.5 & 31.5\\
\hline
\end{longtable}
\endgroup{}

\newpage

\clearpage

\section{Real Data Example - Results}

\begin{figure}[t]
\includegraphics[width=0.99\textwidth,
height=0.4\textheight]{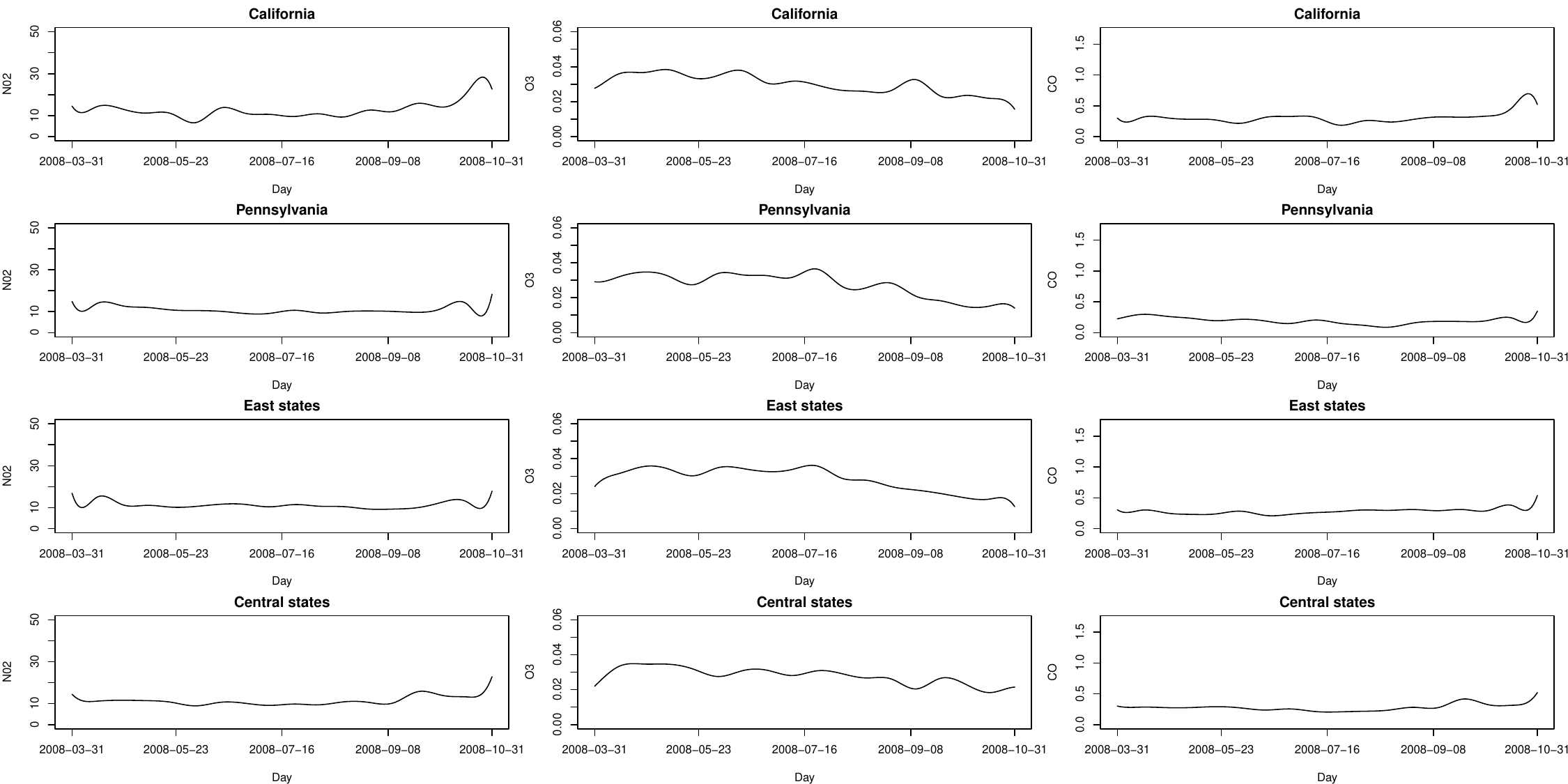}
\caption{Sample mean functions for the air pollution data set.}
\label{fig_rde_2}
\end{figure}

In this section, we present the additional results for the analysis of the U.S. air pollution data set. The sample mean functions for all variables in the this data set are presented in Figure~\ref{fig_rde_2}.

\subsection{Testing Homoscesticity}
\begin{figure}[t]
\includegraphics[width=0.99\textwidth,
height=0.6\textheight]{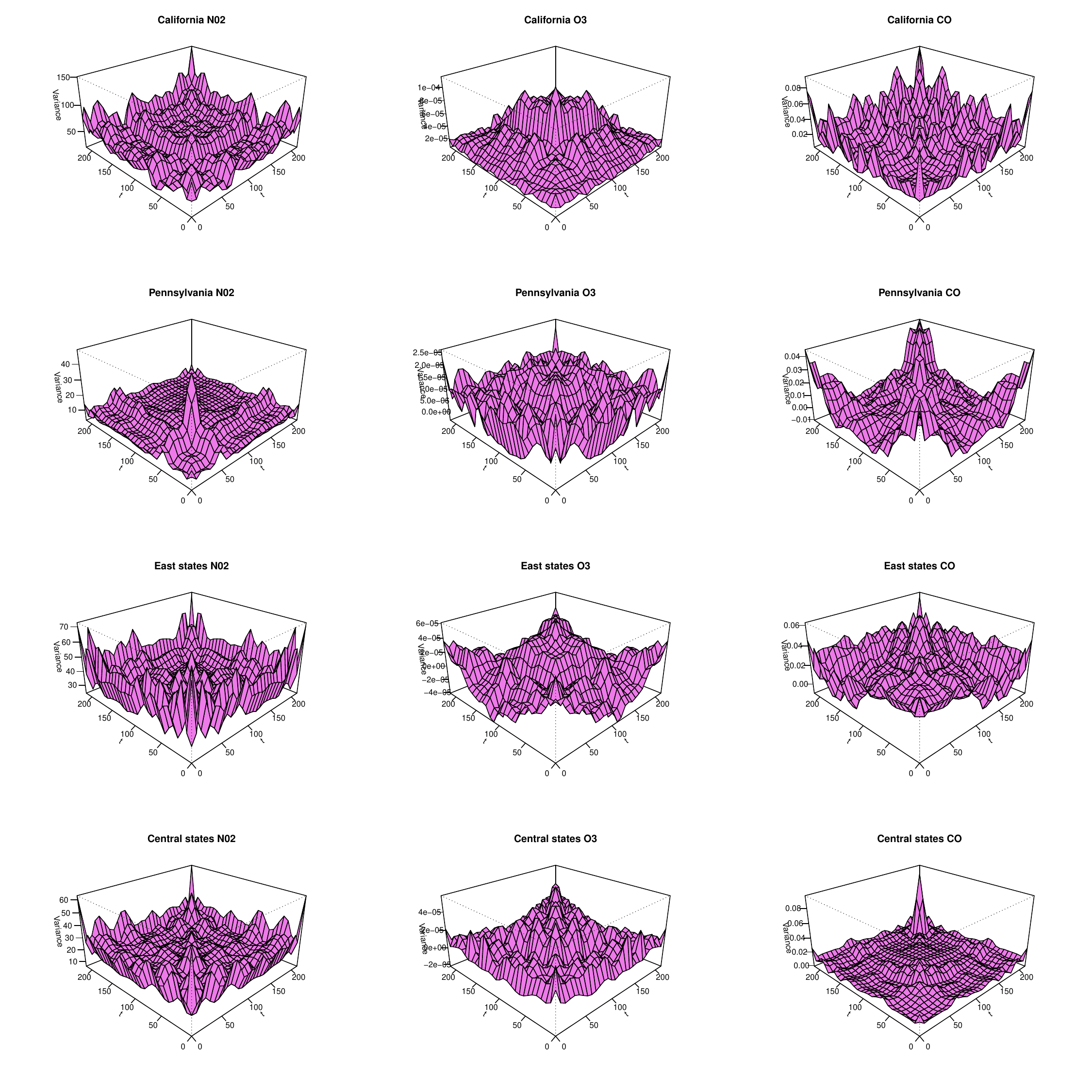}
\caption{Sample covariance functions for the air pollution data set.}
\label{fig_rde_4}
\end{figure}

One would like to check if the air pollution data are homoscedastic or not. To the best of our knowledge, there is no test for equality of covariance functions for multivariate functional data. However, there are a few tests for one-dimensional functional data in \cite{Guoetal2018, Guoetal2019}. We used these tests for each variable separately. The results are presented in Table~\ref{table_cov_fun}. The empirical covariance functions are presented in Figure~\ref{fig_rde_4}. We can observe that the covariance functions for all variables seem to be significantly different. To sum up, it seems that the air pollution data is closer to the heteroscedastic case.

\begin{table}[!h]
\caption[]{$P$-values of tests for equality of covariance functions by \cite{Guoetal2018}, $T_{\max,rp}$, and by  \cite{Guoetal2019}, GPF$_{nv}$, GPF$_{rp}$, $F_{\max,rp}$, for the NO2, O3, and CO variables.}\label{table_cov_fun}
\centering
\begin{tabular}{l|rrrr}\hline
Variable&$T_{\max,rp}$&GPF$_{nv}$&GPF$_{rp}$&$F_{\max,rp}$\\\hline
NO2&0.001&0.038&0.055&0.032\\
O3&0.021& 0.001&0.005&0.093\\
CO&0.122&0.035&0.044&0.014\\
\hline
\end{tabular}
\end{table}

\subsection{Testing Global Hypothesis}

The results for the global hypothesis are given in Table~\ref{tab_rde_1}. See the main paper for description.

\begin{table}[!t]
\caption[$P$-values, empirical size, and power of the tests for global hypothesis in the air pollution data set]{$P$-values (P), empirical size (size), and power (Power) as percentages of the tests for global hypothesis in the air pollution data set. Too liberal results are presented in bold. (We define the test as too liberal, when its empirical size is greater than $6.4\%$, which is the value of the upper limit of $95\%$ binomial confidence interval for empirical size for 1000 simulation runs \citep{DuchesneFrancq2015}.)}
\centering
\begin{tabular}{lrrr}
\hline
Test&$P$&\multicolumn{1}{c}{Size}&\multicolumn{1}{c}{Power}\\\hline
W&\textbf{0.1}&\textbf{10.7}&\textbf{100.0}\\
LH&\textbf{0.2}&\textbf{10.5}&\textbf{100.0}\\
P&\textbf{0.1}&\textbf{10.5}&\textbf{100.0}\\
R&\textbf{0.3}&\textbf{10.9}&\textbf{99.6}\\\addlinespace
Wg&\textbf{0.0}&\textbf{9.1}&\textbf{100.0}\\
LHg&\textbf{0.0}&\textbf{12.7}&\textbf{100.0}\\
Pg&\textbf{0.0}&\textbf{6.7}&\textbf{100.0}\\
Rg&\textbf{0.0}&\textbf{43.8}&\textbf{100.0}\\\addlinespace
Wb&\textbf{0.0}&\textbf{7.2}&\textbf{99.9}\\
LHb&\textbf{0.0}&\textbf{8.5}&\textbf{99.9}\\
Pb&0.0&5.7&99.9\\
Rb&\textbf{0.0}&\textbf{33.6}&\textbf{100.0}\\\addlinespace
ZN&\textbf{0.1}&\textbf{7.5}&\textbf{100.0}\\
ZB&\textbf{0.0}&\textbf{10.9}&\textbf{100.0}\\
Z&0.2&6.4&100.0\\\addlinespace
GPH&0.4&5.1&100.0\\
mGPH&0.0&5.8&99.9\\\addlinespace
SPH&0.0&5.8&100.0\\
mSPH&0.0&5.7&100.0\\
\hline
\end{tabular}
\label{tab_rde_1}
\end{table}

\subsection{Setup of Simulation Study based on Air Pollution Data Set}
To check the correctness of testing results for the air pollution data set, we have conducted a simulation study based on this set. The setup of this study is as follows: We generated the simulation data from the multivariate normal distribution with the following specifications:
\begin{itemize}
\item The sample sizes are equal to those of the samples of cities in California, Pennsylvania, East states, and Central states, i.e., $n_1=13$, $n_2=15$, $n_3=9$, and $n_4=15$.
\item The covariance matrix in the $i$-th group was equal to the sample covariance function for the $i$-th sample from the data set.
\item For checking the type I error level control, in each group, the vector of mean functions was set to the vector of sample mean functions of the pooled data.
\item For power investigation, the vector of mean functions in the $i$-th group was equal to the vector of sample mean functions for the $i$-th sample from the data set.
\end{itemize}
The above setup mimics the data structure to guarantee the appropriateness of the simulation study.

\FloatBarrier
\bibliographystyle{abbrvnat}